\documentclass[prb,aps,twocolumn,superscriptaddress]{revtex4-1}
\usepackage{amsfonts}
\usepackage{amsmath}
\usepackage{amssymb}
\usepackage{amsthm}
\usepackage{bm}
\usepackage{enumerate}
\usepackage{esint}
\usepackage{framed}
\usepackage{graphicx}
\usepackage[pdftex,colorlinks=true]{hyperref}
\usepackage{latexsym}
\usepackage{listings}
\usepackage{mathtools}
\usepackage{psfrag}
\usepackage{subfigure}
\usepackage{xcolor}
\usepackage[all]{xy}
\usepackage{float}
\usepackage{tikz}

\usepackage{amsfonts}
\usepackage{multirow}
\usepackage{psfrag}
\usepackage{array}
\usepackage[english]{babel}
\usepackage[autostyle]{csquotes}

\newcommand{\ket}[1]{|#1\rangle}
\newcommand{\bra}[1]{\langle#1|}
\newcommand{\commute}[2]{\left\lbrack #1,#2 \right\rbrack}
\newcommand{\overlap}[2]{\langle #1|#2 \rangle}

\newcommand{\Tr}{\mathrm{Tr}}

\newcommand{\GS}{\mathrm{GS}}

\begin{document}
\title{Entanglement Entropy From Tensor Network States for Stabilizer Codes}

\author{Huan He}\thanks{These authors contributed equally.}
\affiliation{Physics Department, Princeton University, Princeton, New Jersey 08544, USA}

\author{Yunqin Zheng}\thanks{These authors contributed equally.}
\affiliation{Physics Department, Princeton University, Princeton, New Jersey 08544, USA}

\author{B. Andrei Bernevig}
\affiliation{Physics Department, Princeton University, Princeton, New Jersey 08544, USA}
\affiliation{Donostia International Physics Center, P. Manuel de Lardizabal 4, 20018 Donostia-San Sebastia ́n, Spain}
\thanks{On sabbatical}
\affiliation{
Laboratoire Pierre Aigrain, D\'epartement de physique de l'ENS, \'Ecole normale sup\'erieure, PSL Research University, Universit\'e Paris Diderot, Sorbonne Paris Cit\'e, Sorbonne  Universit\'es, UPMC Univ. Paris 06, CNRS, 75005 Paris, France
}
\thanks{On sabbatical}
\affiliation{Sorbonne Universit\'es, UPMC Univ Paris 06, UMR 7589, LPTHE, F-75005, Paris, France}
\thanks{On sabbatical}

\author{Nicolas Regnault}
\affiliation{
Laboratoire Pierre Aigrain, D\'epartement de physique de l'ENS, \'Ecole normale sup\'erieure, PSL Research University, Universit\'e Paris Diderot, Sorbonne Paris Cit\'e, Sorbonne  Universit\'es, UPMC Univ. Paris 06, CNRS, 75005 Paris, France
}

\date{\today}

\begin{abstract}
In this paper, we present the construction of tensor network states (TNS) for \emph{some} of the degenerate ground states of 3D stabilizer codes. We then use the TNS formalism  to obtain the entanglement spectrum and entropy of these ground states for some special cuts. In particular, we work out the examples of the 3D toric code, the X-cube model and the Haah code. The latter two models belong to the category of ``fracton" models proposed recently, while the first one belongs to the conventional topological phases. We mention the cases for which the entanglement entropy and spectrum can be calculated exactly: for these, the constructed TNS is a singular value decomposition (SVD) of the ground states with respect to particular entanglement cuts. Apart from the area law, the entanglement entropies also have constant and linear corrections for the fracton models, while the entanglement entropies for the toric code models only have constant corrections. For the cuts we consider, the entanglement spectra of these three models are completely flat. We also conjecture that the negative linear correction to the area law is a signature of extensive ground state degeneracy.
Moreover, the transfer matrices of these TNSs can be constructed. We show that the transfer matrices are projectors whose eigenvalues are either $1$ or $0$. The number of nonzero eigenvalues is tightly related to the ground state degeneracy. 
\end{abstract}

\maketitle
\tableofcontents
\clearpage

\section{Introduction}\label{sec.intro}

Since the invention of density matrix renormalization group (DMRG)\cite{white1992density}, tensor network states (TNS) have become important numerical and theoretical tools in quantum many-body physics. At its core, the scaling of the entanglement entropy was the novel physics concept that made possible  the invention of the recent TNS methods \cite{schollwock2011}. The understanding of the behavior of entanglement entropies of different phases of matter fundamentally improved the numerical tools for condensed matter systems and initiated new research directions for both condensed matter and high energy physics \cite{ryu2006holographic,latorre2015holographic,Pastawski2015}. 

\begin{figure}[b]
\centering
\includegraphics[width=0.4\columnwidth]{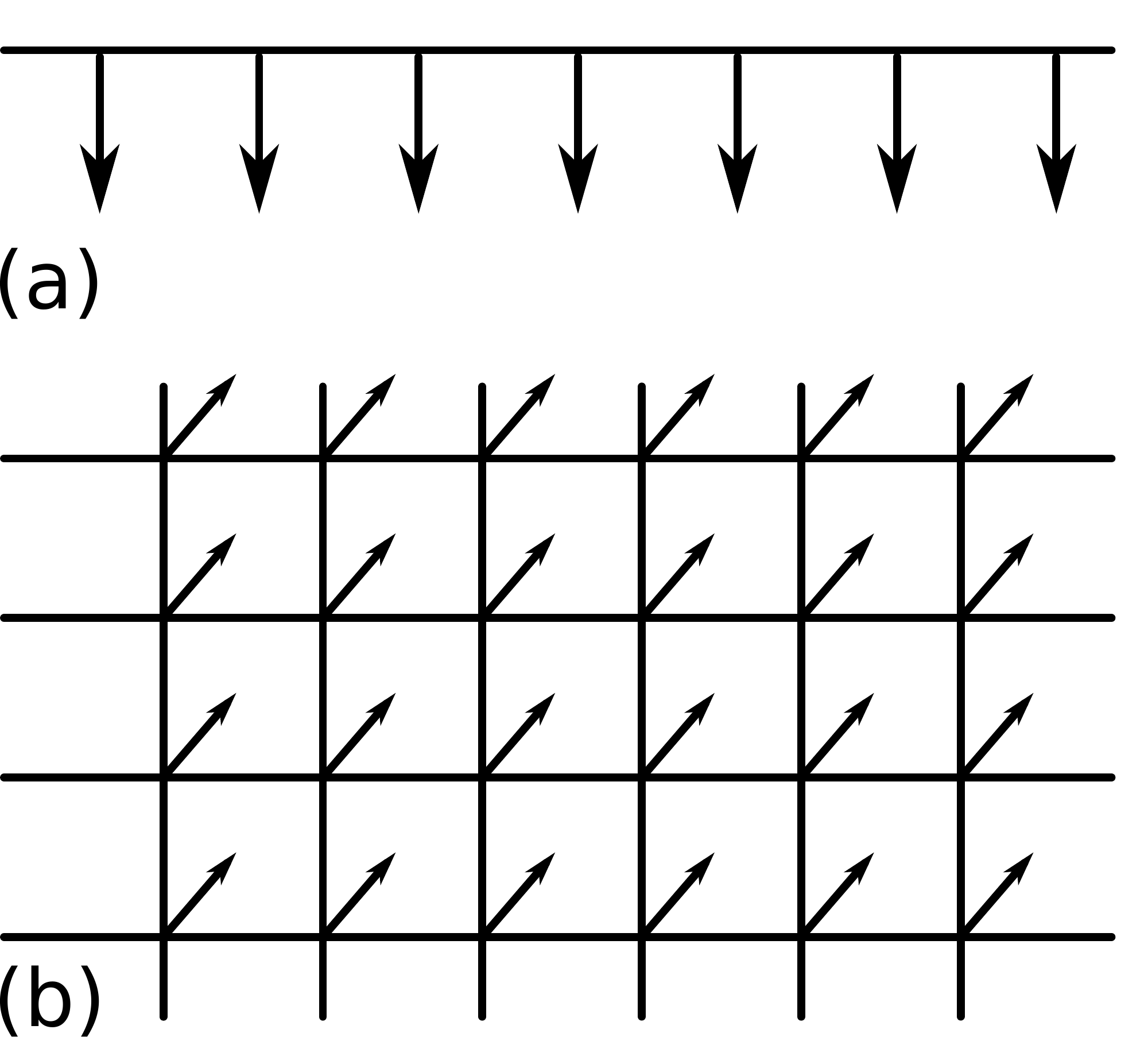}
\caption{Examples of TNS lattice wave functions in 1D and 2D. Each node is a tensor whose indices are the lines connecting to it. The physical indices - of the quantum Hilbert space -  are the lines with arrows, while the lines without any arrows are the virtual indices. Connected lines means the corresponding indices are contracted. Panel (a) is an MPS for 1D systems. Panel (b) is a PEPS on a 2D square lattice.}
\label{fig.TNS_example}
\end{figure}

The study of the entanglement entropy of  one-dimensional (1D)  gapped ground states led to the invention of Matrix Product State (MPS)\cite{perez2006matrix,verstraete2008matrix,orus2014,verstraete2008matrix} in 1D - the simplest but also the most successful example of a TNS. An MPS is a wave function whose coefficients (in some basis decomposition) are represented as matrix products. The immediate generalization of MPS in two dimensions (2D) is the Projected Entangled Pair States (PEPS)\cite{verstraete2008matrix,eisert2013entanglement}   -- the type of TNS we use in this paper. Similar to an MPS, a PEPS is a wave function whose coefficients are tensor contractions. See Fig.~\ref{fig.TNS_example} for a pictorial description of MPS and PEPS. Other types of TNSs include tree Tensor Network (TTN)\cite{shi2006classical}, Multi-Entanglement Renormalization Ansatz (MERA)\cite{vidal2008class}, which are beyond the scope of this paper. In three dimensions (3D), the study of the TNS is not yet well-developed. 

TNS have been heavily used in condensed matter physics in the past decade, especially in the study of 1D and 2D topological phases\cite{wen2016zoo}. Amongst many examples, 
\begin{enumerate}
\item Numerical simulations of the 1D Haldane chain led to the discovery of symmetry protected topological phases (SPT)\cite{pollmann2012symmetry}.
\item Fractional quantum Hall states can be exactly written as MPS\cite{zaletel2012exact,zaletel2013topological,estinne2013matrix,estienne2013fractional,wu2014braiding,estienne2015correlation,wu2015matrix,Grushin2015characterization,zaletel2015infinite,geraedts2015competing,regnault2017entanglement,geraedts2017emergent} which allows performing numerical calculations not accessible by exact diagonalization techniques.
\item A large class of spin liquids wave functions can be constructed using TNS with global spin rotation symmetries and lattice symmetries\cite{schuch2012resonating,wang2013constructing,iqbal2014semionic,poilblanc2014resonating,poilblanc2015critical,mambrini2016systematic,mei2017gapped}. 
\end{enumerate} 

The ground states of gapped local Hamiltonians are conjectured to obey the area law: the entanglement entropy of the ground states with respect to a subsystem $A$ grows linearly with the area of the subsystem's boundary $\mathrm{Area}(\partial A)$:
$$S_A \sim \mathrm{Area}(\partial A).$$
Specifically, in 1D, the entanglement entropy of the subsystem $A$ is a constant:
$$S_l \sim \mathrm{Const},$$ 
since the boundary of $A$ contains only two points.
In 2D, the entanglement entropy of the subsystem $A$ obeys:
$$S_{A} \sim l,$$ 
where $l$ is the perimeter length of the boundary of $A$. For long-range entangled topological phases in 2D, the area law gets supplemented by a leading constant contribution dubbed as topological entanglement entropy \cite{kitaev2006topological,levin2006detecting}, which contains the total quantum dimension. 

In higher spatial dimensions than 2D, more exotic gapped states of matter exist, beyond the paradigm of topological phases.\cite{kitaev2003fault,wen2016zoo}. Recently, 3D so-called fracton models\cite{haah2011local,bravyi2011quantum,haah2013,haah2013lattice,haah2014bifurcation,vijay2015A,haah2016algebraic,Devakul-2017arXiv170910071D,1707.02308,1703.02973,1706.07070,PhysRevB.94.155128,2017arXiv170703838P,2017arXiv170804619S,2017arXiv170910094P,2017arXiv170909673P,2017arXiv171001744M,2017PhRvB..96k5102P,2017arXiv170509300S,2017PhRvD..96b4051P,2017PhRvB..96c5119P,2017EPJST.226..749J,pretko2017fracton,schmitz2017recoverable,slagle2017x,shirley2017fracton,gromov2017fractional} represented by Haah code\cite{haah2011local} and X-cube model have been proposed, attracting the attention of both quantum information\cite{PhysRevA.54.1862} and condensed matter community\cite{BRAVYI2011839,PhysRevLett.94.040402,PhysRevB.95.115139,PhysRevB.93.205406,PhysRevB.95.155133}. They can be realized by stabilizer code  Hamiltonians, whose fundamental property is that they consist solely of sums of terms that commute with each other. They are hence exactly solvable. The defining features of fracton models include (but are not restricted to) that:
\begin{enumerate}
\item Fracton models are gapped, since they can be realized by commuting Hamiltonian terms.
\item The ground state degeneracy on the torus changes as the system size changes. Hence, fracton models seem not to have thermodynamic limits.
\item The low energy excitations can have fractal shapes, other than only points and loops available in conventional topological phases.
\item The excitations of fracton models are not fully mobile: they can only move either along submanifold of the 3D lattice (Type I fracton model), or completely immobile without energy dissipation (Type II fracton model). 
\end{enumerate}

In this paper, we obtain a TNS representation for \emph{some} of the ground states of three stabilizer codes in 3D: the 3D toric code model\cite{kitaev2003fault}, the X-cube model and the Haah code. The two latter ones belong to the catalog of fracton models, while the first one belongs to the conventional topological phases. For instance, the ground state degeneracies on the torus of the X-cube model and the Haah code do not converge to a single number, as the system size increases. In contrast, the ground state degeneracy (GSD) on the torus for 3D toric code model is 8 for all system sizes. Ref.~\onlinecite{vijay2016fracton} treated the X-cube and Haah code models using the idea of lattice gauge theory. 
The gauge symmetry is generally generated by part of the commuting Hamiltonian terms; the rest of the Hamiltonian terms are interpreted as enforcing flat flux conditions. More explicitly, the authors treated the terms only made of Pauli $Z$ operators as the gauge symmetry generators, and the terms only made of Pauli $X$ operators as the flux operators.
The gauge symmetries in the X-cube and the Haah code models are not the conventional $\mathbb{Z}_2$ gauge symmetry such as that in the 3D toric code model, since the gauge symmetry generators, the Pauli $Z$ terms of the X-cube and the Haah code models, are different from those in the 3D toric code model.
Refs.~\onlinecite{ma2017fracton,vijay2017isotropic} derived the X-cube model from ``isotropically" layered 2D toric code models and condensations. The caveat is that this condensation is weaker than the conventional boson condensation in modular tensor category or field theory\cite{eliens2014,kong2014anyon,neupert2016boson,neupert2016nogo}. The authors condense ``composite flux loop" of coupled layers of 2D toric code model. The ``composite flux loop" refers to a composite of four flux excitations near a bond of the lattice. See Ref.~\onlinecite{vijay2017isotropic} for explicit explanations.

Using the TNS representations of some of the ground states, we obtain the entanglement entropy upper bounds for all three models. We then derive the reduced density matrix cuts for which the TNS represents the singular value decomposition (SVD) of the state. For these types of cuts, the entanglement entropy of the three stabilizer codes can be computed exactly. We find that for the fracton models, the entanglement entropy has linear corrections to the area law, corresponding to an exponential degeneracy in the TNS transfer matrix. 

The transfer matrices of TNS of 2D toric code\cite{kitaev2003fault}, whose eigenvalues and eigenstates dominate the correlation functions, have been studied in Ref.~\onlinecite{schuch2013topological,haegeman2015shadows}. The flat entanglement spectra\cite{li2008entanglement} of the 2D toric code were studied in Refs.~\onlinecite{cirac2011entanglement,ho2015edge}. Our TNS construction, when restricted to 2D toric code model, gives the exact results of transfer matrices and entanglement spectra. See App.~\ref{app.toriccode_2D} for explicit calculations and explanations. Beyond the 2D toric code, Refs.~\onlinecite{Zou2016Spurious,2017arXiv171001744M} prove that the reduced density matrix of any stabilizer code is a projector. Hence, the corresponding entanglement spectrum is flat, a property that we will rederive from our TNS.

We will \emph{not} discuss cocycle twisted topological phases, including Dijkgraaf-Witten theories\cite{hu2013twisted,wan2015twisted,he2017field} or (generalized) Walker-Wang models\cite{walker2012,keyserlingk2013three,burnell2013phase,keyserlingk2015walker,zheng2017structure}, even though they can still be realized by commuting Hamiltonians on lattice\cite{wan2015twisted,hu2013twisted,walker2012,levin2005string}. However, the presence of nontrivial cocycles will make the TNS construction very different, based on the experiences in 2D TNS. 
Our construction will
not work for these twisted models. For instance, in 2D, the virtual index dimension using our construction is the same as the physical index dimension. However, when we consider cocycle twisted topological phases, the ``minimal" virtual bond dimension is generally larger than the physical index dimension\cite{buerschaper2014twisted,he2014modular,buerschaper2009explicit,gu2009tensor}. More explicitly, the minimal virtual bond dimension for 2D toric code model is 2, while the minimal virtual bond dimension for 2D double semion model (twisted toric code) is 4\cite{buerschaper2014twisted,he2014modular,buerschaper2009explicit,gu2009tensor}.
The 2D cocycle twisted TNS has been systematically explored in the literature for bosonic\cite{buerschaper2014twisted} and fermionic\cite{wille2017fermionic,williamson2016fermionic} systems respectively.

The organization of this paper is as follows:
In Sec.~\ref{sec.overview}, we set the notations and provide an overview and the general idea of the TNS construction
In Sec.~\ref{sec.TNSSVD}, we present the calculation of the entanglement properties using the developed TNS construction
In Sec.~\ref{sec.toriccode}, we present the TNS construction for the toric code model in 3D. The entanglement entropy is calculated from the obtained TNS. The transfer matrix is constructed afterwards and is proven to be a projector of rank 2.
In Sec.~\ref{sec.Xcube}, we present the TNS construction for X-cube model. The same calculations for the entanglement entropy and the transfer matrix are presented. They are quickly shown to be  very different from the toric code model. Indeed, the entanglement entropies have linear corrections to the area law, and the transfer matrix is exponentially degenerate.
In Sec.~\ref{sec.Haah}, we present the TNS construction for Haah code. The entanglement entropies are calculated for several types of cuts.
In Sec.~\ref{sec.discussion}, we summarize the paper and discuss future directions.

\section{Stabilizer Code Tensor Network States}\label{sec.overview}

In this section, we provide an overview of the stabilizer codes and the tensor network state description of their ground states. In this article, we focus on a few ``main" stabilizer codes in three dimensions : the toric code\cite{kitaev2003fault}, the X-cube model\cite{vijay2016fracton} and the Haah code\cite{haah2011local}. The TNSs for these models have similarities in their derivation and they share several (but importantly not all!) common features. Both aspects are presented in this section. For pedagogical purposes, we discuss the 2D toric code in App.~\ref{app.toriccode_2D}.

\subsection{Notations}\label{subsec.notations}

\begin{figure}[t]
\centering
\includegraphics[width=0.9\columnwidth]{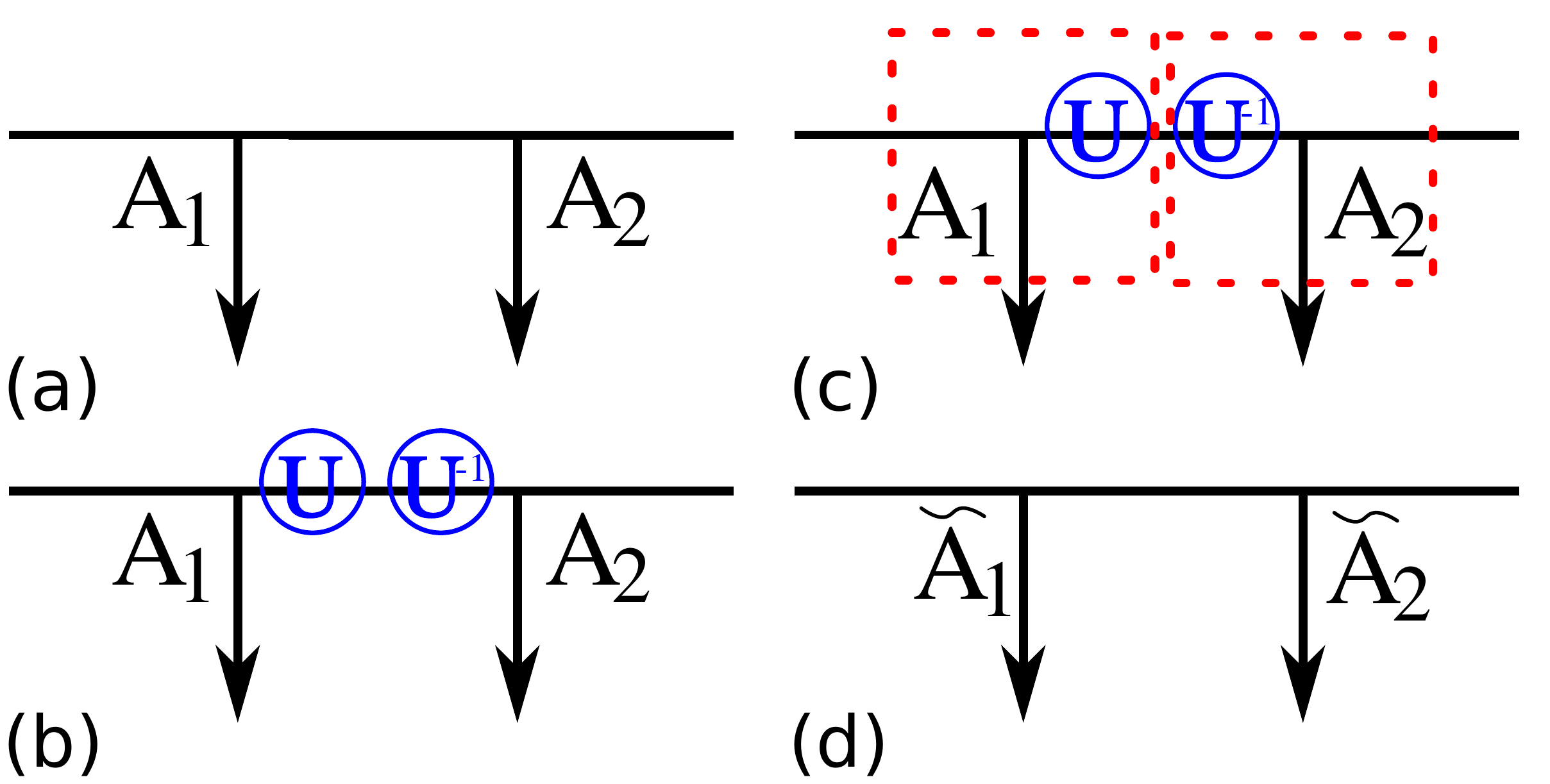}
\caption{An illustration of the TNS gauge in MPS. (a) A part of an MPS. $A_1$ and $A_2$ are two local tensors contracted together. (b) We insert the identity operator $\mathbb{I}=UU^{-1}$ at the virtual level - it acts on the virtual bonds. The tensor contraction of $A_1$ and $A_2$ does not change. (c) We further multiply $U$ with $A_1$ and $U^{-1}$ with $A_2$, resulting in $\tilde{A}_1$ and $\tilde{A}_2$ respectively in Panel (d). The tensor contraction of $A_1$ and $A_2$ is the same as the tensor contraction of $\tilde{A}_1$ and $\tilde{A}_2$. The TNS does not change as well. Similar TNS gauges also appear in other TNS such as PEPS.}
\label{fig.TNSgauge}
\end{figure}

We first fix some of the notations used in the paper, to which we will refer throughout the manuscript:
\begin{enumerate}
\item The Pauli matrices $X$ and $Z$ are defined as:
\begin{equation}
\begin{split}
X = \left(\begin{matrix}
0	&	1	\\
1	&	0
\end{matrix}\right), \;
Z = \left(\begin{matrix}
1	&	0	\\
0	&	-1
\end{matrix}\right).
\end{split}
\end{equation}
\item We introduce a $g$ tensor, which denotes the projector from a physical index to virtual indices. $g$ tensors are essentially the same (up to the number of indices) for all stabilizer codes. $g$ tensors have two virtual indices and one physical index for the 3D toric code model and the X-cube model, while $g$ tensors for the Haah code have four virtual indices and one physical index. They are depicted in Eq.~\eqref{eq.projector2}, \eqref{eq.projector4L} and \eqref{eq.projector4R}.
\item We introduce the $T$ tensor, which denotes the local tensor for each model. It only has virtual indices and thus no physical indices. The specific tensor elements are determined by the Hamiltonian terms.
\item Since we consider mostly models on cubic lattices, the indices of $T$ tensors will be denoted as $x$, $\bar{x}$, $y$, $\bar{y}$, $z$ and $\bar{z}$ in the 3 directions (forward and backward) respectively. The indices will be  collectively denoted using curly brackets. For instance, the physical indices are collectively denoted as $\{s\}$, while the virtual indices are denoted as $\{t\}$. The virtual indices which are not contracted over are called ``open indices". Both the physical indices and the virtual indices are non-negative integer values.
\item Graphically, the physical indices are denoted by arrows, while the virtual indices are not associated with any arrows. See Fig.~\ref{fig.TNS_example}.
\item The contraction of a network of tensors over the virtual indices is denoted as $\mathcal{C}^{\mathcal{M}} \left( \; \right)$ where $\mathcal{M}$ is the spatial manifold that the TNS lives on. The corresponding wave function that arises from the contraction is denoted as $\ket{\mathrm{TNS}}_{\mathcal{M}}$. When evaluating the TNS norms or any other physical quantities, we contract over the virtual indices from both the bra and the ket layer. This contraction is still denoted by $\mathcal{C}^{\mathcal{M}} \left( \; \right)$.
\item $L_x$, $L_y$ and $L_z$ refer to the system sizes in the three directions (the boundary conditions will be specified), while $l_x$, $l_y$ and $l_z$ refer to the sizes of the entanglement cut. Both are measured in units of vertices.
\item The TNS gauge is defined as the gauge degrees of freedom of TNS such that the wave function stays invariant while the local tensors change. One can insert identity operators  $\mathbb{I}=UU^{-1}$ on the  virtual bonds, where $U$ is any invertible matrix acting on the virtual index, multiplying  $U$ and $U^{-1}$ to nearby local tensors respectively. The local tensors then change but the wave function stays invariant. We refer to this gauge degree of freedom as the TNS gauge. The TNS gauge exists in MPS, PEPS etc. See Fig.~\ref{fig.TNSgauge} for an illustration. In our calculations, we only fix the tensor elements up to the TNS gauge.
\end{enumerate}

\subsection{Stabilizer Code and TNS Construction}\label{subsec.TNSingeneral}

\begin{figure*}[t]
\centering
\includegraphics[width=0.6\textwidth]{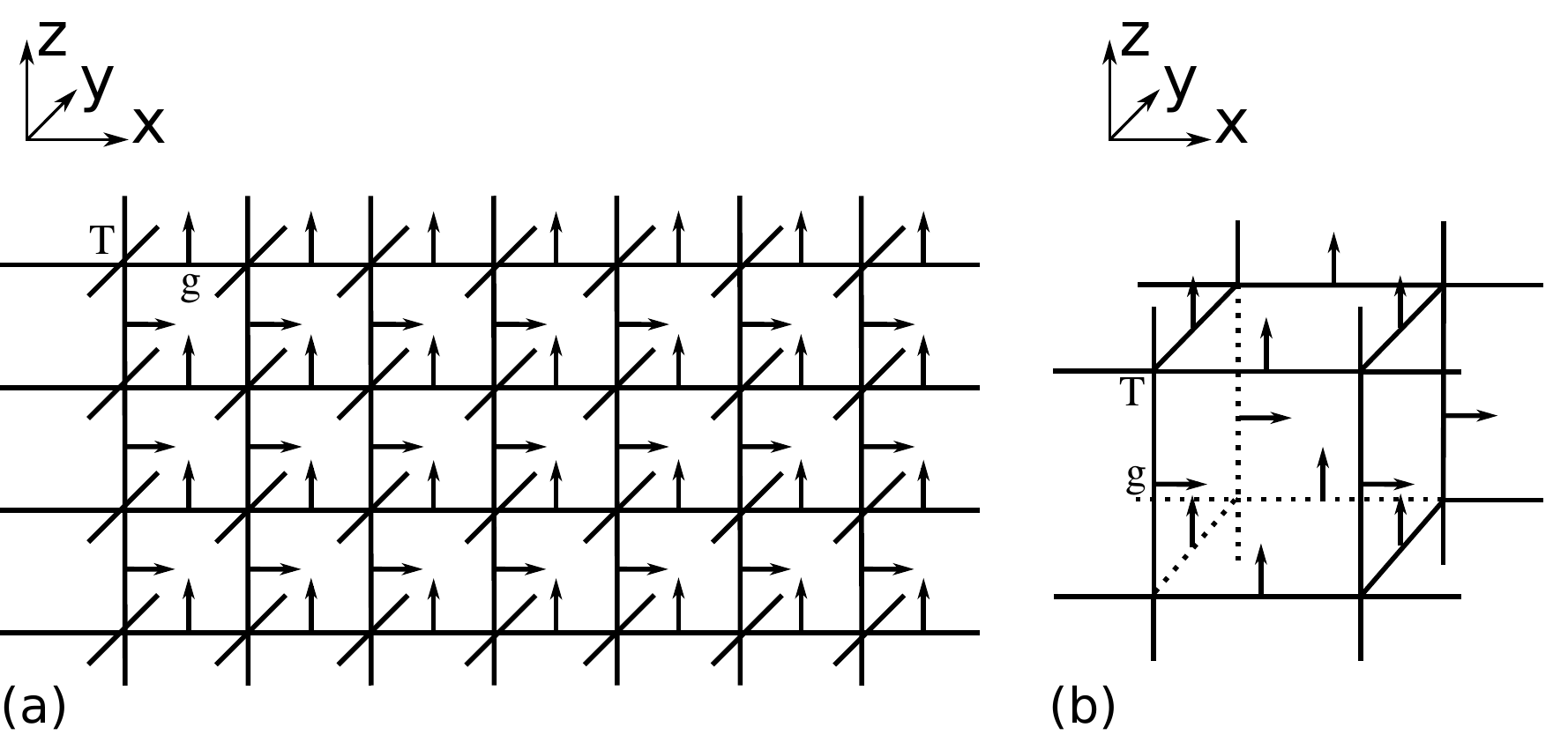}
\caption{(a) A plane of TNS on a cubic lattice. (b) TNS on a cube. The lines with arrows are the physical indices. The connected lines are the contracted virtual indices, while the open lines are not contracted. On each vertex, there lives a $T$ tensor, and on each bond, we have a projector $g$ tensor. }
\label{fig.TNS}
\end{figure*}

We now summarize the general idea of constructing TNSs for stabilizer codes. In App.~\ref{app.toriccode_2D}, we provide the construction of the TNS for the 2D toric code model on a square lattice. In the following, we assume that the physical spins are defined on the bonds of the cubic lattice (such as the 3D toric code and the X-cube models). The cases where the physical spins are defined on vertices can be analyzed similarly.  The generic philosophy of any stabilizer code model is captured by the following exactly solvable Hamiltonian:
\begin{equation}
H = - \sum_{v} A_{v} - \sum_{p} B_p \label{stabilizercodeplaqutteandvertex}
\end{equation} 
where the Hamiltonian is the sum of the $A_v$ terms which are products of only Pauli $Z$ operators, and the $B_p$ terms which are products of only Pauli $X$ operators. 
$v$ and $p$ denotes the positions of the $A_v$ and $B_p$ operators on the lattice. In the 3D toric code, $v$ is the vertex of the cubic lattice while $p$ is the plaquette. In the X-cube model, $v$ is the vertex while $p$ is the cube. In the Haah code, both $v$ and $p$ are cubes. See Sec.~\ref{subsec.toriccode}, \ref{subsec.Xcube} and \ref{subsec.HaahCode} for the definitions of Hamiltonians of these three models.
All these \textit{local} operators commute with each other:
\begin{equation}
\begin{split}
\commute{A_{v}}{A_{v^\prime}} &= 0,	\quad \forall\; v,v^\prime	\\
\commute{B_{p}}{B_{p^\prime}} &= 0,	\quad \forall\; p,p^\prime	\\
\commute{A_{v}}{B_{p}} &= 0,	\quad \forall\; v,p	.\\
\end{split}
\end{equation}
The Hamiltonian eigenstates are the eigenstates of these local terms individually. In particular, any ground state $\ket{\mathrm{GS}}$ should satisfy:
\begin{equation}
\begin{split}
A_{v} \ket{\mathrm{GS}} = \ket{\mathrm{GS}}, \quad \forall\; v	\\
B_{p} \ket{\mathrm{GS}} = \ket{\mathrm{GS}}, \quad \forall\; p	\\
\end{split}
\end{equation}
for all positions labeled by $v$ and $p$. In this paper, we only consider Hamiltonians being of a sum of local terms that are either a product of Pauli $Z$ operators, or a product of Pauli $X$ operators. Thus, we do not include the case of mixed products of Pauli $Z$ and $X$ operators.

The ground states for the stabilizer codes with Hamiltonian as in Eq.~\eqref{stabilizercodeplaqutteandvertex} can be  written exactly in terms of TNS.
Our construction, when restricted to the 2D toric code model, is the same as in the literature\cite{buerschaper2009explicit,gu2009tensor}.
In the following, we provide one possible general construction for such TNSs. We introduce a projector $g$ tensor with one physical index $s$ and two virtual indices $i, j$:
\begin{equation}\label{eq.projector2}
g^{s}_{ij} =
\begin{gathered}
\includegraphics[width=3cm]{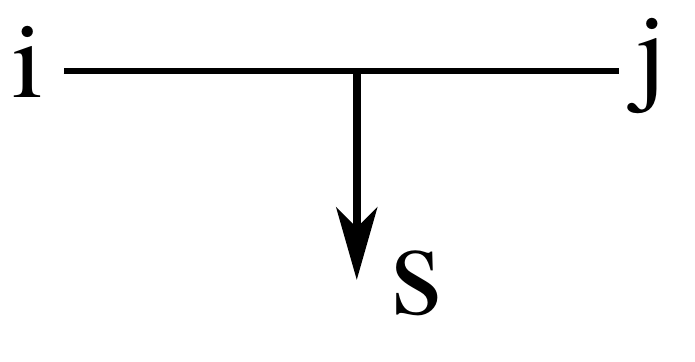}
\end{gathered}
=
\begin{cases}
1	&	s=i=j	\\
0	&	\text{otherwise}
\end{cases}
\end{equation}
where the line with an arrow represents the physical index, and the lines without arrows correspond to the virtual indices. The physical index $s=0, 1$ represents the Pauli $Z$ eigenstates of $|\mathord{\uparrow}\rangle, |\mathord{\downarrow}\rangle$ respectively where $Z|\mathord{\uparrow}\rangle=|\mathord{\uparrow}\rangle$, and $Z|\mathord{\downarrow}\rangle=-|\mathord{\downarrow}\rangle$. The projector $g$ tensor maps the physical spin into the virtual spins exactly. As a result, the virtual index has a bond dimension $2$. 
When a Pauli operator acts on the physical index of a projector $g$ tensor, its action transfers to the virtual indices of $g$. For instance, a Pauli operator $X$ acting on the physical index of a $g$ tensor amounts to two Pauli operators  $X$ acting on \emph{both} virtual indices of the same $g$ tensor, and a Pauli operator $Z$ acting on the physical index of a $g$ tensor amounts to a Pauli operator $Z$ acting on \emph{either} virtual index of the same $g$ tensor. 

To each vertex, we associate a local tensor $T$ which only has virtual indices. To each bond, we associate a projector $g$ tensor. The TNS is obtained by contracting the $g$ and $T$ tensors as depicted in Fig.~\ref{fig.TNS} (a) and (b). We define the TNS as:
\begin{equation}\label{eq.TNS}
\ket{\mathrm{TNS}} = \sum_{\{s\}} \mathcal{C}^{\mathcal{R}^3} \left( g^{s_1}g^{s_2}g^{s_3} \ldots TTT \ldots \right) \ket{\{s\}}
\end{equation}
where $\mathcal{C}^{\mathcal{R}^3}$ denotes the contraction over all virtual indices on $\mathcal{R}^3$ as illustrated in Fig.~\ref{fig.TNS} (b); $\ket{\{s\}}$ is a wave function basis for spin configurations on the cubic lattice in Pauli $Z$ basis.
The TNS can be put on other spatial manifolds such as $\mathcal{T}^3$ and $\mathcal{T}^2 \times \mathcal{R}$. In our notation, they are denoted by changing $\mathcal{C}^{\mathcal{R}^3}$ to $\mathcal{C}^{\mathcal{T}^3}$ and $\mathcal{C}^{\mathcal{T}^2 \times \mathcal{R}}$.
The TNS for the ground states satisfies:
\begin{equation}\label{eq.ABcondition}
\begin{split}
A_v \ket{\mathrm{TNS}} = \ket{\mathrm{TNS}}, \quad \forall \; v	\\
B_p \ket{\mathrm{TNS}} = \ket{\mathrm{TNS}}, \quad \forall \; p	\\
\end{split}
\end{equation}
for all positions labeled by $v$ and $p$. 

The actions of $A_v$ and $B_p$ operators on the TNS can be transferred to the virtual indices, using the definition of the $g$ tensor. Since the virtual indices of projector $g$ tensors are contracted with the virtual indices of $T$ tensors, the actions of $A_v$ and $B_p$ on the physical indices will be transferred to actions on the local tensors $T$. By enforcing the local tensors $T$ to be invariant under $A_v$ and $B_p$ actions, we obtain Eq.~\eqref{eq.ABcondition},  and $\ket{\mathrm{TNS}}$ belongs to the ground state manifold. 
For the three models analyzed in this paper, we have found that up to TNS gauge, the elements of the local tensor $T$ can be reduced to two values, either $1$ or $0$. The first equation of Eq.~\eqref{eq.ABcondition} restricts the local $T$ tensor to be:
\begin{equation}
\begin{split}
T_{x\bar{x}y\ldots}
\begin{cases}
\neq 0	&	\parbox[t]{.6\textwidth}{if the indices $x\bar{x}y\ldots$ satisfy \\ some constraints}\\
= 0	&	\text{otherwise}\\
\end{cases}.
\end{split}
\end{equation}
Applying the second equation of Eq.~\eqref{eq.ABcondition} will further restrict the local $T$ tensor to be:
\begin{equation}
\begin{split}
T_{x\bar{x}y\ldots} = 
\begin{cases}
1	&	\parbox[t]{.6\textwidth}{if the indices $x\bar{x}y\ldots$ satisfy \\ some constraints}\\
0	&	\text{otherwise}\\
\end{cases}.
\end{split}
\end{equation}
For simplicity, we calculate the entanglement entropies of the wave function on $\mathcal{R}^3$ with respect to some specific entanglement cuts, and compute the ground state degeneracy (GSD) of the 3D toric code and X-cube model on $\mathcal{T}^3$. 

We emphasize that in this paper, we are only concerned with the bulk wave functions and their entanglement entropies. In principle, the TNS of Eq.~\eqref{eq.TNS} requires boundary conditions, i.e. the virtual indices at infinity on $\mathcal{R}^3$. The boundary conditions are assumed not to make a difference to the reduced density matrices in the bulk. (Notice that this is true as long as the region considered for the reduced density matrices does not contain any boundary virtual index.) Hence, we do not need to specify the boundary conditions for the TNS in the following calculations of entanglement entropies.

\subsection{TNS Norm}\label{subsec.TNSnorm}

Evaluating the norm of the TNS given by Eq.~\eqref{eq.TNS} (or any scalar product between two TNS) is straightforward. Indeed the $g$ tensors are projectors, and hence greatly simplify the expression of the tensor network norm when we contract over the physical indices.

Given the wave function of Eq.~\eqref{eq.TNS}, we can compute its norm as follows:
\begin{widetext}
\begin{equation}\label{eq.TNSnorm}
\begin{split}
\overlap{\mathrm{TNS}}{\mathrm{TNS}} 
=& \left(\sum_{\{s\}} \mathcal{C}^{\mathcal{R}^3} \left( g^{s_1}g^{s_2}g^{s_3} \ldots TTT \ldots \right) \bra{\{s\}}\right)^\star \left(\sum_{\{s\}} \mathcal{C}^{\mathcal{R}^3} \left( g^{s_1}g^{s_2}g^{s_3} \ldots TTT \ldots \right) \ket{\{s\}}\right) \\
=& \sum_{\{s\}} \mathcal{C}^{\mathcal{R}^3} \left( g^{s_1\star}g^{s_2\star}g^{s_3\star} \ldots T^\star T^\star T^\star \ldots \right) \mathcal{C}^{\mathcal{R}^3} \left( g^{s_1}g^{s_2}g^{s_3} \ldots TTT \ldots \right) \\
=& \sum_{\{s\}} \mathcal{C}^{\mathcal{R}^3} \left( g^{s_1}g^{s_2}g^{s_3} \ldots T^\star T^\star T^\star \ldots \right) \mathcal{C}^{\mathcal{R}^3} \left( g^{s_1}g^{s_2}g^{s_3} \ldots TTT \ldots \right), \\
\end{split}
\end{equation}
\end{widetext}
where $\star$ is the complex conjugation, and we have used the fact that the $g$ tensors are real for our models. Now we specify a contraction order in Eq.~\eqref{eq.TNSnorm}:  we first contract over the physical indices and then we contract over the virtual indices.
If the physical indices of two projector $g$ tensors are contracted over, the four virtual indices will be enforced to be the same following the definition of the projector $g$ tensor:
\begin{equation}\label{eq.projectorcontraction}
\begin{split}
\sum_{s} g^{s}_{ij}g^{s}_{mn}
&=
\begin{gathered}
\includegraphics[width=2.5cm]{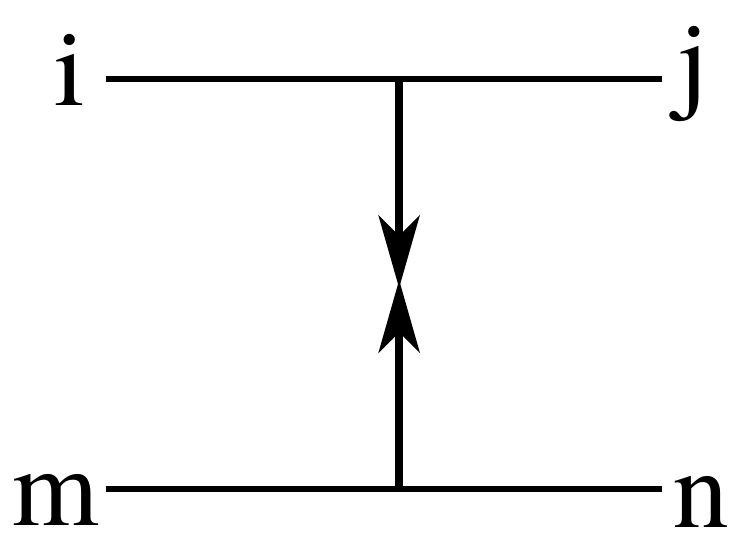}
\end{gathered}	\\
&=
\begin{cases}
1	&	i=j=m=n	\\
0	&	\text{otherwise}	\\
\end{cases}.
\end{split}
\end{equation}
Thus when computing wave function overlap $\overlap{\mathrm{TNS}}{\mathrm{TNS}}$, the virtual indices in the bra layer and the ket layer at the same place are enforced to be same. As a result, we have:
\begin{equation}\label{eq.TNSoverlap}
\overlap{\mathrm{TNS}}{\mathrm{TNS}} = \mathcal{C}^{\mathcal{R}^3} \left( \ldots \mathbb{T}\mathbb{T}\mathbb{T}\ldots\right),
\end{equation}
where $\mathcal{C}^{\mathcal{R}^3}$ stands for the contraction of a network of tensors $\mathbb{T}$ over the virtual indices on $\mathcal{R}^3$. In a slight abuse of notation, $\mathcal{C}^{\mathcal{R}^3}$ in Eq.~\eqref{eq.TNSoverlap} stands for the contraction taken over the virtual indices of both the bra and the ket layer, while the contraction in Eq.~\eqref{eq.TNS} is taken over the virtual indices in only the ket layer. The double tensor $\mathbb{T}$ is defined as
\begin{equation}\label{eq.T2}
\begin{split}
\mathbb{T}_{x\bar{x}y\ldots, x'\bar{x}'y'\ldots} 
&= T^*_{x\bar{x}y\ldots} T_{x'\bar{x}'y'\ldots} \delta_{xx'}\delta_{\bar{x}\bar{x}'}\delta_{yy'}\ldots	\\
&= |T_{x\bar{x}y\ldots}|^2 \delta_{xx'}\delta_{\bar{x}\bar{x}'}\delta_{yy'}\ldots,	\\
\end{split}
\end{equation}
for all the elements of $T$ and $\mathbb{T}$. The indices are not summed over in the above equation. The indices $x\bar{x}y\ldots$ come from the bra layer while the indices $x'\bar{x}'y'\ldots$ come from the ket layer. In a 2D square lattice, a $T$ tensor usually has 4 virtual indices $x,\bar{x},y,\bar{y}$, while in a 3D cubic lattice, a $T$ tensor usually has 6 virtual indices $x,\bar{x},y,\bar{y},z,\bar{z}$.
If the elements of the $T$ tensor are only either $0$ or $1$, we get,
\begin{equation}\label{eq.T2=T1}
\begin{split}
\mathbb{T}_{x\bar{x}y\ldots, x'\bar{x}'y'\ldots}
&= |T_{x\bar{x}y\ldots}|^2\delta_{xx'}\delta_{\bar{x}\bar{x}'}\delta_{yy'}\ldots	\\
&=T_{x\bar{x}y\ldots}\delta_{xx'}\delta_{\bar{x}\bar{x}'}\delta_{yy'}\ldots.	\\
\end{split}
\end{equation}
Then,
\begin{equation}\label{eq.TNSnormContractT}
\begin{split}
\overlap{\mathrm{TNS}}{\mathrm{TNS}}	
=& \mathcal{C}^{\mathcal{R}^3} \left( \ldots \mathbb{T}\mathbb{T}\mathbb{T}\ldots\right)	\\
=& \mathcal{C}^{\mathcal{R}^3} \left( \ldots TTT\ldots \right).
\end{split}
\end{equation}
This result will be frequently used in the following discussions, especially when we compute wave function overlaps or transfer matrices. Eqs.~\eqref{eq.TNSoverlap} and \eqref{eq.TNSnormContractT} hold true on other manifolds as well, such as $\mathcal{T}^3$ and $\mathcal{T}^2\times \mathcal{R}$.

\subsection{Transfer Matrix}\label{subsec.TransferMatrix}

\begin{figure}[t]
\centering
\includegraphics[width=0.4\columnwidth]{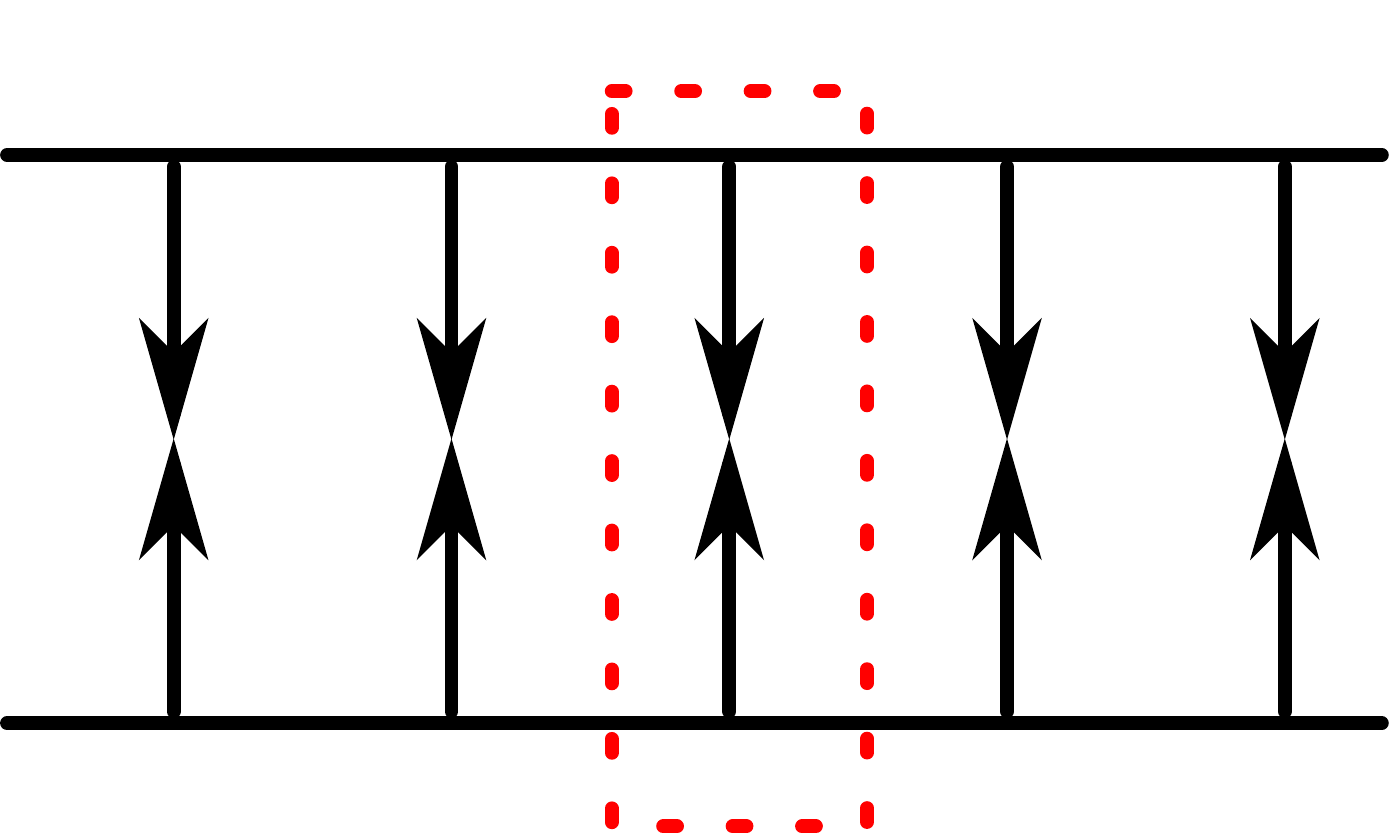}
\caption{Transfer matrix (red dashed square) of a 1D MPS. The connected lines are the contracted virtual indices. The connected arrow lines are the contracted physical indices. The MPS norm (or any other quantities) can be built using the transfer matrix. Higher dimensional transfer matrices are similarly defined for TNS on a cylinder or a torus, by contracting in all directions except one. This leads to a 1D MPS with a bond dimension exponentially larger than the TNS one.}
\label{fig.MPS_TM}
\end{figure}

The transfer matrix method is ubiquitous when using MPS (see Fig.~\ref{fig.MPS_TM} for an illustration of the transfer matrix). It can be generalized to TNS on a 2D cylinder by contracting tensors along the periodic direction of the cylinder. This implies that the bond dimension of the transfer matrix is exponentially large with respect to the cylinder perimeter. In 3D, the TNS norm on $\mathcal{T}^3$ of size $L_x \times L_y \times L_z$ can be written as an MPS using transfer matrices $\mathrm{TM}_{xy}$ in each $xy$-plane:
\begin{equation}\label{eq.TNSnormTransferMatrix}
\begin{split}
\overlap{\mathrm{TNS}}{\mathrm{TNS}} 
&= \mathrm{Tr} \left( \mathrm{TM}_{xy,z=1} \mathrm{TM}_{xy,z=2} \ldots \right)	\\
&= \mathrm{Tr} \left( \left( \mathrm{TM}_{xy,z=1}\right)^{L_z} \right)
\end{split}
\end{equation}
where we have assumed that all transfer matrices in each plane are the same:
\begin{equation}
\mathrm{TM}_{xy,z=1} = \mathrm{TM}_{xy,z=2} = \ldots
\end{equation}
Eq.~\eqref{eq.TNSnormTransferMatrix} is an alternative way of writing the wave function norm and specifies a contraction order of the tensors in Eq.~\eqref{eq.TNSoverlap}: we first contract the virtual indices along $xy$-plane which defines the transfer matrix $\mathrm{TM}_{xy}$, and then contract the virtual indices in $z$-direction which leads to the multiplication and the trace of transfer matrices. 
The transfer matrix $\mathrm{TM}_{xy}$ is defined as:
\begin{widetext}
\begin{equation}\label{eq.TransferMatrix}
\mathrm{TM}_{xy} = \left(\sum_{\{s\}}\mathcal{C}^{\mathcal{T}^2_{xy}} \left( g^{s_1\star}g^{s_2\star}g^{s_3\star}\ldots T^{\star}T^{\star}T^{\star} \ldots \right) \bra{\{s\}}\right) \left(\sum_{\{s\}}\mathcal{C}^{\mathcal{T}^2_{xy}} \left( g^{s_1}g^{s_2}g^{s_3}\ldots TTT \ldots \right) \ket{\{s\}}\right),
\end{equation}
\end{widetext}
where the TNS contraction is performed along the $xy$-plane with periodic boundary conditions, i.e., the 2D torus $\mathcal{T}^2_{xy}$. Denoting $\mathbf{T}_{xy}$ as the TNS in its ket layer, it can depicted as:
\begin{equation}
\begin{gathered}
\includegraphics[width=0.8\columnwidth]{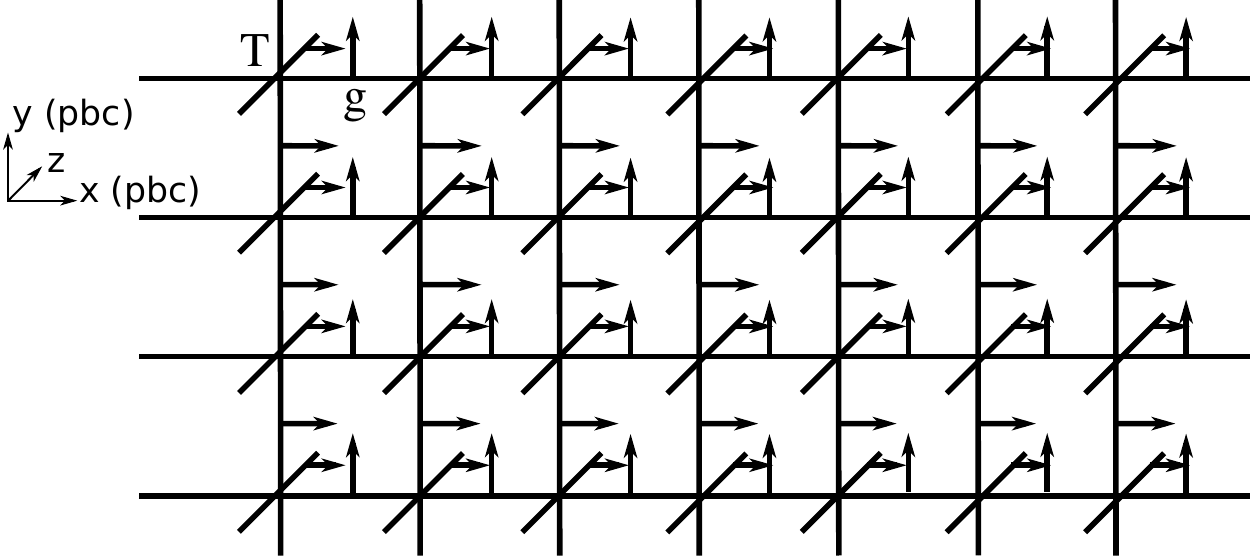}
\end{gathered}
\end{equation}
$\mathrm{TM}_{xy}$ is the overlap of the bra and ket layer of this TNS over the plane with periodic boundary conditions. 

Furthermore, applying Eq.~\eqref{eq.projectorcontraction} to Eq.~\eqref{eq.TransferMatrix}, the virtual indices in the bra layer and the ket layer are identified after the physical indices are contracted in Eq.~\eqref{eq.TransferMatrix}. Hence, we have:
\begin{equation}\label{eq.TransferMatrixContractPhysics}
\mathrm{TM}_{xy} = \mathcal{C}^{\mathcal{T}^2_{xy}} \left( \ldots \mathbb{T}\mathbb{T}\mathbb{T}\ldots \right),
\end{equation}
where the tensors $\mathbb{T}$, defined in Eq.~\eqref{eq.T2}, are in the $xy$-plane with periodic boundary conditions. The indices in the $z$-direction are open. By  Eq.~\eqref{eq.T2=T1}  - which is true when the elements of the $T$ tensor are either $0$ or $1$ -  the transfer matrices is further simplified to:
\begin{equation}\label{eq.TransferMatrixContractT}
\mathrm{TM}_{xy} = \mathcal{C}^{\mathcal{T}^2_{xy}} \left( \ldots TTT \ldots \right).
\end{equation}
Graphically:
\begin{equation}\label{eq.TransferMatrixGraph}
\begin{split}
&\mathrm{TM}_{xy} = 	\\
&\begin{gathered}
\includegraphics[width=0.7\columnwidth]{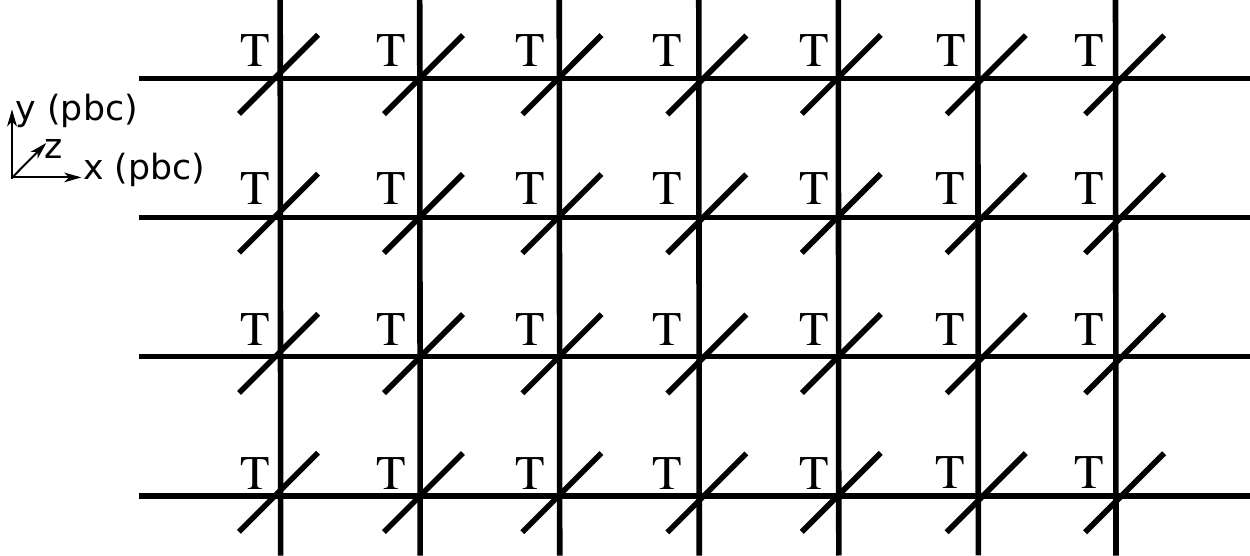}
\end{gathered}.
\end{split}
\end{equation}

Suppose the virtual index is of dimension $D$. Then in Eq.~\eqref{eq.TransferMatrix}, the transfer matrix is of dimension $D^{2L_xL_y} \times D^{2L_xL_y}$. However, in Eq.~\eqref{eq.TransferMatrixContractPhysics}, the transfer matrix reduces to dimension $D^{L_xL_y} \times D^{L_xL_y}$, since the indices in the bra layer and the ket layer are identified due to the contraction over the physical indices of projector $g$ tensors.

\section{Entanglement properties of the stabilizer code TNS}\label{sec.TNSSVD}

The specific structure of the TNS discussed in the previous section allows us to derive its entanglement properties. In this section, we show that for a large class of entanglement cuts, the TNS is already in Schmidt form, i.e. is exactly a singular value decomposition (SVD). We also summarize the main results for the entanglement entropies and the transfer matrices
that we have obtained for the three stabilizer codes.

\subsection{TNS as an exact SVD}\label{subsec.TNSSVD}

We propose a general sufficient condition that the TNS is an SVD with respect to particular entanglement cuts. Let us denote the TNS with open virtual indices $\{t\}$ as:
\begin{equation}
\ket{\{t\}} = \sum_{\{s\}} \mathcal{C}^{\mathcal{M}} \left( TTT \ldots g^{s_1}g^{s_2}g^{s_3}\ldots \right) \ket{\{s\}}, \label{tfunction}
\end{equation}
where $\mathcal{M}$ is an open manifold which the TNS lives on, $\mathcal{C}^{\mathcal{M}}$ stands for the contraction over the virtual indices inside $\mathcal{M}$, but not over the open ones $\{t\}$ that straddle the boundary of $\mathcal{M}$. In Eq.~\eqref{tfunction}, the $T$ tensors and $g$ tensors are the tensors inside $\mathcal{M}$ such that the nodes of the local $T$ tensors and the projector $g$ tensors are inside $\mathcal{M}$. 
For example, when $\mathcal{M}$ is a cube, we have a TNS figure:
\begin{equation}\label{eq.TNSbasis}
\ket{\{t\}} = 
\begin{gathered}
\includegraphics[width=0.3\columnwidth]{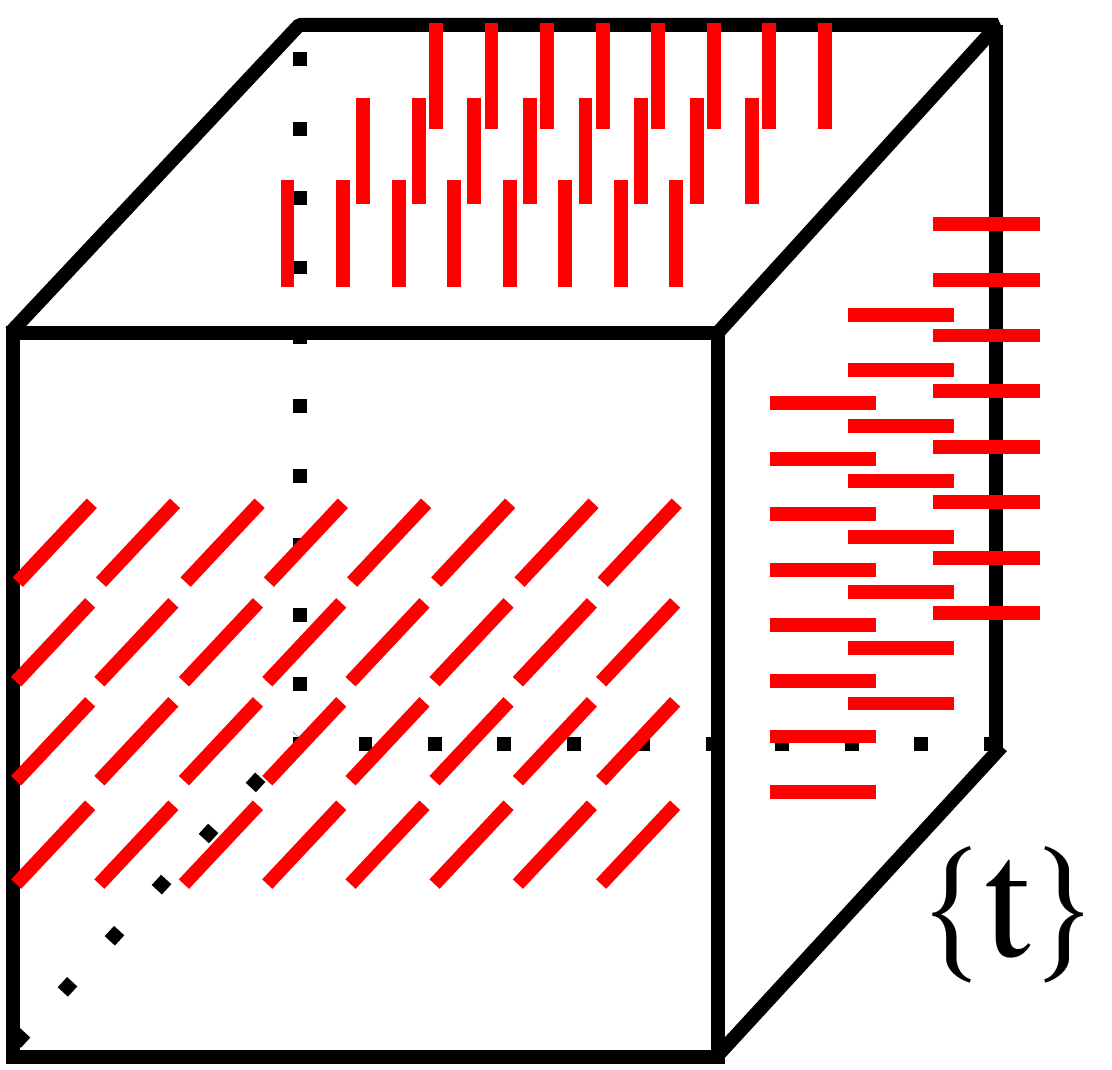}
\end{gathered},
\end{equation}
where inside the cube is a network of contracted tensors which are not explicitly drawn, and the red lines denote the open virtual indices $\{t\}$. With this notation of $\ket{\{t\}}$, the TNS can be written as:
\begin{equation}\label{eq.TNSgeneralDecomposition}
\ket{\mathrm{TNS}} = \sum_{\{t\}} \ket{\{t\}}_{A} \otimes \ket{\{t\}}_{\bar{A}}
\end{equation}
with respect to a region $A$ and its complement $\bar{A}$. $\ket{\{t\}}_{A}$ is the TNS  in region $A$ with open indices $\{t\}$, while $\ket{\{t\}}_{\bar{A}}$ is the TNS in region $\bar{A}$ with the same open indices $\{t\}$ due to tensor contraction. In other words, the TNS naturally induces a bipartition of the wave functions. However, the two partitions do not need to each form orthonormal sets.

We now propose a simple sufficient (but not generally necessary) condition to determine when Eq.~\eqref{eq.TNSgeneralDecomposition} is an exact SVD for the TNS constructed  in this paper. We first have to make an assumption, satisfied by all our TNSs: 

\textit{Local $T$ tensor assumption:} We assume that the indices of the nonzero elements of the local $T$ tensor are constrained: if all the indices of the element $T_{\ldots t\ldots}$ except for $t$ are fixed, then there is only one choice of $t$ such that $T_{\ldots t \ldots}$ is nonzero.

This assumption can be easily verified when the local $T$ tensors are obtained for the three models studied in this paper, such as the 3D toric code model in Eq.~\eqref{eq.ToricCodeTtensor}. We are now ready to express our SVD condition:

\textit{SVD condition:
If there are no two open virtual indices in $\{t\}$ (see Eq.~\eqref{eq.TNSbasis}) of the region $A$ that connect to the same $T$ tensor in the region $A$, then the non-vanishing states $\ket{\{t\}}_{A}$ span an orthogonal basis. Similarly, if there are no two open virtual indices in $\{t\}$ of the region $\bar{A}$ that connect to the same $T$ tensor in the region $\bar{A}$, then the non-vanishing states $\ket{\{t\}}_{\bar{A}}$ form an orthogonal basis. Therefore, Eq.~\eqref{eq.TNSgeneralDecomposition} is an exact SVD.}

\noindent\textbf{Proof}: 

We first prove the statement for the region $A$. Suppose that $\ket{\{t\}}_{A}$ and $\ket{\{t^\prime\}}_{A}$ are two non-vanishing TNSs in the region $A$. Any open index in $\{t\}$ of the region $A$ must connect to either a projector $g$ tensor or a local tensor $T$. We discuss the two situations respectively, and examine the overlap of two different states $_A\overlap{\{t^\prime\}}{\{t\}}_{A}$ as a function of the two indices configurations $\{t^\prime\}$ and $\{t\}$.

(1) If the open virtual index $m$ in the ket layer (i.e. $\ket{\{t\}}_A$) connects to a projector $g$ tensor, then the open virtual index $m^\prime$ in the bra layer (i.e. $_A\bra{\{t^\prime\}}$), at the same place as the index $m$, also connects to a projector $g$ tensor. If we ``zoom in" on the local area of $_A\overlap{\{t^\prime\}}{\{t\}}_{A}$ near the index $m$ and $m^\prime$, we have the following diagram:
\begin{equation}
\begin{gathered}
\includegraphics[width=4cm]{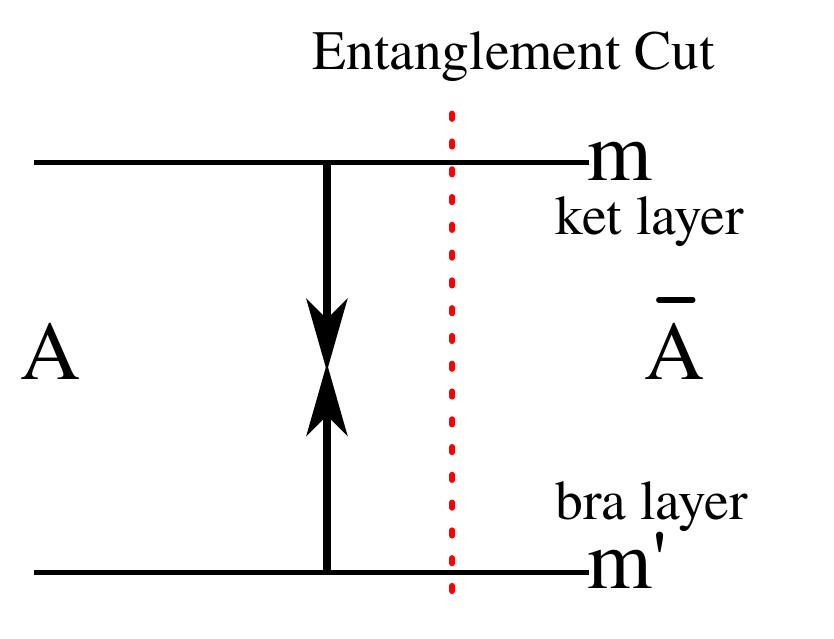}
\end{gathered}
\end{equation}
By using Eq.~\eqref{eq.projectorcontraction}, we can conclude that $m=m^\prime$, otherwise $_A\overlap{\{t^\prime\}}{\{t\}}_{A}=0$.

(2) If the open virtual index $m_0$ in the ket layer connects to a local $T$ tensor, we require by the \textit{SVD condition} that there are no other open virtual indices connecting to this $T$ tensor. Then the other indices of this $T$ tensor are all inside the region $A$. Similarly for the index $m^\prime_0$ in the bra layer. In terms of a diagram, $_A\overlap{\{t^\prime\}}{\{t\}}_{A}$ near the area of the index $m_0$ and $m^\prime_0$ can be represented as:
\begin{equation}
\begin{gathered}
\includegraphics[width=4cm]{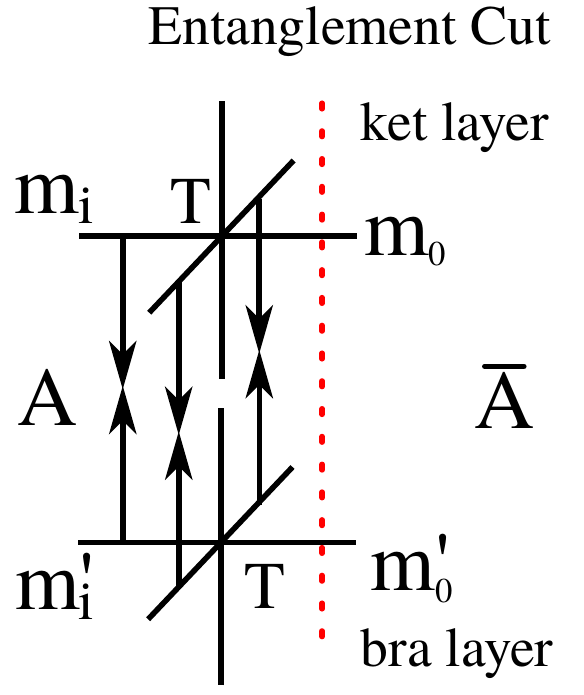}
\end{gathered}
\end{equation}
where $m_i$ and $m_i^\prime$ with $i=1,2,3\ldots$ denote the other virtual indices of the $T$ tensor in the bra and ket layer respectively, except $m_0$ and $m_0^\prime$.
Notice that in the ket layer, the  virtual indices $m_i\;(i=1,2,\ldots)$ of the $T$ tensor (all indices except the index $m_0$) are all connected with contracted projector $g$ tensors inside region $A$. Correspondingly, in the bra layer, the virtual indices $m^\prime_i\;(i=1,2,\ldots)$ are also all connected with the same contracted projector $g$ tensors. Hence, due to these projector $g$ tensors and Eq.~\eqref{eq.projectorcontraction}, all the indices except $m_0$ of the $T$ tensor in the ket layer are equal to their respective analogues in the bra layer:
\begin{equation}\label{eq.indexindentification_SVD}
m_i = m_i^\prime, \quad i=1,2,\ldots
\end{equation}
otherwise the overlap would be $_A\overlap{\{t^\prime\}}{\{t\}}_{A}=0$. The only remaining question is whether the open indices $m_0$ and $m_0^\prime$ should be identified in order to have a non-vanishing overlap $_A\overlap{\{t^\prime\}}{\{t\}}_{A}$.

Using the \textit{local $T$ tensor assumption:}, $m_i\;(i=1,2,\ldots)$ will uniquely determine $m_0$ in order to have nonzero element of the $T$ tensor in the ket layer. Similarly, $m_i^\prime\;(i=1,2,\ldots)$ will uniquely determine $m^\prime_0$ in order for the $T$ tensor in the bra layer to give a nonzero element. Therefore, Eq.~\eqref{eq.indexindentification_SVD} implies that:
\begin{equation}
m_0 = m_0^\prime
\end{equation}
such that the overlap $_A\overlap{\{t^\prime\}}{\{t\}}_{A}$ is nonzero.

Therefore, both situations (1) and (2) lead to the conclusion that the open indices $\{t\}$ and $\{t^\prime\}$ should be identical in order to have a nonzero overlap $_A\overlap{\{t^\prime\}}{\{t\}}_{A}$. The non-vanishing states $\ket{\{t\}}_{A}$ are orthogonal basis. A similar proof can be derived for the region $\bar{A}$. The orthogonality of each set $\ket{\{t\}}_{A}$ and $\ket{\{t\}}_{\bar{A}}$ implies that Eq.~\eqref{eq.TNSgeneralDecomposition} is indeed an SVD. However, the singular values are not clear at this stage since the basis may not be orthonormal (i.e., the states might not be normalized). 
\hfill$\Box$

In the following specific discussions of the 3D toric code model, the X-cube model and the Haah code, we will show that we can select a region $A$ and a cut on the TNS such that $\ket{\{t\}}_{A}$ and $\ket{\{t\}}_{\bar{A}}$ are not only orthogonal, but also normalized. In particular for the 3D toric code model and the X-cube model, we can just select the region $A$ to be a cube which satisfies the \textit{SVD condition} directly. See respectively Sec.~\ref{subsec.ToricCode_Entanglement} and \ref{subsec.Xcube_Entanglement} for detailed discussion of these two models and the SVD condition. However, the Haah code is different: a cubic region $A$ does not fulfill the \textit{SVD condition}, and in Sec.~\ref{subsec:HaahcodeSVDCuts} we generalize the \textit{SVD condition} to the \textit{Generalized SVD Condition} and apply $B_c$ operators to make the TNS an SVD. 

\subsection{Summary of the results}\label{subsec.overviewsummary}

We now summarize the major results derived in this paper for the three stabilizer codes. Fundamentally, our calculations come down to the fact that the indices of the nonzero elements of the local tensor $T$ and $g$ are constrained. More specifically, when we calculate the entanglement entropies with a TNS which is an exact SVD, the only task is to count the number of independent Schmidt states $\ket{\{t\}}_{A}$. The number of independent Schmidt states $\ket{\{t\}}_{A}$ is determined by the \textbf{Concatenation lemma},
i.e., when a network of $T$ tensors and $g$ tensors are concatenated, the open indices of the nonzero elements of the resulting tensors are constrained as well.
\begin{enumerate}
\item The TNS is the exact SVD for the ground states with respect to particular entanglement cuts. The entanglement spectra are flat for models studied in this paper.

\item The entanglement of TNS is bounded by the area law: 
$$S \le \mathrm{Area} \times \log(D),$$ 
where $D$ is the virtual index dimension and $\mathrm{Area}$ is measured in the units of vertices. For the models studied in this paper, the entanglement entropies are strictly smaller than the area law when one is computing in terms of vertices.
For the toric code, the correction is a negative constant, $-\log(2)$. For the X-cube model and Haah code, the correction includes a negative term linear with the system size, presented in Sec.~\ref{subsec.Xcube_Entanglement} and \ref{subsec.Haah_Entanglement}. 

\item The transfer matrices in Eq.~\eqref{eq.TransferMatrix} of the 3D toric code model and the X-cube model are shown to be a projector whose eigenvalues are either $0$ or $1$. For the 3D toric code, the transfer matrix in the $xy$-plane is a projector of rank 2. For the X-cube model, the transfer matrix is a projector of rank $2^{L_x+L_y-1}$ where $L_x$ and $L_y$ are the lattice sizes in $x$- and $y$- directions respectively.

\item We prove that the TNS ground states obtained on the torus using our construction are the $+1$ eigenstates of loop $X$ operators. Hence, our TNS construction does not include all ground states on the torus. The degeneracy of the corresponding transfer matrix is smaller than the GSD on the torus.  We can obtain all the ground states with loop/surface $Z$ operators on the TNS, which generate all the wave functions on the torus. We call the TNS with $Z$ operators, ``twisted TNS". Correspondingly, we also obtain more transfer matrices in the $xy$-plane built from the twisted TNS, and these transfer matrices are all the same projectors.
The same TNS phenomenon in the 2D toric code model has been studied in Ref.~\onlinecite{schuch2013topological}.

\item In our calculations, both the transfer matrix eigenvalue degeneracies, and the corrections to the area law of entanglement entropies are rooted in the \textbf{Concatenation lemma}. Hence, we believe that the two contributions are  related. Specifically, suppose we consider our TNS on a 3D cylinder $\mathcal{T}^2_{xy} \times \mathcal{R}_z$, and the entanglement cut splits the system in two halves $z>0$ and $z<0$. Then, for the toric code model, the transfer matrix $\mathrm{TM}_{xy}$ has the degeneracy 2, and the entanglement entropy correction to the area law is $-\log(2)$. For the X-cube model, the transfer matrix $\mathrm{TM}_{xy}$ has the degeneracy $2^{L_x+L_y-1}$, and the entanglement entropy correction to the area law is $-(L_x+L_y-1)\log(2)$ (See Eq.~\eqref{eq.XcubeCylinderEntropy}). Moreover, the GSD on $\mathcal{T}^3$ is generally larger than the transfer matrix degeneracy. Therefore, given these calculations, we conjecture that the negative linear correction to the area law is a signature of the extensive ground state degeneracy.
\end{enumerate}

\section{3D Toric Code}\label{sec.toriccode}

In this section, we construct the TNS for the 3D toric code model and then calculate the entanglement entropy and GSD, both deriving from the \textbf{Concatenation lemma}. The results are the immediate generalizations of those in the 2D toric code model. We find a topological entanglement entropy in accordance to  that obtained by Ref.~\onlinecite{zheng2017structure} using a field theoretic approach. 

This section is organized as follows:
In Sec.~\ref{subsec.toriccode}, we briefly review the toric code model in a cubic lattice. 
In Sec.~\ref{subsec.ToricCode_TNS}, we construct the TNS for the toric code model.
In Sec.~\ref{subsec.ConcatenationToricCode}, we prove a \textbf{Concatenation lemma} for toric code TNS, which is useful in the following calculations.
In Sec.~\ref{subsec.ToricCode_Entanglement}, we calculate the entanglement entropies on $\mathcal{R}^3$.
In Sec.~\ref{subsec.ToricCode_TransferMatrix}, we construct the transfer matrix and prove it is a projector of rank 2.
In Sec.~\ref{subsec.ToricCode_GSD}, we show how to construct $8$ ground states on torus by twisting the TNS.

\subsection{Hamiltonian of 3D Toric Code Model}\label{subsec.toriccode}

\begin{figure}[t]
\centering
\includegraphics[width=0.6\columnwidth]{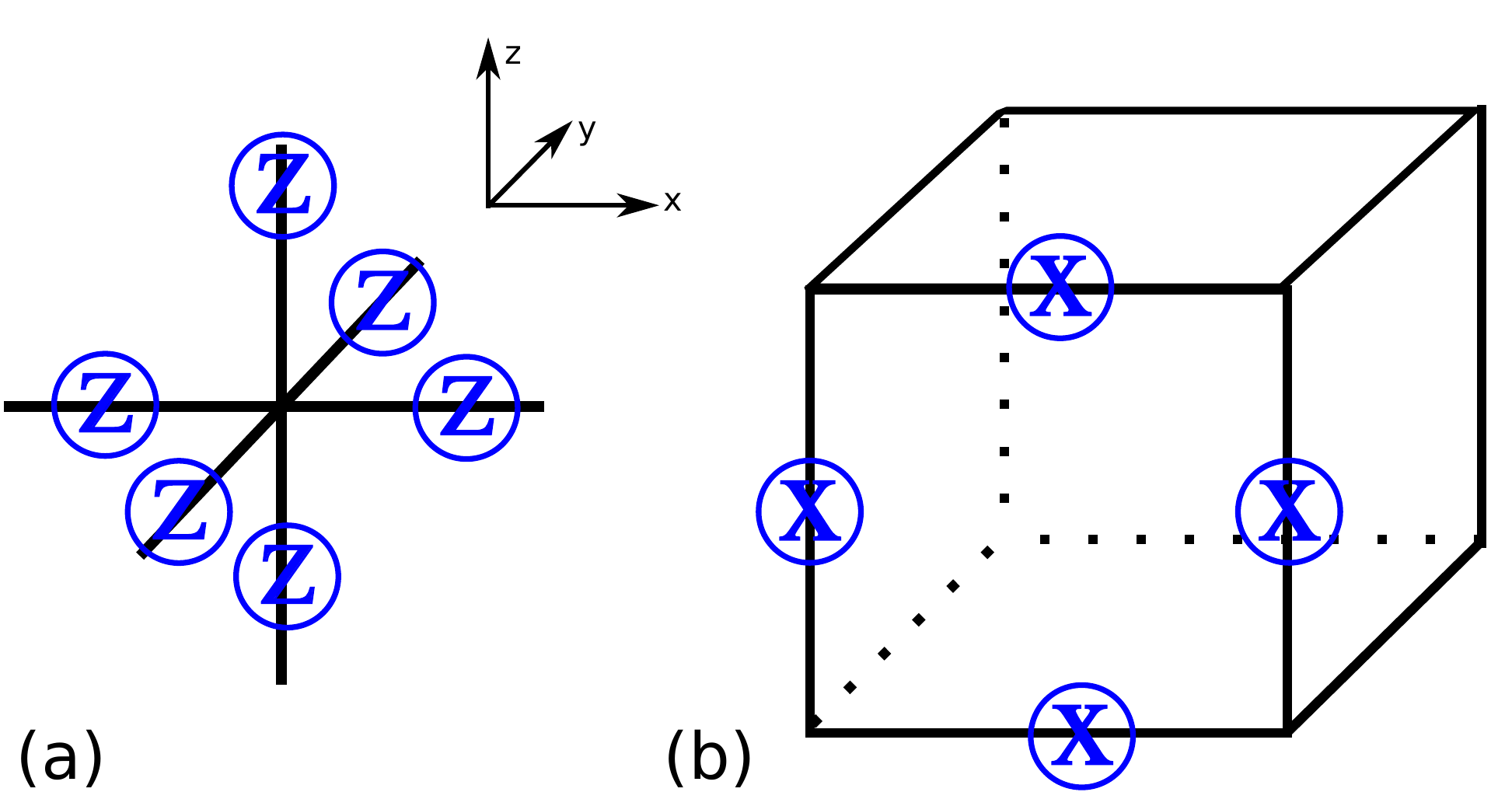}
\caption{The Hamiltonian terms of the 3D toric code model. Panel (a) is $A_{v}$ which is a product of 6 $Z$ operators, and Panel (b) is $B_{p}$ which is a product of 4 $X$ operators. The circled $X$ and $Z$ represent the Pauli matrices acting on the spin-$1/2$'s. The toric code Hamiltonian includes $A_v$ terms on all vertices $v$ and $B_p$ terms on all plaquettes $p$.}
\label{fig.ToricCodeHamiltonian}
\end{figure}

The 3D toric code model can be defined on any random lattice. However, for simplicity, we only work on the cubic lattice. On a cubic lattice, the physical spins are defined on the bonds of the lattice, and the Hamiltonian is built from two types of terms:
\begin{equation}
H = -\sum_{v} A_v - \sum_{p} B_p.
\end{equation}
where $A_v$ is defined around a vertex $v$, and $B_p$ is defined on a plaquette $p$:
\begin{equation}
A_v = \prod_{i \in v} Z_i, \quad
B_p = \prod_{i \in p} X_i,
\end{equation}
where $Z_i$ and $X_i$ are Pauli matrices for the $i$-th spin. On a cubic lattice, $A_v$ is composed of $6$ Pauli $Z$ operators while $B_p$ is composed of $4$ Pauli $X$ operators. These two terms are depicted in Fig.~\ref{fig.ToricCodeHamiltonian}. In the 2D toric code, $A_v$ is composed of $4$ Pauli $Z$ operators on a square lattice. The Hamiltonian is the sum of  $A_v$ operators on all vertices $v$ and $B_p$ operators on all plaquettes $p$.

It is easy to verify that all the Hamiltonian terms commute:
\begin{equation}
\begin{split}
\commute{A_{v}}{A_{v^\prime}} &= 0, \quad \forall\; v,v^\prime	\\
\commute{B_{p}}{B_{p^\prime}} &= 0, \quad \forall\; p,p^\prime	\\
\commute{A_{v}}{B_{p}} &= 0, \quad \forall\; v,p,	\\
\end{split}
\end{equation}
and their eigenvalues are $\pm 1$:
\begin{equation}
A_{v}^2 = 1,
\quad
B_{p}^2 = 1.
\end{equation}
The ground states $\ket{\mathrm{GS}}$ should satisfy:
\begin{equation}\label{eq.ToricCodeGS}
\begin{split}
A_v \ket{\mathrm{GS}} &= \ket{\mathrm{GS}},	\quad \forall\; v	\\
B_p \ket{\mathrm{GS}} &= \ket{\mathrm{GS}}.	\quad \forall\; p
\end{split}
\end{equation}
These two sets of equations are enough to derive the local $T$ tensor and to construct TNS for the toric code model. In particular, one of the ground states on the torus that we will find is 
\begin{eqnarray}\label{GSoftoriccode}
|\psi\rangle= \prod_{v} \frac{1+A_v}{2} |0_x\rangle,
\end{eqnarray}
where $|0_x\rangle$ is the tensor product of all $X=1$ eigenstates defined on each link. See App.~\ref{app.TNS_projected} for a proof that Eq.~\eqref{GSoftoriccode} is indeed the TNS that we will now construct.

\subsection{TNS for 3D Toric Code}\label{subsec.ToricCode_TNS}

We first introduce a projector $g$ tensor Eq.~\eqref{eq.projector2} on each bond of the lattice. Both the virtual indices and the physical indices take two values, $0$ and $1$.
The projector $g$ tensor satisfies:
\begin{equation}\label{eq.projectorcondition ToricCode}
\begin{gathered}
\includegraphics[width=\columnwidth]{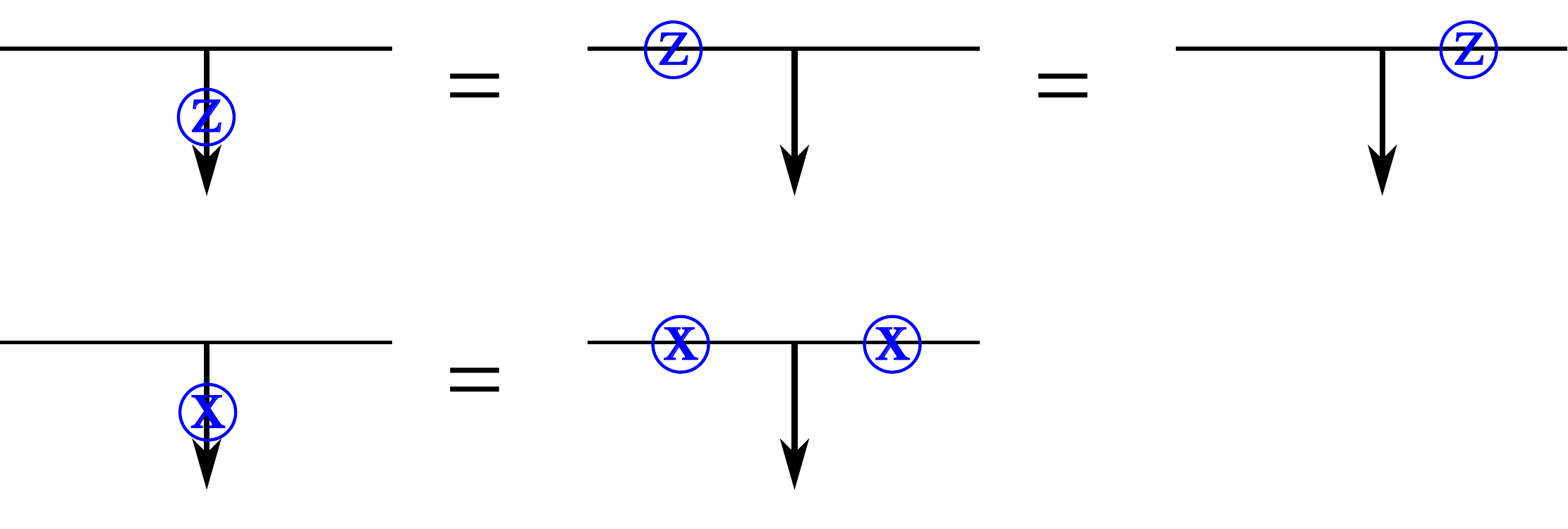}
\end{gathered}.
\end{equation}
In terms of algebraic equations, these diagrams correspond to:
\begin{equation}
\begin{split}
&g^{s}_{i,j}(-1)^{s}=g^{s}_{i,j}(-1)^{i}=g^{s}_{i,j}(-1)^{j}	\\
&g^{1-s}_{i,j}=g^{s}_{1-i,1-j}.	\\
\end{split}
\end{equation}
These two sets of equations are true, because (1) the indices $s$, $i$ and $j$ are identified for nonzero $g^{s}_{i,j}$, (2) the nonzero $g^{s}_{i,j}$ are always $1$ according to Eq.~\eqref{eq.projector2}. We can use these conditions to transfer the action of the physical operators to the virtual operators. 
Now we introduce an additional $T$ tensor on each vertex of the cubic lattice, which has six virtual indices. Graphically, we represent such a $T$ tensor as:
\begin{equation}
\begin{gathered}
\includegraphics[width=2cm]{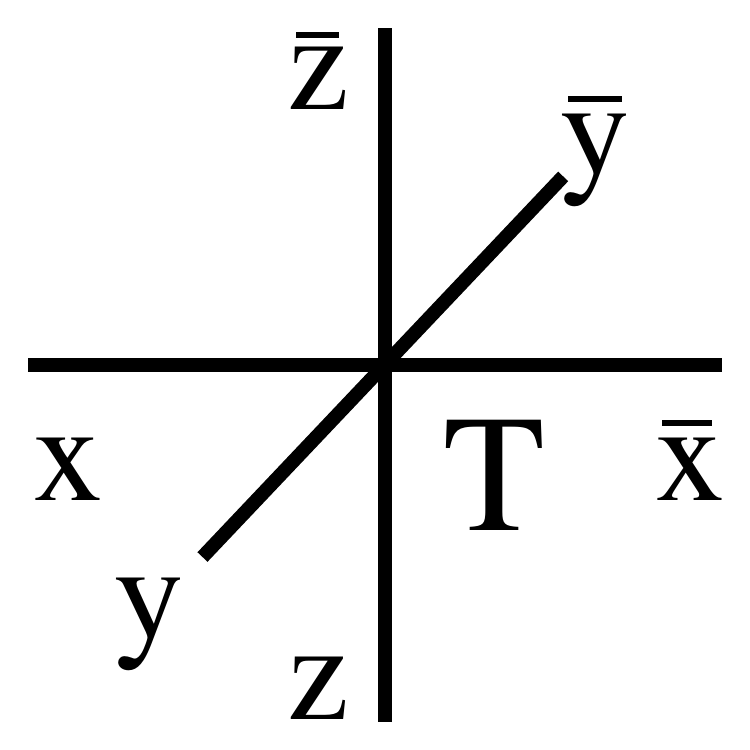}
\end{gathered}.
\end{equation}
Next we need to fix the elements of the $T$ tensor, up to the TNS gauge freedom. The method to fix the $T$ tensor is to make it invariant under the actions of $A_{v}$ and $B_{p}$ operators, in order to implement the local conditions for the ground states in Eq.~\eqref{eq.ToricCodeGS}. The actions of $A_v$ and $B_p$ operators on the local tensors are:
\begin{equation}\label{eq.3DToricCode_DericeT}
\begin{gathered}
\includegraphics[width=0.8\columnwidth]{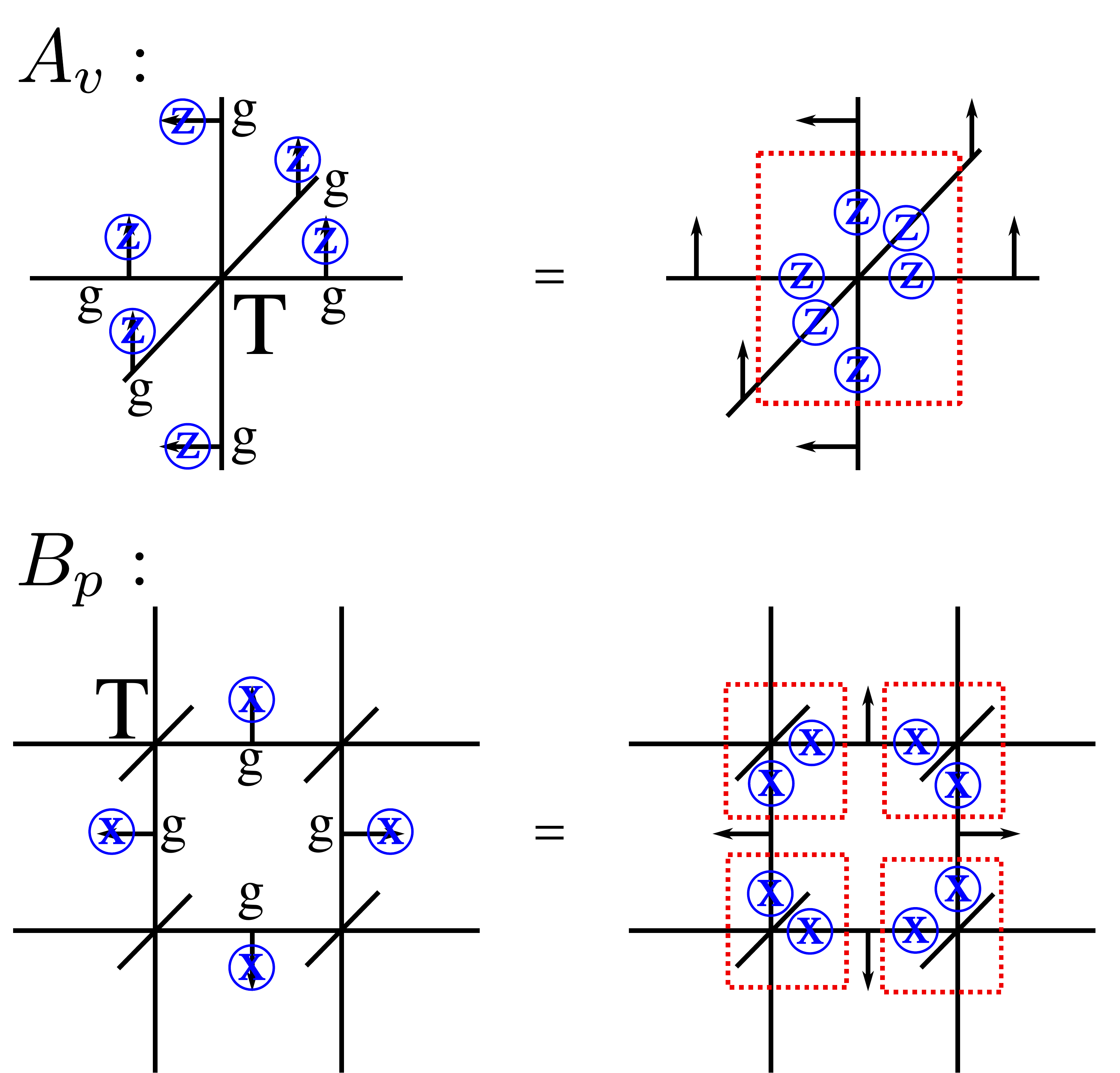}
\end{gathered}.
\end{equation}
where we have used Eq.~\eqref{eq.projectorcondition ToricCode} to transfer the physical operators to the virtual ones. We require a strong version of the solution to the above equations. We want the tensors in the dashed red rectangles  to be invariant under the actions of any of the $A_v$ and $B_p$ (this is a sufficient constraint which guarantees that the tensors form the ground state), which leads to the following equations:
\begin{equation}\label{eq.ToricCodeTcondition}
\begin{gathered}
\includegraphics[width=0.9\columnwidth]{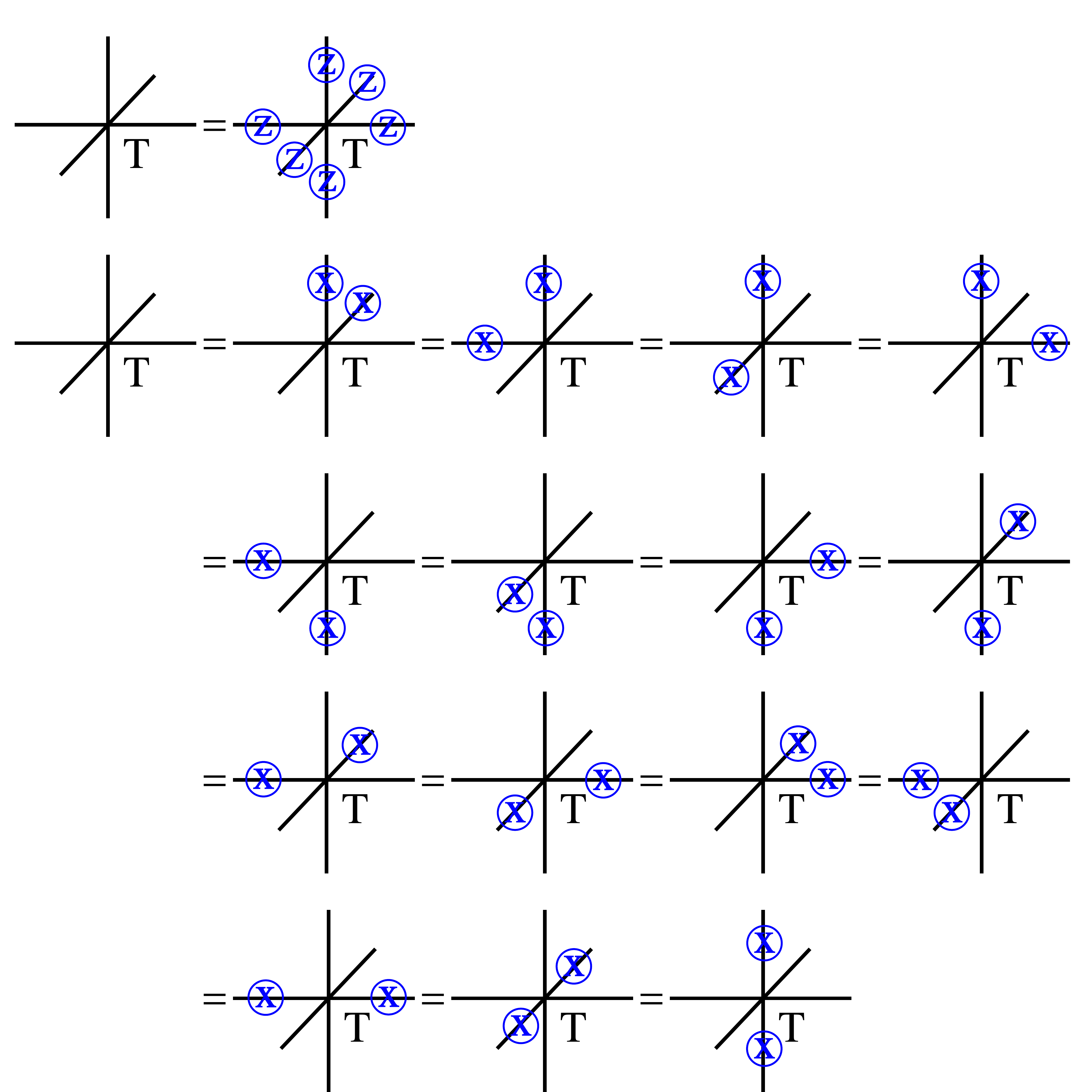}
\end{gathered}
\end{equation}
In the second set of equations, the first 12 equalities are obvious from the red dashed squares, and the last 3 equalities can be derived from the first 12 ones. 
Expanding the first set of conditions by using $Z_{ij}=\delta_{ij} (-1)^{i}$, we have:
\begin{equation}\label{eq.3Dtoriccode.symmetric}
\begin{split}
&T_{x\bar{x},y\bar{y},z\bar{z}} = (-1)^{x+\bar{x}+y+\bar{y}+z+\bar{z}} T_{x\bar{x},y\bar{y},z\bar{z}} \\
&\Leftrightarrow	\\
&T_{x\bar{x},y\bar{y},z\bar{z}} 
\begin{cases}
=0, &\;\text{if}\;\; x+\bar{x}+y+\bar{y}+z+\bar{z} = 1 \mod{2}	\\
\neq 0, &\;\text{if}\;\; x+\bar{x}+y+\bar{y}+z+\bar{z} = 0 \mod{2},	\\
\end{cases}
\end{split}
\end{equation}
where $x,\bar{x},y,\bar{y},z,\bar{z}$ are the six indices of $T$ in the three directions respectively. 
We emphasize for clarity that $\bar{x}$ is not $-x$; these are notations for different indices. The second set of conditions in Eq.~\eqref{eq.ToricCodeTcondition} further enforces that an even number of index flipping of the virtual indices of a tensor does not change the value of the tensor elements. For instance, in terms of components, we have:
\begin{equation}
\begin{split}
T_{x\bar{x},y\bar{y},z\bar{z}} 
=& T_{(1-x)(1-\bar{x}),y\bar{y},z\bar{z}}	\\
=& T_{(1-x)\bar{x},(1-y)\bar{y},z\bar{z}}	\\
=& T_{x\bar{x},y\bar{y},(1-z)(1-\bar{z})}	\\
=& \ldots
\end{split}.
\end{equation} 
Hence, the nonzero elements of the $T$ tensor are all equal. Up to an overall normalization, we have the unique solution:
\begin{equation}\label{eq.ToricCodeTtensor}
T_{x\bar{x},y\bar{y},z\bar{z}} =
\begin{cases}
0, &\;\text{if}\; x+\bar{x}+y+\bar{y}+z+\bar{z} = 1 \mod{2}	\\
1, &\;\text{if}\; x+\bar{x}+y+\bar{y}+z+\bar{z} = 0 \mod{2}.	\\
\end{cases}
\end{equation}
The TNS is then Eq.~\eqref{eq.TNS} with the local $T$ being Eq.~\eqref{eq.ToricCodeTtensor}. The local $T$ tensors are the same on other spatial manifolds, such as $\mathcal{T}^3$. 

A similar set of conditions as the first equality in Eq.~\eqref{eq.ToricCodeTcondition} have been introduced by several other names in tensor network literature: $\mathbb{Z}_2$-injectivity\cite{schuch2010peps}, MPO-injectivity\cite{csahinouglu2014characterizing}, $\mathbb{Z}_2$ gauge symmetry\cite{he2014modular} etc. The previous studies were in 2D, and our condition is the 3D generalization. Notice that the first equation in Eq.~\eqref{eq.ToricCodeTcondition} alone will not necessarily lead to topological order. It only implies that the ground state is $\mathbb{Z}_2$ symmetric. The state which only satisfies the first condition in  Eq.~\eqref{eq.ToricCodeTcondition} could also be a topological trivial state by tuning the relative strength of the nonzero elements of $T$ tensor.  This can be interpreted as a condensation transition from topological phases to trivial phases. See Refs.~\onlinecite{he2014modular,shukla2016boson,marien2016condensation,duivenvoorden2017entanglement,garre2017symmetry} for explanations and examples in the case of 2D TNS. 

\subsection{Concatenation Lemma}\label{subsec.ConcatenationToricCode}

In this section, we consider the contraction of a network of local $T$ tensors with open virtual indices. One example of such a contraction is the tensor network norm Eq.~\eqref{eq.TNSnormContractT} or the transfer matrix Eq.~\eqref{eq.TransferMatrixContractT}. Since the elements of a local $T$ tensor are 0 for the odd sector and 1 for the even sector (see Eq.~\eqref{eq.ToricCodeTtensor}), we will show that, in general, a network of contracted $T$ tensors obeys a similar rule: some elements are zeros while the others are nonzero and identical. A \textbf{Concatenation lemma} is proposed to derive the rule for the contraction of several tensors in general and will be frequently used in the following discussions. For example, we will use this lemma to show in Sec.~\ref{subsec.ToricCode_TransferMatrix} that the transfer matrix $\mathrm{TM}_{xy}$ for the 3D toric code model
is a projector of rank 2. 

\begin{figure}[t]
\centering
\includegraphics[width=0.4\columnwidth]{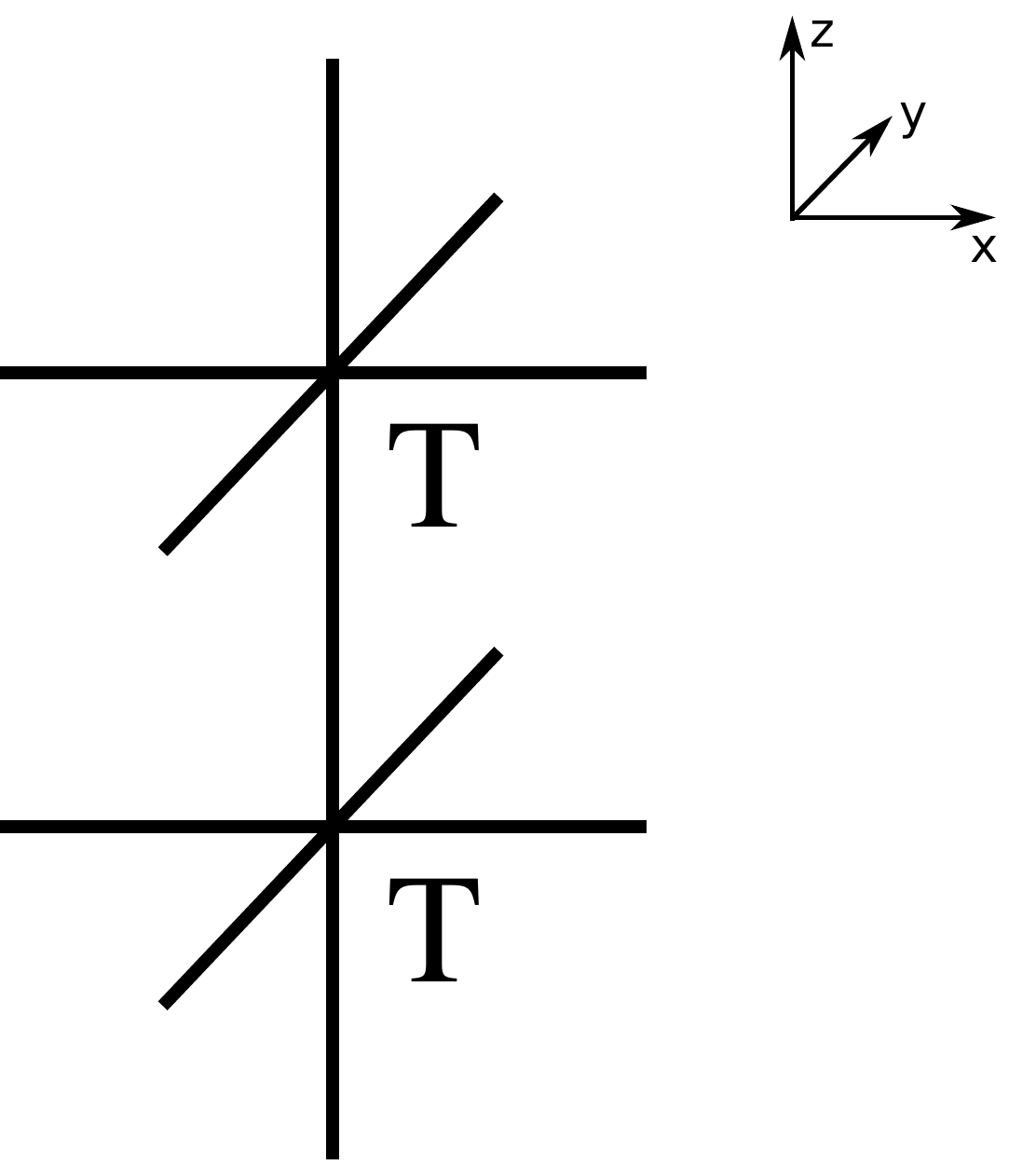}
\caption{Contraction of two local $T$ tensors in the $z$-direction. We emphasize that there is no projector $g$ tensor in this figure.}
\label{fig.Contract2T}
\end{figure}

\begin{framed}
\textbf{Concatenation Lemma:} For a network of contracted $T$ tensors Eq.~\eqref{eq.ToricCodeTtensor} with open indices, the open indices need to sum to $0 \mod{2}$, otherwise the element of the network tensor is zero. Moreover, if nonzero, the elements of the network tensor are constants, independent of open indices.
\end{framed}

This lemma can be easily proved by using $\mathbb{Z}_2$ symmetry Eq.~\eqref{eq.ToricCodeTtensor} and induction. The proof is in App.~\ref{app.ToricCode_Concatenation}. 
We explain this lemma by a simple example. Suppose we have two $T$ tensors contracted over a pair of indices:
\begin{equation}
\begin{split}
&\mathbf{T}_{x_1,\bar{x}_1,y_1,\bar{y}_1,z_1,x_2,\bar{x}_2,y_2,\bar{y}_2,\bar{z}_2}	\\
=&\sum_{\bar{z}_1,z_2} T_{x_1\bar{x}_1,y_1\bar{y}_1,z_1\bar{z}_1} T_{x_2\bar{x}_2,y_2\bar{y}_2,z_2\bar{z}_2} \delta_{\bar{z}_1z_2}.	\\
\end{split}
\end{equation}
Graphically, the tensor $\mathbf{T}$ is represented by Fig.~\ref{fig.Contract2T}. The open indices of the tensor $\mathbf{T}$ need to sum to an even number in order for the elements of the $\mathbf{T}$ tensor to be nonzero. This comes out of writing the constraints of each of the $T$ tensors:
\begin{equation}
\begin{split}
&\begin{cases}
&x_1+\bar{x}_1+y_1+\bar{y}_1+z_1+\bar{z}_1 = 0, \mod{2}	\\	
&x_2+\bar{x}_2+y_2+\bar{y}_2+z_2+\bar{z}_2 = 0, \mod{2}	\\
&\bar{z}_1=z_2
\end{cases}
\\
\Rightarrow\;
&x_1+\bar{x}_1+y_1+\bar{y}_1+z_1+ x_2+\bar{x}_2+y_2+\bar{y}_2+\bar{z}_2	\\
=& 0, \mod{2}.
\end{split}
\end{equation}
Otherwise, the tensor element of $\mathbf{T}$ is zero. Moreover, the elements of the contracted tensor are 1, if nonzero:
\begin{widetext}
\begin{equation}
\mathbf{T}_{x_1,\bar{x}_1,y_1,\bar{y}_1,z_1,x_2,\bar{x}_2,y_2,\bar{y}_2,\bar{z}_2}
= \begin{cases}
0	&	\text{if}\;\; x_1+\bar{x}_1+y_1+\bar{y}_1+z_1+ x_2+\bar{x}_2+y_2+\bar{y}_2+\bar{z}_2 = 1, \mod{2}	\\
1	&	\text{if}\;\; x_1+\bar{x}_1+y_1+\bar{y}_1+z_1+ x_2+\bar{x}_2+y_2+\bar{y}_2+\bar{z}_2 = 0, \mod{2}.
\end{cases}
\end{equation}
\end{widetext}
For a more complicated contraction of $T$ tensors, we have:
\begin{equation}
\mathbf{T}_{\{t\}} = \begin{cases}
0	&	\text{if}\;\; \sum_i t_i = 1, \mod{2}	\\
\mathrm{Const}	&	\text{if}\;\; \sum_i t_i = 0, \mod{2}	\\
\end{cases}
\end{equation}
where $\{t\}$ denotes all the indices of the tensor $\mathbf{T}$. We emphasize that the nonzero constant does not depend on $\{t\}$ .

\subsection{Entanglement}\label{subsec.ToricCode_Entanglement}

\begin{figure}[b]
\centering
\includegraphics[width=0.6\columnwidth]{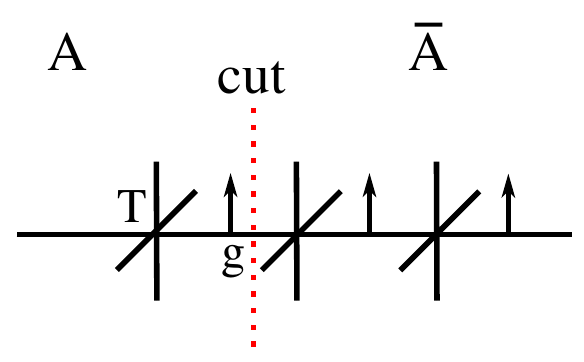}
\caption{The splitting of tensors near the entanglement cut.}
\label{fig.cut}
\end{figure}

We now show that Eq.~\eqref{eq.TNS} is exactly an SVD for the wave function with respect to the entanglement cut illustrated in Fig.~\ref{fig.cut}.
For simplicity, suppose that the TNS is defined on infinite $\mathcal{R}^3$. As we have emphasized at the end of Sec.~\ref{subsec.TNSingeneral}, we do not specify the boundary conditions of the TNS, since we are only concerned with the bulk wave functions whose reduced density matrices are assumed not to be influenced by the boundary conditions. If we put the wave function on a large but finite $\mathcal{R}^3$, we have to specify the boundary conditions of the TNS by fixing the indices on the boundary. Suppose the open indices on the boundary are denoted as $\{t^b\}$. The norm of the TNS on open $\mathcal{R}^3$, which can be expressed as a network of contracted $T$ tensors with open virtual indices $\{t^b\}$, is zero when $\sum_i t^b_i=1 \mod{2}$ and nonzero when $\sum_i t^b_i=0 \mod{2}$, according to the \textbf{Concatenation lemma} of the 3D toric code model. Hence, we can only fix the boundary indices $\{t^b\}$ to be $\sum_{i} t^b_i = 0 \mod{2}$.
Calculating the entanglement on a nontrivial manifold is ambiguous since  multiple degenerate ground states, which cannot be distinguished locally, appear. Their superpositions have different entanglement entropies.

We rewrite Eq.~\eqref{eq.TNS} by separating the tensor contractions to a spatial region $A$ and its complement region $\bar{A}$. Region $A$ contains the $g$ tensors near the entanglement cut as illustrated in Fig.~\ref{fig.cut}:
\begin{equation}\label{eq.ToricCodeSVD}
\begin{split}
\ket{\mathrm{TNS}}_{\mathcal{R}^3} = \sum_{\{t\}} \ket{\{t\}}_{A} \otimes \ket{\{t\}}_{\bar{A}}
\end{split}
\end{equation}
where 
\begin{widetext}
\begin{equation}\label{eq.SVDbasisA}
\begin{split}
|\{t\}\rangle_A
=\sum_{\{s\}\in A}\sum_{\{i\} \in A} \mathcal{C}^{A} (g^{s_1}_{t_1 i_1} g^{s_2}_{t_2 i_2}\ldots
g^{s_3}_{i_3i_4}g^{s_4}_{i_5i_6} T_{i_7\ldots}T_{i_8\ldots }\ldots )|\{s\}\rangle.
\end{split}
\end{equation}
\end{widetext}
Indices denoted by $s$ are the physical indices; indices denoted by $t$ are the open virtual indices straddling the entanglement cut from the region $A$; indices denoted by $i$ are the contracted virtual indices inside the region $A$. 
The tensors  $g^{s_1}_{t_1i_1}$ and $g^{s_2}_{t_1i_2}$ etc are the projector $g$ tensors near the entanglement cut on the region $A$ side as illustrated in Fig.~\ref{fig.cut}; $g^{s_3}_{i_3i_4}$ and $g^{s_4}_{i_5i_6} $ are the projector $g$ tensors inside the region $A$; for this cut, all the $T$ tensors are inside the region $A$. The summation is over all physical indices $\{s\}$ inside the region A.

Thereby, $\ket{\{t\}}$ is the TNS for region $A$ with open virtual indices $\{t\}$. We choose a convention of splitting tensors whereby $g$ tensors near the entanglement cut belong to the region $A$, as illustrated in Fig.~\ref{fig.cut}. For instance, when the region $A$ is a cube, we can graphically denote the basis $\ket{\{t\}}$ as Eq.~\eqref{eq.TNSbasis},
where in the bulk of this cube is a TNS, and the red lines are the outgoing virtual indices $\{t\}$. 
The $g$ tensors connecting with these red lines are inside the cube.
Similarly for the region $\bar{A}$:
\begin{equation}
\begin{split}
|\{t\}\rangle_{\bar{A}}=\sum_{\{s\}\in \bar{A}}\sum_{\{i\} \in \bar{A}}\mathcal{C}^{\bar{A}}( g^{s_1}_{i_1i_2}g^{s_2}_{i_3i_4} T_{t_1i_5\ldots}T_{t_2i_6\ldots }\ldots )|\{s\}\rangle.
\end{split}
\end{equation}
Since the TNSs for region $A$ and $\bar{A}$ share the same boundary virtual indices $\{t\}$, then in Eq.~\eqref{eq.ToricCodeSVD} the two basis for region $A$ and $\bar{A}$ have the same label $\{t\}$. For the TNS of Eq.~\eqref{eq.TNS}, the boundary virtual indices $\{t\}$ of the regions $A$ and $\bar{A}$ are contracted over, and thus in Eq.~\eqref{eq.ToricCodeSVD} $\{t\}$ are summed over.

We now show that $\ket{\{t\}}_{A}$ and $\ket{\{t\}}_{\bar{A}}$ are an orthonormal basis (normalized up to constant)
for the region $A$ and the region $\bar{A}$ respectively. Therefore, Eq.~\eqref{eq.ToricCodeSVD} is exactly the SVD for the ground state wave function, i.e.,
\begin{equation}\label{eq.ToricCodeOrthoNormal}
_A\overlap{\{t^\prime\}}{\{t\}}_A \propto \delta_{\{t^\prime\},\{t\}} \delta(\sum_i t_i =0 \mod{2}).
\end{equation}

\noindent \textbf{Proof:}

Applying the \textit{SVD condition} to the toric code TNS, we can immediately conclude that the $\ket{\{t\}}_{A}$ span an orthogonal basis, and the TNS is exactly an SVD. However, the \textit{SVD condition} does not tell us whether the basis is orthonormal. In the following, we show that $\ket{\{t\}}_{A}$ is not only orthogonal, but also orthonormal with a norm independent on ${t}$,
which leads to the flat singular values.
Following the definition of our basis:
\begin{widetext}
\begin{equation}
\begin{split}
_A\overlap{\{t^\prime\}}{\{t\}}_{A}=	
& \left(\sum_{\{s'\}\in A}\sum_{\{j\} \in A}\mathcal{C}^{A}(g^{s'_1\star}_{t'_1 j_1} g^{s'_2\star}_{t'_2 j_2}\ldots g^{s'_3\star}_{j_3j_4}g^{s'_4\star}_{j_5j_6} T^{\star}_{j_7\ldots}T^{\star}_{j_8\ldots }\ldots )\langle \{s'\}| \right) \\
&\left( \sum_{\{s\}\in A}\sum_{\{i\} \in A}\mathcal{C}^{A}(g^{s_1}_{t_1 i_1} g^{s_2}_{t_2 i_2}\ldots g^{s_3}_{i_3i_4}g^{s_4}_{i_5i_6} T_{i_7\ldots}T_{i_8\ldots }\ldots )|\{s\}\rangle \right).
\end{split}
\end{equation}
\end{widetext}
When the open virtual indices $\{t^\prime\} \neq \{t\}$, the overlap is clearly zero, as the spin configurations on the boundary are different due to the projector $g$ tensors. Hence, the basis $\ket{\{t\}}_{A}$ are orthogonal. 

Next we show that $_A\overlap{\{t\}}{\{t\}}_{A}$ is zero when $\left( \sum_{t_i \in \{t\}} t_i \right)$ is odd. Following the same derivations in Sec.~\ref{subsec.TNSnorm}, we have:
\begin{equation}
_A\overlap{\{t\}}{\{t\}}_{A} = \mathcal{C}^{A} \left( \ldots TTT \ldots \right)
\end{equation}
with the open virtual indices $\{t\}$. The contraction $\mathcal{C}^{A}$ is over the $T$ tensors in the region $A$. Applying the \textbf{Concatenation lemma}, $_A\overlap{\{t\}}{\{t\}}_{A}$ is zero if the open indices $\{t\}$ are summed to be $1\mod{2}$:
\begin{equation}
\sum_i t_i = 1 \mod{2} \;\Rightarrow\; _A\overlap{\{t\}}{\{t\}}_{A}=0.
\end{equation}
Moreover,
\begin{equation}
_A\overlap{\{t\}}{\{t\}}_{A}=\mathrm{Const},	\quad\text{when}\; \sum_i t_i = 0 \mod{2}.
\end{equation}
Hence $\ket{\{t\}}$ is orthonormal basis up to an overall normalization factor that can be obtained by the normalization of $\ket{\mathrm{TNS}}$.
\hfill$\Box$\\

The same proof works for the region $\bar{A}$ and $\ket{\{t\}}_{\bar{A}}$. Therefore, we can conclude that Eq.~\eqref{eq.ToricCodeSVD} is indeed an SVD, and the singular values are all identical. Hence, for a entanglement cut, we only need to count the number of singular vectors in Eq.~\eqref{eq.ToricCodeSVD}. For a connected entanglement surface with $N$ open virtual indices, the number of singular vectors in Eq.~\eqref{eq.ToricCodeSVD} is $2^{N-1}$, because the open virtual indices need to sum to be $0\mod{2}$. Hence, the entanglement entropy for a region whose entanglement surface is singly connected is:
\begin{equation}
S = N\log(2)-\log(2).
\end{equation}
If the entanglement surface still has $N$ open virtual indices but is separated into $n$ disconnected surfaces, then the entanglement entropy is:
\begin{equation}
\begin{split}
S 
&= N\log(2) - n\log(2)	\\
&= \mathrm{Area}\times \log(2) - n\log(2).
\end{split}
\end{equation}
The above is true because the condition that the open indices need to have an even summation holds true for each component of the entanglement cut. Furthermore, if we place our TNS ground state  on a 3D cylinder $\mathcal{T}^2_{xy} \times \mathcal{R}_z$, and the entanglement cut splits the cylinder into two halves $z>0$ and $z<0$, then the entanglement entropy of either side is also 
$S = \text{Area}\times \log(2) - \log(2)$.  The results can be easily generalized to $\mathbb{Z}_K$ lattice gauge models on $\mathcal{R}^3$:
\begin{equation}
S = \mathrm{Area} \times \log(K) - n\log(K)
\end{equation} 
with the same equation holding on a cylinder $\mathcal{T}^2_{xy} \times \mathcal{R}_z$. The entanglement spectrum is also flat.
The area is measured by the number of open virtual indices straddling the entanglement cut. 

Following the same logic, for the toric code in $(d+1)$ dimensions, all the open virtual indices of region $A$, $\{t_i\}$,  have to satisfy a single constraint $\sum_i t_i=0\mod 2$, because they have to obey the \textbf{Concatenation lemma}. If there are $N$ open virtual indices on the surface of region $A$, there are $N-1$ independent open virtual indices.  Hence the rank of the reduced density matrix is still $2^{N-1}$, because each independent open index can take 2 values. The entanglement entropy is 
\begin{equation}
S=N\log(2)-\log(2).
\end{equation}
The topological entanglement entropy $S_{\mathrm{topo}}[\mathcal{T}^{d-1}]$ is independent of the dimensionality, and it obeys the conjecture presented in Ref.~\onlinecite{zheng2017structure}:
\begin{equation}
\exp(-d S_{\mathrm{topo}}[\mathcal{T}^{d-1}])=\mathrm{GSD}[\mathcal{T}^d]
\end{equation}
where $\mathrm{GSD}[\mathcal{T}^d]=2^{d}$.

\subsection{Transfer Matrix as a Projector}\label{subsec.ToricCode_TransferMatrix}

The $z$-direction transfer matrix $\mathrm{TM}_{xy}$ in 3D is defined as a tensor network overlap in the $xy$-plane, with periodic boundary conditions. The indices in the $z$ direction are open and not contracted over (see Eq.~\eqref{eq.TransferMatrix} to Eq.~\eqref{eq.TransferMatrixGraph}). In this section, we will show that $\mathrm{TM}_{xy}$ for the 3D toric code model is a projector of rank 2. Let us denote the indices of the transfer matrix as 
\begin{equation}\label{eq.TransferMatrixIndex}
\begin{split}
&\left(\mathrm{TM}_{xy}\right)_{\{z\},\{\bar{z}\}}= \begin{gathered}
\includegraphics[width=0.5\columnwidth]{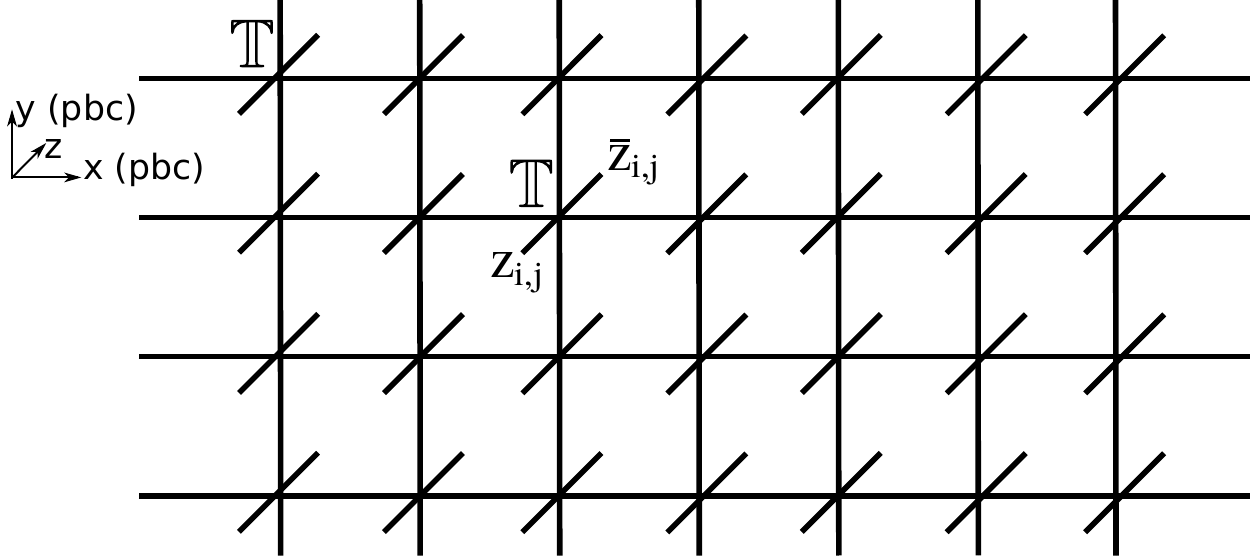}
\end{gathered},
\end{split}
\end{equation}
where $z_{i,j}$ and $\bar{z}_{i,j}$ are the indices at the position $(i,j)$ on the $xy$-plane.
The vector space of this transfer matrix is of dimension $2^{L_xL_y}$. Suppose the vector space is spanned by the basis $e_{\{z\}}$, where
\begin{equation}\label{eq.TMbasis}
e_{\{z\}} = \bigotimes_{i=1}^{L_x}\bigotimes_{j=1}^{L_y} e_{z_{i,j}} = e_{z_{1,1}} \otimes e_{z_{1,2}} \otimes \ldots	e_{z_{L_x,L_y}}.
\end{equation}
$e_{z_{i,j}}$ is the local ``virtual bond Hilbert space" spin $\ket{0}=\ket{\mathord{\uparrow}}$, $\ket{1}=\ket{\mathord{\downarrow}}$ basis for the index $z_{i,j}$, where $i$ and $j$ are the coordinates of $z_{i,j}$ in the $x$- and $y$-directions respectively. 
We can consider the matrix multiplication of the transfer matrix $\mathrm{TM}_{xy}$ with an element of the basis $e_{\{\bar{z}\}}$:
\begin{equation}\label{eq.TMactBasis}
\mathrm{TM}_{xy} \cdot e_{\{\bar{z}\}} = \sum_{\{z\}} \left( \mathrm{TM}_{xy} \right)_{\{z\},\{\bar{z}\}}
\end{equation}
where $\{\bar{z}\}$ is fixed for the both LHS and RHS.
Applying the \textbf{Concatenation lemma} to $\textstyle \left( \mathrm{TM}_{xy} \right)_{\{z\},\{\bar{z}\}}$, we conclude that the terms satisfying 
\begin{equation}
\sum_{i=1}^{L_x}\sum_{j=1}^{L_y} z_{i,j} + \bar{z}_{i,j} = 0 \mod{2}
\end{equation} 
contribute equally to the RHS of Eq.~\eqref{eq.TMactBasis}, while the terms satisfying
\begin{equation}
\sum_{i=1}^{L_x}\sum_{j=1}^{L_y} z_{i,j} + \bar{z}_{i,j} = 1 \mod{2}
\end{equation} 
do not contribute. Therefore, we can rewrite the summation more precisely:
\begin{equation}
\begin{split}
\mathrm{TM}_{xy} \cdot e_{\{\bar{z}\}} 
&= \sum_{\{z\} \text{ with } \sum_i z_i+\bar{z}_i \text{ even } } \left( \mathrm{TM}_{xy} \right)_{\{z\},\{\bar{z}\}}	\\
&\propto \sum_{\{z\} \text{ with } \sum_i z_i+\bar{z}_i \text{ even } } e_{\{z\}}.
\end{split}
\end{equation}
When the $\{\bar{z}\}$ satisfies $\sum_{i,j} \bar{z}_{i,j}=0 \mod{2}$, we have
\begin{equation}\label{eq.ToricCode_projector_even}
\begin{split}
&\mathrm{TM}_{xy} \cdot e_{\{\bar{z}\}} \propto \sum_{\{z\} \text{ with } \sum_i z_i \text{ even } } e_{\{z\}},	\\
\end{split}
\end{equation}
while when the $\{\bar{z}\}$ satisfies $\sum_{i,j} \bar{z}_{i,j}=1 \mod{2}$, we have
\begin{equation}\label{eq.ToricCode_projector_odd}
\begin{split}
&\mathrm{TM}_{xy} \cdot e_{\{\bar{z}\}} \propto \sum_{\{z\} \text{ with } \sum_i z_i \text{ odd } } e_{\{z\}}.	\\
\end{split}
\end{equation}
Therefore, $\mathrm{TM}_{xy}$ is a projector of rank 2. 
Hence, it has only two eigenvalues of $1$, and the corresponding unnormalized eigenvectors are:
\begin{equation}\label{eq.ToricCodeTMeigenstate}
\begin{split}
\sum_{\{z\} \text{ with } \sum_i z_i \text{ even } } e_{\{z\}}	\\
\sum_{\{z\} \text{ with } \sum_i z_i \text{ odd } } e_{\{z\}}.
\end{split}
\end{equation}

\subsection{GSD and Transfer Matrix}\label{subsec.ToricCode_GSD}

We know that the 3D toric code model has three pairs of nonlocal operators along the cycles of 3D torus:
\begin{equation}\label{eq.ToricCodeNonLocal}
\begin{split}
&W_X[C_x]=\prod_{i\in C_x} X_i, \quad W_Z[\tilde{C}_{yz}]=\prod_{i\in \tilde{C}_{yz}} Z_i; \\
&W_X[C_y]=\prod_{i\in C_y} X_i, \quad W_Z[\tilde{C}_{xz}]=\prod_{i\in \tilde{C}_{xz}} Z_i; \\ 
&W_X[C_z]=\prod_{i\in C_z} X_i,	\quad W_Z[\tilde{C}_{xy}]=\prod_{i\in \tilde{C}_{xy}} Z_i. \\ 
\end{split}
\end{equation}
where $C_x$ is a loop along the cycle of $x$ direction on lattice, $\tilde{C}_{yz}$ is a closed surface along the cycles of $yz$ directions on dual lattice, and similarly for the other directions. The figures for these operators are:
\begin{equation}\label{eq.figureoperatorsToricCode}
\begin{gathered}
\includegraphics[width=0.7\columnwidth]{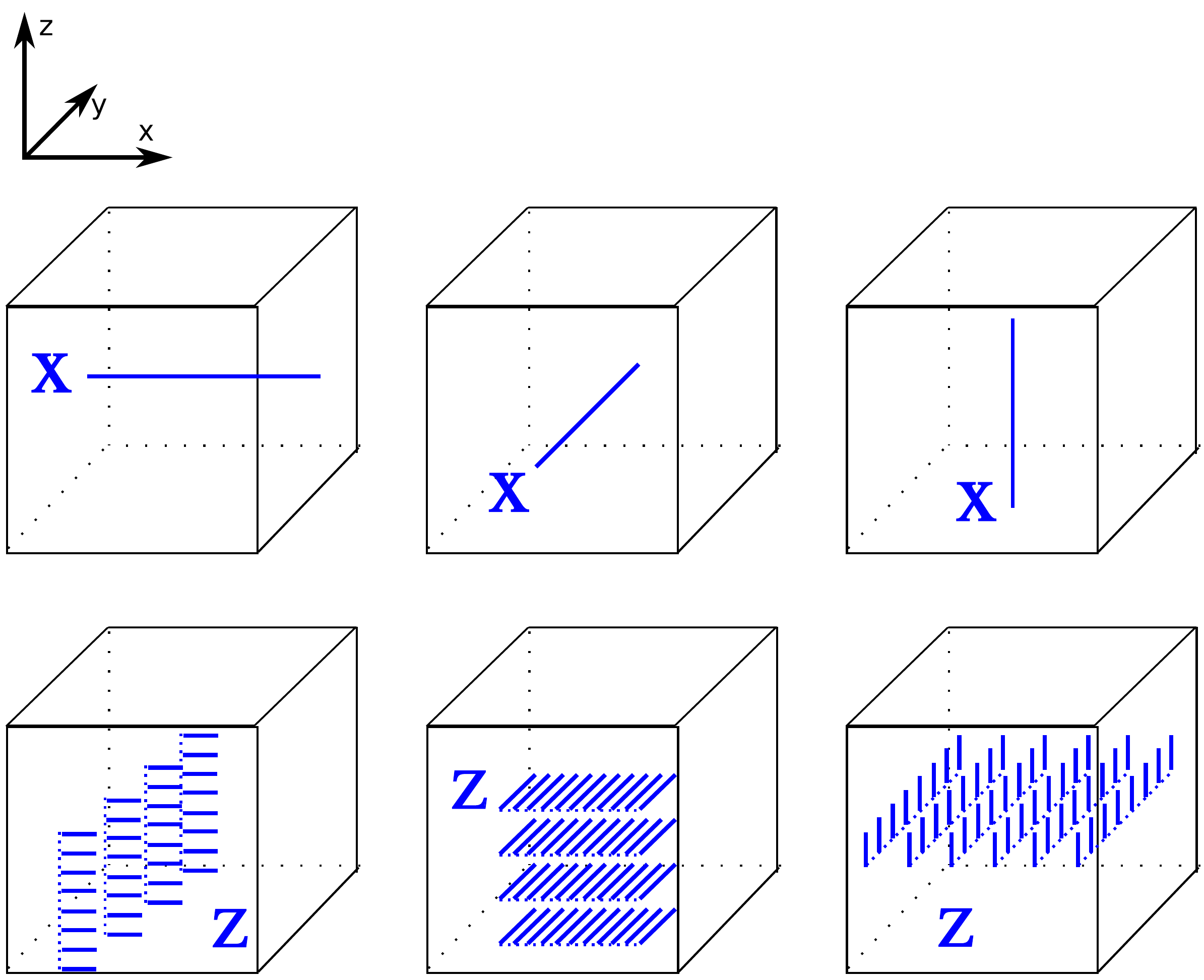}
\end{gathered}
\end{equation}
The commutation relations include:
\begin{equation}
\begin{split}
W_X[C_x] W_Z[\tilde{C}_{yz}] &= -W_Z[\tilde{C}_{yz}] W_X[C_x],	\\
W_X[C_y] W_Z[\tilde{C}_{xz}] &= -W_Z[\tilde{C}_{xz}] W_X[C_y],	\\
W_X[C_z] W_Z[\tilde{C}_{xy}] &= -W_Z[\tilde{C}_{xy}] W_X[C_z].
\end{split}
\end{equation}
All other combinations of operators commute. Hence, there are 8 degenerate ground states in total on the torus, assuming that Eq.~\eqref{eq.ToricCodeNonLocal} has exhausted all nonlocal operators.

We can also put our TNS on the 3-torus, i.e., $\ket{\mathrm{TNS}}_{\mathcal{T}^3}$.
It is not hard to verify using Eq.~\eqref{eq.projectorcondition ToricCode} and \eqref{eq.ToricCodeTcondition} that:
\begin{equation}
\begin{split}
W_X[C_x] \ket{\mathrm{TNS}}_{\mathcal{T}^3} &= \ket{\mathrm{TNS}}_{\mathcal{T}^3},	\\
W_X[C_y] \ket{\mathrm{TNS}}_{\mathcal{T}^3} &= \ket{\mathrm{TNS}}_{\mathcal{T}^3},	\\
W_X[C_z] \ket{\mathrm{TNS}}_{\mathcal{T}^3} &= \ket{\mathrm{TNS}}_{\mathcal{T}^3}.	\\
\end{split}
\end{equation}
As already mentioned in Sec.~\ref{subsec.ToricCode_TNS}, $|\mathrm{TNS}\rangle_{\mathcal{T}^3}=|\psi\rangle$ where $|\psi\rangle$ is defined in Eq.~\eqref{GSoftoriccode}(see App.~\ref{app.TNS_projected} for a proof). Both $|\mathrm{TNS}\rangle_{\mathcal{T}^3}$ and $|\psi\rangle$ are $+1$ eigenstates of $W_X$ operators. 
However, the transfer matrix defined by $\ket{\mathrm{TNS}}_{\mathcal{T}^3}$ does not provide 8 fold degenerate eigenvalues, but only 2, as shown in Sec.~\ref{subsec.ToricCode_TransferMatrix}. 

We can act with the $W_Z[\tilde{C}_{yz}]$ and $W_Z[\tilde{C}_{xz}]$ on the TNS by using Eq.~\eqref{eq.projectorcondition ToricCode} and \eqref{eq.ToricCodeTcondition} to generate all the ground states. The TNSs obtained by this action in terms of a $xy$-plane of tensors are depicted as below:
\begin{equation}\label{eq.figuresToricCodeTransferMatrix}
\begin{gathered}
\includegraphics[width=0.6\columnwidth]{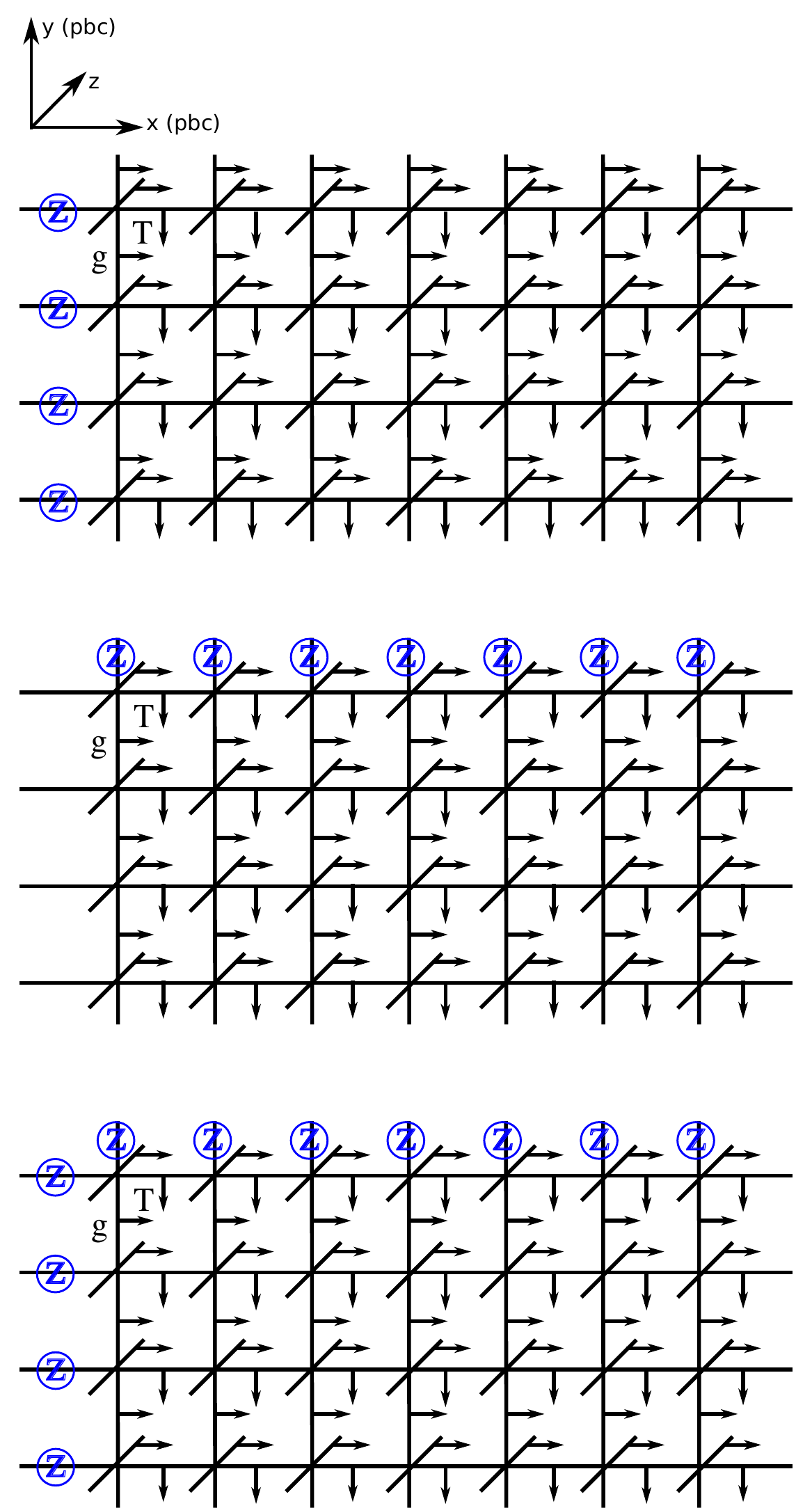}
\end{gathered}
\end{equation}
The intersection of $W_Z[\tilde{C}_{yz}]$ and $W_Z[\tilde{C}_{xz}]$ with the $xy$-plane is the line $Z$ operators, illustrated by the blue circle $Z$ in Eq.~\eqref{eq.figuresToricCodeTransferMatrix}. $W_Z[\tilde{C}_{xy}]$ acts on the $xy$-plane on the dual lattice, and thus does not change the transfer matrix at all. We denote these $xy$-planes of TNSs in Eq.~\eqref{eq.figuresToricCodeTransferMatrix} as $\mathbf{T}_{xy}^{\alpha,\beta}$ (with open indices along the $z$ direction), where $\alpha,~\beta\in \{0,1\}$ label whether we have inserted the $Z$ operators in the $x$ and $y$ direction respectively. The subindex $xy$ in $\mathbf{T}_{xy}^{\alpha,\beta}$ means that the TNS is on a $xy$-plane.
Clearly, $\mathbf{T}_{xy}^{\alpha,\beta}$ are different since they support different holonomies of $W_X[C_x]$ operators and $W_X[C_y]$ operators. After obtaining $\mathbf{T}_{xy}^{\alpha,\beta}$, we can define four twisted transfer matrices correspondingly by contracting the physical indices between bra $\mathbf{T}_{xy}^{\alpha,\beta}$ and ket $\mathbf{T}_{xy}^{\alpha,\beta}$. The twisted transfer matrices are denoted as $\mathrm{TM}_{xy}^{\alpha,\beta}$. For instance, $\mathrm{TM}_{xy}^{0,0}$ is the untwisted transfer matrix in Eq.~\eqref{eq.TransferMatrixIndex}.

Each of these transfer matrices $\mathrm{TM}_{xy}^{\alpha,\beta}$ is also a projector of rank 2. The reasons are that (1) the contraction of the projector $g$ tensors between the bra $\mathbf{T}_{xy}^{\alpha,\beta}$ and ket $\mathbf{T}_{xy}^{\alpha,\beta}$ makes the indices in the bra layer and ket layer identical; (2) the $Z$ operators in the bra and ket layer will cancel each other and produce an identity operator. Hence, the transfer matrices built from the twisted TNS $\mathrm{TM}_{xy}^{1,0}$, $\mathrm{TM}_{xy}^{0,1}$ and $\mathrm{TM}_{xy}^{1,1}$ are equal to that built from the untwisted TNS $\mathrm{TM}_{xy}^{0,0}$:
\begin{equation}\label{eq.ToricCode_TwistedTM}
\mathrm{TM}_{xy}^{\alpha,\beta} = \mathrm{TM}_{xy}^{0,0}, 	\quad\forall\; \alpha,\beta\in \{0,1\}
\end{equation}
The transfer matrix has degeneracy 2, and it is the same for each of 4 TNSs which are different. We have $4\times 2 =8$ degenerate eigenstates.
We have to emphasize that Eq.~\eqref{eq.ToricCode_TwistedTM} holds true, when there are no physical operators in between the bra and ket layer of $\mathbf{T}_{xy}^{\alpha,\beta}$, and the physical indices of projector $g$ tensors are directly contracted. If a physical operator is inserted, for instance $W_X[C_x]$, then the twisted transfer matrices in the presence of $W_X[C_x]$ are NOT the same as the untwisted one sandwiching the same $W_X[C_x]$.

In constructing the TNS on torus, we need to choose the same $\mathbf{T}_{xy}^{\alpha,\beta}$ for each $xy$-plane. If we use different $\mathbf{T}_{xy}^{\alpha,\beta}$ in each $xy$-plane to construct a TNS on torus, the corresponding wave function is no longer a ground state for the 3D toric code model. 
More specifically, in the 3D toric code model, if we contract $L_z-1$  $\mathbf{T}_{xy}^{0,0}$'s with one twisted $\mathbf{T}_{xy}^{\alpha,\beta}$ Eq.~\eqref{eq.figuresToricCodeTransferMatrix}, then the corresponding wave function will support a pair of loop excitations because the $B_p$ operators up and below $Z$ operators are violated, and is no longer a ground state. Graphically, the contraction of these $L_z-1$ number of $\mathbf{T}_{xy}^{0,0}$ and one $\mathbf{T}_{xy}^{1,0}$ is:
\begin{equation}
\begin{gathered}
\includegraphics[width=0.4\columnwidth]{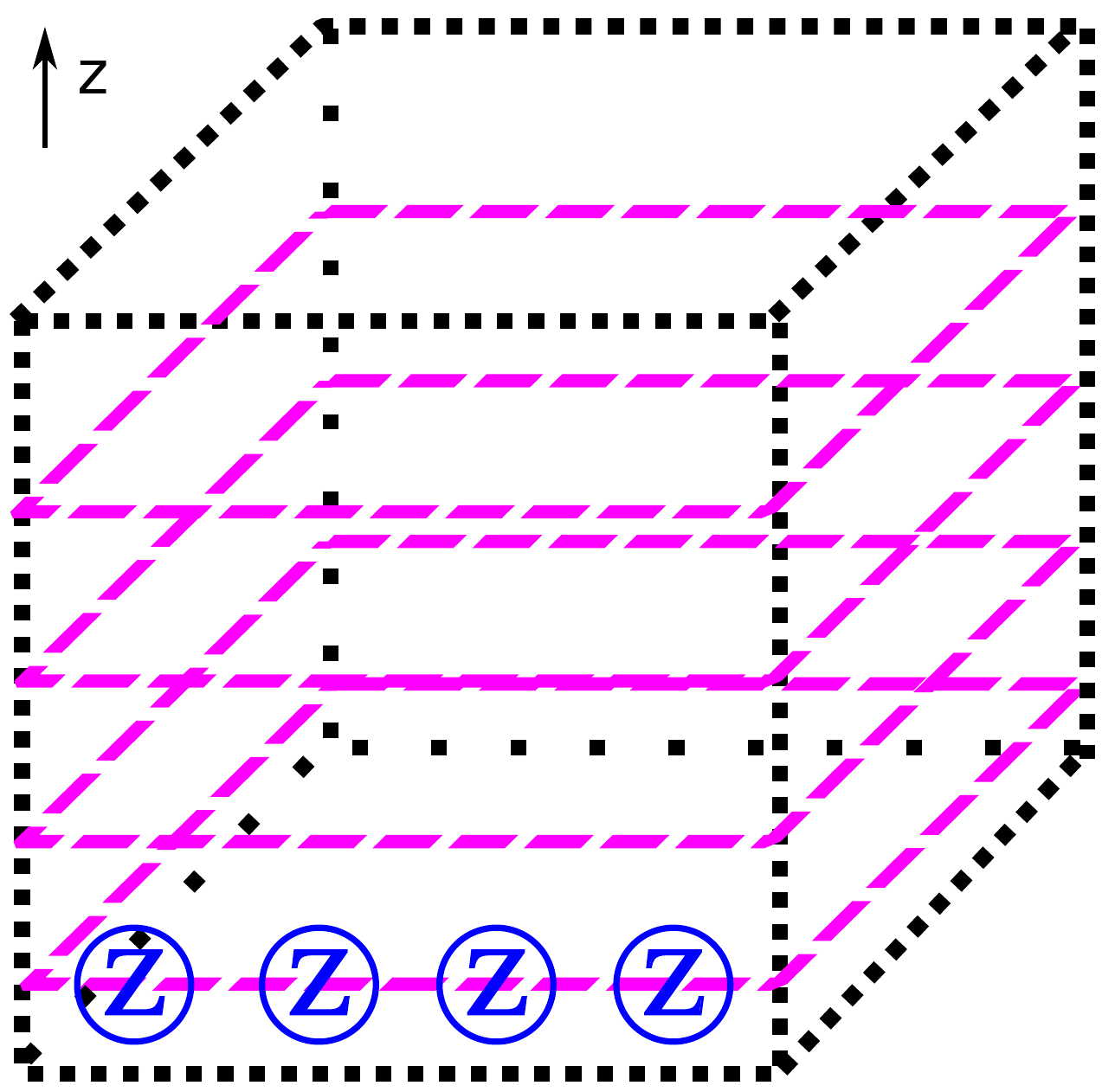}
\end{gathered}.
\end{equation}
The reason for this energy costing is that there is a line of $Z$ operators. However, the nonlocal operators in 3D toric code do not have a line $Z$ operator, but only have surface $Z$ operators. 
See Eq.~\eqref{eq.figureoperatorsToricCode}. 
On the other hand, the TNS made of all the same $\mathbf{T}_{xy}^{\alpha,\beta}$, for instance $\mathbf{T}_{xy}^{1,0}$:
\begin{equation}
\begin{gathered}
\includegraphics[width=0.4\columnwidth]{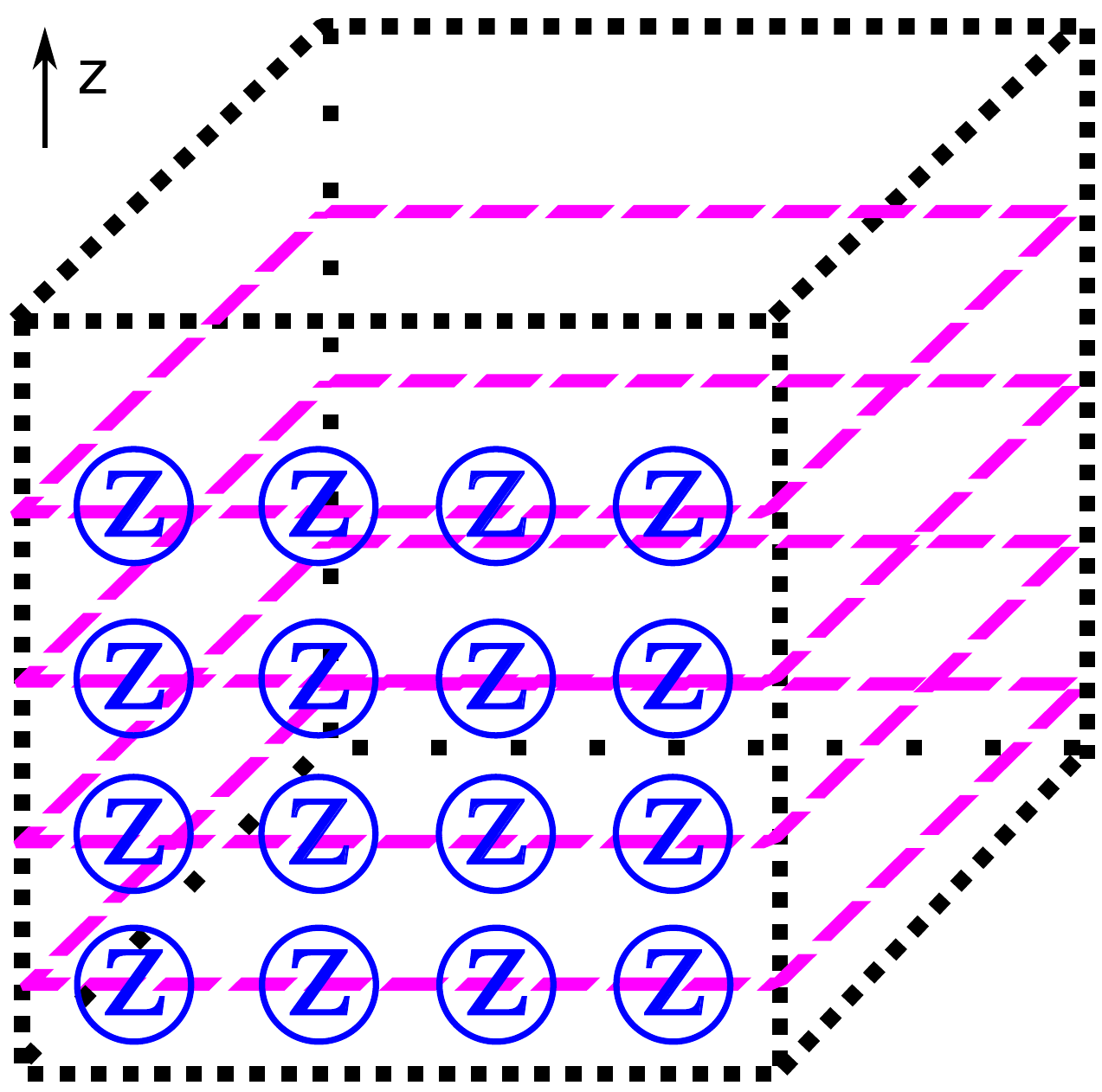}
\end{gathered}
\end{equation}
is allowed, because it corresponds to a closed surface operator included in Eq.~\eqref{eq.figureoperatorsToricCode}. We will come back to this point in Sec.~\ref{subsec.Xcube_GSD} where this issue is subtle and makes a difference when we count GSD from transfer matrices.

\begin{figure*}[t]
\centering
\includegraphics[width=0.6\textwidth]{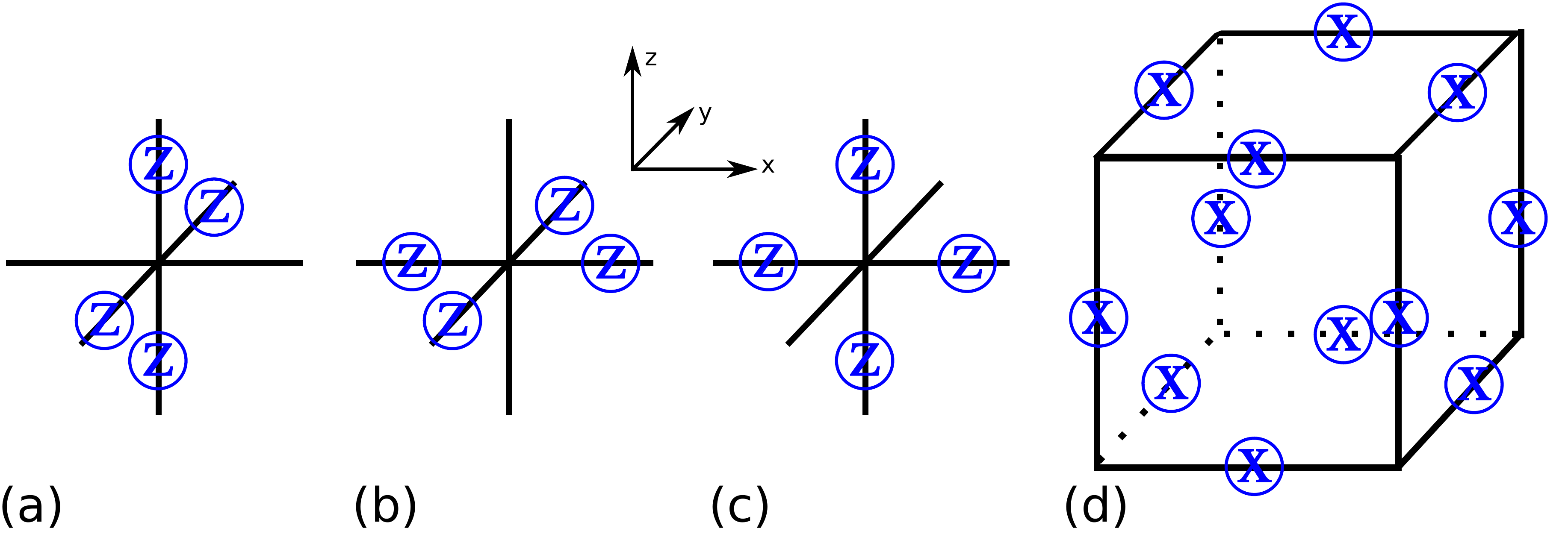}
\caption{The Hamiltonian terms of the X-cube model: (a) $A_{v,yz}$, (b) $A_{v,xy}$, (c) $A_{v,xz}$ and (d) $B_{c}$. The circled $X$ and $Z$ represent the Pauli matrices acting on the physical spin-$1/2$'s.}
\label{fig.XCubeHamiltonian}
\end{figure*}

\section{X-cube Model}\label{sec.Xcube}

In this section, we construct the TNS for the X-cube model, one of the simplest fracton models. Using it, we then calculate the entanglement entropy and the GSD of this model. The results are generally different from those in conventional topological phases. The entanglement entropy has a linear correction to the area law, and the GSD grows exponentially with the size of the system.
This section is organized as follows:
In Sec.~\ref{subsec.Xcube}, we briefly review the X-cube model in a cubic lattice. 
In Sec.~\ref{subsec.Xcube_TNS}, we construct the TNS for the X-cube model.
In Sec.~\ref{subsec.ConcatenationXcube}, we prove a \textbf{Concatenation lemma} for the X-cube TNS.
In Sec.~\ref{subsec.Xcube_Entanglement}, we calculate the entanglement entropies on $\mathcal{R}^3$.
In Sec.~\ref{subsec.Xcube_TransferMatrix}, we construct the transfer matrix and prove that it is a projector of rank $2^{L_x+L_y-1}$.
In Sec.~\ref{subsec.Xcube_GSD}, we show how the transfer matrix gives rise to an extensive GSD on torus.
The App.~\ref{app.transfermatrixnumerics} is also related to this section. It includes the numerical results of the transfer matrix eigenvalue degeneracy, which further confirm the analytical results in Sec.~\ref{subsec.Xcube_TransferMatrix}.

\subsection{Hamiltonian of X-cube Model}\label{subsec.Xcube}

The model is defined on a cubic lattice. All the spin $1/2$'s are associated with the bonds of the cubic lattice. The Hamiltonian is:
\begin{equation}
H = -\sum_{v} \left( A_{v,xy}+A_{v,yz}+A_{v,xz} \right) - \sum_{c} B_c
\end{equation}
where the summation is taken over all vertices and cubes respectively. Each term is depicted in Fig.~\ref{fig.XCubeHamiltonian}. More specifically, $A_{v,xy}$ is the product of four $Z$ operators around the vertex $v$ in the $xy$ plane. Similarly for $A_{v,yz}$ and $A_{v,xz}$. $B_{c}$ flips the $12$ spins of a cube $c$. 
It is trivial to show that all the Hamiltonian terms commute:
\begin{equation}
\begin{split}
\commute{A_{v,xy}}{A_{v^\prime,xy}}=&
\commute{A_{v,yz}}{A_{v^\prime,yz}}=
\commute{A_{v,xz}}{A_{v^\prime,xz}}=0,	\\
\commute{A_{v,xy}}{A_{v^\prime,yz}}=&
\commute{A_{v,xy}}{A_{v^\prime,xz}}=	
\commute{A_{v,yz}}{A_{v^\prime,xz}}=0,	\\
\commute{B_{c}}{A_{v^\prime,xy}}=&	
\commute{B_{c}}{A_{v^\prime,yz}}=
\commute{B_{c}}{A_{v^\prime,xz}}=0	\\
\commute{B_{c}}{B_{c^\prime}}=&0,	
\quad\forall\; v,v^\prime,c,c^\prime. \\
\end{split}
\end{equation}
Hence, this model can be exactly solved. The ground state $\ket{\mathrm{GS}}$ (on $\mathcal{R}^3$) needs to satisfy that:
\begin{equation}\label{eq.GScondition}
\begin{split}
A_{v,xy}\ket{\mathrm{GS}}=&\ket{\mathrm{GS}},	\\
A_{v,yz}\ket{\mathrm{GS}}=&\ket{\mathrm{GS}},	\\
A_{v,xz}\ket{\mathrm{GS}}=&\ket{\mathrm{GS}},	\\
B_{c}\ket{\mathrm{GS}}=&\ket{\mathrm{GS}},	\quad \forall\; v,c.
\end{split}
\end{equation}
This set of equations will be used to derive the local $T$ tensor for the X-cube model.

The nonlocal operators of the X-cube model which are required to commute with all Hamiltonian terms on torus include 9 loop operators\cite{vijay2016fracton}:
\begin{equation}\label{eq.operatorsXcube}
\begin{split}
&W_X[C_x] = \prod_{i \in C_x} X_i,
W_X[C_y] = \prod_{i \in C_y} X_i,
W_X[C_z] = \prod_{i \in C_z} X_i,	\\
&W_Z[\tilde{C}_{x,z}] = \prod_{i \in \tilde{C}_x} Z_i,
W_Z[\tilde{C}_{y,z}] = \prod_{i \in \tilde{C}_y} Z_i,
W_Z[\tilde{C}_{z,x}] = \prod_{i \in \tilde{C}_z} Z_i,	\\
&W_Z[\tilde{C}_{x,y}] = \prod_{i \in \tilde{C}_x} Z_i,
W_Z[\tilde{C}_{y,x}] = \prod_{i \in \tilde{C}_y} Z_i,
W_Z[\tilde{C}_{z,y}] = \prod_{i \in \tilde{C}_z} Z_i,	\quad
\end{split}
\end{equation}
where $C_x$ is a straight line along the cycle of the $x$ direction on lattice, and $\tilde{C}_{x,z}$ is a line along the cycle of the $x$ direction on dual lattice while the lattice bonds of $\tilde{C}_{x,z}$ are in the $z$-direction, and similarly for the other directions. The figures for them are:
\begin{equation}\label{eq.figureoperatorsXcube}
\begin{gathered}
\includegraphics[width=0.77\columnwidth]{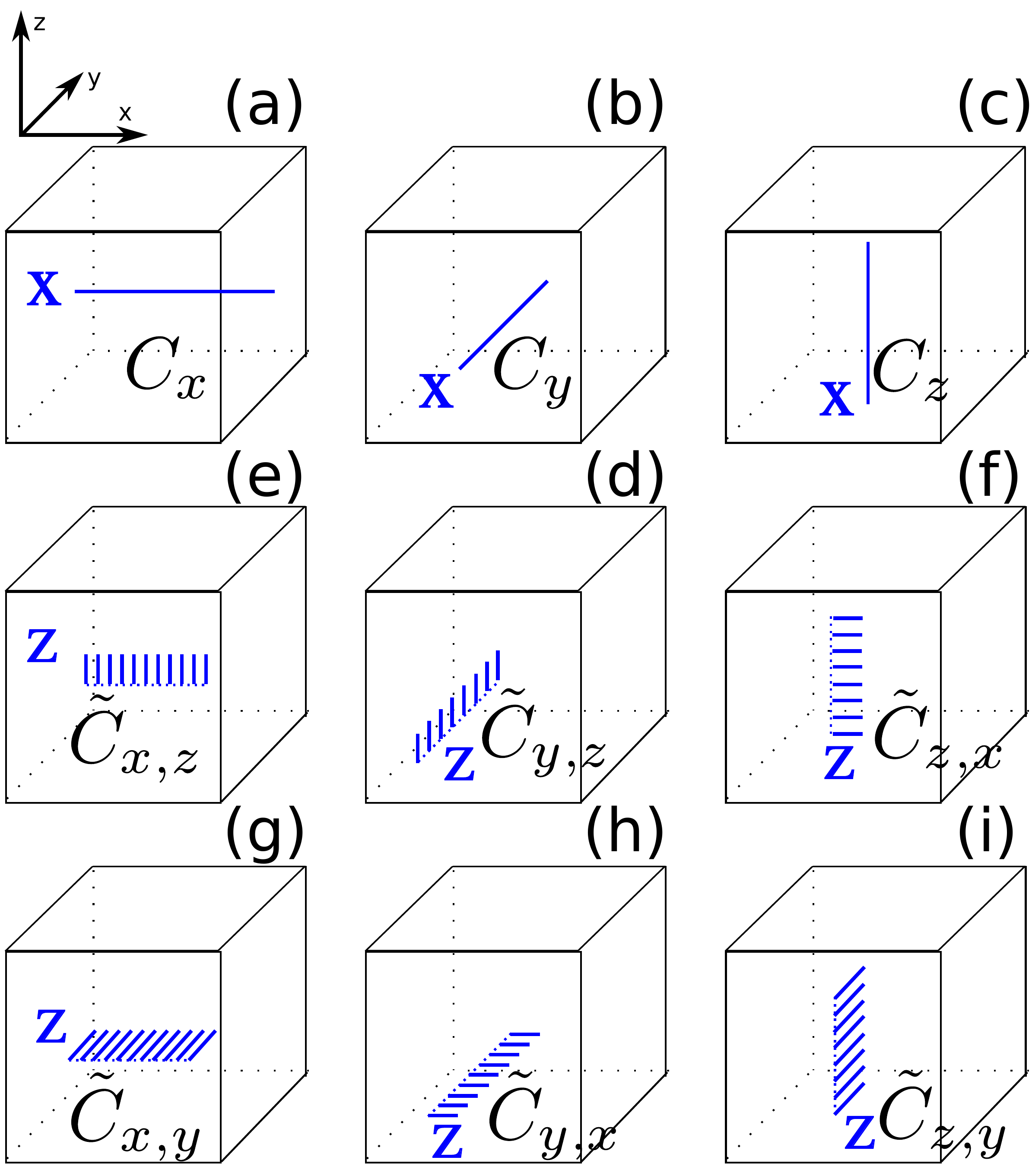}
\end{gathered}.
\end{equation}

The algebras of these loop operators are grouped into three independent sets:
\begin{enumerate}
\item The operator (a) in Eq.~\eqref{eq.figureoperatorsXcube} anti-commutes with the operator (f) and the operator (h) when they overlap at one spin. Thus, $W_X[C_x]$ anti-commutes with $W_Z[\tilde{C}_{z,x}]$ and $W_Z[\tilde{C}_{y,x}]$ when they overlap at one spin. 
\item The operator (b) in Eq.~\eqref{eq.figureoperatorsXcube} anti-commutes with the operator (g) and the operator (i) when they overlap at one spin. Thus, $W_X[C_y]$ anti-commutes with $W_Z[\tilde{C}_{x,y}]$ and $W_Z[\tilde{C}_{z,y}]$ when they overlap at one spin.
\item The operator (c) in Eq.~\eqref{eq.figureoperatorsXcube} anti-commutes with the operator (d) and the operator (e) when they overlap at one spin. Thus, $W_X[C_z]$ anti-commutes with $W_Z[\tilde{C}_{x,z}]$ and $W_Z[\tilde{C}_{y,z}]$ when they overlap at one spin.
\end{enumerate}
All other combinations of operators commute, because they do not overlap at the same physical spin.
Each of the three algebras gives a representation of dimension\cite{2017arXiv170804619S} 
\begin{equation}
2^{L_y+L_z-1},	\quad
2^{L_z+L_x-1},	\quad
2^{L_x+L_y-1}.	\quad
\end{equation}
respectively.
Hence the total dimension of the ground state space is $2^{2L_x+2L_y+2L_z-3}$. The derivations of the GSD in terms of these operator algebras are explained in App.~\ref{app.XcubeGSD}.

\subsection{TNS for X-cube Model}\label{subsec.Xcube_TNS}

Following the same prescription in Sec.~\ref{sec.toriccode}, we can write down the TNS for the X-cube model. First, we introduce a projector $g$ tensor on the bonds of the lattice (see Eq.~\eqref{eq.projector2}). The virtual index is either $0$ or $1$. The $g$ tensor is a projector which projects the physical spin to the virtual indices. The tensor $g$ satisfies Eq.~\eqref{eq.projectorcondition ToricCode}.

The TNS is not only composed of the $g$ tensor on the bonds of the lattice, but also the $T$ tensors on the vertices. The $T$ tensor has six virtual indices and no physical index. Unlike the $g$ tensor, the $T$ tensor will be specified by the Hamiltonian terms. We now implement Eq.~\eqref{eq.GScondition} on the TNS. Using the condition Eq.~\eqref{eq.projectorcondition ToricCode}, we can transfer the operators in Hamiltonian terms from physical indices to virtual indices:
\begin{equation}
\begin{gathered}
\includegraphics[width=0.8\columnwidth]{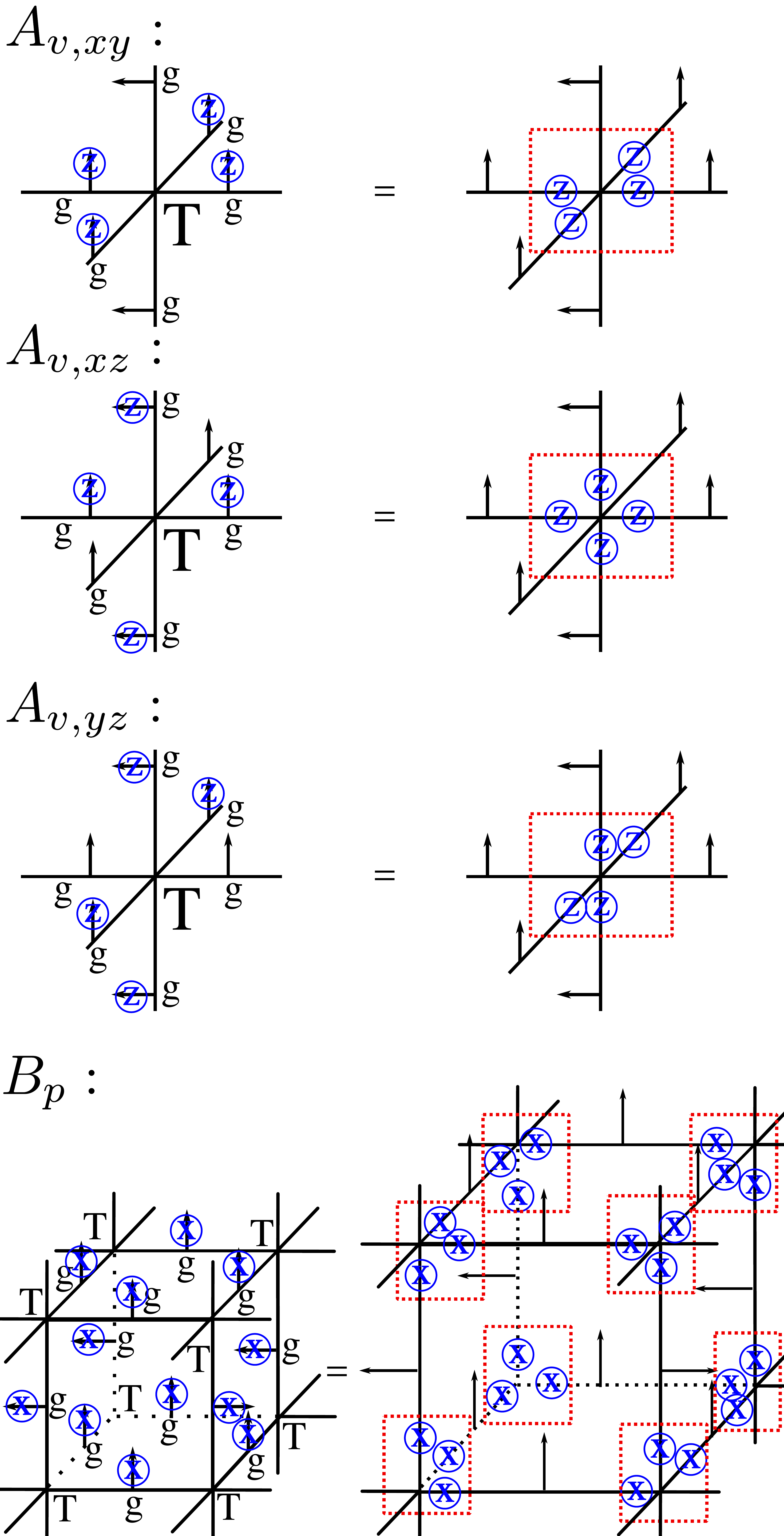}
\end{gathered}.
\end{equation}
Requiring that the tensors in the dashed red rectangles are invariant will lead to the following (strong) conditions on the  $T$ tensor:
\begin{equation}\label{eq.XcubeTtensor}
\begin{gathered}
\includegraphics[width=0.6\columnwidth]{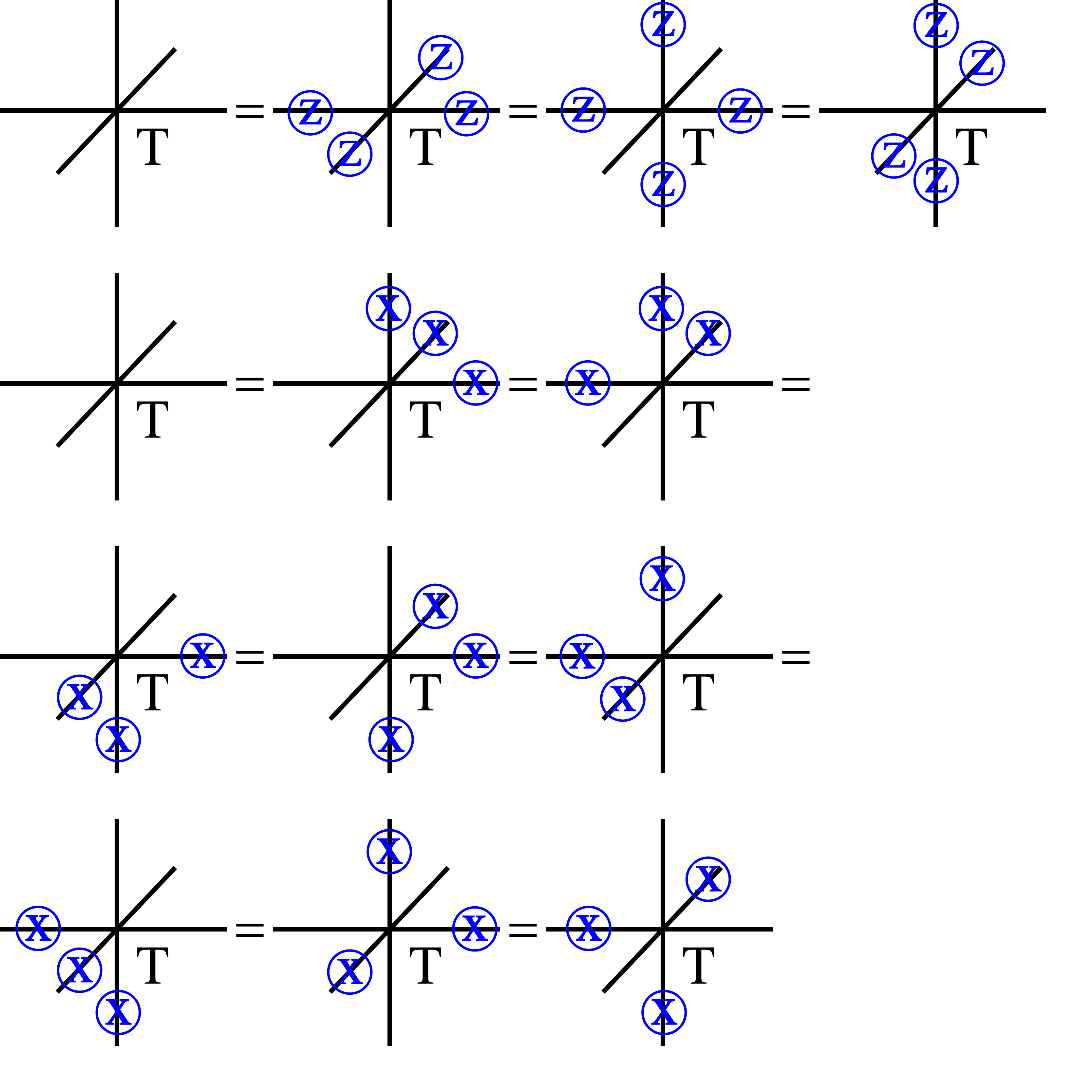}
\end{gathered}.
\end{equation}
The first set of conditions is required by the operators $A_{v,xy}$, $A_{v,yz}$ and $A_{v,xz}$ around the vertex $v$. The second set of equations is required by the operators $B_{c}$ around the cube $c$. The $X$ operators acting on the 12 spins of the cube $c$ will be transferred to the virtual spins of the eight $T$ tensors around the cube $c$, using Eq.~\eqref{eq.projectorcondition ToricCode}. The $X$ operators act on the eight quadrants of a $T$ tensor. Clearly, these two sets of the conditions are not independent. The solution to these conditions is:
\begin{equation}\label{eq.localT}
\begin{split}
&T_{x\bar{x},y\bar{y},z\bar{z}} =\begin{cases}
1	&	\text{if } \begin{cases}
x+\bar{x}+y+\bar{y}=0 \mod{2}, 	\\
x+\bar{x}+z+\bar{z}=0 \mod{2},	\\
y+\bar{y}+z+\bar{z}=0 \mod{2}.
\end{cases}	\\
0	&	\text{otherwise}.
\end{cases}
\end{split}
\end{equation}
A useful consequence of Eq.~\eqref{eq.XcubeTtensor} is:
\begin{equation}\label{eq.XcubeStraightX}
\begin{gathered}
\includegraphics[width=0.7\columnwidth]{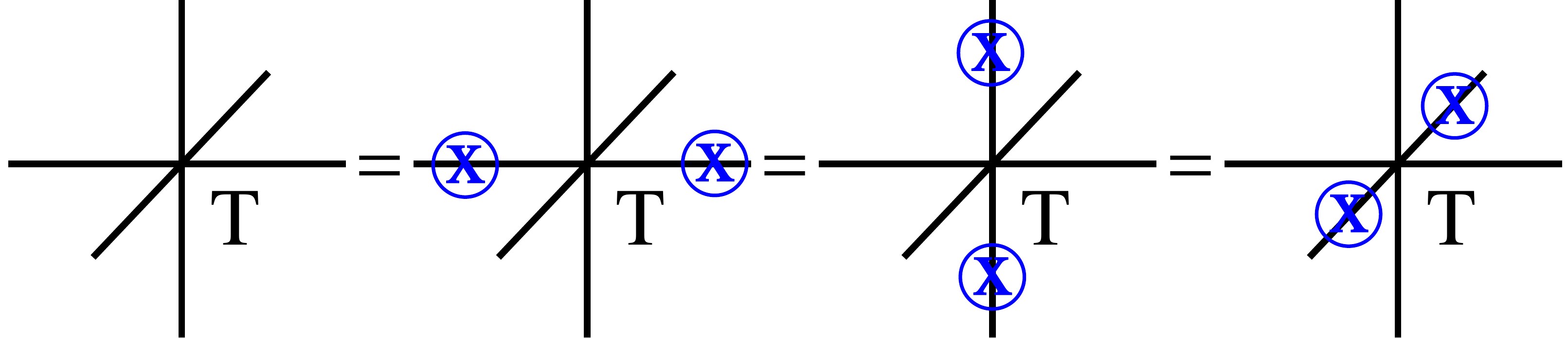}
\end{gathered}.
\end{equation}
We now have fixed the TNS for the X-cube model using the local conditions Eq.~\eqref{eq.GScondition}. The wave function on $\mathcal{R}^3$ can also be represented as a tensor contraction of Eq.~\eqref{eq.TNS}.

\subsection{Concatenation Lemma}\label{subsec.ConcatenationXcube}

In this section, we consider the contraction of a network of local $T$ tensors with open virtual indices for the X-cube model, similar to the idea developed in Sec.~\ref{subsec.ConcatenationToricCode}. However, here the situation is more complicated than the 3D toric code model. The elements of a local $T$ tensor are either $0$ or $1$ in Eq.~\eqref{eq.XcubeTtensor}, depending on the even/odd sector in three directions. The elements of the contracted $T$ tensors will also be either $0$ or $1$ with a similar criterion.

\begin{framed}
\textbf{Concatenation Lemma:} For a network of the contracted $T$ tensors in Eq.~\eqref{eq.localT}, the sums of the open indices along each $xy$, $yz$ and $xz$ planes have to be even. Otherwise, the tensor element of this network is zero. The nonzero elements are constants independent of the virtual indices.
\end{framed}

This lemma is a consequence of Eq.~\eqref{eq.localT}. See App.~\ref{app.Xcube_Concatenation} for an induction proof. In this section, we explain this result by considering a simple example. Suppose that we have two $T$ tensors contracted along the $z$ direction:
\begin{equation}
\begin{split}
&\mathbf{T}_{x_1,\bar{x}_1,y_1,\bar{y}_1,z_1,x_2,\bar{x}_2,y_2,\bar{y}_2,\bar{z}_2} \\
=&\sum_{\bar{z}_1,z_2} T_{x_1\bar{x}_1,y_1\bar{y}_1,z_1\bar{z}_1} T_{x_2\bar{x}_2,y_2\bar{y}_2,z_2\bar{z}_2} \delta_{\bar{z}_1z_2}.	\\
\end{split}
\end{equation}
Graphically, $\mathbf{T}$ is the same as depicted in Fig.~\ref{fig.Contract2T}. For each of the two $T$ tensors, we have
\begin{equation}
\begin{cases}
x_1+\bar{x}_1+y_1+\bar{y}_1 &= 0 \mod{2} \\
x_1+\bar{x}_1+z_1+\bar{z}_1 &= 0 \mod{2} \\
y_1+\bar{y}_1+z_1+\bar{z}_1 &= 0 \mod{2} \\
\end{cases}	
\end{equation}
and
\begin{equation}
\begin{cases}
x_2+\bar{x}_2+y_2+\bar{y}_2 &= 0 \mod{2} \\	
x_2+\bar{x}_2+z_2+\bar{z}_2 &= 0 \mod{2} \\
y_2+\bar{y}_2+z_2+\bar{z}_2 &= 0 \mod{2}. \\
\end{cases}
\end{equation}
Therefore, setting $\bar{z}_1 = z_2$ due to the tensor contraction, the open indices of $\mathbf{T}$ need to satisfy:
\begin{equation}
\begin{split}
\begin{cases}
x_1+\bar{x}_1+y_1+\bar{y}_1 &= 0, \mod{2}	\\
x_2+\bar{x}_2+y_2+\bar{y}_2 &= 0, \mod{2}	\\
z_1+\bar{z}_2+x_1+\bar{x}_1+x_2+\bar{x}_2 &= 0, \mod{2}	\\
z_1+\bar{z}_2+y_1+\bar{y}_1+y_2+\bar{y}_2 &= 0, \mod{2}	\\
\end{cases}
\end{split}
\end{equation} 
in order for the elements of the tensor $\mathbf{T}$ to be nonzero. The last set of equations intuitively means that the open indices of the tensor $\mathbf{T}$ (Fig.~\ref{fig.Contract2T}) in each $xy$, $yz$ and $xz$ plane need to have an even summation. Moreover, the elements of the contracted $T$ tensor are 1 independent of indices:
\begin{widetext}
\begin{equation}\label{eq.Xcube_Ttensor_Example}
\mathbf{T}_{x_1,\bar{x}_1,y_1,\bar{y}_1,z_1,x_2,\bar{x}_2,y_2,\bar{y}_2,\bar{z}_2} =
\begin{cases}
1	&	\text{if}\;\;
\begin{cases}
x_1+\bar{x}_1+y_1+\bar{y}_1 &= 0, \mod{2}	\;\;\text{and}\\
x_2+\bar{x}_2+y_2+\bar{y}_2 &= 0, \mod{2}	\;\;\text{and}\\
z_1+\bar{z}_2+x_1+\bar{x}_1+x_2+\bar{x}_2 &= 0, \mod{2}	\;\;\text{and}\\
z_1+\bar{z}_2+y_1+\bar{y}_1+y_2+\bar{y}_2 &= 0, \mod{2}	\\
\end{cases}	\\
0	&	\text{otherwise}.
\end{cases}
\end{equation}
\end{widetext}

Generally for a complicated contraction of local $T$ tensors denoted by $\mathbf{T}$, we have:
\begin{equation}
\mathbf{T}_{\{t\}} = \begin{cases}
\mathrm{Const}\neq 0	&	\text{\textbf{Concatenation lemma} is satisfied}	\\
0	&	\text{else}.	\\
\end{cases}
\end{equation} 
This is the notation that we will use when computing the entanglement entropies or the transfer matrix degeneracies.

\subsection{Entanglement}\label{subsec.Xcube_Entanglement}

\begin{figure*}[t]
\centering
\includegraphics[width=0.7\textwidth]{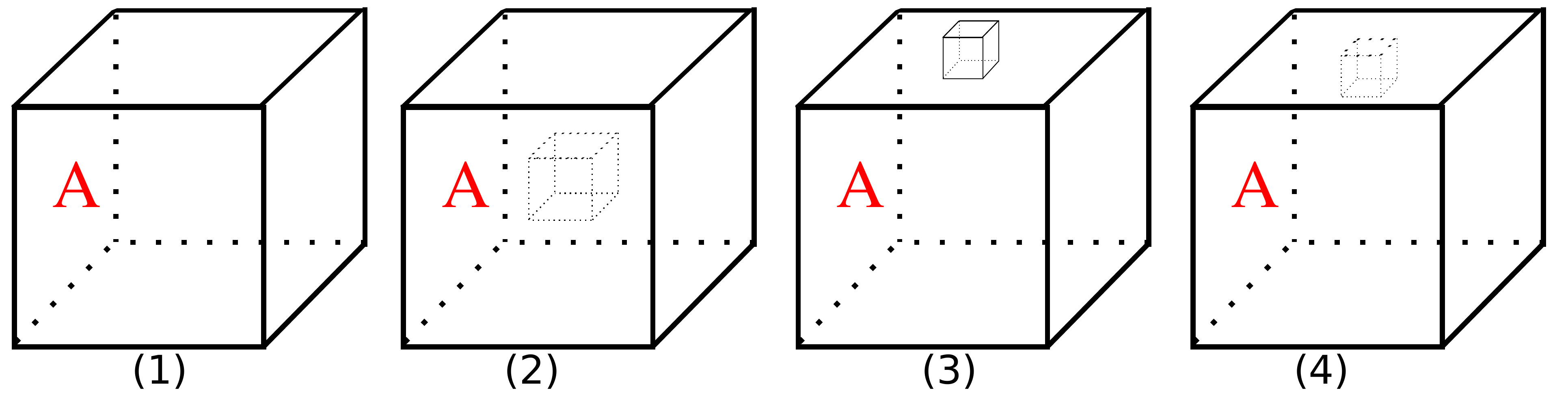}
\caption{Figures for several regions $A$ for which we calculate the entanglement entropies. (1) Region A is a cube of size $l_x \times l_y \times l_z$. (2) Region A is a cube of $l_x \times l_y \times l_z$ with a hole of size $l_x^\prime \times l_y^\prime \times l_z^\prime $ in it. (3) Region A is a cube of size $l_x \times l_y \times l_z$ and a small cube of height $l_z^\prime$ on top of it. (4) Region A is a cube of size $l_x \times l_y \times l_z$ carved on top of it a small cube of height $l_z^\prime$. 
}
\label{fig.RegionA}
\end{figure*}

We can show that a cubic region $A$ and its deformations are the exact SVD of $\ket{\mathrm{TNS}}$. See Fig.~\ref{fig.RegionA} for some examples of deformed $A$ regions. We can read out the singular values of $\ket{\mathrm{TNS}}$ for these entanglement cut. 
Suppose we consider the wave function on $\mathcal{R}^3$ for simplicity (i.e., without dealing with the multiple ground states on $\mathcal{T}^3$). We rewrite the wave function:
\begin{equation}\label{eq.XcubeSVD}
\ket{\mathrm{TNS}} = \sum_{\{t\}} \ket{\{t\}}_{A} \otimes \ket{\{t\}}_{\bar{A}}.
\end{equation}
where $\{t\}$ represent the tensor virtual indices connecting the region $A$ and $\bar{A}$. $\ket{\{t\}}_{A}$ is the TNS for the region $A$ with open virtual indices $\{t\}$. Similarly for $\ket{\{t\}}_{\bar{A}}$ for the complement region $\bar{A}$. Because every virtual bond is contracted over for $\ket{\mathrm{TNS}}$, $\ket{\{t\}}_{A}$ and $\ket{\{t\}}_{\bar{A}}$ have to share the same $\{t\}$, and $\{t\}$ is summed over in Eq.~\eqref{eq.XcubeSVD}.

Next we show that this set of basis $\ket{\{t\}}_{A}$ is orthonormal:
\begin{equation}
\begin{split}
&_A\langle \{t^\prime \}| \{t\} \rangle_{A} \propto \delta_{\{t^\prime\},\{t\}},
\end{split}
\end{equation}
when the \textbf{Concatenation lemma} is satisfied for both of $\{t^\prime\}$, $\{t\}$, and the basis $\ket{\{t\}}_{A}$ are not null vectors.

\noindent\textbf{Proof:}

The open virtual indices $\{t\}$ satisfy the \textit{SVD condition} in Sec.~\ref{subsec.TNSSVD}. Hence, we can conclude that $\ket{\{t\}}_{A}$ and $\ket{\{t\}}_{\bar{A}}$ are orthogonal basis for the region $A$ and $\bar{A}$, and thus Eq.~\eqref{eq.XcubeSVD} is exactly SVD. In order to calculate the entanglement entropies, we need to show that $\ket{\{t\}}_{A}$ and $\ket{\{t\}}_{\bar{A}}$ are not only orthogonal, but also orthonormal.

The proof is essentially the same as in Sec.~\ref{subsec.ToricCode_Entanglement}. Here we briefly repeat it. We use the same conventions of SVD basis $\ket{\{t\}}_{A}$ and $\ket{\{t^\prime\}}_{\bar{A}}$ as in Sec.~\ref{subsec.ToricCode_Entanglement}.
Suppose $\{t^\prime\} $ and $ \{t\}$ both satisfy \textbf{Concatenation lemma}. Hence, $\ket{\{t\}}_A$ and $\ket{\{t^\prime\}}_A$ are not null vectors.
More specifically, the wave function $\ket{\{t\}}_A$ is the same as in Eq.~\eqref{eq.SVDbasisA}. All the virtual indices except $\{t\}$ are contracted over.

When $\{t^\prime\} \neq \{t\}$, the basis overlap is zero, because the spins on the boundary of $A$ are different. When $\{t^\prime\} = \{t\}$, the overlap is nonzero. Moreover, the overlaps $_A\langle \{t \}| \{t\} \rangle_A$ are constants independent of $\{t\}$, due to the \textbf{Concatenation lemma}.
Hence, $\ket{\{t\}}_{A}$ is an orthonormal basis for the region $A$, up an overall normalization factor.
\hfill$\Box$\\

Therefore, using the orthonormal basis $\ket{\{t\}}_{A}$ and $\ket{\{t\}}_{\bar{A}}$, Eq.~\eqref{eq.XcubeSVD} is indeed the SVD. Furthermore, the singular values are all identical. As a result, the entanglement of the region $A$ is determined by the number of basis states $\ket{\{t\}}_A$ which are involved in Eq.~\eqref{eq.XcubeSVD}, i.e., the rank of the contracted tensor for the region $A$. The rank of the contracted tensor can be counted by the \textbf{Concatenation lemma}. We only need to count the number of indices that satisfy the \textbf{Concatenation lemma}. We now list a few simple examples of entanglement entropies. The entanglement cuts are displayed in Fig.~\ref{fig.RegionA}. Correspondingly, their entanglement entropies are:
\begin{enumerate}
\item When region $A$ is a cube of size $l_x \times l_y \times l_z$:
\begin{equation}
\begin{split}
\frac{S_{A}}{\log2} 
&= 2l_xl_y+2l_yl_z+2l_xl_z - l_x - l_y - l_z + 1	\\
&= \mathrm{Area} - l_x - l_y - l_z + 1.	\\
\end{split}\label{entropy1xcode}
\end{equation}
The calculation details are the following:

The number of indices straddling this entanglement cut is $2l_xl_y+2l_yl_z+2l_xl_z$. This is the maximum possible number of basis states in the SVD of Eq.~\eqref{eq.XcubeSVD}. However, these indices are not free. They are subject to certain constraints, in order for the singular vectors to have non-vanishing norms. Using the \textbf{Concatenation lemma}, we know that the open indices in each $xy$, $yz$, and $xz$ plane must have even summations. We denote the indices to be $t_{i,j,k}$ where $i,j,k$ are the coordinates of such a index. Then, we have; 
\begin{equation}
\begin{split}
\sum_{i,j} t_{i,j,k} = 0 \mod{2}, \quad\forall~ k	\\
\sum_{i,k} t_{i,j,k} = 0 \mod{2}, \quad\forall~ j	\\
\sum_{j,k} t_{i,j,k} = 0 \mod{2}, \quad\forall~ i	\\
\end{split}
\end{equation}
where the summation is only taken over open virtual indices near the entanglement cut in each $xy$, $yz$ and $xz$ plane.
Therefore, we have $l_x$, $l_y$, $l_z$ number of constraints respectively. However, these constraints are not independent. Only $l_x+l_y+l_z-1$ of them are independent, because the three sets of constraints sum to be an even number. Hence, the number of free indices is $$2l_xl_y+2l_yl_z+2l_xl_z - l_x - l_y - l_z + 1.$$ The number of singular vectors in Eq.~\eqref{eq.XcubeSVD} is: $$2^{2l_xl_y+2l_yl_z+2l_xl_z - l_x - l_y - l_z + 1},$$ which leads to the entropy written in Eq.~\eqref{entropy1xcode}.
\hfill$\Box$

\item When the region $A$ is a cube of size $l_x \times l_y \times l_z$ with an empty hole of size $l_x^\prime \times l_y^\prime \times l_z^\prime$:
\begin{equation}
\begin{split}
\frac{S_{A}}{\log2} 
=& 2l_xl_y+2l_yl_z+2l_xl_z+ \\
&
2 l_x^\prime l_y^\prime + 
2 l_y^\prime l_z^\prime +
2 l_x^\prime l_z^\prime \\
&
- l_x^\prime - l_y^\prime - l_z^\prime
- l_x - l_y - l_z
+2	\\
=& \mathrm{Area} - l_x^\prime - l_y^\prime - l_z^\prime - l_x - l_y - l_z + 2.
\end{split}
\end{equation}

\item  When the region $A$ is a cube of size $l_x \times l_y \times l_z$ with an additional convex cube of height $l_z^\prime$ on top (Fig.~\ref{fig.RegionA} (3)):
\begin{equation}
\begin{split}
\frac{S_{A}}{\log2} 
=& \mathrm{Area} - l_x - l_y - l_z - l_z^\prime + 1.
\end{split}
\end{equation}

\item  When the region $A$ is a cube of size $l_x \times l_y \times l_z$ with an additional concave cube of height $l_z^\prime$ on top (Fig.~\ref{fig.RegionA} (4)):
\begin{equation}
\begin{split}
\frac{S_{A}}{\log2} 
=& \mathrm{Area} - l_x - l_y - l_z - l_z^\prime + 1.
\end{split}
\end{equation}
\end{enumerate}
The area part of the entanglement is measured by the number of indices straddling the entanglement cut.
The constant contribution to the entanglement entropy is universal\cite{casini2009entanglement}, as it counts the number of connected components of the entanglement surface. As opposed to the toric code case, the constants are positive numbers. We emphasize that the linear corrections to the area law in the entanglement entropies states have not been observed in quantum field theories.

Furthermore, if we put the TNS on a cylinder $\mathcal{T}^2_{xy} \times \mathcal{R}_z$ and the entanglement cut splits the cylinder into two halves $z>0$ and $z<0$, then the entanglement entropy of either side is:
\begin{equation}\label{eq.XcubeCylinderEntropy}
\frac{S_{A}}{\log2} = \mathrm{Area} - L_x-L_y + 1.
\end{equation} We emphasize that the entanglement spectrum is flat, because all singular values are identical in Eq.~\eqref{eq.XcubeSVD}.

\subsection{Transfer Matrix as a Projector}\label{subsec.Xcube_TransferMatrix}

\begin{figure*}[t]
\centering
\includegraphics[width=0.7\textwidth]{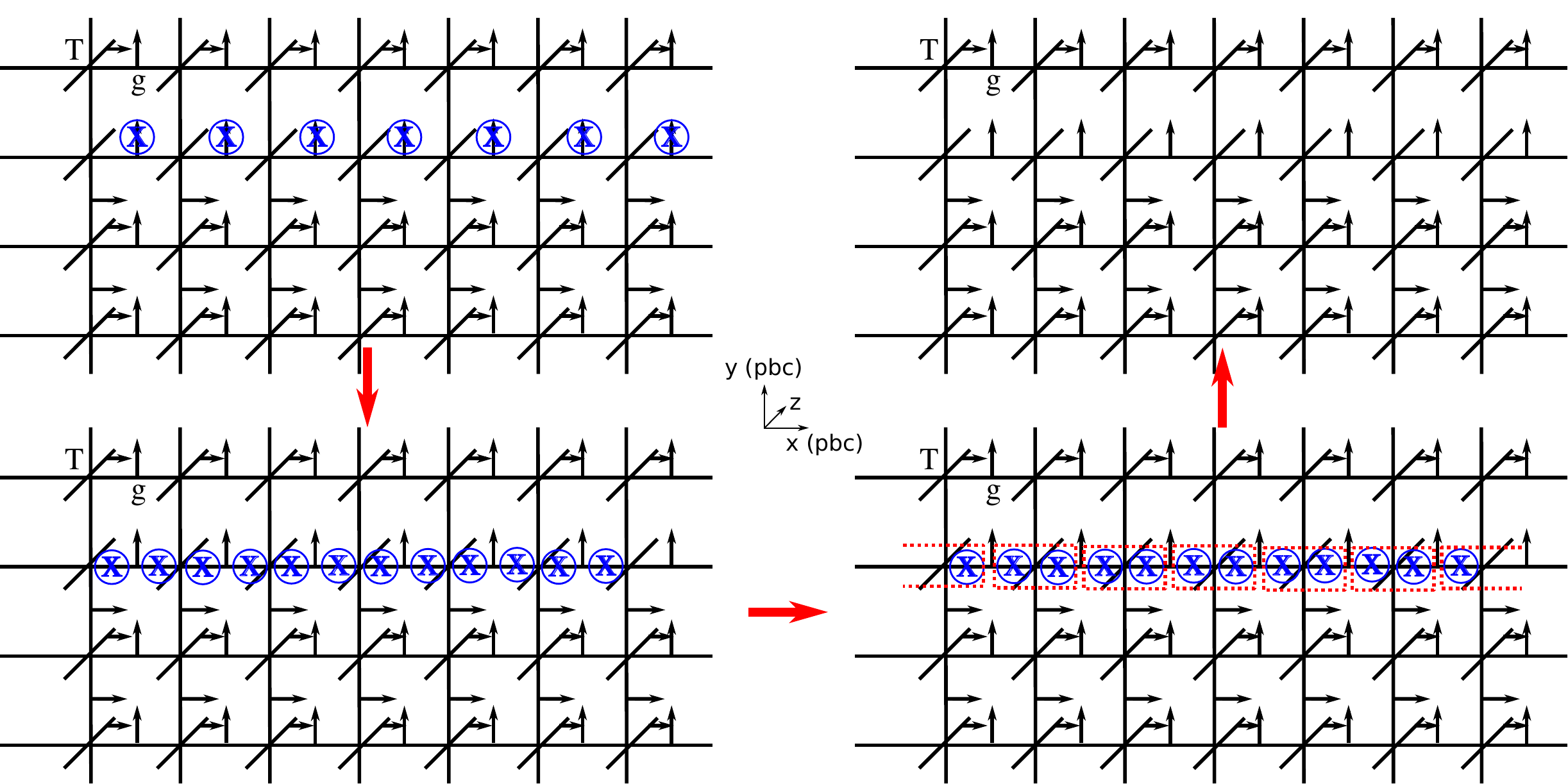}
\caption{Derivations for the first equation in Eq.~\eqref{eq.TNS_WX_eigenstate}. The rest two equations can be proved similarly. In the first step, the physical $X$ operators can be transferred to the virtual level by using Eq.~\eqref{eq.projectorcondition ToricCode}, and in the third step, all the virtual $X$ operators are exactly canceled in pairs (dashed red rectangles in the third figure) due to Eq.~\eqref{eq.XcubeStraightX}. }
\label{fig.TNS_WX_eigenstate}
\end{figure*}

Following the same reasoning explained in Sec.~\ref{subsec.TransferMatrix}, the transfer matrix of the X-cube model in the $xy$-plane is:
\begin{equation}
\mathrm{TM}_{xy} = \mathcal{C}^{\mathcal{T}^2_{xy}} \left( \ldots TTT \ldots \right)
\end{equation}
with open virtual indices in the $z$-direction. Graphically, see Eq.~\eqref{eq.TransferMatrixGraph} and Eq.~\eqref{eq.TransferMatrixIndex}.
In this section, we will show that the $\mathrm{TM}_{xy}$ for the X-cube model is also a projector. However, the projection is more complicated than in the 3D toric code example.
Using the transfer matrix basis $e_{\{\bar{z}\}}$ defined in Eq.~\eqref{eq.TMbasis}, we show that:
\begin{equation}\label{eq.XcubeTransferMatrixMultiplication}
\begin{split}
\mathrm{TM}_{xy} \cdot e_{\{\bar{z}\}} 
&= \sum_{\text{Concatenation Lemma}} \left(\mathrm{TM}_{xy}\right)_{\{z\},\{\bar{z}\}}	\\
&\propto \sum_{\text{Concatenation Lemma}} e_{\{z\}}	\\
\end{split}
\end{equation}
where the notation $\{z\}$ and $\{\bar{z}\}$ collectively denote all $z$ indices perpendicular to the $xy$-plane.
The summation with the Concatenation lemma means that:
\begin{equation}\label{eq.XcubeTMcondition}
\begin{split}
\sum_{i} z_{i,j} + \bar{z}_{i,j} = 0 \mod{2}, \quad\forall~ j	\\
\sum_{j} z_{i,j} + \bar{z}_{i,j} = 0 \mod{2}, \quad\forall~ i	\\
\end{split}
\end{equation}
where the subindex $i,j$ of $z_{i,j}$ denote the positions of $z_{i,j}$ in the $x$- and $y$-direction respectively.
These two equations mean that in each $xz$ and $yz$ plane, the indices have an even summation. For instance, the red dashed rectangles below:
\begin{equation}
\begin{gathered}
\includegraphics[width=\columnwidth]{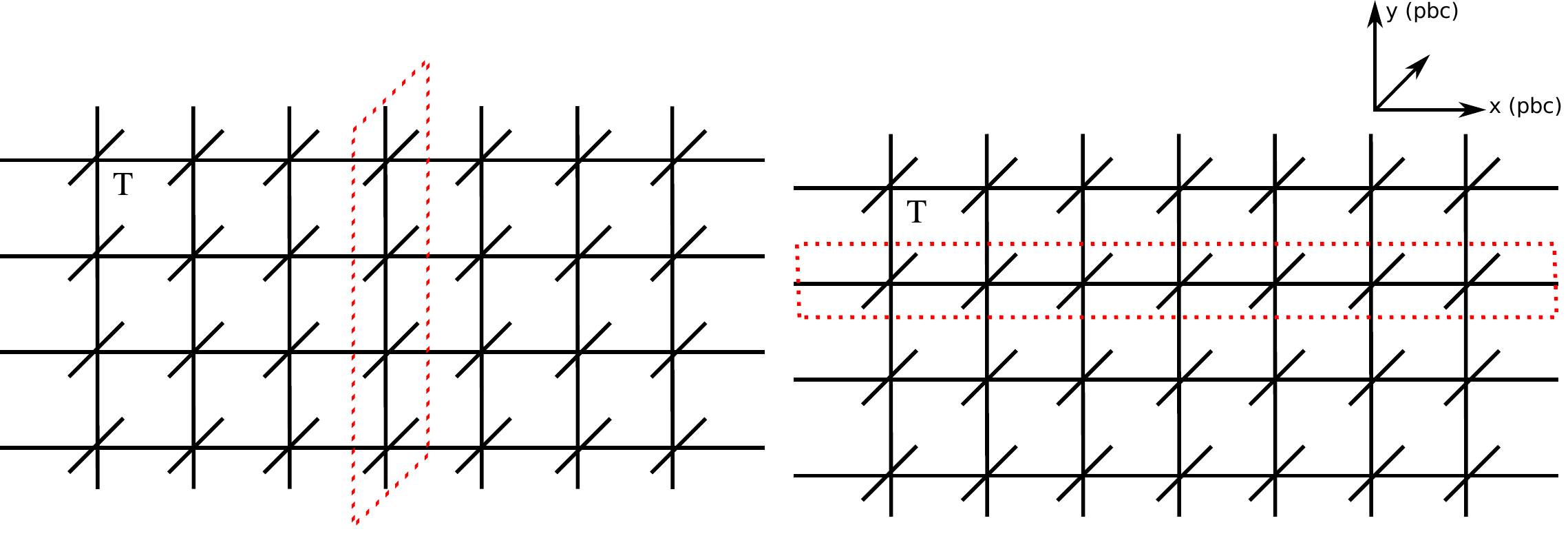}
\end{gathered}.
\end{equation}
Among the $L_x+L_y$ equations in Eq.~\eqref{eq.XcubeTMcondition}, only $L_x+L_y-1$ are linearly independent, because the summations of the two sets of constraints are the same:
\begin{equation}
\begin{split}
&\sum_{j}\left( \sum_{i} z_{i,j} + \bar{z}_{i,j} \right) = 0 \mod{2}	\\
\Leftrightarrow
&\sum_{i}\left( \sum_{j} z_{i,j} + \bar{z}_{i,j} \right) = 0 \mod{2}.	\\
\end{split}
\end{equation}
The summation in Eq.~\eqref{eq.XcubeTransferMatrixMultiplication} can be separated into $L_x+L_y-1$ number of different ``parity" sectors, similar to the 3D toric code case Eq.~\eqref{eq.ToricCodeTMeigenstate}. Hence, $\mathrm{TM}_{xy}$ is a projector of rank $2^{L_x+L_y-1}$. It has $2^{L_x+L_y-1}$ degenerate nonzero eigenvalues. 

\subsection{GSD and Transfer Matrix}\label{subsec.Xcube_GSD}

The TNS, which gives us the single ground state on $\mathcal{R}^3$, has the minimum energy by construction. If we contract these tensors on the torus $\mathcal{T}^3$ with periodic boundary conditions, then we still yield only one ground state. Moreover, this ground state is the $+1$ eigenstate of all $W_X$ operators in Eq.~\eqref{eq.operatorsXcube}:
\begin{equation}\label{eq.TNS_WX_eigenstate}
\begin{split}
W_{X}[C_x] \ket{\mathrm{TNS}}_{\mathcal{T}^3} &= \ket{\mathrm{TNS}}_{\mathcal{T}^3},	\\
W_{X}[C_y] \ket{\mathrm{TNS}}_{\mathcal{T}^3} &= \ket{\mathrm{TNS}}_{\mathcal{T}^3},	\\
W_{X}[C_z] \ket{\mathrm{TNS}}_{\mathcal{T}^3} &= \ket{\mathrm{TNS}}_{\mathcal{T}^3}.	\\
\end{split}
\end{equation}
These three equations can be proved by using Eq.~\eqref{eq.projectorcondition ToricCode} and Eq.~\eqref{eq.XcubeTtensor}; the derivations are  summarized in Fig.~\ref{fig.TNS_WX_eigenstate}.

Other ground states on the torus can be obtained by acting with the nonlocal operators $W_{Z}[\tilde{C}_{z,x}]$ and $W_{Z}[\tilde{C}_{z,y}]$ in Eq.~\eqref{eq.operatorsXcube} on the TNS $\ket{\mathrm{TNS}}_{\mathcal{T}^3}$. The physical operator $W_{Z}[\tilde{C}_{z,x}]$ and $W_{Z}[\tilde{C}_{z,y}]$ can be transferred to the virtual indices using Eq.~\eqref{eq.projectorcondition ToricCode}. 

\begin{figure*}[t]
\centering
\includegraphics[width=0.8\textwidth]{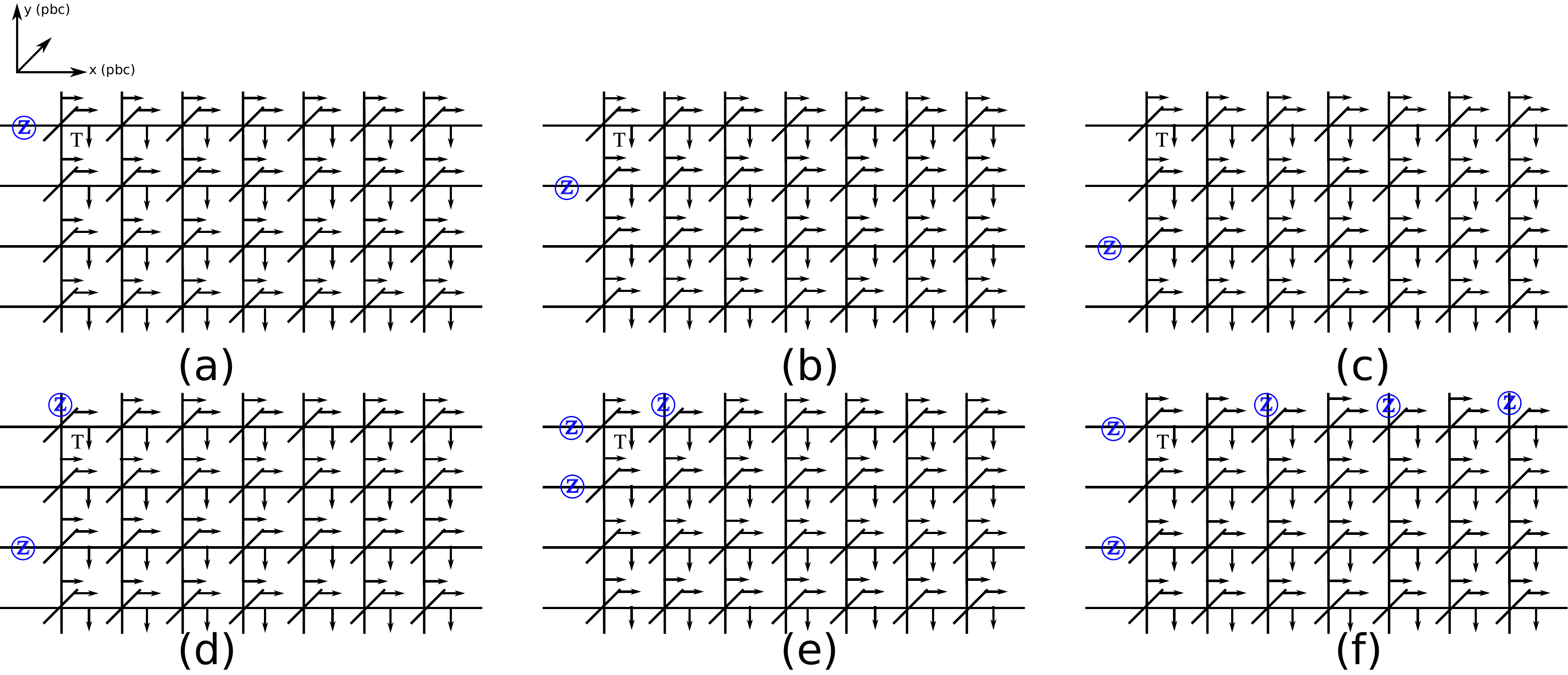}
\caption{Examples for the X-cube TNS in a $xy$-plane, obtained by acting $W_{Z}[\tilde{C}_{z,x}]$ and $W_{Z}[\tilde{C}_{z,y}]$ on the constructed TNS. The intersection of one $W_{Z}[\tilde{C}_{z,x}]$ operator and one $W_{Z}[\tilde{C}_{z,y}]$ operator with the $xy$-plane is only one Pauli $Z$ operator, i.e., the circled blue $Z$ in this figure.}
\label{fig.XcubeTwistedTM}
\end{figure*}

After applying $W_{Z}[\tilde{C}_{z,x}]$ and $W_{Z}[\tilde{C}_{z,y}]$ in Eq.~\eqref{eq.operatorsXcube} on TNS, we can generate $2^{L_x+L_y}$ TNSs exemplified by Fig.~\ref{fig.XcubeTwistedTM}. The intersections of $W_{Z}[\tilde{C}_{z,x}]$ and $W_{Z}[\tilde{C}_{z,y}]$ with the $xy$-plane are the blue circled $Z$ operators in Fig.~\ref{fig.XcubeTwistedTM}. We denote these planes of tensors in Fig.~\ref{fig.XcubeTwistedTM} as $\mathbf{T}_{xy}^{\vec{\alpha},\vec{\beta}}$ where $\vec{\alpha}$ and $\vec{\beta}$ are binary vectors (values in $\{0,1\}$) of length $L_x$ and $L_y$ representing the absence or presence of Pauli $Z$ operators. For instance, the untwisted plane of TNS is $\mathbf{T}_{xy}^{\vec{0},\vec{0}}$ using this convention.
This notation $\mathbf{T}_{xy}^{\vec{0},\vec{0}}$ is similar to that in Sec.~\ref{subsec.ToricCode_GSD}.
Inserting a $Z$ operator at the virtual level will change the holonomy of $W_X$ in the $xy$-plane. For instance, for Panel (a) in Fig.~\ref{fig.XcubeTwistedTM}, the $W_{X}$ operator along the first row has a $-1$ eigenvalue, while the $W_{X}$ operators along the second, the third and the fourth row have a $+1$ eigenvalue. 

Each of these $\mathbf{T}_{xy}^{\vec{\alpha},\vec{\beta}}$ will generate a transfer matrix $\mathrm{TM}_{xy}^{\vec{\alpha},\vec{\beta}}$ which has $2^{L_x+L_y-1}$ degenerate eigenvalues. 
The reasons are that (1) when building the transfer matrices from the twisted $\mathbf{T}_{xy}^{\vec{\alpha},\vec{\beta}}$, the contraction over the physical indices of the projector $g$ tensors makes the virtual indices from the bra layer and ket layer identical; (2) the $Z$ operators in the bra layer cancel their analogues respectively in the ket layer. Hence, the transfer matrices $\mathbf{TM}_{xy}^{\vec{\alpha},\vec{\beta}}$ obtained from the twisted $\mathbf{T}_{xy}^{\vec{\alpha},\vec{\beta}}$ are equal to that obtained from the untwisted one $\mathbf{T}_{xy}^{\vec{0},\vec{0}}$. 
\begin{equation}
\mathrm{TM}_{xy}^{\vec{\alpha},\vec{\beta}} = \mathrm{TM}_{xy}^{\vec{0},\vec{0}}, 	\quad \forall\; \vec{\alpha}, \vec{\beta}
\end{equation}
Thus, the transfer matrices $\mathrm{TM}_{xy}^{\vec{\alpha},\vec{\beta}}$ built from the twisted $\mathbf{T}_{xy}^{\vec{\alpha},\vec{\beta}}$ are also projectors of dimension $2^{L_x+L_y-1}$.

\begin{figure*}[t]
\centering
\includegraphics[width=0.6\textwidth]{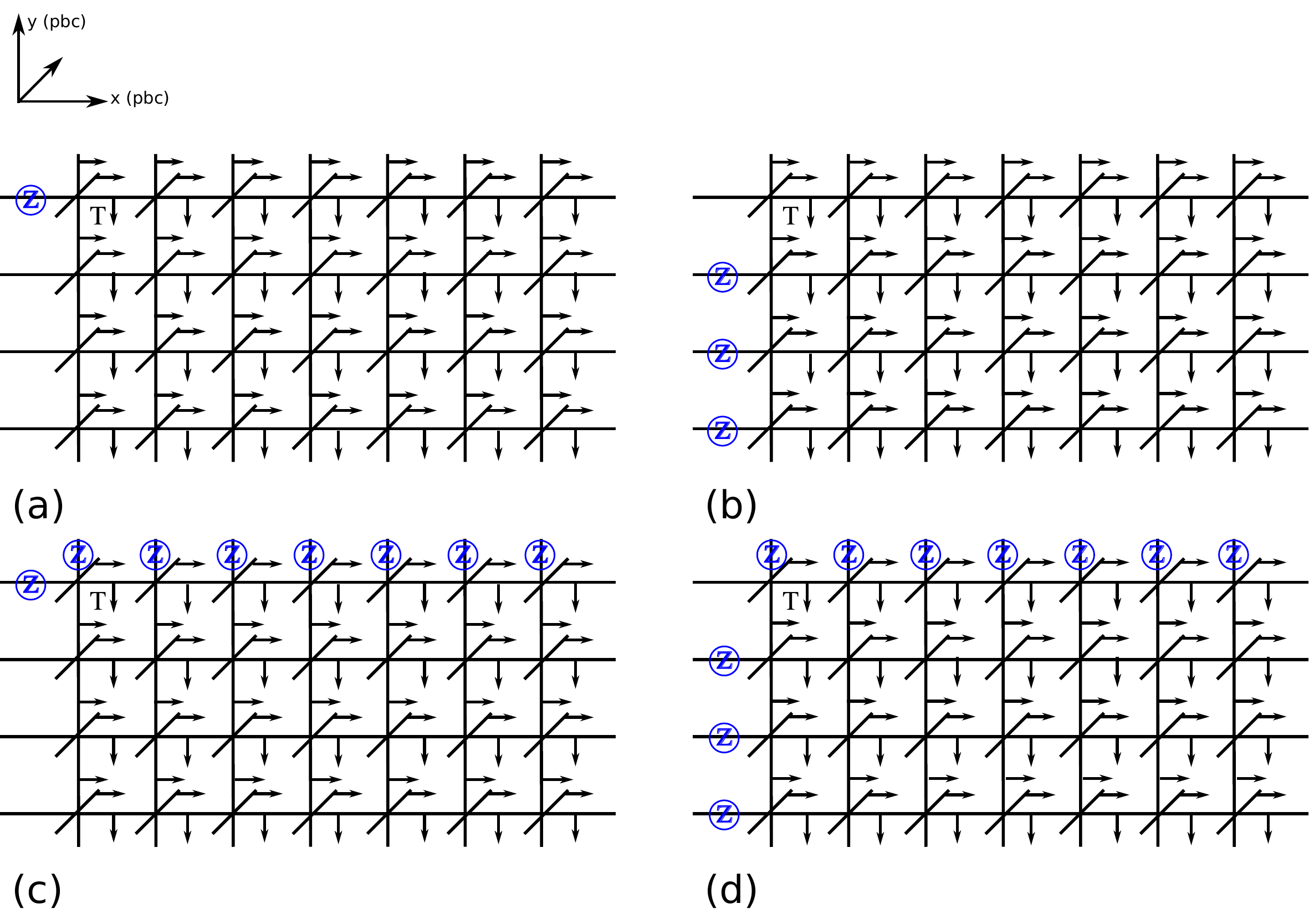}
\caption{We act $W_{Z}[\tilde{C}_{x,y}]$ and $W_{Z}[\tilde{C}_{y,x}]$ operators on Panel (a) in Fig.~\ref{fig.XcubeTwistedTM}. Hence, we have four TNSs in this $xy$-plane that can be related to each other. See the text for the explanations.}
\label{fig.twistedTNS}
\end{figure*}

The TNS on torus is then constructed as the contraction of $\mathbf{T}_{xy}^{\vec{\alpha},\vec{\beta}}$ on each $xy$-plane.
However, the subtlety of the X-cube model is that in constructing TNS on the torus, there are still degree of freedom we can play with. In the 3D toric code case Sec.~\ref{subsec.ToricCode_GSD}, as we evolve the state in the $z$ direction,  we have to use the same $\mathbf{T}_{xy}^{\alpha,\beta}$ in each plane, otherwise the corresponding wave function is no longer a ground state. However, we have more choices for the X-cube model. In each plane of $\mathbf{T}_{xy}^{\vec{\alpha},\vec{\beta}}$, we have four choices that do not change the energy: we can still act $W_{Z}[\tilde{C}_{x,y}]$ and $W_{Z}[\tilde{C}_{y,x}]$ on TNS in the $xy$-plane without affecting other $xy$-planes. These choices do not raise the energy because $W_{Z}[\tilde{C}_{x,y}]$ and $W_{Z}[\tilde{C}_{y,x}]$ are the nonlocal operators of the X-cube model Eq.~\eqref{eq.operatorsXcube}: they do not cost any energy.
For each of the $\mathbf{T}_{xy}^{\vec{\alpha},\vec{\beta}}$ built from Fig.~\ref{fig.XcubeTwistedTM}, we can find 3 others: all $4$ planes of tensors $\mathbf{T}_{xy}^{\vec{\alpha},\vec{\beta}}$ can be used in the $z$ direction.  Take Panel (a) of Fig.~\ref{fig.XcubeTwistedTM} as an example at one point in the $z$ direction. The 4 $\mathbf{T}_{xy}^{\vec{\alpha},\vec{\beta}}$ are depicted in Fig.~\ref{fig.twistedTNS}. Their expressions are:
\begin{enumerate}
\item For Panel (a), we do not apply any operators.
\item For Panel (b), we apply $W_{Z}[\tilde{C}_{y,x}]$ on TNS.
\item For Panel (c), we apply $W_{Z}[\tilde{C}_{x,y}]$ on TNS.
\item For Panel (d), we apply both $W_{Z}[\tilde{C}_{x,y}]$ and $W_{Z}[\tilde{C}_{y,x}]$ on TNS.
\end{enumerate}
Due to this choice, four twisted $\mathbf{T}_{xy}^{\vec{\alpha},\vec{\beta}}$ will be grouped together, and there are 
\begin{equation}\label{eq.ambiguity}
\frac{2^{L_x+L_y} }{4}
\end{equation} 
number of groups of twisted $\mathbf{T}_{xy}^{\vec{\alpha},\vec{\beta}}$. 
Hence, the total GSD that we can obtain from the transfer matrices built from $\mathbf{T}_{xy}^{\vec{\alpha},\vec{\beta}}$ is:
\begin{equation}
2^{L_x+L_y-1}  \times \frac{2^{L_x+L_y} }{4}\times 4^{L_z} = 2^{2L_x+2L_y+2L_z-3}.
\end{equation}
Each factor in the above formula has an explanation:
\begin{enumerate}
\item $2^{L_x+L_y-1}$ is the degeneracy of each transfer matrix.
\item $\frac{2^{L_x+L_y} }{4}$ is the number of groups of the twisted $\mathbf{T}_{xy}^{\vec{\alpha},\vec{\beta}}$ due to the ``ambiguity" explained in the paragraph before Eq.~\eqref{eq.ambiguity}.
\item $4^{L_z}$ is the number of combinations for $\mathbf{T}_{xy}^{\vec{\alpha},\vec{\beta}}$, since for each $xy$ plane we can pick any of the 4 $\mathbf{T}_{xy}^{\vec{\alpha},\vec{\beta}}$ belonging to the same group.
\end{enumerate}
This is the total GSD of X-cube model on torus.

\section{Haah Code}\label{sec.Haah}

In this section, we derive the TNS for the Haah code following a similar prescription as that in Sec.~\ref{subsec.ToricCode_TNS} and Sec.~\ref{subsec.Xcube_TNS}. We then compute the entanglement entropies using the TNS for several types of entanglement cuts. This section is organized as follows. In Sec.~\ref{subsec.HaahCode}, we review the Hamiltonian of the Haah code. In Sec.~\ref{subsec.TNS_Haah}, we present the construction of TNS for the Haah code. In Sec.~\ref{subsec:HaahcodeSVDCuts}, we discuss the entanglement cuts for which the TNS is an exact SVD. In Sec.~\ref{subsec.Haah_Entanglement}, we discuss the cubic entanglement cut, where the TNS is not an exact SVD. The calculation of the entanglement entropies proceeds in the same way as that of the 3D toric code model or X-cube model: one counts the number of constraints for open indices. 

\subsection{Hamiltonian of Haah code}\label{subsec.HaahCode}

The Haah code is defined on a cubic lattice. As opposed to the 3D toric code and the X-cube model discussed in Sec.~\ref{sec.toriccode} and \ref{sec.Xcube}, there are two spin-$1/2$'s defined on each \emph{vertex} of a cubic lattice. The Hamiltonian of the Haah code is a sum of commuting operators where each term is the product of Pauli $X$ or $Z$ operators. Specifically, there are two types of the Hamiltonian terms:
\begin{eqnarray}
	H= -\sum_{a,b,c}A_{abc}-\sum_{a,b,c} B_{abc}.
\end{eqnarray}
The $A$ and $B$ operators are defined on each cube in the cubic lattice, and the indices $a,b,c$ represent the vertex coordinates. If we choose the space to be $\mathcal{R}^3$, then $a, b, c\in \mathbb{Z}$. If we choose the space to be a 3D torus of the size $L_x\times L_y\times L_z$ with periodic boundary condition on each side, then $a\in \mathbb{Z}_{L_x}$, $b\in \mathbb{Z}_{L_y}$ and $c\in \mathbb{Z}_{L_z}$. The operators defined on $a=0, b=0, c=0$ are
\begin{eqnarray}
	\begin{split}
		A_{000}&=Z^{L}_{110}Z^L_{101}Z^L_{011}Z^{L}_{111}Z^{R}_{100}Z^R_{010}Z^R_{001}Z^{R}_{111}\\
		B_{000}&=X^{L}_{000}X^L_{110}X^L_{101}X^L_{011}X^R_{000}X^R_{100}X^R_{010}X^R_{001}.
	\end{split}
\end{eqnarray}
The superscripts $L/R$ represent the left or the right spin on a vertex where the Pauli operators act on. The subscripts $(ijk)\in \mathbb{Z}_2\times\mathbb{Z}_2\times \mathbb{Z}_2$ represent the coordinate of vertices (on a cube). The operators $A_{abc}$ and $B_{abc}$ on all other cubes can be obtained by translation. Pictorially the two types of terms are:
\begin{equation}\label{HaahHamiltonian}
	\begin{gathered}
		\includegraphics[width=0.8\columnwidth]{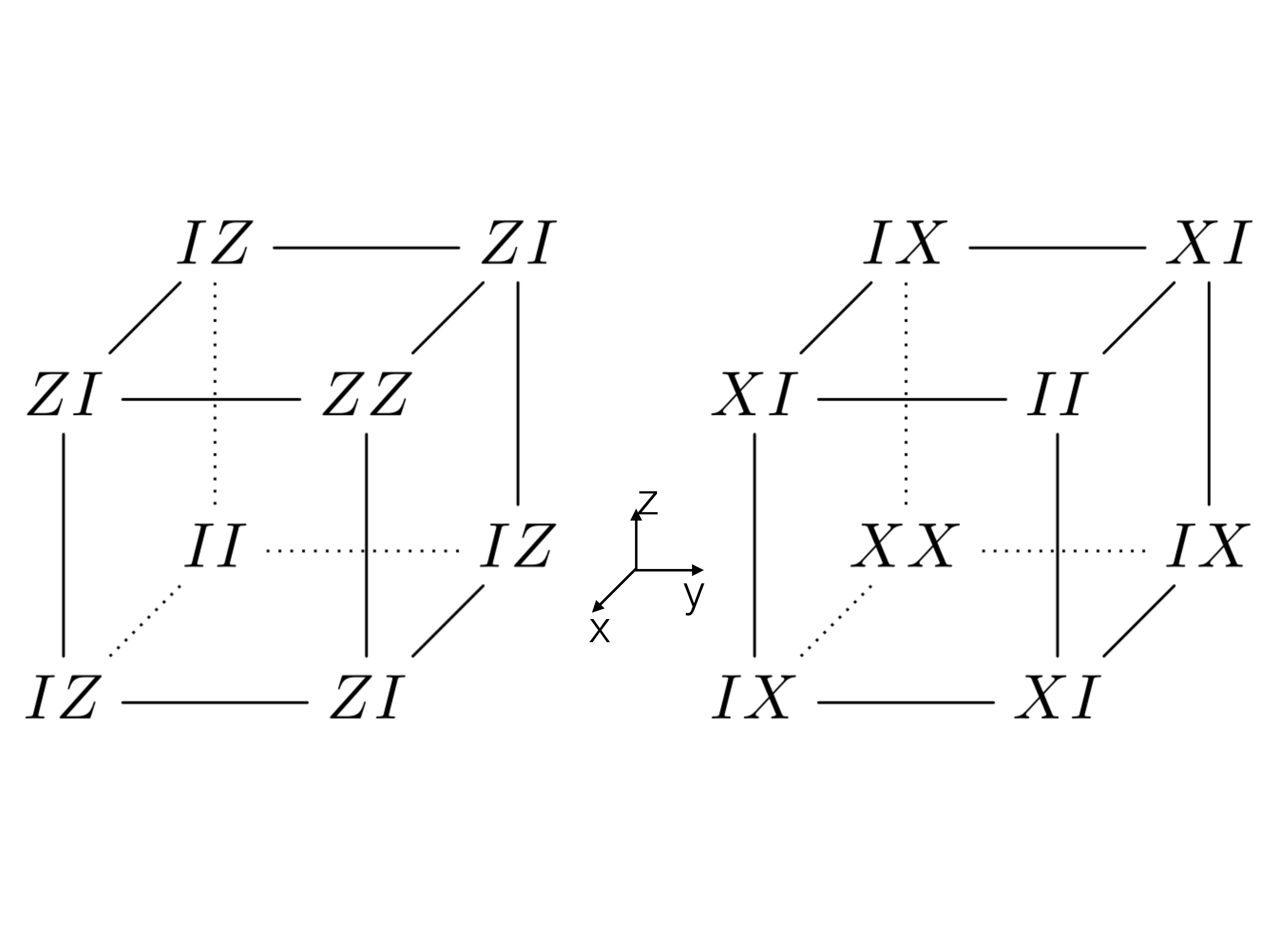}
	\end{gathered}
\end{equation}
It is straightforward to check that all the Hamiltonian terms commute. 

\subsection{TNS for Haah Code}\label{subsec.TNS_Haah}

\begin{figure*}[t]
	\centering
	\includegraphics[width=0.6\columnwidth]{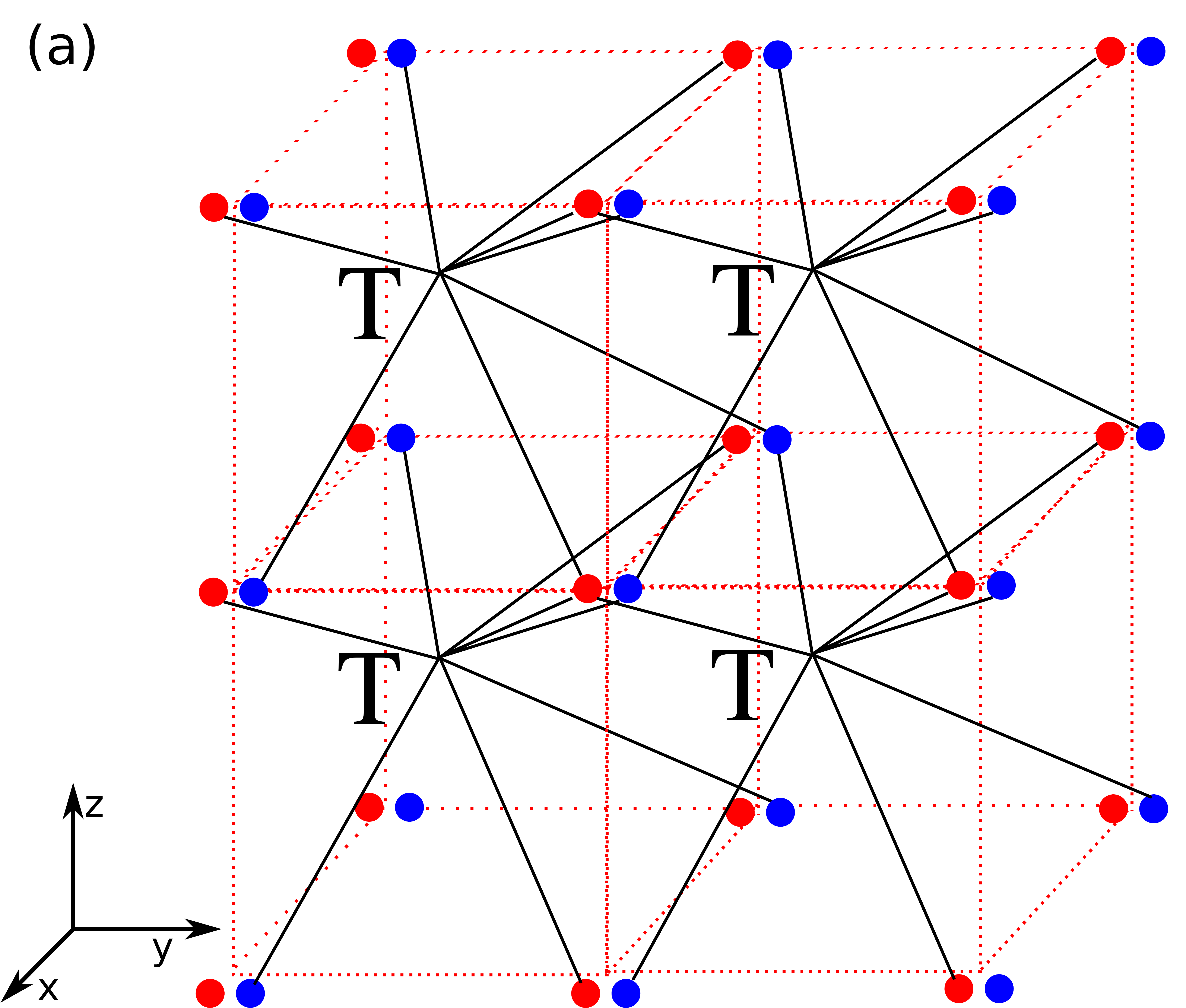}
	\includegraphics[width=0.5\columnwidth]{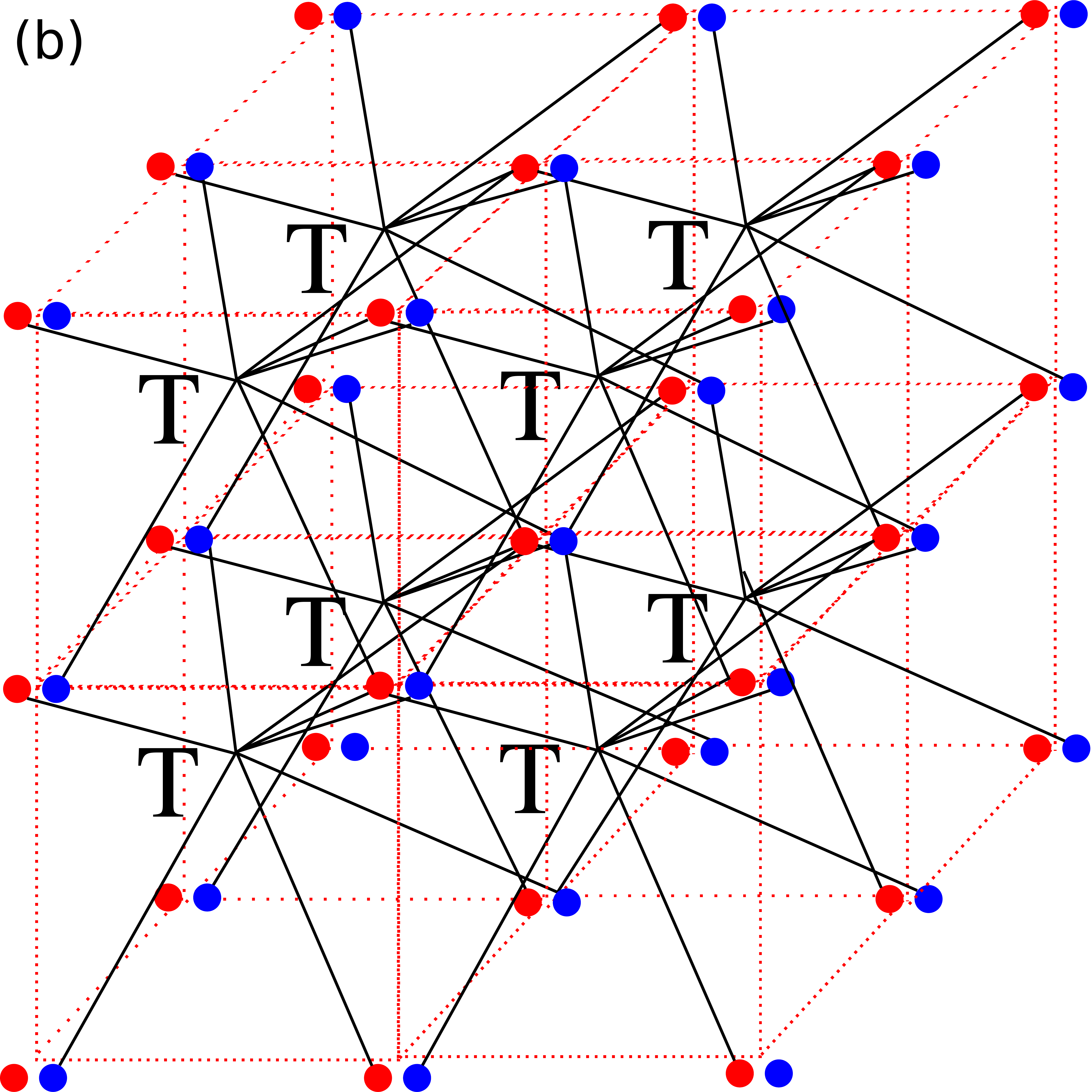}
	\caption{Tensor contraction for the Haah code TNS. (a) The lattice size is $2 \times 3 \times 3$. (b) The lattice size is $3 \times 3 \times 3$ }
	\label{fig.HaahTNS}
\end{figure*}

The ground state $|\mathrm{\GS}\rangle$ is obtained by requiring
\begin{eqnarray}
	A_{abc}|\GS\rangle=|\GS\rangle\label{AconditionHaah}	\\
	B_{abc}|\GS\rangle=|\GS\rangle\label{BconditionHaah}
\end{eqnarray}
for every $a,b,c$. We can solve these two equations similarly to the 3D toric code model in Sec.~\ref{subsec.ToricCode_TNS} and the X-cube model in Sec.~\ref{subsec.Xcube_TNS} to obtain a TNS representation. 
We now specify the projector $g$ tensor and the local $T$ tensor. 

There are 2 types of $g$ tensors $g^{L}$ and $g^{R}$ associated with the left and right physical spins on each vertex. Each $g$ tensor has 1 physical index $s$ and 4 virtual indices $i,j,k,l$. The reason for these 4 virtual indices (rather than 2 virtual indices as in the toric code and the X-cube examples) is that, for each vertex, the virtual indices from $T$ tensors (to be defined below) in the neighboring 8 octants need to be fully contracted; this requires the $g$ tensor to have $4$ virtual indices. The index assignment of the left and right $g$ tensors, $g^{Ls}_{ijkl}$ and $g^{Rs}_{ijkl}$, are:
\begin{equation}\label{eq.projector4L}
	\begin{split}
		&g^{Ls}_{ijkl}=	\begin{gathered}
			\includegraphics[width=0.65\columnwidth]{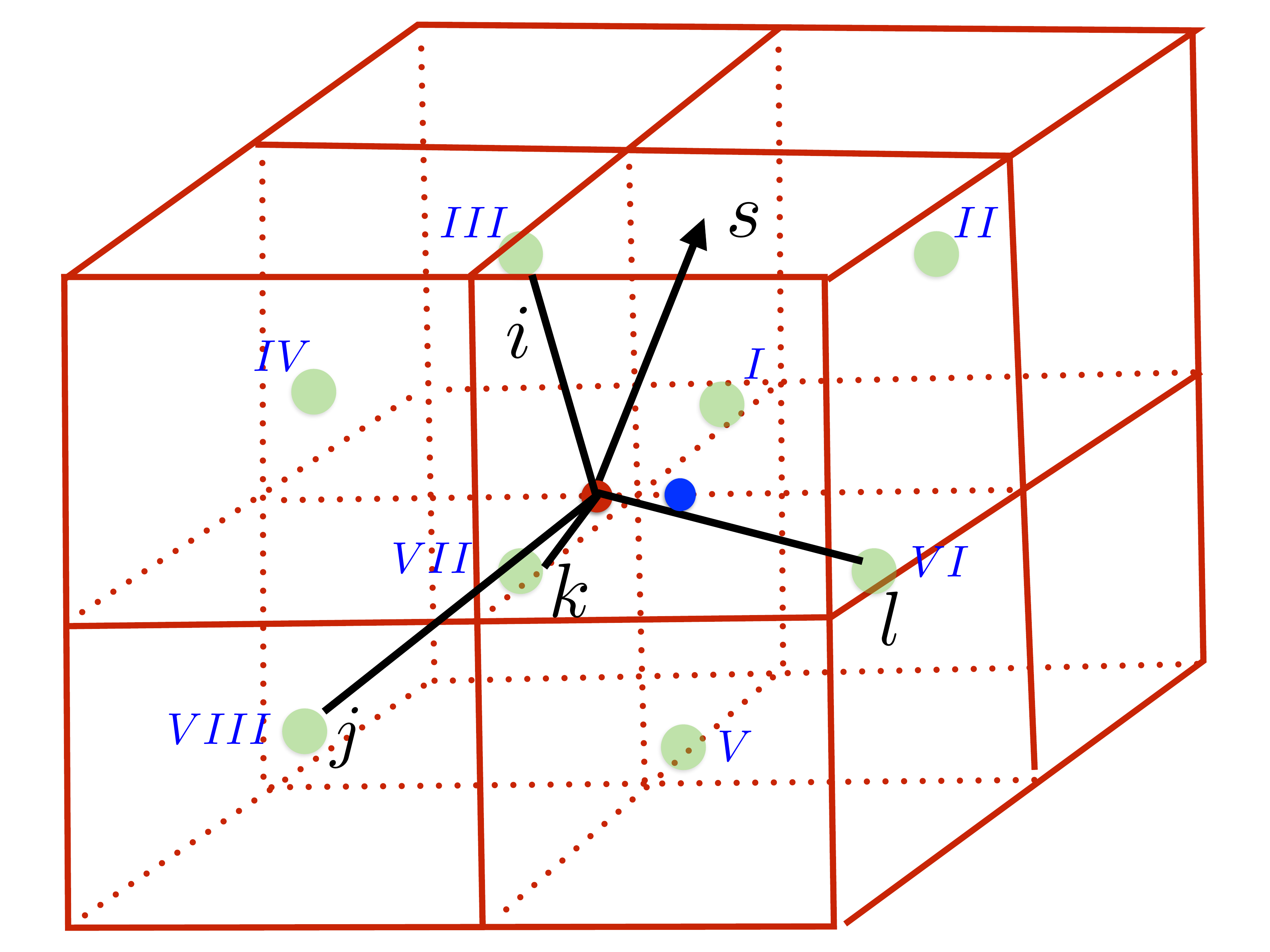}
		\end{gathered}
	\end{split}
\end{equation}
and 
\begin{equation}\label{eq.projector4R}
	\begin{split}
		&g^{Rs}_{ijkl}=\begin{gathered}
			\includegraphics[width=0.65\columnwidth]{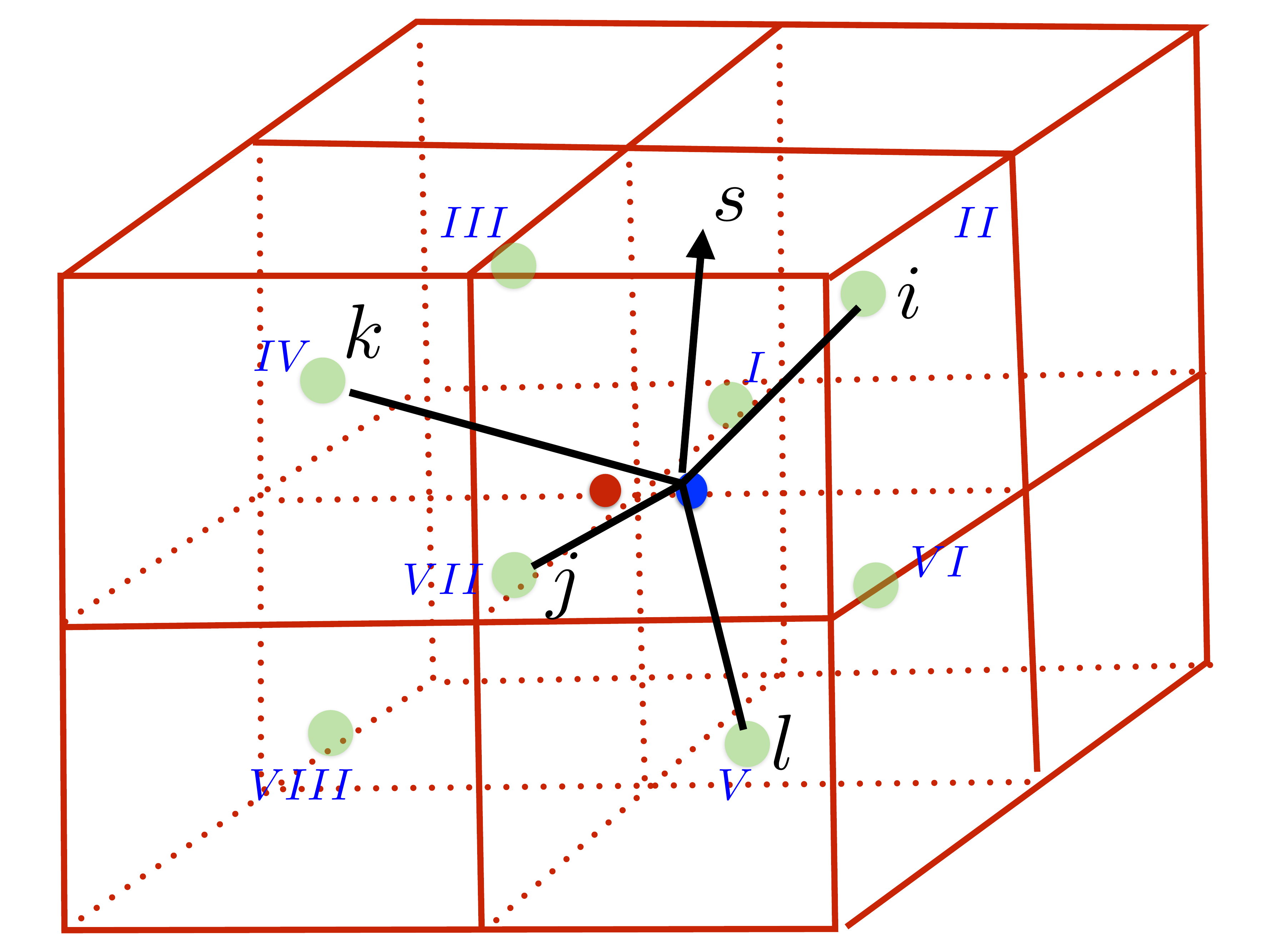}
		\end{gathered}
	\end{split}
\end{equation} where $s$ is the physical index in $\{\ket{0}=\ket{\mathord{\uparrow}},\ket{1}=\ket{\mathord{\downarrow}}\}$, and $ijkl$ are virtual indices. We use a blue dot for the $g$ tensor on the right spin and a red dot for the $g$ tensor on the left spin. The green dots at the center of each cube represent $T$ tensors (which we define below). Similar to the toric code model and the X-cube model, we require that the $g$ tensor acts as a projector from the physical index to the four virtual indices: 
\begin{equation}\label{Haahgtensorprojectioncond}
	\begin{split}
		g^{Ls}_{ijkl} &= \begin{cases}
			1	&	i=j=k=l=s	\\
			0	&	\text{otherwise}
		\end{cases},	\\
		g^{Rs}_{ijkl} &= \begin{cases}
			1	&	i=j=k=l=s	\\
			0	&	\text{otherwise}
		\end{cases}.
	\end{split}
\end{equation}
The four virtual indices of $g^{Ls}_{ijkl}$ extend along the III, VIII, VII, VI octants (as shown in Eq.~\eqref{eq.projector4L}), and the four virtual indices of $g^{Rs}_{ijkl}$ extend along the II, VII, IV, V octants (as shown in Eq.~\eqref{eq.projector4R}). 

The tensor $T_{\{i\}}$ is defined at the center of each cube, and every $T$ tensor has 8 virtual indices. Graphically, the $T$ tensor is:
\begin{equation}\label{HaahTtensor}
	\begin{split}
		&T_{i_1i_2i_3i_4i_5i_6i_7i_8}=
		\begin{gathered}
			\includegraphics[width=0.35\columnwidth]{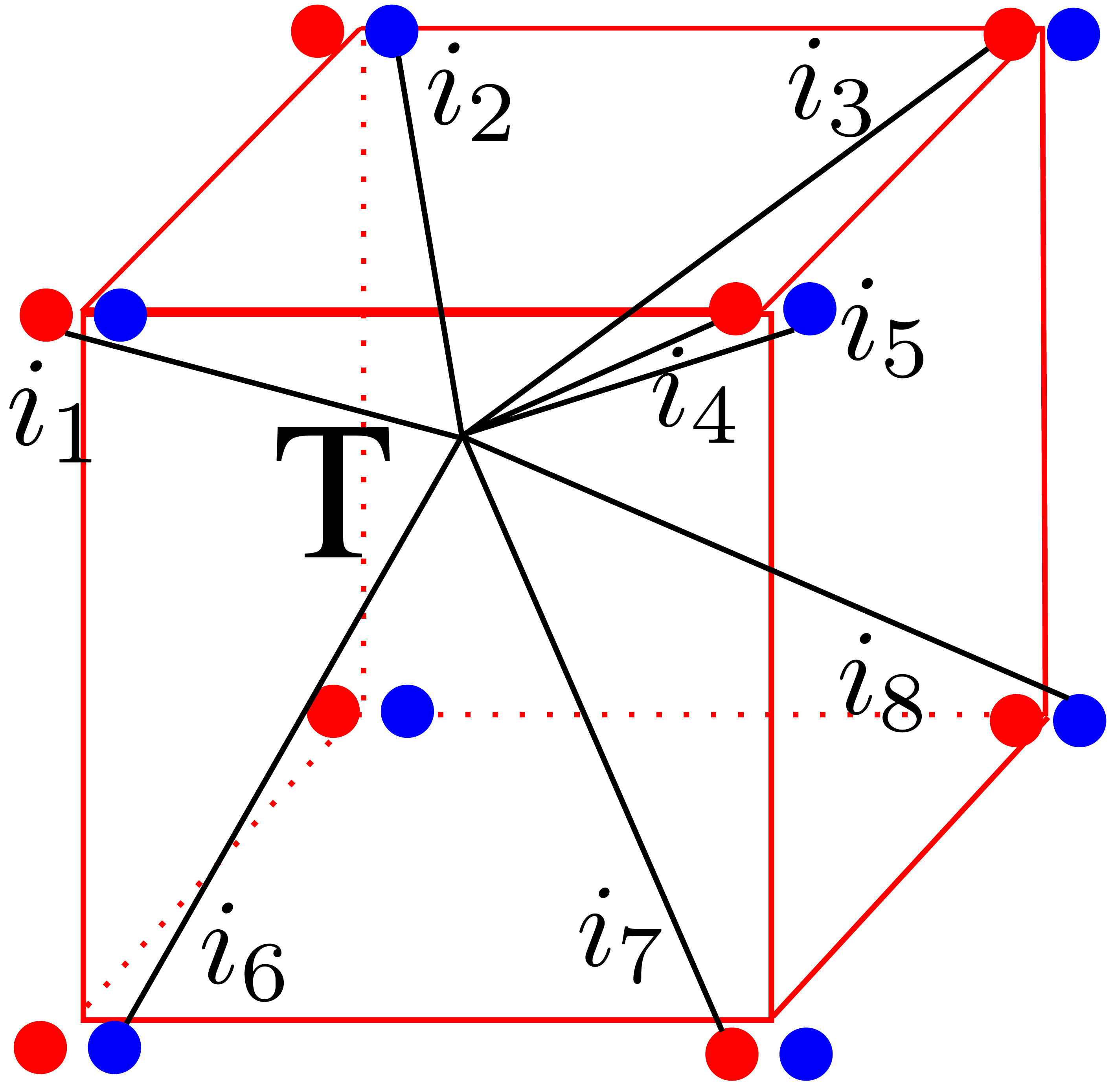}
		\end{gathered}.
	\end{split}
\end{equation}
The $T$ tensor is contracted to 8 of the total 16 (8 vertices times 2 degrees of freedom per vertex) $g$ tensors located at the cube corners via the virtual indices. The reason for only 8 virtual indices (instead of 16 virtual indices) in the $T$ tensor is that among 16 spins around the cube $(a,b,c)$ only eight of them are addressed by the Pauli $Z$ operators in the $A_{abc}$ term of the Hamiltonian. The elements of the $T$ tensor for a given set of virtual indices $i_1i_2i_3i_4i_5i_6i_7i_8$ are determined by solving Eq.~\eqref{AconditionHaah} and Eq.~\eqref{BconditionHaah}. Imposing the condition Eq.~\eqref{AconditionHaah} and transferring the physical $Z$ operators to the virtual level, we find that:
\begin{equation}
\begin{gathered}
\includegraphics[width=0.35\columnwidth]{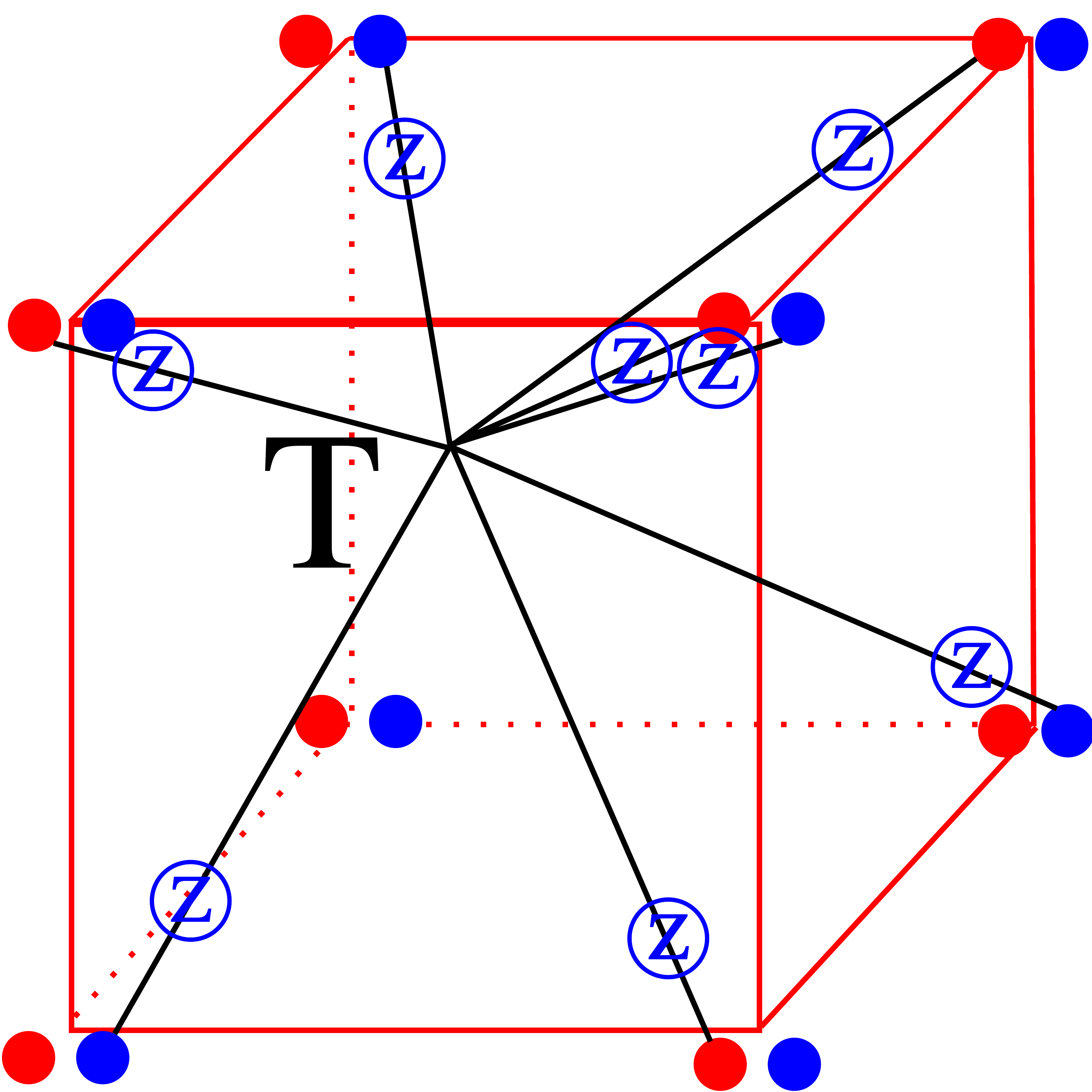}
\end{gathered}
=
\begin{gathered}
\includegraphics[width=0.35\columnwidth]{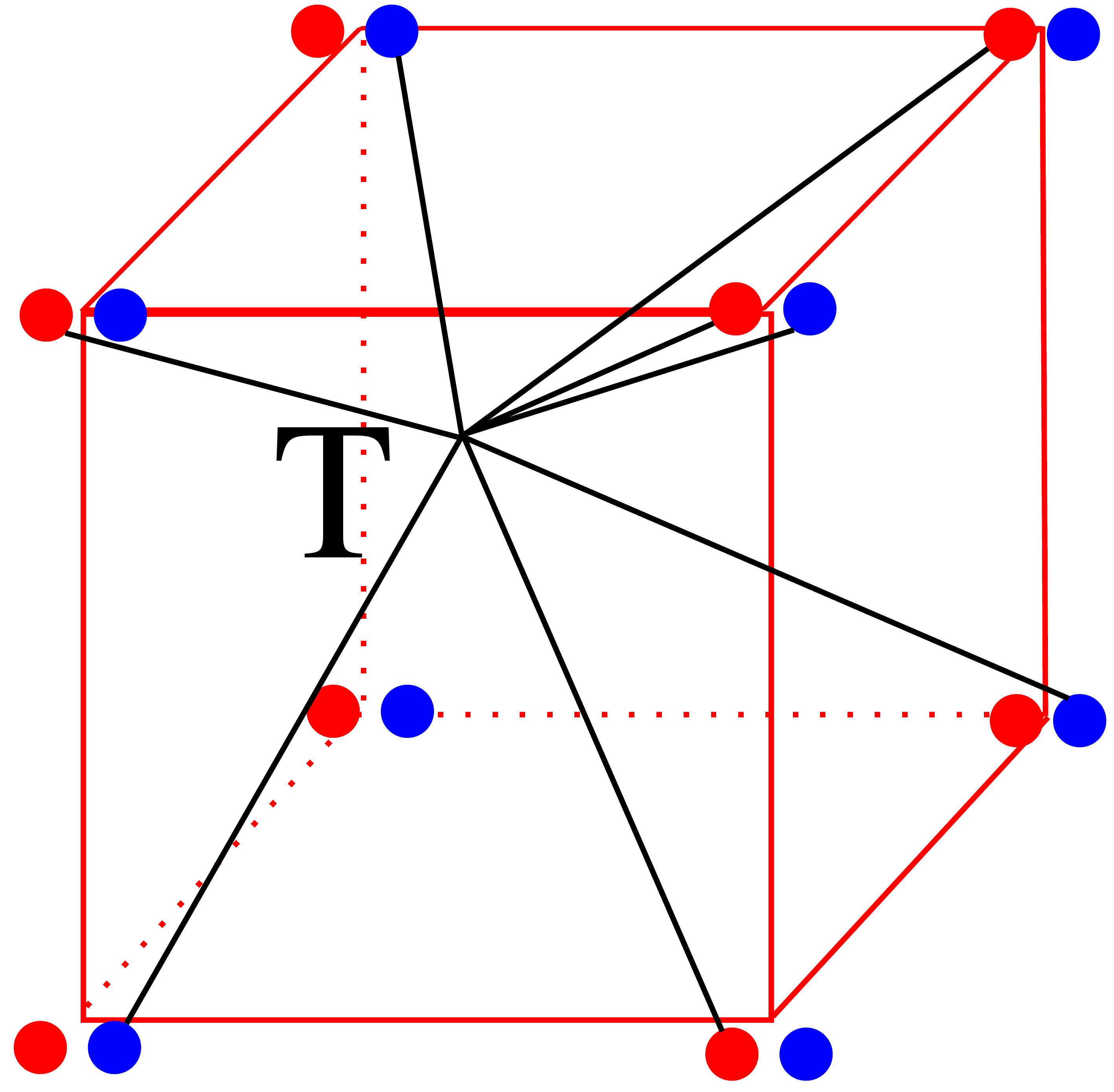}
\end{gathered}
\end{equation}
which amounts to 
\begin{equation}\label{HaahTtensorconstraint}
	T_{i_1i_2i_3i_4i_5i_6i_7i_8}= (-1)^{\sum_{n=1}^{8} i_n} T_{i_1i_2i_3i_4i_5i_6i_7i_8},
\end{equation}
where $i_1, \cdots, i_8$ are the eight virtual indices of the $T$ tensor defined in Eq.~\eqref{HaahTtensor}. 
Hence, 
\begin{equation}
	\begin{split}
		T_{i_1i_2i_3i_4i_5i_6i_7i_8}=0,\;\mathrm{if}\; \sum_{n=1}^{8}i_n=1\mod 2.
	\end{split}
\end{equation} 
Imposing the condition Eq.~\eqref{BconditionHaah} and transferring the physical $X$ operators to the virtual level, we find that 
\begin{equation}\label{HaahXtensorconstruction}
\begin{split}
&\begin{gathered}
\includegraphics[width=0.25\columnwidth]{figures/HaahT_noindex.pdf}
\end{gathered}
=
\begin{gathered}
\includegraphics[width=0.25\columnwidth]{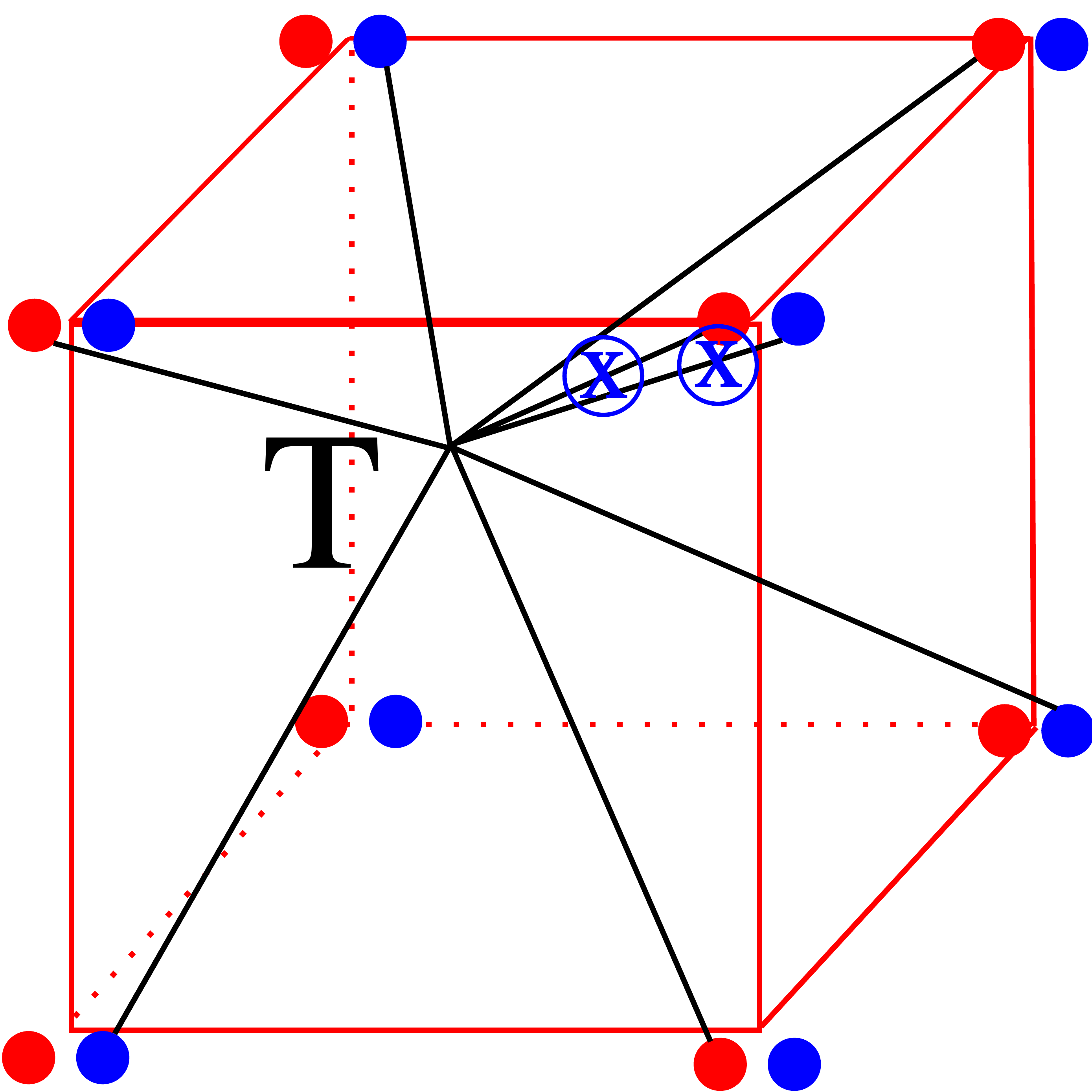}
\end{gathered}
=
\begin{gathered}
	\includegraphics[width=0.25\columnwidth]{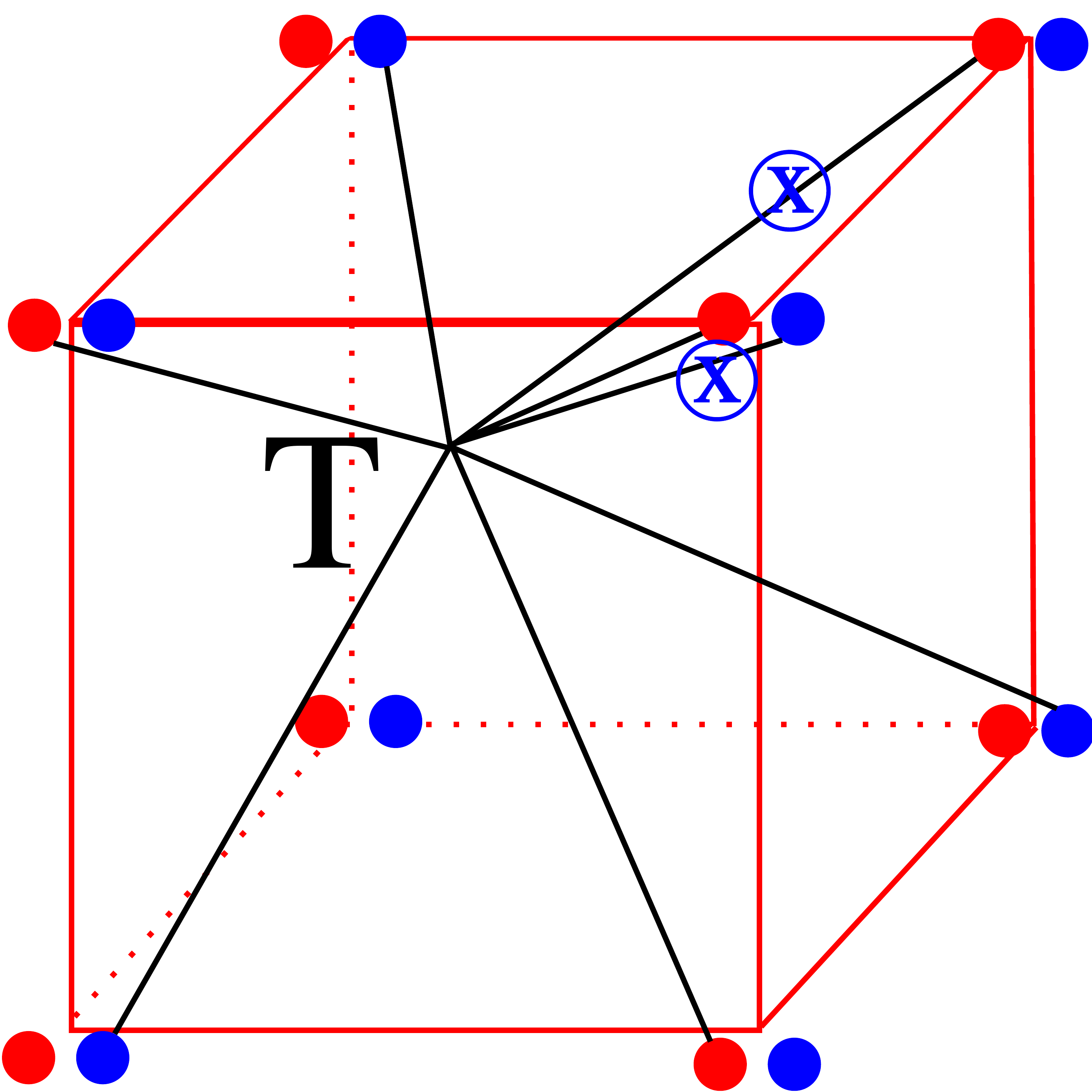}
\end{gathered}\\
=&
\begin{gathered}
\includegraphics[width=0.25\columnwidth]{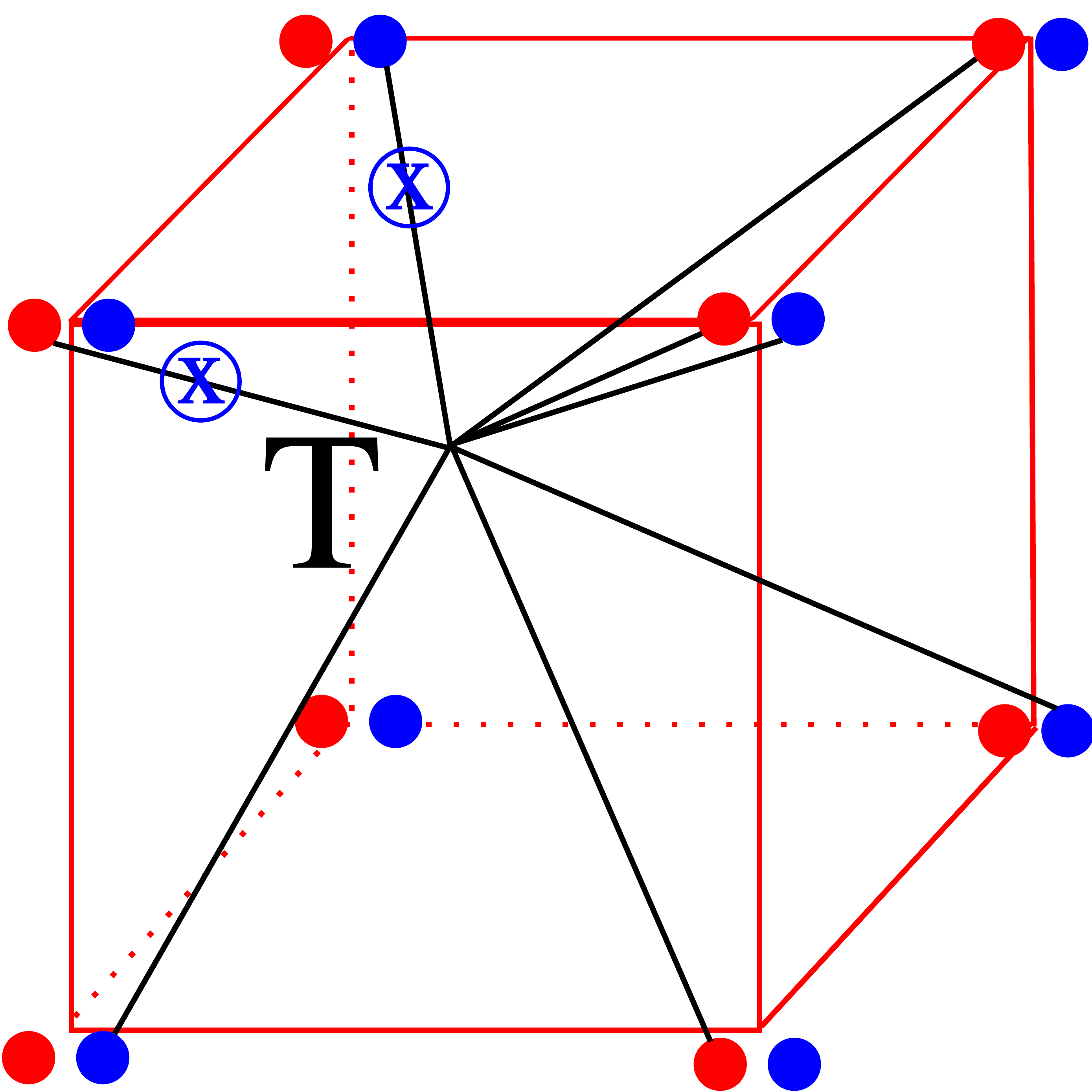}
\end{gathered}
=
\begin{gathered}
\includegraphics[width=0.25\columnwidth]{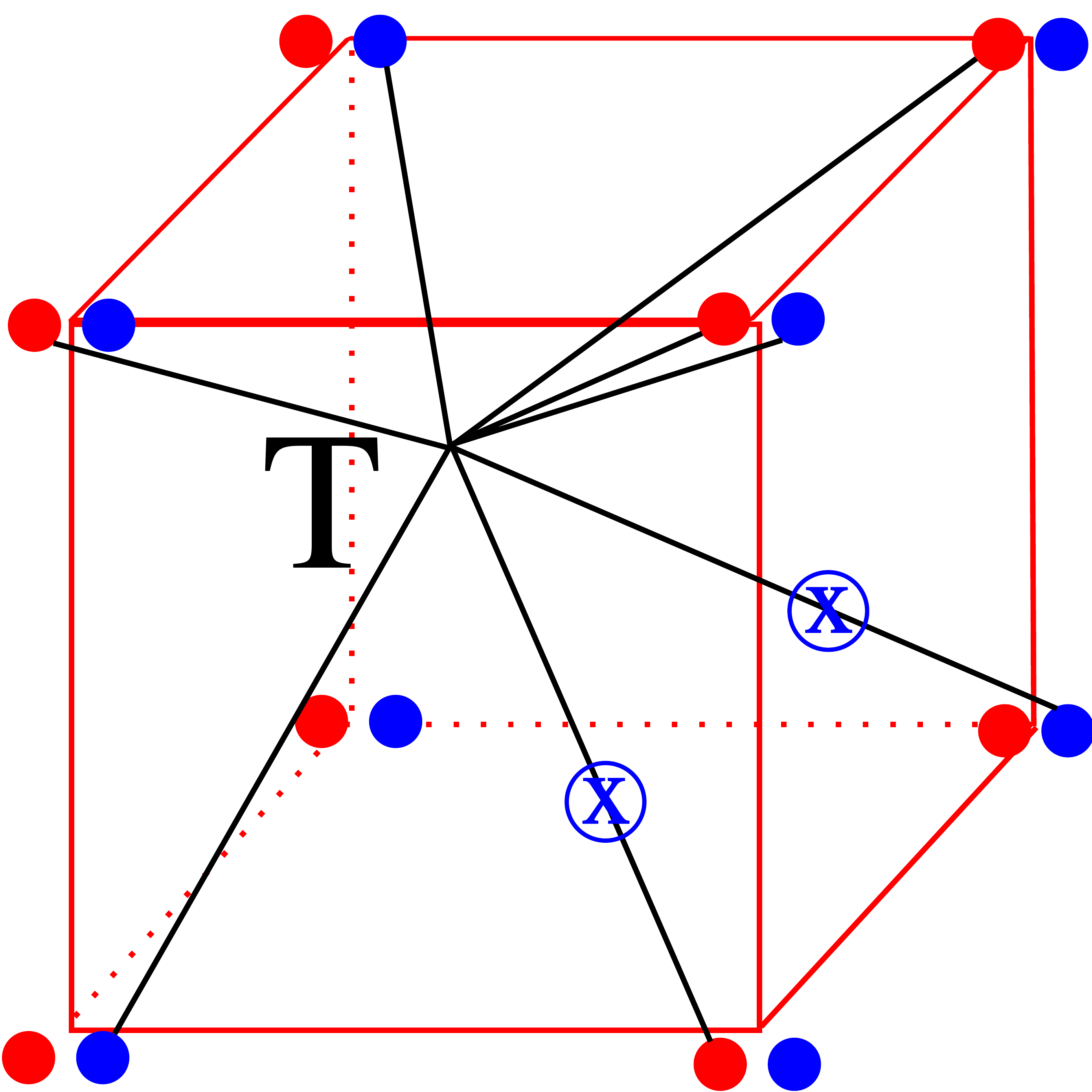}
\end{gathered}
=
\begin{gathered}
\includegraphics[width=0.25\columnwidth]{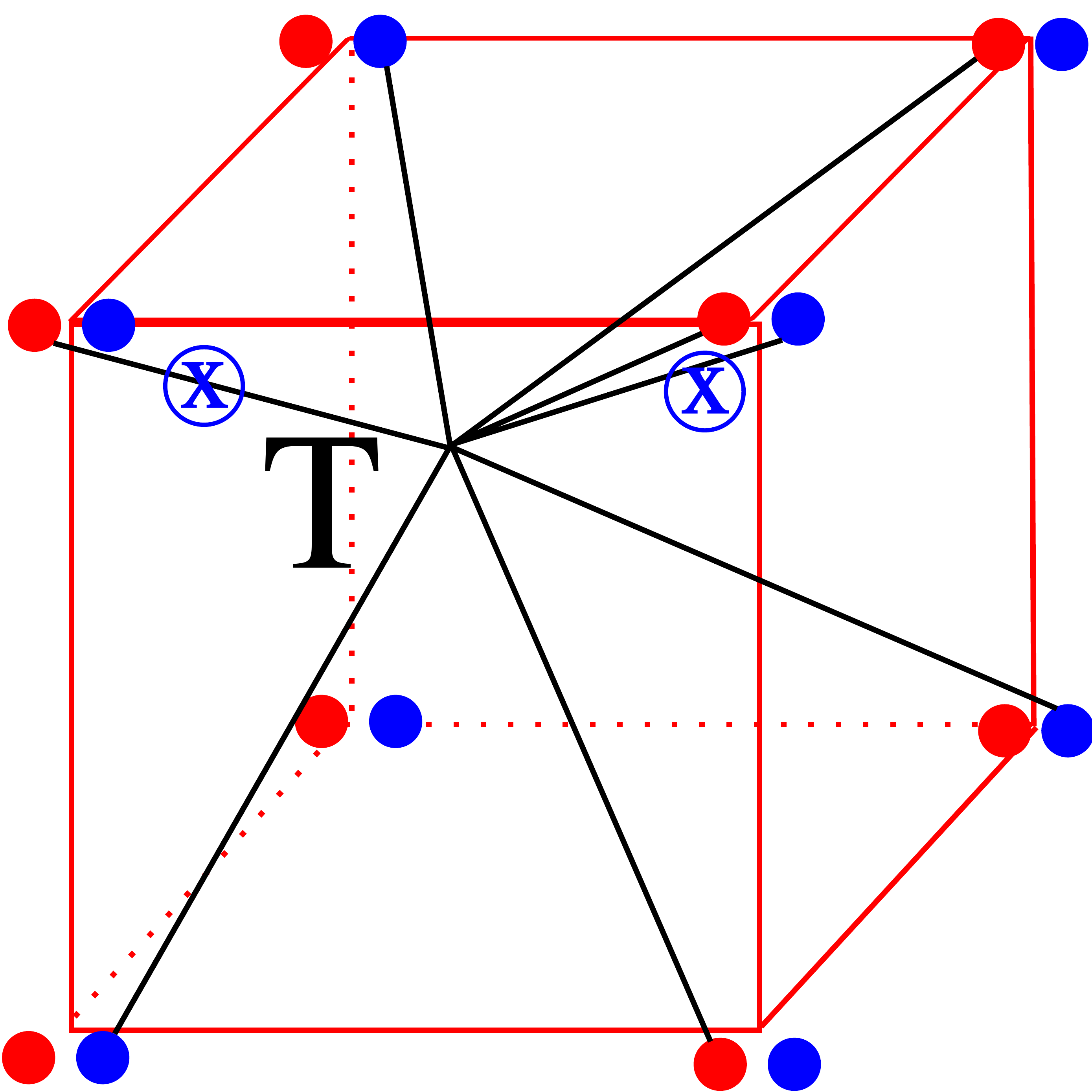}
\end{gathered}\\
=&
\begin{gathered}
\includegraphics[width=0.25\columnwidth]{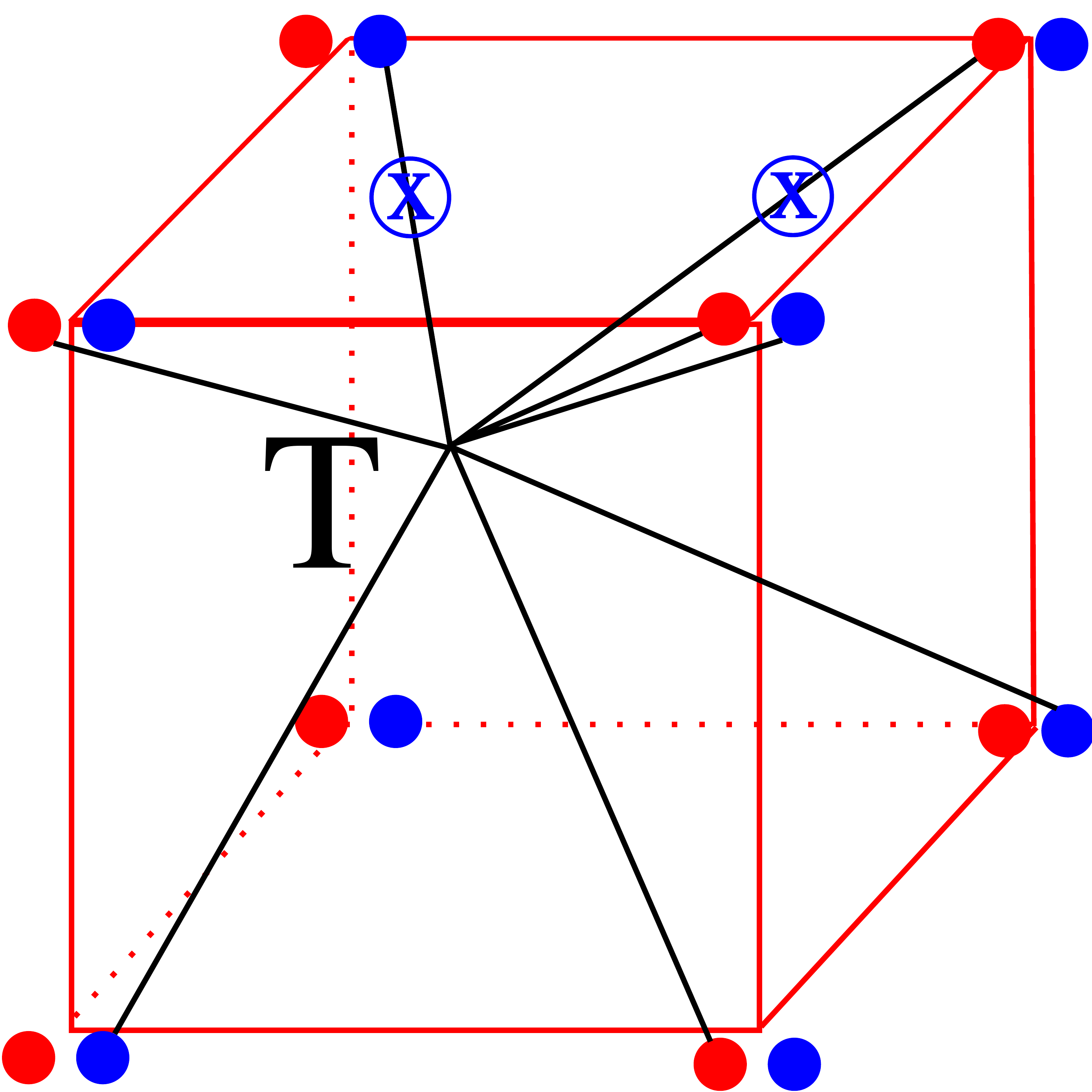}
\end{gathered}
=
\begin{gathered}
\includegraphics[width=0.25\columnwidth]{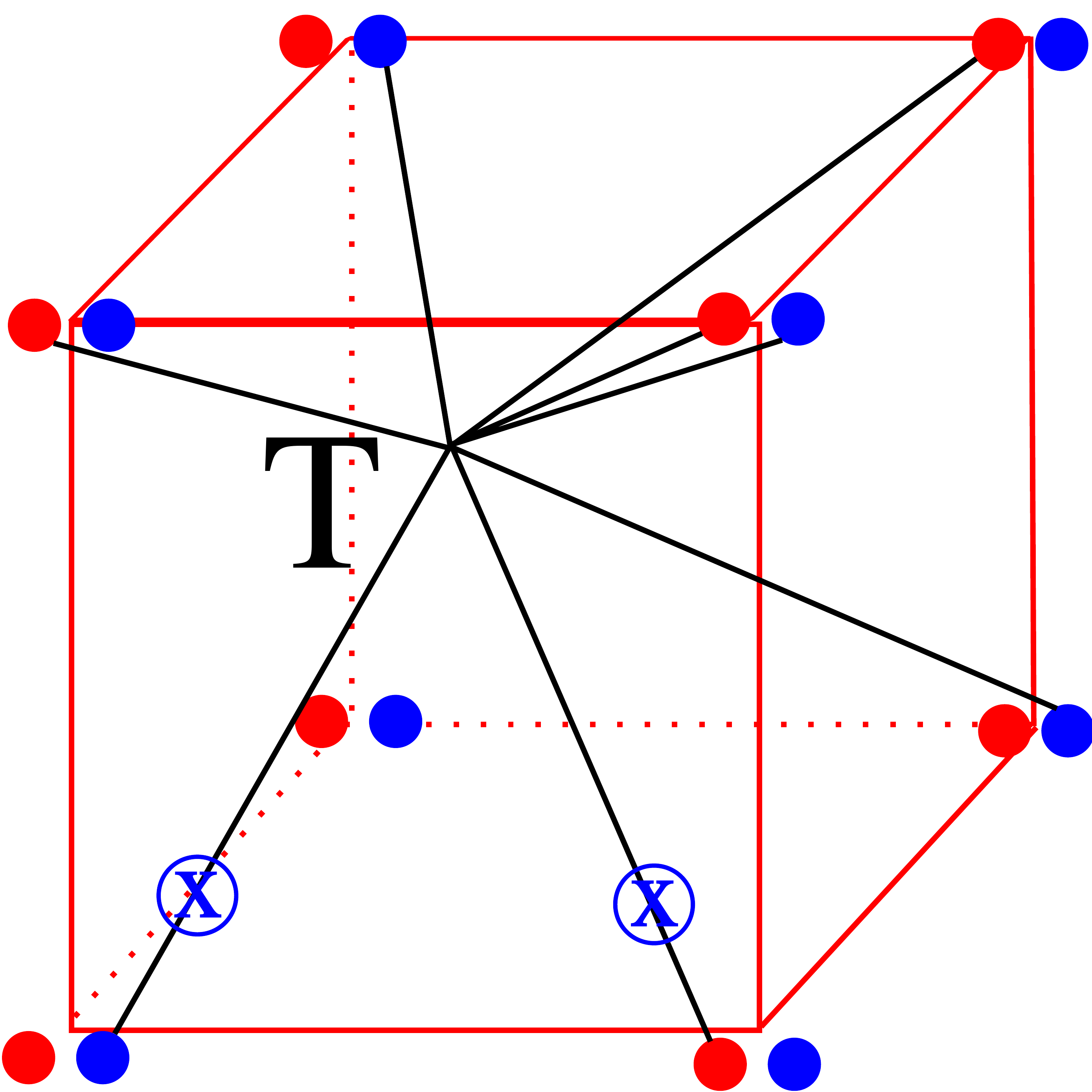}
\end{gathered}
=
\begin{gathered}
\includegraphics[width=0.25\columnwidth]{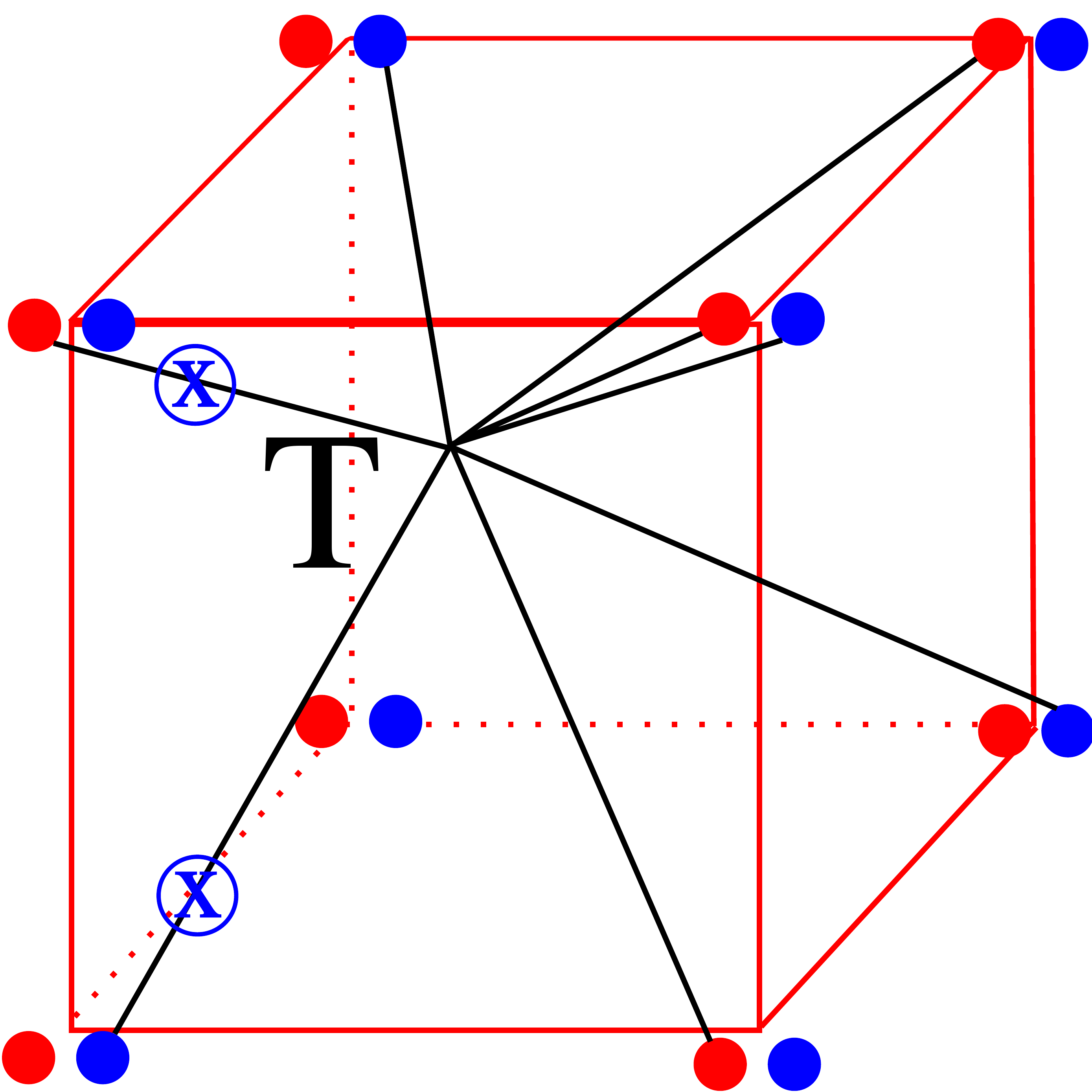}
\end{gathered}\\
=&
\begin{gathered}
\includegraphics[width=0.25\columnwidth]{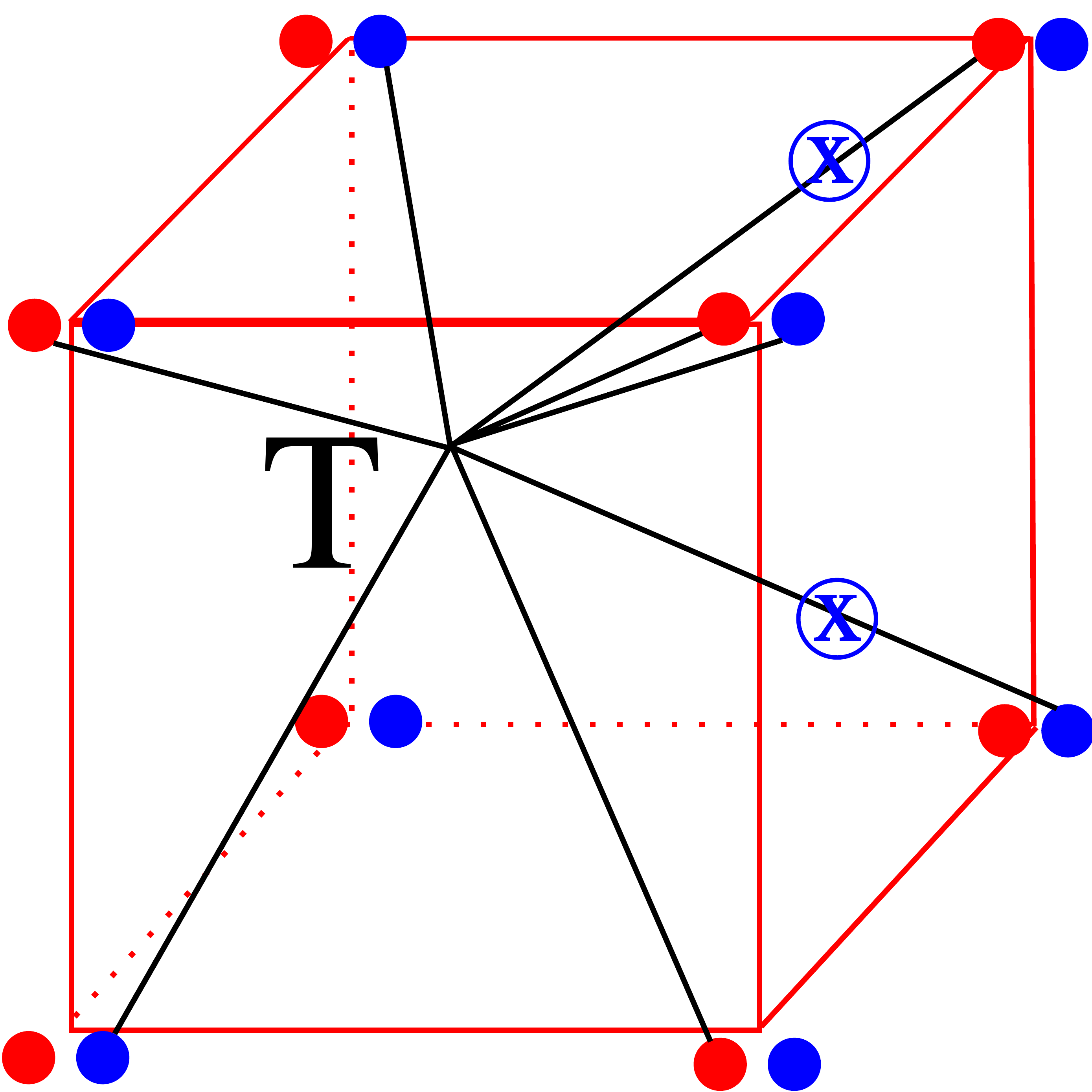}
\end{gathered}
=
\begin{gathered}
\includegraphics[width=0.25\columnwidth]{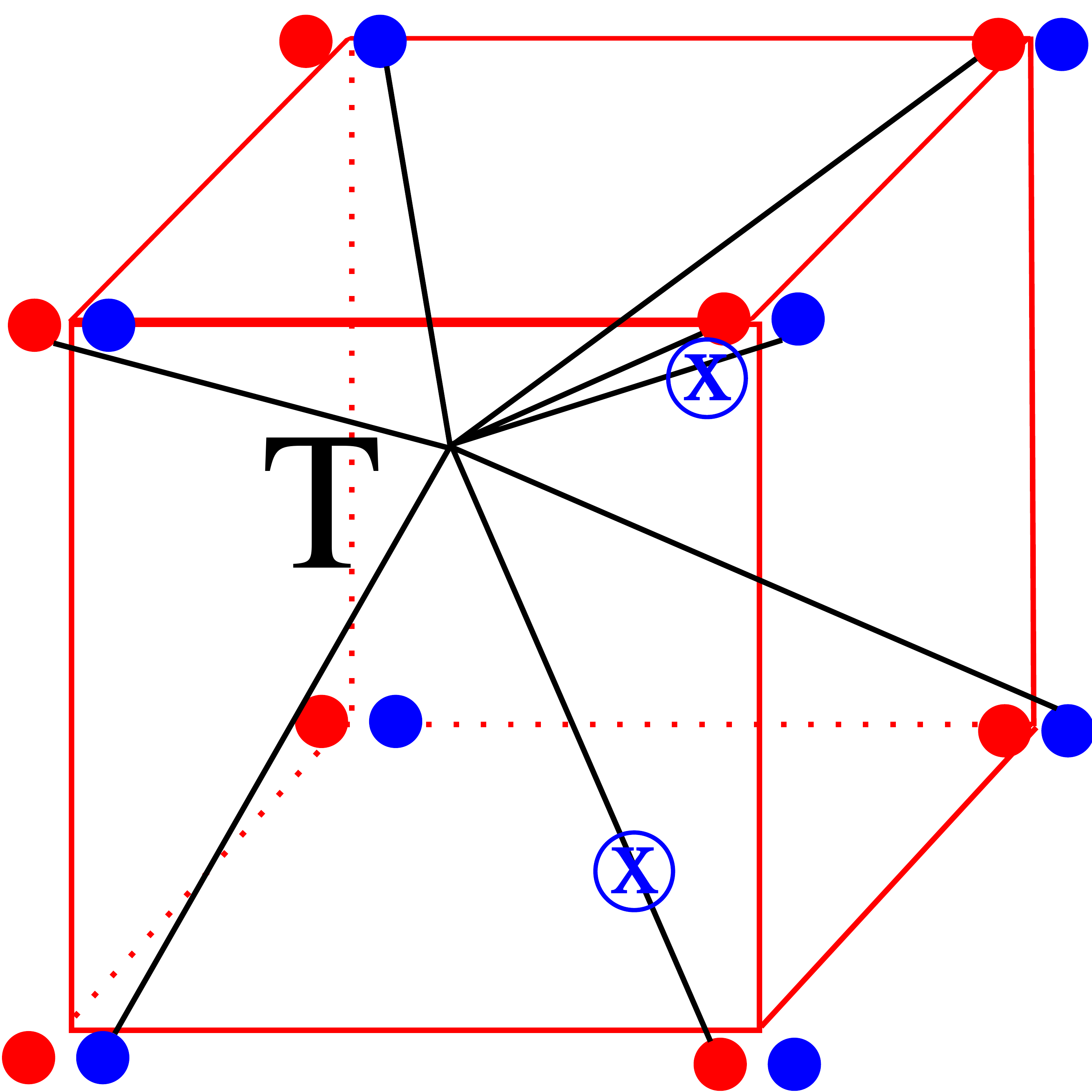}
\end{gathered}
=
\begin{gathered}
\includegraphics[width=0.25\columnwidth]{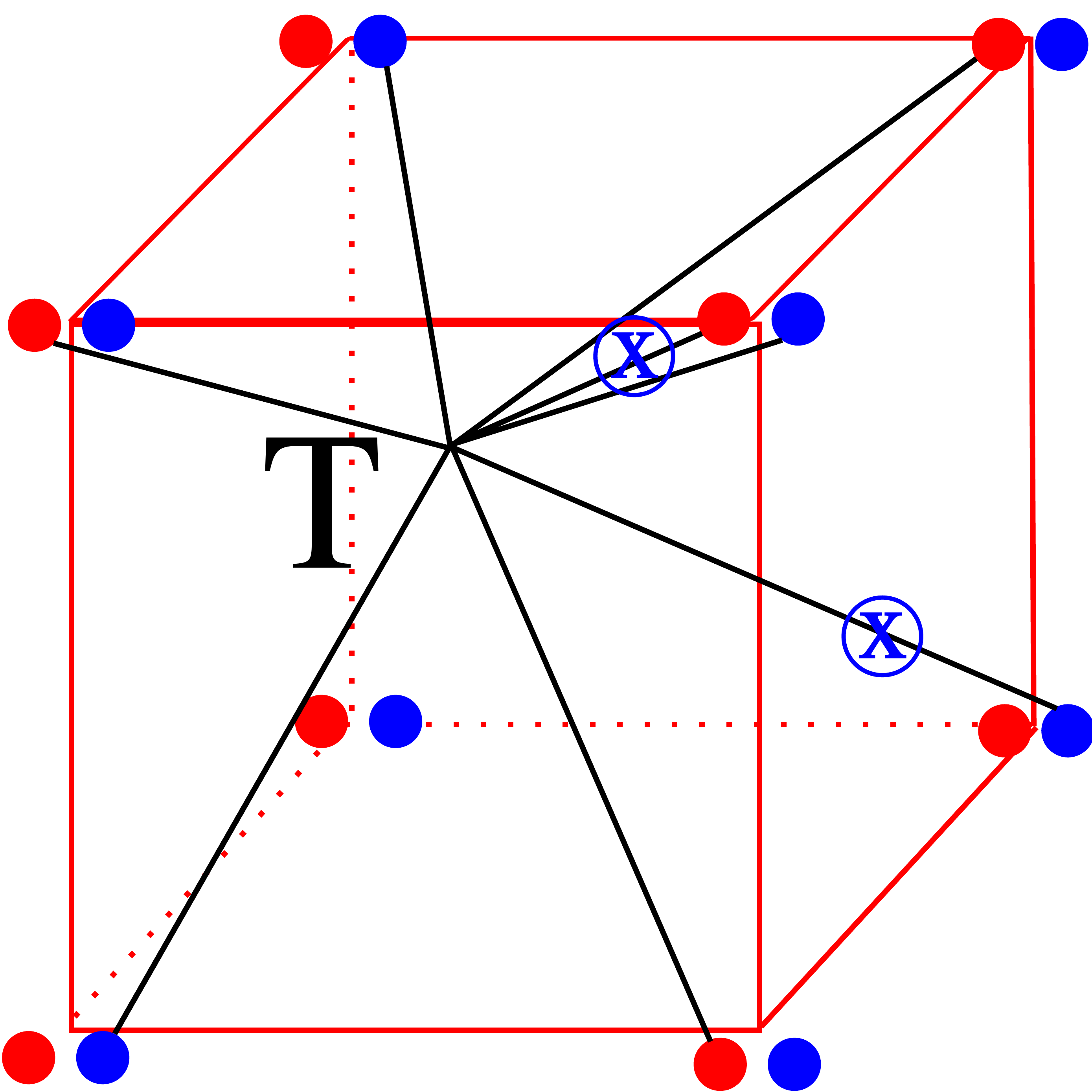}
\end{gathered}\\
=&
\begin{gathered}
\includegraphics[width=0.25\columnwidth]{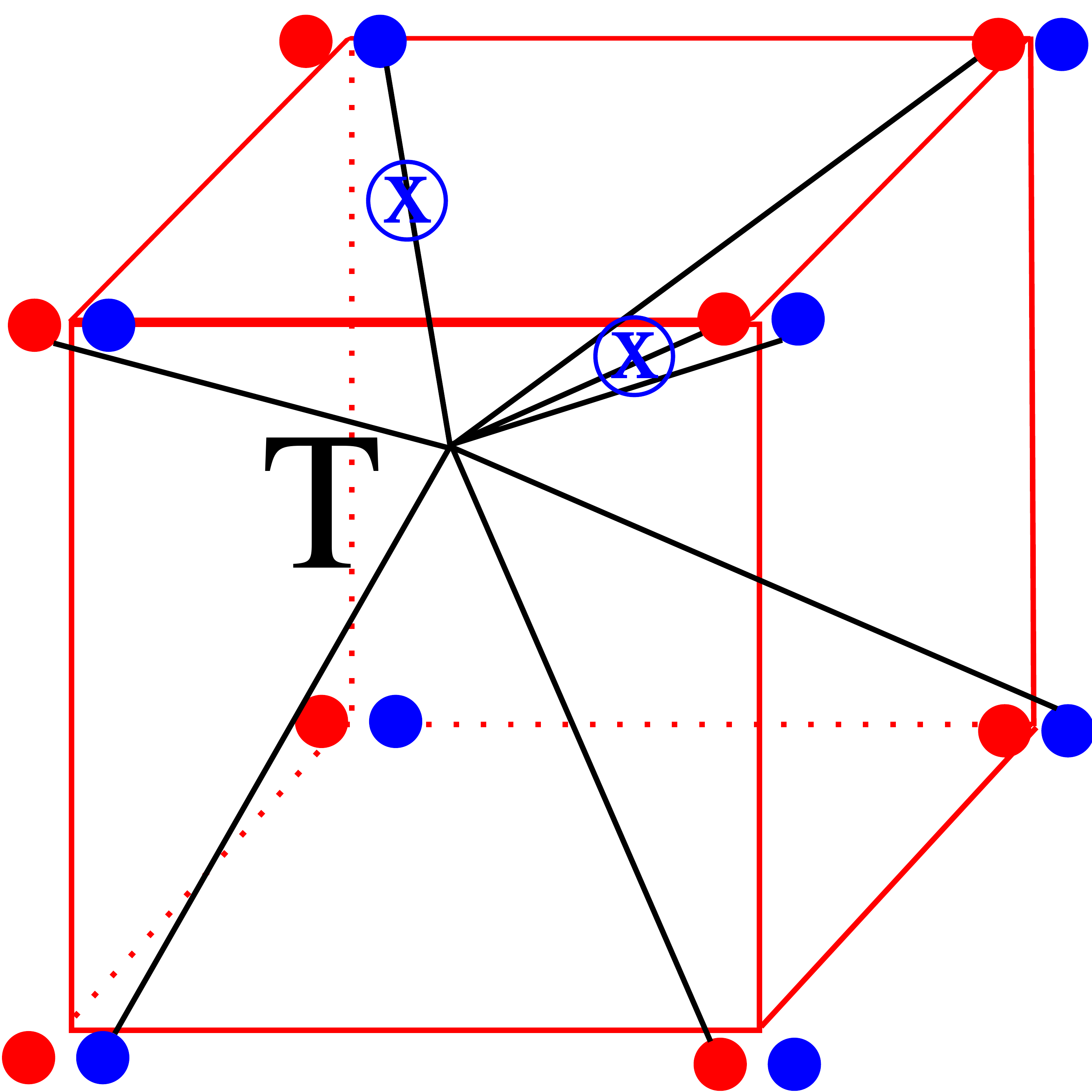}
\end{gathered}
=
\begin{gathered}
\includegraphics[width=0.25\columnwidth]{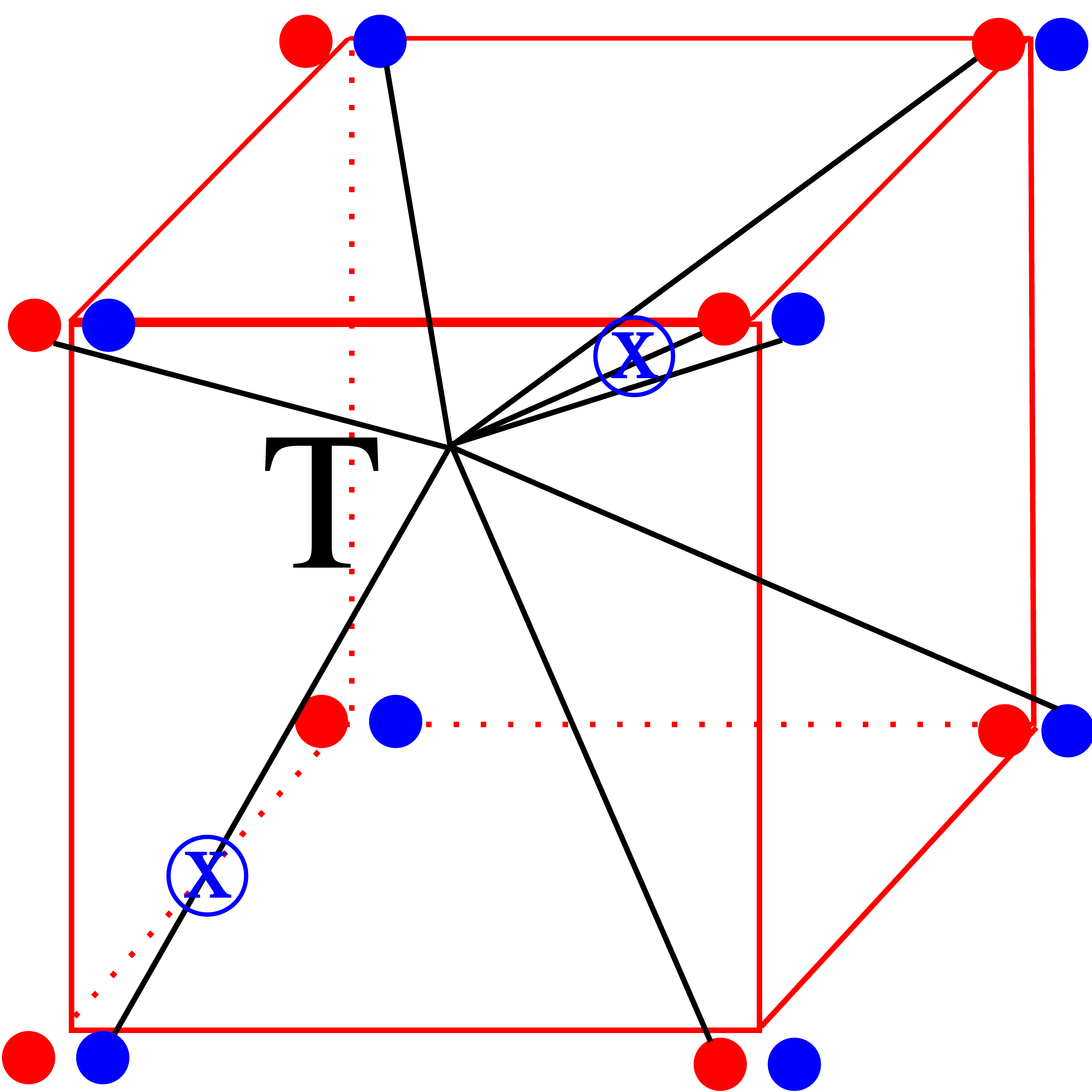}
\end{gathered}
=
\begin{gathered}
\includegraphics[width=0.25\columnwidth]{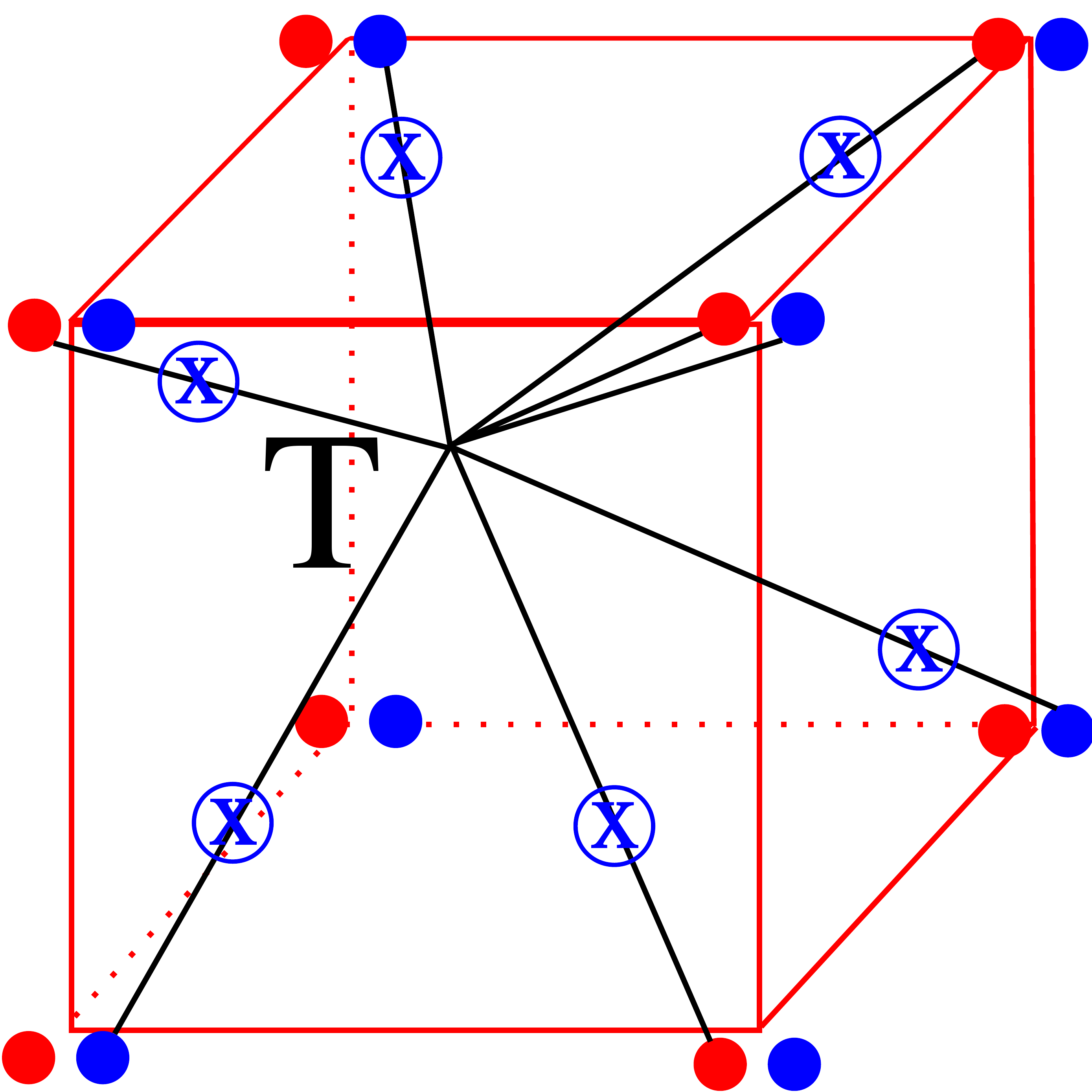}
\end{gathered}.\\
\end{split}
\end{equation}
In terms of components, Eq.~\eqref{HaahXtensorconstruction} means that $T_{i_1i_2i_3i_4i_5i_6i_7i_8}=T_{i'_1i'_2i'_3i'_4i'_5i'_6i'_7i'_8}$ where $i'_1i'_2i'_3i'_4i'_5i'_6i'_7i'_8$ are obtained by flipping arbitrary pairs of indices from $i_1i_2i_3i_4i_5i_6i_7i_8$. For example, 
\begin{eqnarray}\label{HaahXoperatorconstraints}
\begin{split}
T_{i_1i_2i_3i_4i_5i_6i_7i_8}&=T_{(1-i_1)(1-i_2)i_3i_4i_5i_6i_7i_8}\\
&=T_{i_1(1-i_2)(1-i_3)i_4i_5i_6i_7i_8}\\
&=T_{i_1i_2(1-i_3)(1-i_4)i_5i_6i_7i_8}\\
&=...
\end{split}
\end{eqnarray}
Combining Eq.~\eqref{HaahTtensorconstraint} and \eqref{HaahXoperatorconstraints}, we find that any configuration of $T_{i_1i_2i_3i_4i_5i_6i_7i_8}$ satisfying the condition  $\sum_{k=1}^8 i_k=0\mod 2$ are equal. 
We can rescale the $T$ tensor such that $T_{i_1i_2i_3i_4i_5i_6i_7i_8}=1$ for $\sum_{k=1}^8 i_k=0\mod 2$, i.e., 
\begin{equation}\label{HaahTtensor2}
	T_{i_1i_2i_3i_4i_5i_6i_7i_8}=
	\begin{cases}
		1 & \sum_{n=1}^{8}i_n=0 \mod{2}\\
		0 & \sum_{n=1}^{8}i_n=1 \mod{2}.\\
	\end{cases}
\end{equation}
For simplicity, we consider the space to be $\mathcal{R}^3$ where the Haah code has a unique ground state.
\begin{equation}\label{eq.HaahTNS}
	\ket{\mathrm{TNS}} = \sum_{\{s\}} \mathcal{C}^{R^3} \left( g^{L,s_1}g^{R,s_2}g^{L,s_3}g^{R,s_4} \ldots TTT \ldots \right) \ket{\{s\}}.
\end{equation}

We emphasize that the contraction of the Haah code TNS is quite different from that of the 3D toric code model and the X-cube model. The main difference is that the $g$ tensor has 4 virtual indices for the Haah code, while it has only 2 virtual indices for the 3D toric code and the X-cube model. As an example of contraction,  we take two blocks of size $2 \times 2 \times 1$ and $2 \times 2 \times 2$  in Fig.~\ref{fig.HaahTNS}. The $T$ tensors with their virtual indices are drawn explicitly. Each red or blue node in the two figures is a projector $g$ tensor, whose physical index is not drawn; we only draw the virtual legs that are connected to the $T$ tensors inside the blocks.  In the block $2\times 2 \times 2$, all the 8 virtual indices of the two $g$ tensors (4 per each $g$ tensor) in the middle of all the cubes are contracted with $T$ tensors, while other $g$ tensors have open virtual indices (which are not explicitly drawn).

\subsection{Entanglement Entropy for SVD Cuts}
\label{subsec:HaahcodeSVDCuts}

In this section, we compute the entanglement entropies for two types of cuts for which the TNS is an SVD. 

\subsubsection{Two types of SVD Cuts}

\begin{figure}[H]
	\centering
	\includegraphics[width=0.25\columnwidth]{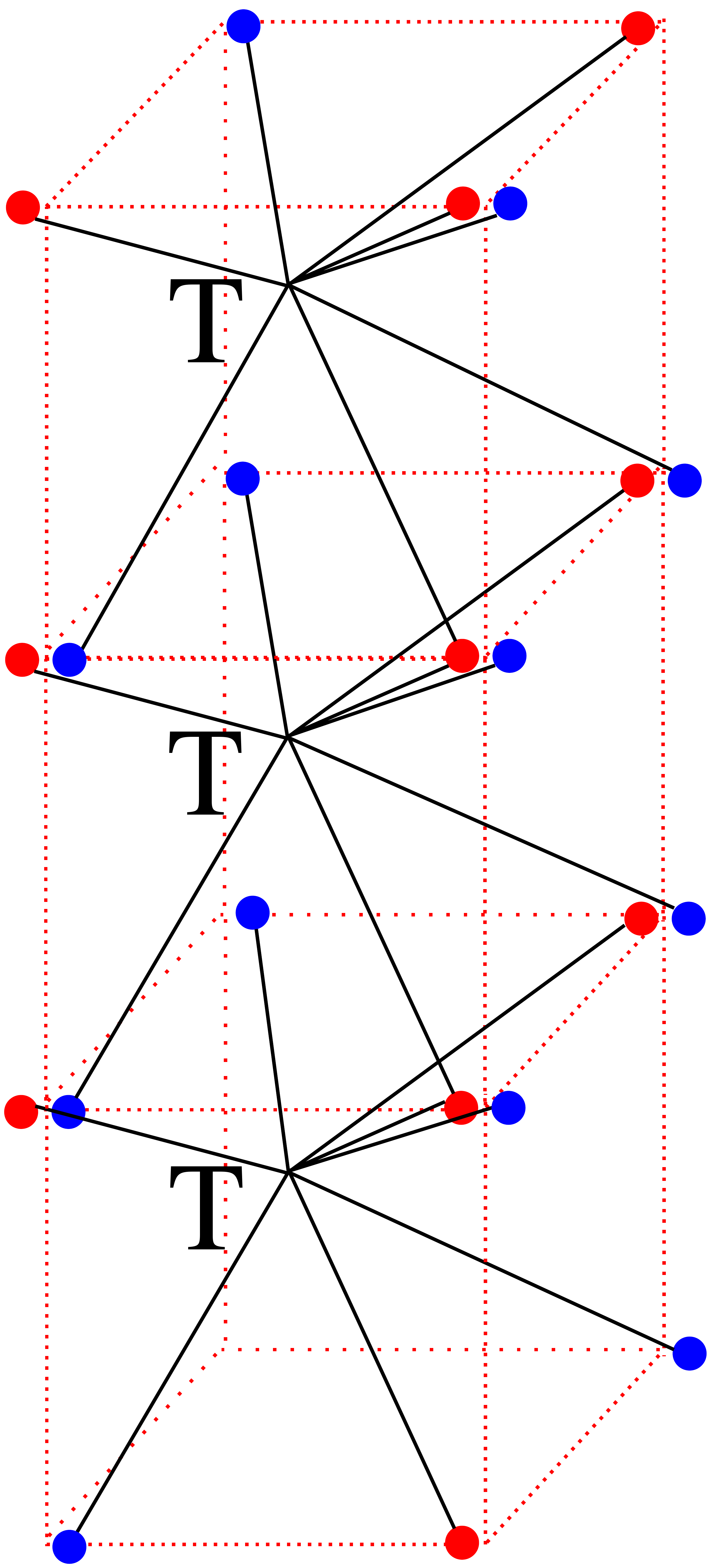}
	\caption{Region $A$ contains all the spins connecting with $l-1$ $T$ tensors which are contracted along z direction. The figure shows an example with $l=4$.}
	\label{fig.Haah_SVD2}
\end{figure}

\begin{figure}[H]
	\centering
	\includegraphics[width=0.4\columnwidth]{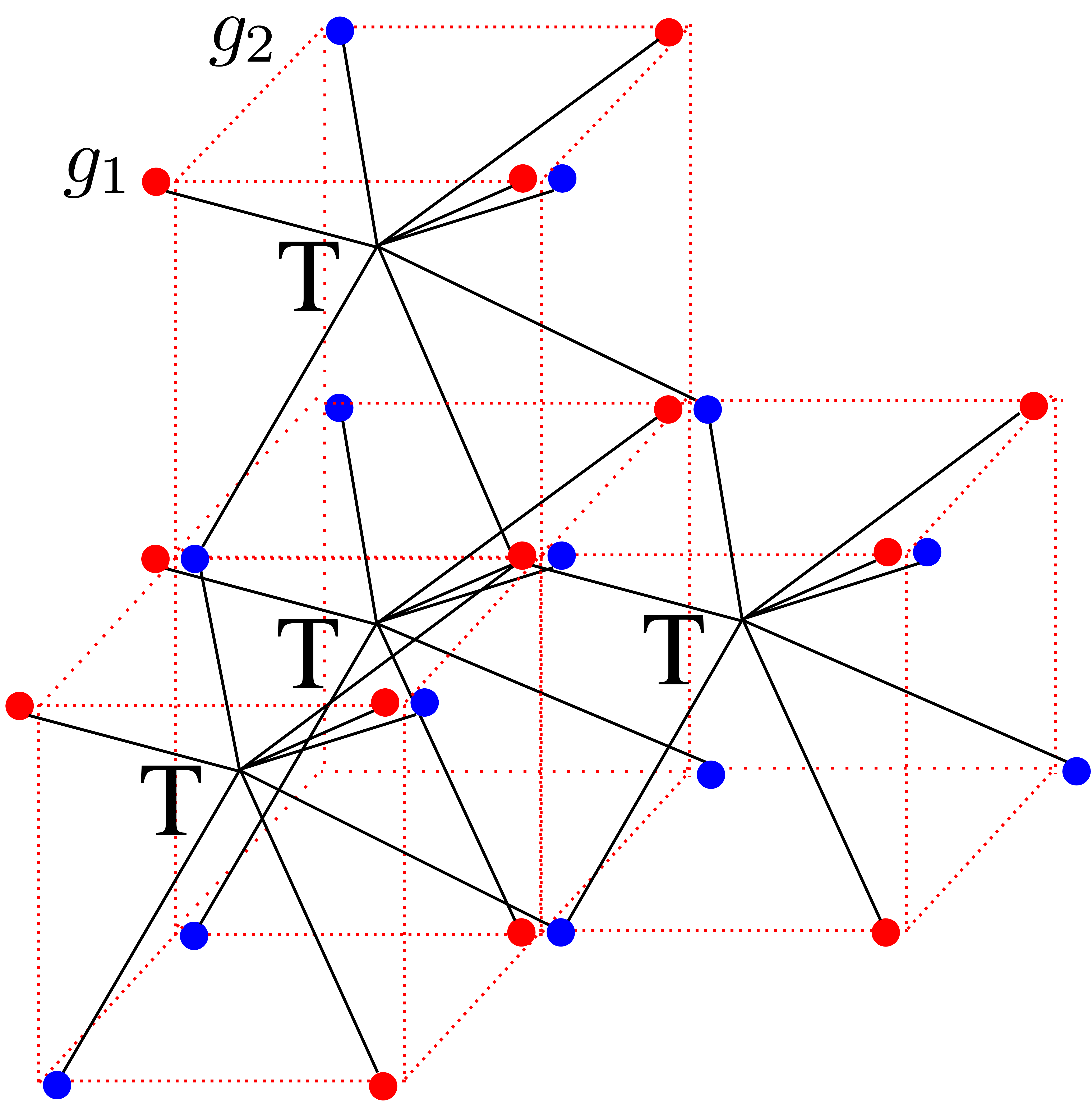}
	\caption{Region $A$ contains all the spins connecting with $T$ tensors which are contracted in a ``tripod-like" shape, where three legs extend along $x, y, z$ directions. There are three legs extending along $x, y, z$ directions respectively. In general, three legs can have different length, each with $l_x-1, l_y-1, l_z-1$ cubes along three directions.  
		This figure shows an example where $l_x=l_y=l_z=3$. }
	\label{fig.Haah_SVD1}
\end{figure}

To compute the entanglement entropy, we use the same convention adopted in the discussion of the 3D toric code (in Sec.~\ref{sec.toriccode}) and the X-cube model (in Sec.~\ref{sec.Xcube}): the open virtual  indices of the region $A$ connect directly to the $g$ tensors inside $A$ while the open virtual indices of the region $\bar{A}$ connect with $T$ tensors inside $\bar{A}$. We further choose a region $A$ such that the TNS is an SVD, and compute the entanglement entropy. We find two types of entanglement cuts for which the Haah code TNS is an exact SVD. For the general cubic region $A$, we need an extra step to perform the SVD of the TNS. This derivation will be presented in Sec.~\ref{subsubsec.HaahCodeSVD}. We now specify the two types of regions that the Haah code TNS is an exact SVD.
\begin{enumerate}
	\item Region $A$ only consists of the spins connecting to a set of $(l-1)$ $T$ tensors which are contracted along a certain direction. Figure \ref{fig.Haah_SVD2} shows an example with $l-1=3$ contracted along the $z$ direction. (Since in Sec.~\ref{sec.toriccode} and \ref{sec.Xcube}, we used $l$ as the number of vertices along each side of region $A$, there are $l-1$ bonds (or cubes) along each side.)
	
	\item Region $A$ contains all the spins connecting with $T$ tensors which are contracted in a ``tripod-like" shape, where three legs extend along $x, y, z$ directions. If there are $l_x-1$ cubes in the $x$ leg, $l_y-1$ cubes in the $y$ leg, and $l_z-1$ cubes in the $z$ leg, then there are $1+(l_x-2)+(l_y-2)+(l_z-2)=l_x+l_y+l_z-5$ cubes (or $T$ tensors) region $A$. Figure \ref{fig.Haah_SVD1} shows an example with $l_x=l_y=l_z=3$. 
\end{enumerate}
In the first case and for $l=2,3$, we use brute-force numerics to find that the reduced density matrix is diagonal (see App.~\ref{app.HaahcodeBruteforce} for details), which shows that the TNS is an exact SVD. 

In order to show that the above cuts correspond to an SVD, we follow the arguments developed in Sec.~\ref{subsec.TNSSVD}. In Sec.~\ref{subsec.TNSSVD}, we proposed a \emph{SVD Condition}. However, we find that the region $A$ of both types, shown in Fig.~\ref{fig.Haah_SVD2} and \ref{fig.Haah_SVD1}, do not satisfy the \emph{SVD Condition}: Two open virtual indices in region $\bar{A}$ connects with the same $T$ tensor, which violates the \emph{SVD Condition}. For instance, the $g_1$ and $g_2$ in Fig.~\ref{fig.Haah_SVD1} connects to the same $T$ tensor in their upper-left cube which is in the region $\bar{A}$.
Here, we propose a \emph{Generalized SVD Condition} which suffices to prove that the entanglement cut corresponding to Fig.~\ref{fig.Haah_SVD2} and \ref{fig.Haah_SVD1} are SVD. 

\textit{Generalized SVD condition:
	Let $\{t\}$ be the set of open virtual indices. Given a set of physical indices $\{s\}$ inside region $\bar{A}$, if $\{t\}$ can be uniquely determined by the $\{s\}$ inside region $\bar{A}$ via the $g$ tensor projection condition Eq.~\eqref{Haahgtensorprojectioncond} and $T$ tensor constraints Eq.~\eqref{HaahTtensor2}, then $\ket{\{t\}}_{\bar{A}}$ is orthogonal. Since $\ket{\{t\}}_A$ is orthogonal because all the open virtual indices are connected with $g$ tensors, the TNS $|TNS\rangle=\sum_{\{t\}}\ket{\{t\}}_A\otimes \ket{\{t\}}_{\bar{A}}$ is SVD. }

To prove the \emph{Generalized SVD Condition}, we notice that if we have two different sets of open virtual indices $\{t\}_{\bar{A}}$ and $\{t'\}_{\bar{A}}$, the physical indices $\{s\}_{\bar{A}}$ and $\{s'\}_{\bar{A}}$ which connect (via $g$ tensors) to the $T$ tensors on the boundary of region $\bar{A}$ cannot be the same. Otherwise, if $\{s\}_{\bar{A}}=\{s'\}_{\bar{A}}$, since the physical indices $\{s\}_{\bar{A}}$ and $\{s'\}_{\bar{A}}$ in the region $\bar{A}$ uniquely determine the open virtual indices $\{t\}_{\bar{A}}$ and $\{t'\}_{\bar{A}}$, $\{t\}_{\bar{A}}=\{t'\}_{\bar{A}}$, hence it is in contradiction with our assumption $\{t\}_{\bar{A}}\neq \{t'\}_{\bar{A}}$. Therefore, $\{t\}_{\bar{A}}\neq \{t'\}_{\bar{A}}$ implies $\{s\}_{\bar{A}}\neq \{s'\}_{\bar{A}}$, and hence  $_{\bar{A}} \langle\{t\}|\{t'\}\rangle_{\bar{A}}=0$. This is in the same spirit of the proof in Sec.~\ref{subsec.TNSSVD}. The proof of normalization of the wave function is independent of $\{t\}$ is also the same as in Sec.~\ref{subsec.TNSSVD}. Furthermore, $_A \langle\{t\}|\{t'\}\rangle_A=0$ for $\{t\}\neq \{t'\}$ is the straightforward because $\{t\}_A$ are connected with $g$ tensors. In summary, if the entanglement cut satisfies the \emph{Generalized SVD Condition}, we have 
\begin{enumerate}
	\item $_A\langle \{t\}|\{t'\}\rangle_A \propto \delta_{\{t\},\{t'\}}$ when $\ket{\{t\}}_{A}$ and $\ket{\{t^\prime\}}_{A}$ are not null vectors;
	\item $_{\bar{A}}\langle \{t\}|\{t'\}\rangle_{\bar{A}} \propto \delta_{\{t\},\{t'\}}$ when $\ket{\{t\}}_{\bar{A}}$ and $\ket{\{t^\prime\}}_{\bar{A}}$ are not null vectors.
\end{enumerate}
This shows that the TNS is an SVD. 

We explain the \emph{Generalized SVD Condition} in the simplest example,  i.e., $l=2$ in case 1. There is only one $T$ tensor, and the region A contains $8$ physical spins. 
\begin{center}
	\includegraphics[width=0.5\columnwidth]{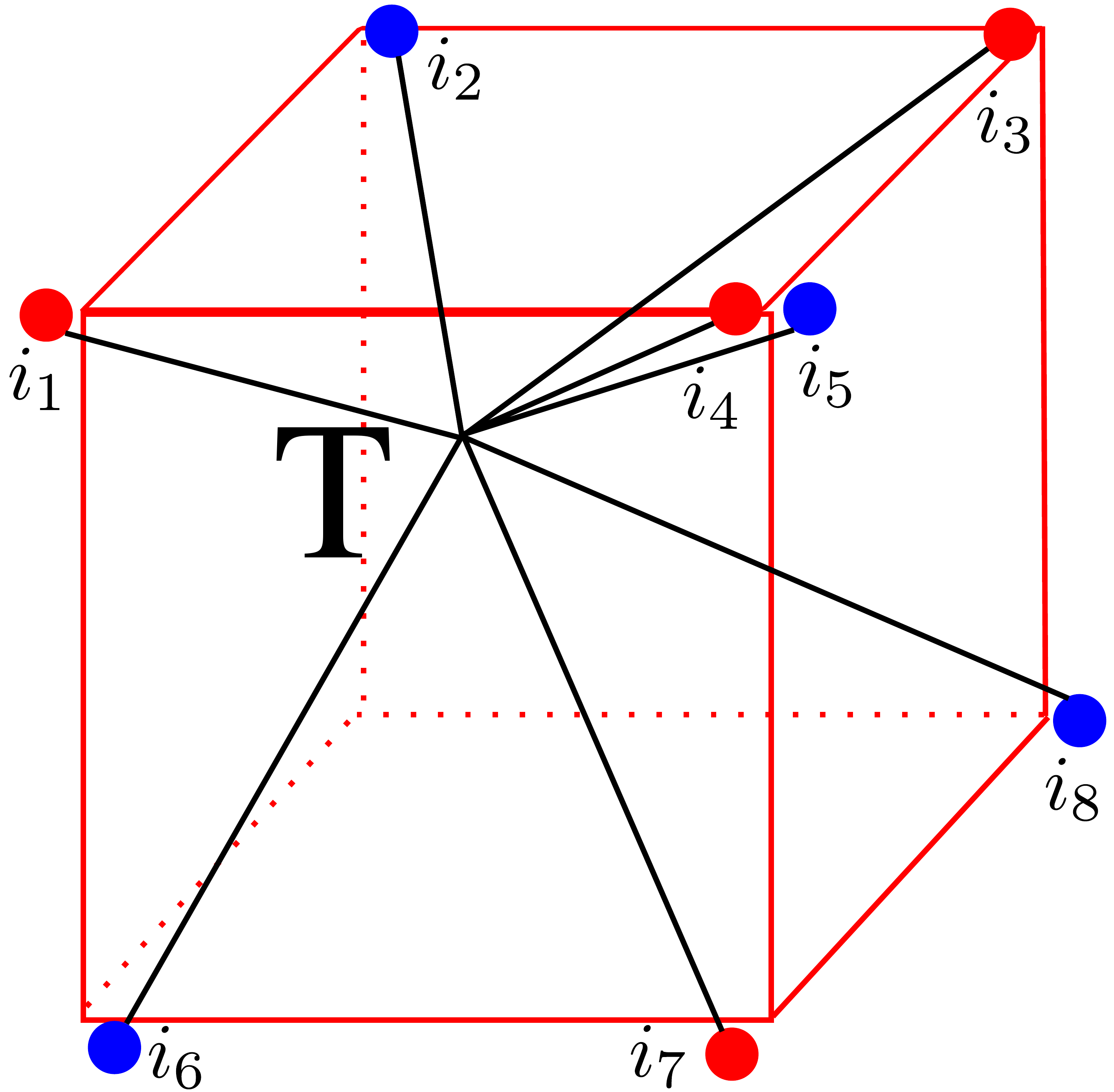}
\end{center}
All other spins apart from the eight connecting with the $T$ tensor belong to region $\bar{A}$.  Because the virtual indices and physical indices are related by the $g$ tensor which is a projector, we use $i_1$ to denote the values of both virtual indices and physical indices connecting with left $g$ tensor located at $(x,y,z)=(0,0,1)$. Here, we use the coordinate convention where the $(x,y,z)=(0,0,0)$ is located at the left down frontmost corner as in Fig.~\ref{fig.HaahTNS}. Similarly we use $ i_2, i_3, i_4, i_5, i_6, i_7, i_8$ to label the values of the virtual/physical indices on the remaining seven nodes connecting with the same $T$ tensor. Hence the set of open indices is effectively $\{i_1,  i_2, i_3, i_4, i_5, i_6, i_7, i_8\}$ (after identified by the $g$ tensors). We further consider how the physical indices from the region $\bar{A}$ constrain the open indices. Consider the $T$ tensor in the region $\bar{A}$ (which we denote by $T'$) which shares two spins $i_7, i_8$ with the region $A$ (The $T^\prime$ tensor lives in the lower right corner): 
\begin{equation}
\begin{gathered}
\includegraphics[width=5cm]{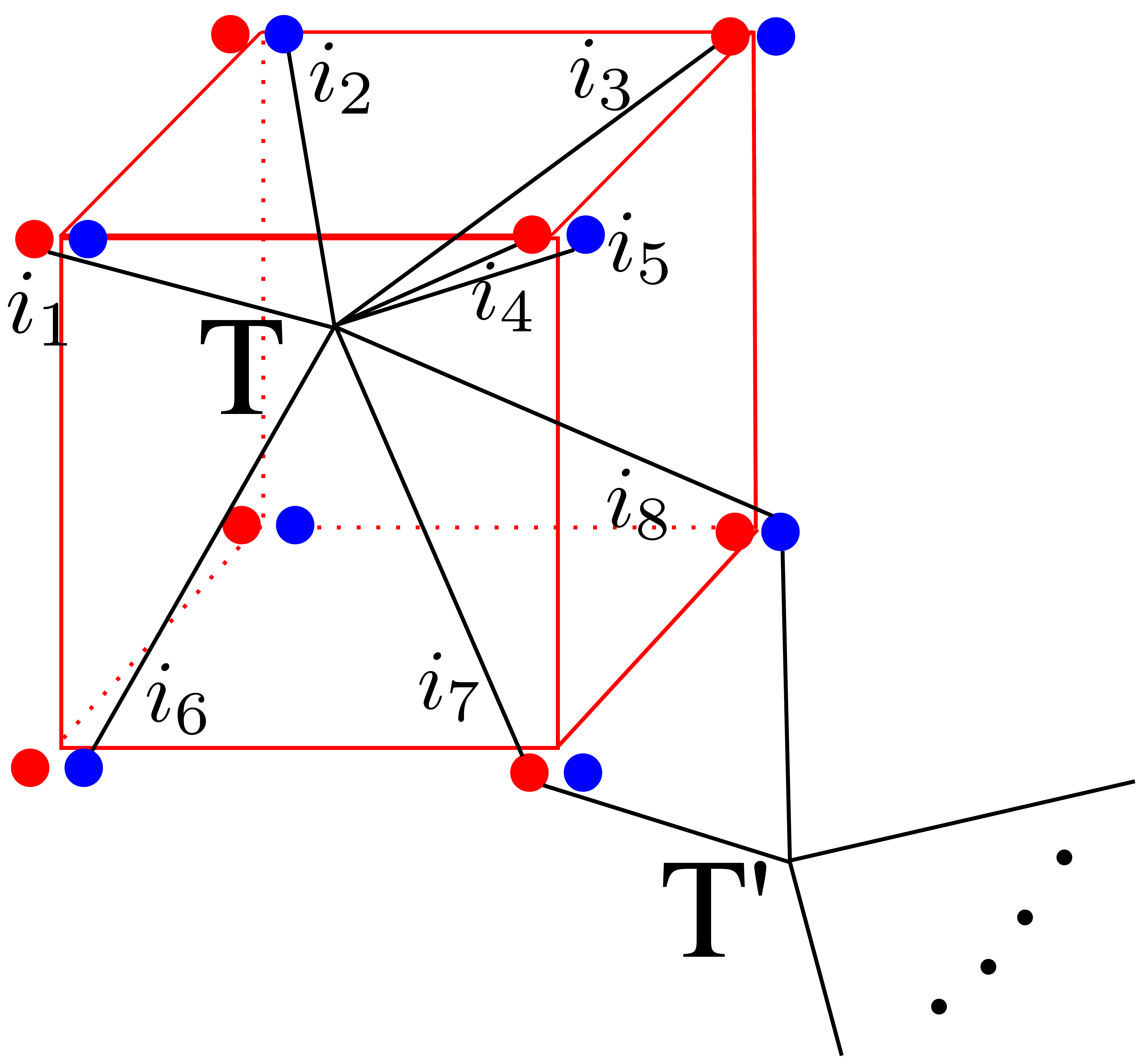}
\end{gathered}
\end{equation}
Since six among the eight virtual indices of $T'$ are contracted with $g$ tensors inside region $\bar{A}$, the remaining two open virtual indices, i.e., $i_7$ and $i_8$ are subject to one constraint from the $T'$ tensor:
\begin{eqnarray}
i_7+i_8=\mathrm{fixed},
\end{eqnarray}
where ``fixed" means that the sum is fixed by the physical indices inside the region $\bar{A}$. We can similarly consider the constraints coming from other $T$ tensors in region $\bar{A}$. The whole set of constraints are listed as follows:   
\begin{eqnarray}
i_7+i_8&=&\mathrm{fixed}\nonumber\\
i_1+i_2&=&\mathrm{fixed}\nonumber\\
i_5&=&\mathrm{fixed}\nonumber\\
i_6&=&\mathrm{fixed}\nonumber\\
i_6+i_7&=&\mathrm{fixed}\nonumber\\
i_2+i_3&=&\mathrm{fixed}\nonumber\\
i_5&=&\mathrm{fixed}\nonumber\\
i_8&=&\mathrm{fixed}\nonumber\\
i_1&=&\mathrm{fixed}\nonumber\\
i_4&=&\mathrm{fixed}\nonumber\\
i_3&=&\mathrm{fixed}\nonumber\\
i_7&=&\mathrm{fixed}.
\end{eqnarray}
The ``fixed" on the right hand side of the equations means that the virtual indices or the sum of the virtual indices are fixed by the physical indices in the region $\bar{A}$. All variables and equations are defined module 2. The above equations uniquely determine all the open virtual indices $i_1...i_8$. Therefore, such a choice of region $A$ of the entanglement cut satisfies the \emph{Generalized SVD Condition}.

For the first type of the region $A$ with general $l$, and the second type of region $A$ with general $l_x, l_y, l_z$, we can similarly check that the TNS satisfies the \emph{Generalized SVD Condition}. Numerically, we checked that the Haah code TNS indeed satisfies the \emph{Generalized SVD Condition} for $2\leq l\leq 9$ for the first type, and $3\leq l_x\leq 8, 3\leq l_y\leq 8, 3\leq l_z\leq 8$ for the second type. The numerical procedure for this check is to list all the constraints for the indices in the region $A$ and find how many solutions exist for these constraints.

\subsubsection{Entanglement entropy}

We now compute the entanglement entropy for the exact SVD TNSs. We first consider the case 1 with general $l$, such as in Fig.~\ref{fig.Haah_SVD2}. All the spins connecting with $l-1$ contracted $T$ tensors along the $z$ directions are in region $A$, and the remaining belong to region $\bar{A}$.  
The number of open virtual indices, after identified by the local $g$ tensors, is $8+7(l-2)=7l-6$. The number of constraints from the local $T$ tensors is simply the number of $T$ tensors $l-1$, because they are all independent. Hence the number of independent open virtual indices is $7l-6-(l-1)=6l-5$. Therefore, the entanglement entropy is
\begin{eqnarray}\label{HaahexactSVDtype1}
\frac{S(A)}{\log 2}=6l-5.
\end{eqnarray}
In appendix.~\ref{app.HaahcodeBruteforce}, we numerically brute-force compute the reduced density matrix for $l=2$ and $l=3$, and find that the results match the general formula Eq.~\eqref{HaahexactSVDtype1}.

We further consider the case 2 --- the region $A$ of tripod shape. The legs in the $x,y,z$ direction contains $l_x-1, l_y-1, l_z-1$ $T$ tensors respectively. We first count the total  number of open virtual indices. When $l_x=l_y=l_z=3$ as shown in Fig.~\ref{fig.Haah_SVD1}, there are 26 physical spins (or $g$ tensors) in total. However, there is one $g$ tensor (at the left spin of $(x,y,z)=(1,1,1)$) whose four virtual indices are all contracted by the $T$ tensors within region $A$. Hence the number of open virtual indices, after identified by the local $g$ tensor, is 25. Moreover, we notice that adding one $T$ tensor in one of the three legs of region $A$ brings 7 extra spins. Therefore the total number of open virtual indices (after identified by the $g$ tensor) is 
$(26-1)+7(l_x-3)+7(l_y-3)+7(l_z-3)=7l_x+7l_y+7l_z-38$. We further numerically count the number of constraints that these open virtual indices satisfy. We find the number of constraints is the number of cubes minus 1, i.e., $(l_x+l_y+l_z-5)-1=l_x+l_y+l_z-6$. Therefore the number of  independent open virtual indices is $(7l_x+7l_y+7l_z-38)-(l_x+l_y+l_z-6)=6l_x+6l_y+6l_z-32$. The entanglement entropy is 
\begin{eqnarray}
\frac{S(A)}{\log 2}=6l_x+6l_y+6l_z-32.
\end{eqnarray}

\subsection{Entanglement Entropy for Cubic Cuts}\label{subsec.Haah_Entanglement}

In this section, we consider the case where the region $A$ is a cube of size $l \times l \times l$, where $l$ is the number of vertices in each direction of the cube. The cut is chosen such that all the open virtual indices straddling the region A are connected to $g$ tensors in the region $A$ (i.e., all the physical spins near the boundary belong to the region $A$). For example, for $l=2$ as shown in \eqref{HaahTtensor}, all 16 physical spins belong to the region $A$. For $l=3$ as shown in Fig.~\ref{fig.HaahTNS} (b), all 54 physical spins belong to the region $A$. 
For the simplicity of notations, in this section, we denote the Hamiltonian terms as $A_c$ and $B_c$ where the subscript refers to a cube $c$.

\subsubsection{SVD for TNS}\label{subsubsec.HaahCodeSVD}

For the cubic region $A$, we find that the TNS for the Haah code is different from that for the toric code and X-cube model: the TNS for the Haah code is \emph{not} an exact SVD. The TNS basis in the region $A$, $\ket{\{t\}}_{A}$, are orthonormal, since the open virtual indices are connected with $g$ tensors. However, the TNS basis $\ket{\{t\}}_{\bar{A}}$ in the region $\bar{A}$ are \emph{not} orthogonal. In other words, the basis $\ket{\{t\}}_{\bar{A}}$ is over complete.

The subtlety that the TNS bipartition is not an exact SVD manifests as follows: the singular vectors in the region $A$ for the ground states of the Haah code have to be the eigenvectors of all $A_c$ and $B_c$ operators that actually lie in the region $A$, and the corresponding eigenvalues should all be $1$. Notice that our TNS basis state $\ket{\{t\}}_{A}$, if not null, are the eigenvectors of all $A_c$ operators inside the region $A$ with eigenvalues $1$, and are also the eigenvectors of $B_c$ operators with eigenvalues $1$ when $B_c$ operators are deep inside the region $A$, i.e., when they do not act on any spin at the boundary of $A$. However, $\ket{\{t\}}_{A}$ are \emph{not} the eigenvectors of $B_c$ operators, when $B_c$ operators are inside the region $A$ but also adjacent to the region $A$'s boundary. The reason is that the $B_c$ operators adjacent to the region $A$'s boundary, when acting on the TNS basis $\ket{\{t\}}_{A}$, will flip the physical spins on the boundary, and thus flip the open virtual indices $\{t\}$ due to the projector $g$ tensors. Therefore, the basis $\ket{\{t\}}_{A}$ is no longer the singular vectors for the Haah code. This is not an \textit{a priori} problem, but a result of the geometry of the Haah code, whose spins cannot be written on bonds but have to be written on sites. A similar situation would occur if the 2D toric code model would be re-written to have its spins on sites.

The method to find the correct SVD for the TNS is to use the $\ket{\{t\}}_{A}$ to construct the eigenvectors of $B_c$ operators by projection. We prove the following statement:

\textit{If $\ket{\{t^\prime\}}_{A}=B_c \ket{\{t\}}_{A}$ when $B_c$ is inside the region $A$ and also adjacent to the region $A$'s boundary, then $_A\langle \{t'\}|\{t\}\rangle_A=0$ and  $\ket{\{t^\prime\}}_{\bar{A}}=\ket{\{t\}}_{\bar{A}}$.}

\begin{figure}[t]
\centering
\includegraphics[width=0.7\columnwidth]{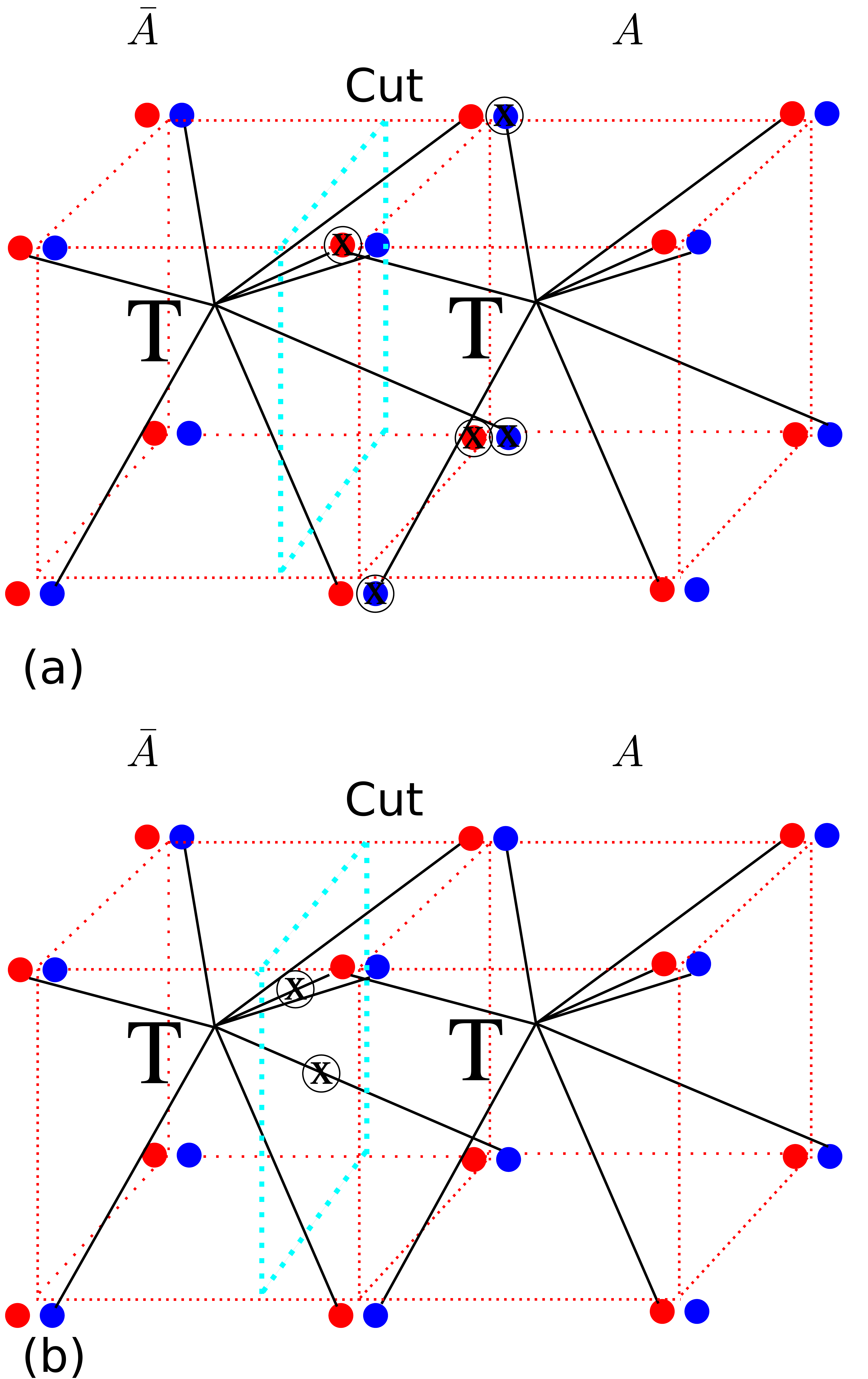}
\caption{Transferring the Pauli $X$ operators of the $B_c$ operator from the region $A$ (a) to the region $\bar{A}$ (b).}
\label{fig.Haah_transfer}
\end{figure}

The proof is as follows. The first part of the statement is a consequence of the $\ket{\{t\}}_A$ basis state orthogonality. Indeed, $B_c$ flips physical spins located at the region $A$'s boundary. Thus the two sets $\{t\}$ and $\{t'\}$ are distinct. The second part of the statement is more involved. Suppose for simplicity that we consider two nearest neighbor $T$ tensors for the region $A$ and $\bar{A}$ in Fig.~\ref{fig.Haah_transfer}. The $B_c$ operator acts on the right cube Fig.~\ref{fig.Haah_transfer} (a). The physical spins on the boundary of the region $A$ which are flipped by $B_c$ are those covered by circled $X$ in Fig.~\ref{fig.Haah_transfer} (a). Then these Pauli $X$ operators can be transferred to the virtual indices due to projector $g$ tensors, and the virtual indices of the $T$ tensor in the region $\bar{A}$ obtain two $X$ operators as in Fig.~\ref{fig.Haah_transfer} (b). Notice that the $T$ tensor for the Haah code is invariant under this action (see the 12th cube in  Eq.~\eqref{HaahXtensorconstruction}). This is also true for other $T$ tensors in the region $\bar{A}$ that are affected by $B_c$.
The transfer of $X$ operators from the region $A$ to the region $\bar{A}$ gives exactly the same equations in Eq.~\eqref{HaahXtensorconstruction} when we solve for the $T$ tensors. Hence, the X operators transferred to the open virtual indices in the region $\bar{A}$ do not change the state at all, i.e., $\ket{\{t^\prime\}}_{\bar{A}}=\ket{\{t\}}_{A}$. As a consequence, we can perform the following factorization
\begin{equation}
\begin{split}
&\ket{\{t\}}_{A} \otimes \ket{\{t\}}_{\bar{A}} + \ket{\{t^\prime\}}_{A} \otimes \ket{\{t^\prime\}}_{\bar{A}}	\\
=& \Big[(1+B_c) \ket{\{t\}}_{A}\Big] \otimes \ket{\{t\}}_{\bar{A}}.
\end{split}
\end{equation}

The left part of the tensor product is an eigenstate of $B_c$ with eigenvalue 1. 
Therefore, in the TNS decomposition Eq.~\eqref{eq.TNSgeneralDecomposition}, we can group the basis state $\ket{\{t\}}_{A}$ which are connected by this $B_c$ operator. This factorization can be extended to any product of $B_c$ operators inside the region $A$ and also adjacent to the region $A$'s boundary. Notice that any such product has at least one $X$ operator belonging to only one $B_c$ and so is different from the identity. When acting with all the possible products of these $B_c$ operator (including the identity) on a given $|\{t\}\rangle_A$ will generate as many unique states as there are $B_c$'s. The TNS can be brought to the following form
\begin{eqnarray}
|\mathrm{TNS}\rangle=\sum_{\{t\}'}\bigg[\prod_{c} \bigg(\frac{1+B_c}{2}\bigg)\ket{\{t\}}_A\bigg]\otimes |\{t\}\rangle_{\bar{A}},
\end{eqnarray}
where the product over $c$ only involves the $B_c$ operators inside the region A and also adjacent to the region $A$'s boundary and the sum over $\{t\}'$ is over the open virtual index configurations that are not related by the action of these $B_c$ operators. 

\subsubsection{Counting the number of TNS basis in region $A$: Notations}

To compute the upper bound of the entanglement entropy, we need to find the number of singular vectors in the region $A$ that are also eigenstates of any  $B_c$ operators fully lying in the region $A$. This number that we denote as  $\mathrm{basis}(\mathrm{TNS}(A))$ is 
\begin{equation}
	\mathrm{basis}(\mathrm{TNS}(A))= 2^{N-N_B},
\end{equation}
where $N$ is the number of independent open virtual indices and $N_B$ is the number of $B_c$ operators inside the region $A$ and also adjacent to the region $A$'s boundary. Every open virtual index connected to a $g$ tensor located in $A$ and at the boundary of this region. Since each $g$ tensor has a unique independent virtual index, we have $N=N_g-N_c$ where $N_g$ is the number of $g$ tensors in $A$ and at the boundary of this region and $N_c$ is the number of constraints on the open indices coming from the $T$ tensors within the region $A$. We thus get
\begin{equation}
	\log_2 (\mathrm{basis}(\mathrm{TNS}(A)))= N_{g}-N_c-N_{B}
\end{equation}
and the upper bound on the entanglement entropy reads
\begin{eqnarray}
S(A) = (N_{g}-N_c-N_{B})\log 2.
\end{eqnarray}

\subsubsection{Counting $N_{g}$ and $N_B$}

We first count $N_{g}$. The number of $g$ tensors can be computed by looking at Fig.~\ref{fig.HaahTNS} (b). We consider the region $A$ with size $l_x\times l_y\times l_z$ (Notice that $l_x, l_y, l_z$ are the number of vertices in each direction). In eight corners, there are $8\times 2=16$ vertices. On the four hinges along $x$ direction, there are $2\times 4\times (l_x-2)$ vertices, where 2 means there are two spins on each vertex, and 4 means four hinges. And similar for $2\times 4\times (l_y-2)$ and $2\times 4\times (l_z-2)$ in the $y$ and $z$ directions respectively. For the $xy$-plane, there are $2\times 2\times (l_x-2)(l_y-2)$, where the first 2 comes from two spins per vertex, and the second 2 comes from two $xy$-planes. Similarly $2\times 2\times (l_x-2)(l_z-2)$ and $2\times 2\times (l_y-2)(l_z-2)$ for $xz$ and $yz$ plane respectively. Therefore, the total number of $g$ tensors is 
\begin{eqnarray}
\begin{split}
N_g=&16+8(l_x-2)+8(l_y-2)+8(l_z-2)\\
&+4(l_x-2)(l_y-2)+4(l_x-2)(l_z-2)\\
&+4(l_y-2)(l_z-2)\\
=&4l_xl_y+4l_xl_z+4l_yl_z-8l_x-8l_y-8l_z+16.
\end{split}
\end{eqnarray}

We further count $N_B$. As explained in Sec.~\ref{subsubsec.HaahCodeSVD}, $N_B$ is the number of $B_c$ operators inside the region $A$ and adjacent to the boundary of the region $A$. For a cubic region $A$ with size $l\times l\times l$ (which is the case we consider below), the number of such  $B_c$ operators are
\begin{equation}
\begin{split}
N_B=(l-1)^3-(l-3)^3=6l^2-24l+26, \forall l\geq 3.
\end{split}
\end{equation}
For $l=2$, we just have one $B_c$ operator. Hence we have 
\begin{equation}
N_B=6l^2-24 l +26-\delta_{l,2}, \forall l\geq 2.
\end{equation}

\subsubsection{Counting $N_c$: Contribution from the $T$ tensors}
\label{subsec:countingNcfromTtensors}

The open indices may be constrained by the $T$ tensors fully inside the region $A$.
In the following, we will discuss the specific entanglement cuts where $l_x=l_y=l_z=l$.
We rely on numerical calculations to evaluate $N_c$. We first consider the examples $l=2$ and $l=3$ in detail, and then we describe our algorithm to search the number of linearly independent constraints. 

For $l=2$, as shown in Eq.~\eqref{HaahTtensor}, no $g$ tensor has all virtual indices contracted. There is only one $T$ tensor within region $A$. The element of the $T$ tensor is
\begin{equation}
	T_{i_1i_2i_3i_4i_5i_6i_7i_8}
\end{equation}
where $i_1,i_2,i_3,i_4,i_5,i_6,i_7,i_8$ are all contracted virtual indices. Because they are contracted with $g$ tensors where at least one virtual index is open, all the contracted virtual indices $i_1,i_2,i_3,i_4,i_5,i_6,i_7,i_8$ are equal to some open indices, and we denote them as 
\begin{equation}
	\begin{split}
		i_1=t_1,\; i_2=t_2,\; i_3=t_3,\; i_4=t_4, \\
		i_5=t_5,\; i_6=t_6,\; i_7=t_7,\; i_8=t_8.
	\end{split}
\end{equation}
The constraints on $\{i\}$'s are hence equivalent to the constraints on $\{t\}$'s, i.e., 
\begin{equation}
	t_1+t_2+t_3+t_4+t_5+t_6+t_7+t_8=0\mod 2.
\end{equation}
There is only one constraint from the $T$ tensor. Hence $N_c=1$ for $l=2$.

For $l=3$, as shown in the Fig.~\ref{fig.HaahTNS} (b), we have eight constraints from eight $T$ tensors which involve the open indices via the $g$ tensors. The eight equations are
\begin{equation}
	\begin{split}
		\sum_{n=1}^8 i_n^{(x,y,z)}=0\mod 2, \; x,y,z\in \{0,1\}
	\end{split}
\end{equation}
where the up-index $(x,y,z)$ represents the position of the $T$ tensor, and $n$ counts the eight indices of each cube in the $2\times 2 \times 2$ cut. All the $i$'s are contracted virtual indices. However, except the virtual indices that are connected with the central two $g$ tensors (which are defined on the two spins at the vertex $(x,y,z)=(1,1,1)$), all other indices (which are defined on two spins at vertices $(x,y,z)$, $x,y,z\in\{0,1,2\}$ except $(x,y,z)=(1,1,1)$) 
are equal to some open indices via $g$ tensors. Specifically, the virtual indices that are connected with the two center $g$ tensors are 
\begin{equation}
	\begin{split}
		i^{000}_{4}=i^{100}_{3}=i^{010}_{1}=i^{001}_{7}\mod 2\\
		i^{000}_{5}=i^{110}_{2}=i^{101}_{8}=i^{011}_{6}\mod 2.
	\end{split}
\end{equation}
Since we only count the number of constraints for the open indices, we need to Gauss-eliminate all these eight virtual indices $i^{000}_{4}, i^{100}_{3}, i^{010}_{1}, i^{001}_{7}, i^{000}_{5}, i^{110}_{2}, i^{101}_{8}, i^{011}_{6}$ from the above 8 equations. 
Therefore, we obtain $8-2=6$ independent equations in terms of open indices only. Hence there are 6 constraints for the open indices. 

For the general $l$, we apply the same principle. We first enumerate all possible constraints from the $T$ tensors, and then we Gauss-eliminate all the virtual indices that are contracted within the region $A$. 
Hence we obtain a set of equations purely in terms of the open indices. The number of constraints is the rank of this set of equations. We list the number of linear independent constraints for the open indices as follows:
\begin{widetext}
	\begin{equation}
		\begin{array}{@{}*{15}{c}@{}}
			l(\ge 3)\quad	&	3	&	4	&	5	&	6	&	7	&	8	&	9	&	10	&	11	&	12	&	13	&	14	&	15 & 16	\\
			N_c\quad	&	6	&	12	&	18	&	24	&	30	&	36	&	42	&	48	&	54	&	60	&	66	&	72	&	78	&	84	\\
		\end{array}
	\end{equation}
\end{widetext}
Hence, for $l \geq 3$, there are 
\begin{equation}
	6l-12
\end{equation} 
linearly independent constraints for the open indices.
Taking into account the fact that when $l=2$ the  number of constraints is $1$, we infer that the number of constraints for a generic $l$ is:
\begin{equation}
	6l-12+\delta_{l,2}.
\end{equation}

\subsubsection{Entanglement entropy}

We are ready to collect all the data we have obtained and compute the entanglement entropy for the cubic cut. 
For the entanglement cut of size $l\times l\times l$, the total number of $g$ tensors is 
\begin{equation}
	\begin{split}
		N_g
		&=12l^2-24l+16.
	\end{split}
\end{equation}
The number of of $T$ tensor constraints is
\begin{equation}
	\begin{split}
		N_{c}
		=6l-12+\delta_{l,2}, \forall l\geq 2.
	\end{split}
\end{equation}
The number of $B_c$ operators is
\begin{eqnarray}
N_B=6l^2-24l+26-\delta_{l,2}, \forall l\geq 2.
\end{eqnarray}
Therefore the upper bound of the entanglement entropy reads
\begin{equation}\label{Haahlllentanglemententropy}
\begin{split}
\frac{S}{\log 2}
=& N_{g}-N_c - N_B \\
=& 6l^2 - 6l + 2, \forall l\geq 2.
\end{split}
\end{equation}
The entanglement entropies also have negative linear corrections. 

If the region $\bar{A}$ is much larger than the region $A$, we conjecture that the region $\bar{A}$ will not impose any additional constraint. In that case, the upper bound would be saturated. The numerical calculations in App.~\ref{app.HaahcodeBruteforce} also support this conjecture.

\section{Conclusion and Discussion}\label{sec.discussion}

In this paper, we present our TNS construction for three stabilizer models in 3D. The ground states of these stabilizer codes are the eigenstates of all local Hamiltonian terms with $+1$ eigenvalues. The constructions of these TNSs share the same general idea and work in other dimensions as well:
\begin{enumerate}
\item We introduce a projector $g$ tensor for each physical spin which identifies the physical index with the virtual indices.
\item The physical operators acting on the TNS can be transferred to the virtual indices using Eq.~\eqref{eq.projectorcondition ToricCode}.
\item The local $T$ tensors contracted with the projector $g$ tensors are specified by the local Hamiltonian terms.
\end{enumerate}

After we obtain the TNS for the ground state, we can prove that the TNS is an exact SVD for the ground state with some specific entanglement cuts. The entanglement spectra are completely flat for the models studied in this paper. The entanglement entropies can be computed by counting the number of singular vectors. For the 3D toric code model, the entanglement entropies have a constant correction to the area law, $-\log(2)$. For the X-cube model and the Haah code, the entanglement entropies have linear corrections to the area law as shown in Sec.~\ref{subsec.Xcube_Entanglement} and \ref{subsec.Haah_Entanglement}. The analytical calculation of the entanglement entropies is rooted in the \textbf{Concatenation lemma}, since the \textbf{Concatenation lemma} is introduced to count the number of singular vectors. The \textbf{Concatenation lemmas} are rooted the symmetry properties of the local tensors. For instance, Eq.~\eqref{eq.ToricCodeTcondition} and \eqref{eq.ToricCodeTtensor} for the 3D toric code model.

The transfer matrices can also be constructed. For the 3D toric code and the X-cube models, we prove that the transfer matrix is a projector whose dimension is counted by the \textbf{Concatenation lemma} as well. For the 3D toric code model, the transfer matrix is of dimension 2. For the X-cube model, the transfer matrix is of dimension $2^{L_x+L_y-1}$ where $L_x$ and $L_y$ are the sizes of the torus in the $x$ and $y$ directions respectively. The GSD on the torus is generally larger than the degeneracy of the transfer matrix.

Since both the entanglement entropies and the transfer matrix degeneracies are rooted in the \textbf{Concatenation lemma} (or more fundamentally the symmetry properties of the local $T$ tensors), we believe that these two phenomena are related. Moreover, we conjecture that the negative linear correction to the area law is a signature of fracton models. This is similar to the negative constant correction (i.e., the topological entanglement entropy\cite{kitaev2006topological,levin2006detecting}) in 2D. 

In this paper, the TNSs are all the ground states of some exactly solvable local models. If we move away from these fine-tuned points without going through phase transitions, we expect the transfer matrix degeneracies to be still robust, since these degeneracies give rise to the GSD. In Ref.~\onlinecite{haegeman2015shadows}, this statement has been numerically verified in the 2D toric code model and its phase transitions to the trivial phases. 

If we move away from the fine-tuned points, we also expect that the linear term of the entanglement entropies for the fracton models does not vanish, although the specific coefficients of the linear terms might change. An important result is about the topological entanglement entropy. The topological entanglement entropy for the fracton models was first introduced in Ref.~\onlinecite{2017arXiv171001744M}, and is defined as the linear combinations of the entanglement entropies of different regions, in order to exactly cancel the area law. See Ref.~\onlinecite{2017arXiv171001744M} for the definition details. Importantly, the topological entanglement entropies of fracton models are linear with respect to the sizes of the entanglement cuts. Furthermore, Ref.~\onlinecite{2017arXiv171001744M} argues using perturbation theories that the topological entanglement entropies, of the same three models as in our paper are robust to adiabatic perturbations. Hence, Ref.~\onlinecite{2017arXiv171001744M} indicates that there should be a linear correction to the area law which does not vanish, even when moving away from the fine-tuned wave functions.

However, we also have to admit that the rigorous statements, about the entanglement spectra, entropies and the transfer matrix degeneracies of a generic fracton model ground state, need to be verified by the numerical studies for the 3D fracton models in the future.

\section*{Acknowledgments}

H. He and Y. Zheng wish to thank the support from the physics department of Princeton University.
N. R. acknowledges M. Hermanns and O. Petrova for fruitful discussions. 
The authors are grateful to G.~Sierra and B. Bradlyn for numerous and enlightening discussions about related topics of fracton models.
B. A. B. wishes to thank Ecole Normale Superieure, UPMC Paris, and the Donostia International Physics Center for their generous sabbatical hosting during some of the stages of this work. 
B. A. B. acknowledges support for the analytic work from NSF EAGER grant DMR - 1643312, ONR - N00014-14-1-0330,  NSF-MRSEC DMR-1420541.  The computational part of the Princeton work was performed under department of Energy de-sc0016239, Simons Investigator Award, the Packard Foundation, and the Schmidt Fund for Innovative Research.
N. R. was supported by Grants No. ANR-17-CE30-0013-01 and No. ANR-16-CE30-0025.

\emph{Note added}: A week before the submission of this manuscript, a very interesting paper Ref.~\onlinecite{2017arXiv171001744M} appeared which computes, by a completely different method, the entanglement entropies of the X-Cube model and the Haah code. When particularized to their cut, our approach gives the same results for these two models. Furthermore, we thank Rahul Nandkishore and Siddharth Parameswaran for discussions that made us realize that the bound obtained in an earier version of our manuscript for the entanglement entropy for the Haah code could be tightened and saturated so as to give matching results to their paper. 

\appendix
\clearpage
\numberwithin{figure}{section}
\numberwithin{table}{section}

\section{2D Toric Code TNS}\label{app.toriccode_2D}

\begin{figure}[b]
\centering
\includegraphics[width=0.6\columnwidth]{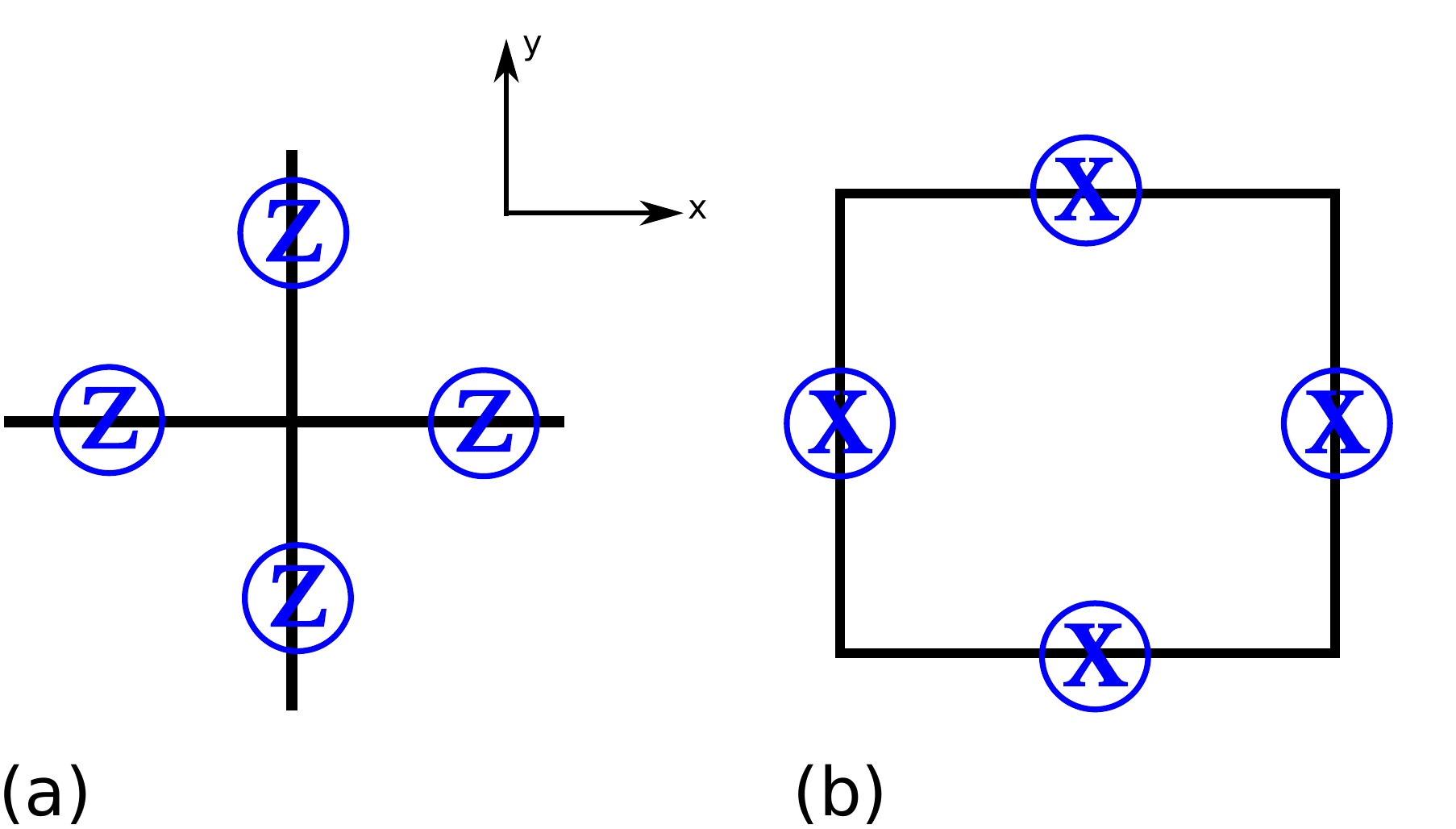}
\caption{The Hamiltonian terms of the 2D toric code model. Panel (a) is $A_{v}$ which is a product of 4 $Z$ operators around the vertex $v$, and Panel (b) is $B_{p}$ which is a product of 4 $X$ operators around the plaquette $p$.}
\label{fig.ToricCodeHamiltonian_2D}
\end{figure}

In this section, we construct the TNS for the 2D toric code model by using the same recipe in Sec.~\ref{sec.toriccode}, \ref{sec.Xcube} and \ref{sec.Haah}. In particular, the 2D toric code is defined on a square lattice with physical spins on all bonds, and the Hamiltonian consists of a sum of vertex terms and plaquette terms:
\begin{equation}
H = -\sum_{v} A_v - \sum_{p} B_p.
\end{equation}
Here, $A_v$ is the product of four Pauli $Z$ matrices around a vertex $v$ and $B_p$ is the product of four Pauli $X$ matrices around a plaquette $p$. 
\begin{equation}
A_v = \prod_{i \in v} Z_i, \quad
B_p = \prod_{i \in p} X_i.
\end{equation}
These two terms are illustrated in Fig.~\ref{fig.ToricCodeHamiltonian_2D}. The ground states of the 2D toric code model need to satisfy
\begin{equation}\label{eq.localcondition2Dtoriccode}
\begin{split}
A_v \ket{\mathrm{GS}} =& \ket{\mathrm{GS}}	\\
B_p \ket{\mathrm{GS}} =& \ket{\mathrm{GS}},	\\
\end{split}
\end{equation}
for all vertices $v$ and plaquettes $p$.

\begin{figure}[b]
\centering
\includegraphics[width=0.6\columnwidth]{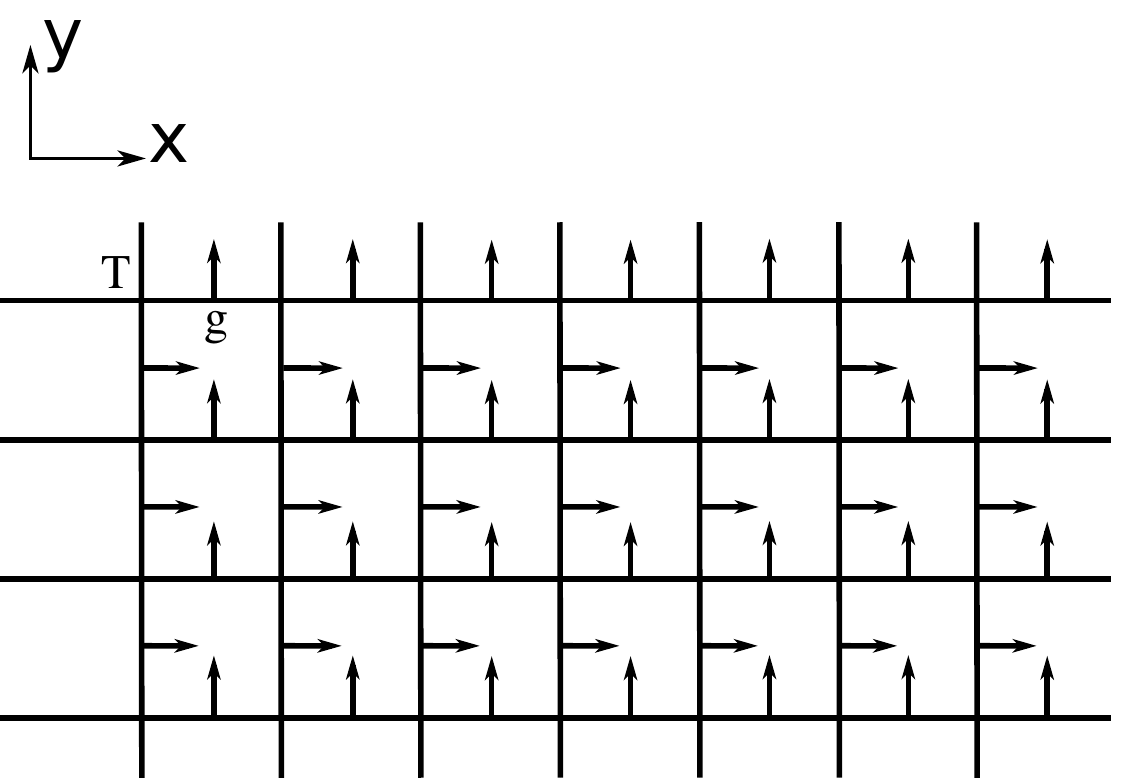}
\caption{The TNS for the 2D toric code model on a square lattice. On each bond, we associate a projector $g$ tensor, and on each vertex, we associate a local $T$ tensor. The connected lines are contracted virtual indices. The lines with arrows are the physical indices.}
\label{fig.TNS_2D}
\end{figure}

In order to construct the TNS for the 2D toric code model, we introduce projector $g$ tensors on each bond of the square lattice and local $T$ tensors on each vertex of the square lattice. The TNS is depicted in Fig.~\ref{fig.TNS_2D}. The projector $g$ tensor is the same as in Eq.~\eqref{eq.projector2}. We can use the same strategy as in Sec.~\ref{subsec.ToricCode_TNS} to implement the ground state conditions in Eq.~\eqref{eq.localcondition2Dtoriccode} on the TNS. The actions of $X$ and $Z$ Pauli matrices on the physical spins are transferred to the virtual indices using Eq.~\eqref{eq.projectorcondition ToricCode}:
\begin{equation}
\begin{gathered}
\includegraphics[width=0.9\columnwidth]{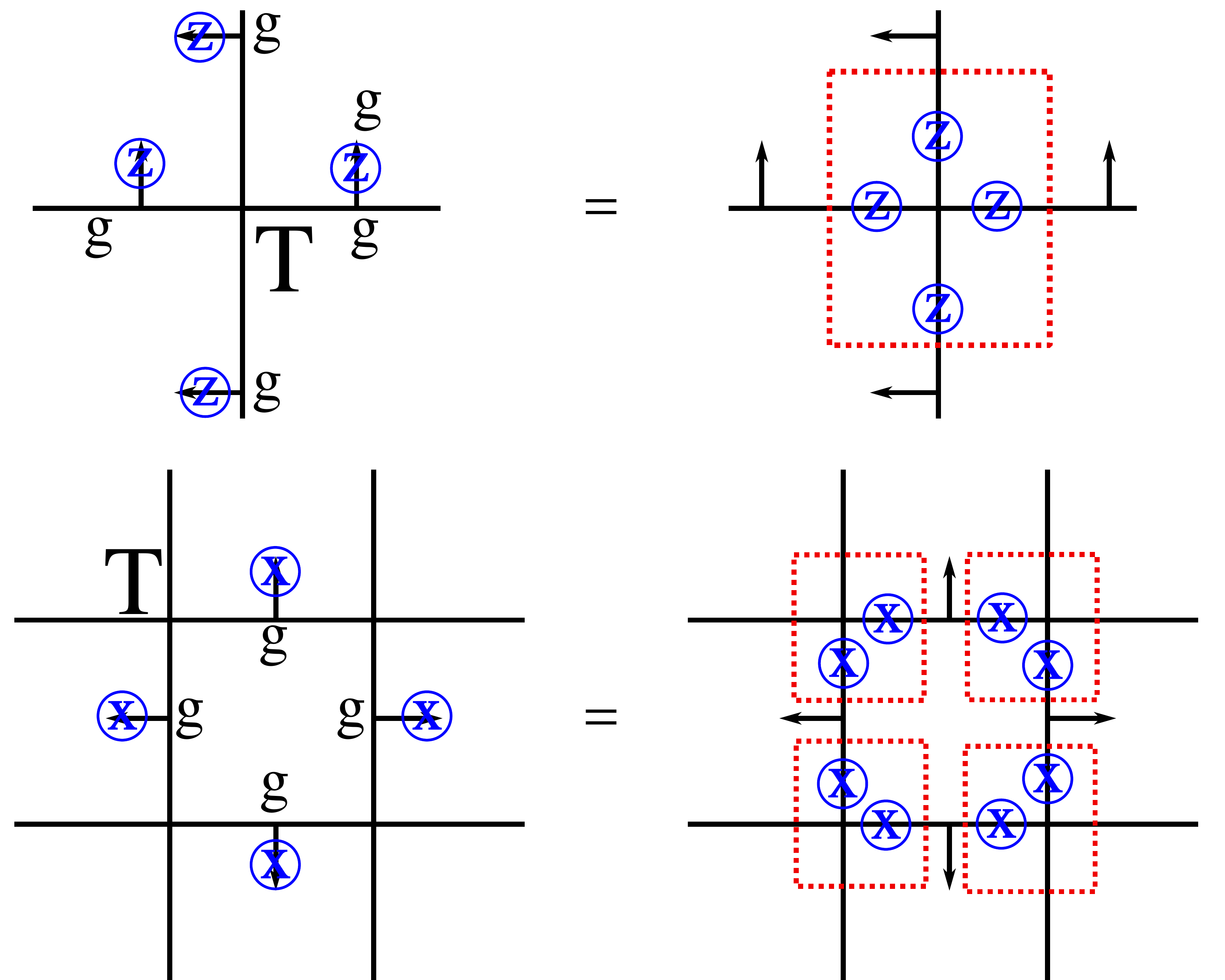}
\end{gathered}.
\end{equation}
In order to implement Eq.~\eqref{eq.localcondition2Dtoriccode}, we require the four Pauli $Z$ matrices (in the first line) and two Pauli $X$ matrices (in the second line) acting on the virtual indices in the dashed red squares to be identity operations. Then we have two (strong) conditions respectively:
\begin{equation}
\begin{split}
T_{x\bar{x},y\bar{y}} &= (-1)^{x+\bar{x}+y+\bar{y}} T_{x\bar{x},y\bar{y}}	\\
T_{x\bar{x},y\bar{y}} &= T_{(1-x)\bar{x},(1-y)\bar{y}} = T_{(1-x)\bar{x},y(1-\bar{y})}	\\
&= T_{x(1-\bar{x}),(1-y)\bar{y}} = T_{x(1-\bar{x}),y(1-\bar{y})}.
\end{split}
\end{equation}
As a result, the local $T$ tensors are fixed up to an overall factor:
\begin{equation}
T_{x\bar{x},y\bar{y}} = \begin{cases}
0 & \text{if } x+\bar{x}+y+\bar{y} = 1 \mod{2}	\\
1 & \text{if } x+\bar{x}+y+\bar{y} = 0 \mod{2}.	\\
\end{cases}
\end{equation}
The transfer matrix properties of such a TNS were studied in Ref.~\onlinecite{schuch2013topological}.

\section{Proof for the Concatenation Lemma for the 3D Toric Code Model}\label{app.ToricCode_Concatenation}

In this section, we prove the \textbf{Concatenation lemma} for the 3D toric code model by induction. First of all, we propose and prove two lemmas:
\begin{description}
\item[(A)] Let $\mathbf{T}_{t_1t_2t_3...}$ be a contraction of a network of local $T$ tensors, whose (i.e. open) indices $\{t_1t_2t_3...\}$ are un-contracted virtual indices. If $\mathbf{T}_{t_1t_2t_3...}$ satisfies the \textbf{Concatenation lemma} of the 3D toric code model, then the contraction of $\mathbf{T}_{t_1t_2t_3...}$ over a subset of its open virtual indices, say contracting over $t_1$ and $t_2$, i.e., $\sum_{t_1t_2} \mathbf{T}_{t_1t_2t_3...}\delta_{t_1t_2}$ still satisfies the \textbf{Concatenation lemma} of the 3D toric code model.
\item[(B)] If $\mathbf{T}_{t_1t_2t_3...}$ and $\tilde{\mathbf{T}}_{\tilde{t}_1\tilde{t}_2\tilde{t}_3...}$ are two networks of contracted local $T$ tensors both of which satisfying the \textbf{Concatenation lemma} of the 3D toric code model, then the contraction over one pair of indices, say $\sum_{t_1\tilde{t}_1} \mathbf{T}_{t_1t_2t_3...}\tilde{\mathbf{T}}_{\tilde{t}_1\tilde{t}_2\tilde{t}_3...}\delta_{t_1\tilde{t}_1}$, still satisfies the \textbf{Concatenation lemma} of the 3D toric code model.
\end{description}

\noindent\textbf{Proof:}

(A): Since $\mathbf{T}$ satisfies the \textbf{Concatenation lemma}, its elements $\mathbf{T}_{\{t\}}$ are:
\begin{equation}
\mathbf{T}_{\{t\}} = \begin{cases}
0	&	\text{if } \sum_{i} t_i = 1 \mod{2}	\\
N	&	\text{if } \sum_{i} t_i = 0 \mod{2},	\\
\end{cases}
\end{equation}
where $N$ is a constant independent of the open virtual indices in the \textbf{Concatenation lemma}.
Suppose that we contract two indices of $\mathbf{T}$, $t_m,\; t_n\in\{t\}$, and we denote the contraction as $\mathbf{T}^\prime$ and the remaining open virtual indices after the contraction $\{t^\prime\}$. Then we have:
\begin{equation}
\begin{split}
\mathbf{T}^\prime_{\{t^\prime\}} 
=& \sum_{t_m,t_n} \mathbf{T}_{\{t\}} \delta_{t_m,t_n}	\\
=& \sum_{t_m,t_n} \mathbf{T}_{\ldots t_m \ldots t_n \ldots} \delta_{t_m,t_n}	\\
=& \sum_{t_m} \mathbf{T}_{\ldots t_m \ldots t_m \ldots}	\\
=& \mathbf{T}_{\ldots 0 \ldots 0 \ldots}+\mathbf{T}_{\ldots 1 \ldots 1 \ldots}	\\
\end{split}
\end{equation}
Hence, the contraction still satisfies the \textbf{Concatenation lemma}:
\begin{equation}
\mathbf{T^\prime}_{\{t^\prime\}} = \begin{cases}
0	&	\text{if } \sum_{i} t^\prime_i = 1 \mod{2}	\\
2N	&	\text{if } \sum_{i} t^\prime_i = 0 \mod{2}.	\\
\end{cases}
\end{equation}

(B): Since $\mathbf{T}$ and $\tilde{\mathbf{T}}$ satisfy the \textbf{Concatenation lemma}, their elements $\mathbf{T}_{\{t\}}$ and $\tilde{\mathbf{T}}_{\{\tilde{t}\}}$ are:
\begin{equation}
\begin{split}
\mathbf{T}_{\{t\}} = \begin{cases}
0	&	\text{if } \sum_{i} t_i = 1 \mod{2}	\\
N	&	\text{if } \sum_{i} t_i = 0 \mod{2}	\\
\end{cases}	\\
\tilde{\mathbf{T}}_{\{\tilde{t}\}} = \begin{cases}
0	&	\text{if } \sum_{i} \tilde{t}_i = 1 \mod{2}	\\
\tilde{N}	&	\text{if } \sum_{i} \tilde{t}_i = 0 \mod{2},	\\
\end{cases}
\end{split}
\end{equation}
where $N$ and $\tilde{N}$ are the constants independent of the indices $\{t\}$ and $\{\tilde{t}\}$ respectively. Suppose that we contract two indices $t_m\in \{t\}$ and $\tilde{t}_n\in \{\tilde{t}\}$, and we denote the contraction as $\mathbf{T}^\prime$ and the remaining open virtual indices after the contraction $\{t^\prime\}$. Then we have:
\begin{equation}
\begin{split}
\mathbf{T}^\prime_{\{t^\prime\}} 
=& \sum_{t_m,\tilde{t}_n} \mathbf{T}_{\{t\}}\tilde{\mathbf{T}}_{\{\tilde{t}\}} \delta_{t_m,\tilde{t}_n}	\\
=& \sum_{t_m,\tilde{t}_n} \mathbf{T}_{\ldots t_m \ldots}\tilde{\mathbf{T}}_{\ldots \tilde{t}_n \ldots} \delta_{t_m,\tilde{t}_n}	\\
=& \mathbf{T}_{\ldots 0 \ldots}\tilde{\mathbf{T}}_{\ldots 0 \ldots} + \mathbf{T}_{\ldots 1 \ldots}\tilde{\mathbf{T}}_{\ldots 1 \ldots},
\end{split}
\end{equation}
The last line is nonzero if and only if $\sum_{i\ne m} t_i$ and $\sum_{j\ne n} \tilde{t}_n$ have the same parity. If this parity is even (resp. odd), only $\mathbf{T}_{\ldots 0 \ldots}\tilde{\mathbf{T}}_{\ldots 0 \ldots}$ (resp. $\mathbf{T}_{\ldots 1 \ldots}\tilde{\mathbf{T}}_{\ldots 1 \ldots}$) is nonzero and equal to $N \tilde{N}$. Since $\sum_i t^{\prime}_i = \sum_{i\ne m} t_i + \sum_{j\ne n} \tilde{t}_n$, we conclude that:
\begin{equation}
\mathbf{T}^\prime_{\{t^\prime\}} = 
\begin{cases}
0	&	\text{if } \sum_{i} t^\prime_i = 1 \mod{2}	\\
N\tilde{ N}	&	\text{if } \sum_{i} t^\prime_i = 0 \mod{2}.	\\
\end{cases}
\end{equation}
$\mathbf{T}^\prime_{\{t^\prime\}}$ still satisfies the \textbf{Concatenation lemma}.
\hfill$\Box$

Having proved Lemma (A) and (B), we can further prove that:
\begin{description}
\item[(C)] If $\mathbf{T}$ and $\tilde{\mathbf{T}}$ are two networks of contracted local $T$ tensors which both satisfy the \textbf{Concatenation lemma} of the 3D toric code model, then their contraction over any pairs of indices still satisfies the \textbf{Concatenation lemma} of the 3D toric code model.
\end{description}

\noindent\textbf{Proof:}

We can decompose the contraction process into two steps: (1) contract $\mathbf{T}$ and $\tilde{\mathbf{T}}$ over one pair of indices; (2) contract the rest of the indices. Lemma (B) guarantees that the outcome tensor of the contraction (1) still satisfies the \textbf{Concatenation lemma}. Lemma (A) guarantees that the outcome tensor of the contraction (2) also satisfies the \textbf{Concatenation lemma}. Hence, Lemma (C) is proved.
\hfill$\Box$

Now we can complete the induction proof for the \textbf{Concatenation lemma} of the 3D toric code model: First of all, we point out the a single local $T$ tensor satisfies the \textbf{Concatenation lemma}. Next, we assume that two networks of contracted local $T$ tensors satisfy the \textbf{Concatenation lemma}, and prove that their contraction also satisfies the \textbf{Concatenation lemma}. This induction step is, in fact, Lemma (C). Therefore, we have completed the induction proof for the \textbf{Concatenation lemma} of the 3D toric code model.

\section{Proof for the Concatenation Lemma for the X-cube Model}\label{app.Xcube_Concatenation}

In this section, we prove the \textbf{Concatenation lemma} for the X-cube model by induction. We point out that the subtlety of the \textbf{Concatenation lemma} of the X-cube model is that the number of constraints for the open virtual indices is linear with respect to the system size. More precisely, each constraint corresponds to a set of open virtual indices along a $xy$, $yz$ or $xz$ plane to have an even summation. See Eq.~\eqref{eq.localT} for an example. Hence, we need to keep track of the constraints when we contract more tensors. For clarity, we denote the sets of constraints of $\{t\}$ as:
\begin{equation}
\begin{split}
f_{xy}(\{t\}): \left\{ \sum_{i \in \text{$xy$-plane}} t_i,  \text{for all $xy$-planes} \right\}	\\
f_{yz}(\{t\}): \left\{ \sum_{i \in \text{$yz$-plane}} t_i,  \text{for all $yz$-planes} \right\}	\\
f_{xz}(\{t\}): \left\{ \sum_{i \in \text{$xz$-plane}} t_i,  \text{for all $xz$-planes} \right\}.	\\
\end{split}
\end{equation}
Notice that these equations are not linearly independent. Their summation is automatically true. Physically, we can view these quantities $f_{xy}(\{t\})$, $f_{yz}(\{t\})$ and $f_{xz}(\{t\})$ as the ``parities" in each $xy$, $yz$ and $xz$ plane.
To begin with, we propose and prove Lemma (D) which is the induction step:
\begin{description}
\item[(D)] If $\mathbf{T}$ is a network of contracted local $T$ tensors of the X-cube model which satisfies the \textbf{Concatenation lemma} of the X-cube model, then the contraction of $\mathbf{T}$ and a $T$ tensor still satisfies the \textbf{Concatenation lemma} of the X-cube model.
\end{description}

\noindent\textbf{Proof:}

We first fix the notations: 
the elements of $\mathbf{T}$ are $\mathbf{T}_{\{t\}}$;
$\{t^c\}=\{t^c_1,t^c_2,\ldots\}\subset \{t\}$ are the indices that contract with the indices of the local $T$ tensor;
and the outcome tensor is denoted as $\mathbf{T}^{\prime}_{\{t^\prime\}}$ where $\{t^\prime\}$ denotes the open virtual indices after contraction. Since $\mathbf{T}$ is assumed to satisfy the \textbf{Concatenation lemma}, we have:
\begin{equation}
\mathbf{T}_{\{t\}} = \begin{cases}
N	&	\text{if } 
\begin{cases}
f_{xy} (\{t\})=0 \mod{2}	\\
f_{yz} (\{t\})=0 \mod{2}	\\
f_{xz} (\{t\})=0 \mod{2}	\\
\end{cases}\\
0	&	\text{otherwise},	\\
\end{cases}
\end{equation}
where $N$ is the constant independent of the open virtual indices $\{t\}$, and $f_{xy} (\{t\})$, $f_{yz} (\{t\})$ and $f_{xz} (\{t\})$ denote the set of summations over the open virtual indices in each $xy$, $yz$ and $xz$ plane. See Eq.~\eqref{eq.Xcube_Ttensor_Example} for an example when $\mathbf{T}$ is a contraction of two local $T$ tensors. Notice that the elements of $T_{x\bar{x}y\bar{y}z\bar{z}}$ also satisfy the \textbf{Concatenation lemma} for the X-cube model, as shown by Eq.~\eqref{eq.localT}.

Using these notations, the tensor contraction is just:
\begin{equation}\label{eq.Xcube_induction_contraction}
\mathbf{T}^\prime_{\{t^\prime\}} = \sum_{\{t^c\}} \mathbf{T}_{\ldots t^c_1 \ldots t^c_2 \ldots } T_{\ldots t^c_1 \ldots t^c_2 \ldots }.
\end{equation}
We now discuss one particular way of contraction: contraction over one pair of indices. Other contractions can be proved using the exact same method.

Suppose $\{t^c\}$ contains only one index. Without loss of generality, we assume that this index is the $x$ index of $T_{x\bar{x}y\bar{y}z\bar{z}}$. 
Then the tensor contraction is:
\begin{equation}
\mathbf{T}^\prime_{\{t^\prime\}} 
= \sum_{x} \mathbf{T}_{\ldots x \ldots } T_{x\ldots}	
\end{equation}
Graphically, 
\begin{equation}
\begin{gathered}
\includegraphics[width=3cm]{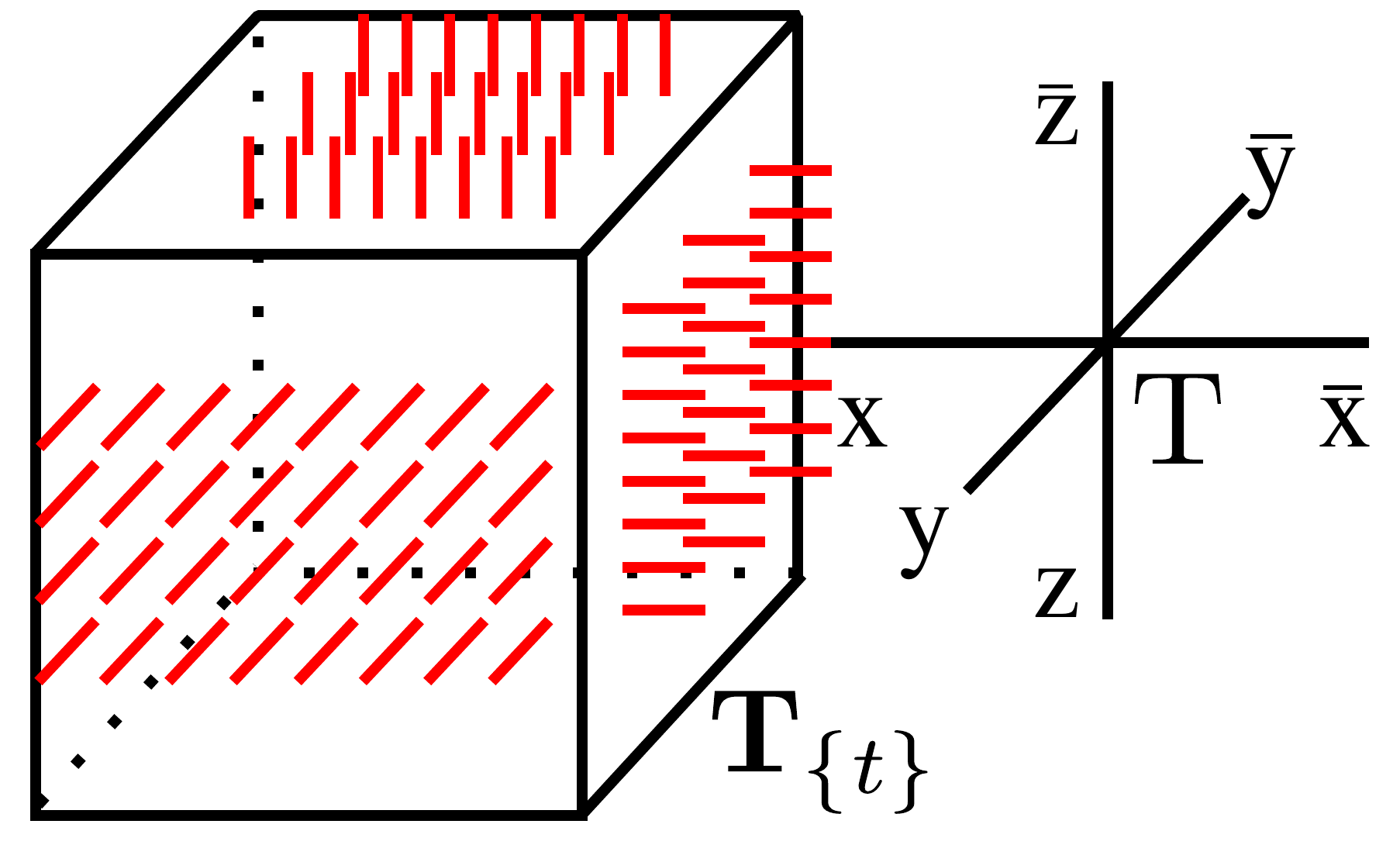}
\end{gathered}.
\end{equation}
When $\mathbf{T}_{\ldots x \ldots}$ and $T_{x\ldots}$ are both nonzero, the indices satisfy that:
\begin{equation}
\begin{split}
f_{xy}(\{t\}/{x},x) =& 0 \mod{2}	\\
f_{xz}(\{t\}/{x},x) =& 0 \mod{2}	\\
x+\bar{x}+y+\bar{y} =& 0 \mod{2}	\\
x+\bar{x}+z+\bar{z} =& 0 \mod{2}	\\
\Rightarrow&\\
f_{xy}(\{t\}/{x},\bar{x},y,\bar{y}) =& 0 \mod{2}	\\
f_{xz}(\{t\}/{x},\bar{x},z,\bar{z}) =& 0 \mod{2},	\\
\end{split}
\end{equation}
where $\{t\}/x$ denotes the $t$ indices excluding the $x$ index. We only list the equations whose variables include the index $x$. Then, we include the $f_{yz}$ constraints from $\mathbf{T}$ and $T$, which can be concatenated into $f_{yz}(\{t^\prime\})$. We find:
\begin{equation}
\mathbf{T}^\prime_{\{t^\prime\}} = \begin{cases}
N	&	\text{if } 
\begin{cases}
f_{xy} (\{t^\prime\})=0 \mod{2}	\\
f_{yz} (\{t^\prime\})=0 \mod{2}	\\
f_{xz} (\{t^\prime\})=0 \mod{2}	\\
\end{cases}\\
0	& 	\text{otherwise}.
\end{cases}
\end{equation}
Hence, $\mathbf{T}^\prime$ still satisfies the \textbf{Concatenation lemma} of the X-cube model. We can further contract other indices of $T$ tensor with $\mathbf{T}$. For instance, the index $y$ of $T$ tensor with another index of $\mathbf{T}$ in the same plane. Then the outcome tensor still satisfies the \textbf{Concatenation lemma} of the X-cube model, because (1) the ``parities" $f_{xy}$, $f_{yz}$ and $f_{xz}$ do not change after contraction, (2) the contraction is the same for the open indices of the same parities. Therefore, Lemma (D) is proved.
\hfill$\Box$

Having proven Lemma (D), now we can complete the induction proof for the \textbf{Concatenation lemma} of the X-cube model: First of all, we point out that a single local $T$ tensor of the X-cube model satisfies the \textbf{Concatenation lemma}. Next, as the induction step, we assume that one network of contracted local $T$ tensors satisfies the \textbf{Concatenation lemma}, and prove that contracting one more local $T$ tensor also satisfies the \textbf{Concatenation lemma}. This induction step is, in fact, Lemma (D). Therefore, we have completed the induction proof for the \textbf{Concatenation lemma} of the X-cube model.

\section{GSD for the X-cube Model}\label{app.XcubeGSD}

In this section, we work out the representation dimension of the operators in Eq.~\eqref{eq.figureoperatorsXcube}. In particular, the first group of algebras is that $W_X[C_x]$ anti-commutes with $W_Z[\tilde{C}_{z,x}]$ and $W_Z[\tilde{C}_{y,x}]$ when they have overlaps. In the projected $yz$ plane, the operators $W_X[C_x]$, $W_Z[\tilde{C}_{z,x}]$ and $W_Z[\tilde{C}_{y,x}]$ can be depicted as:
\begin{equation}
\begin{gathered}
\includegraphics[width=0.5\columnwidth]{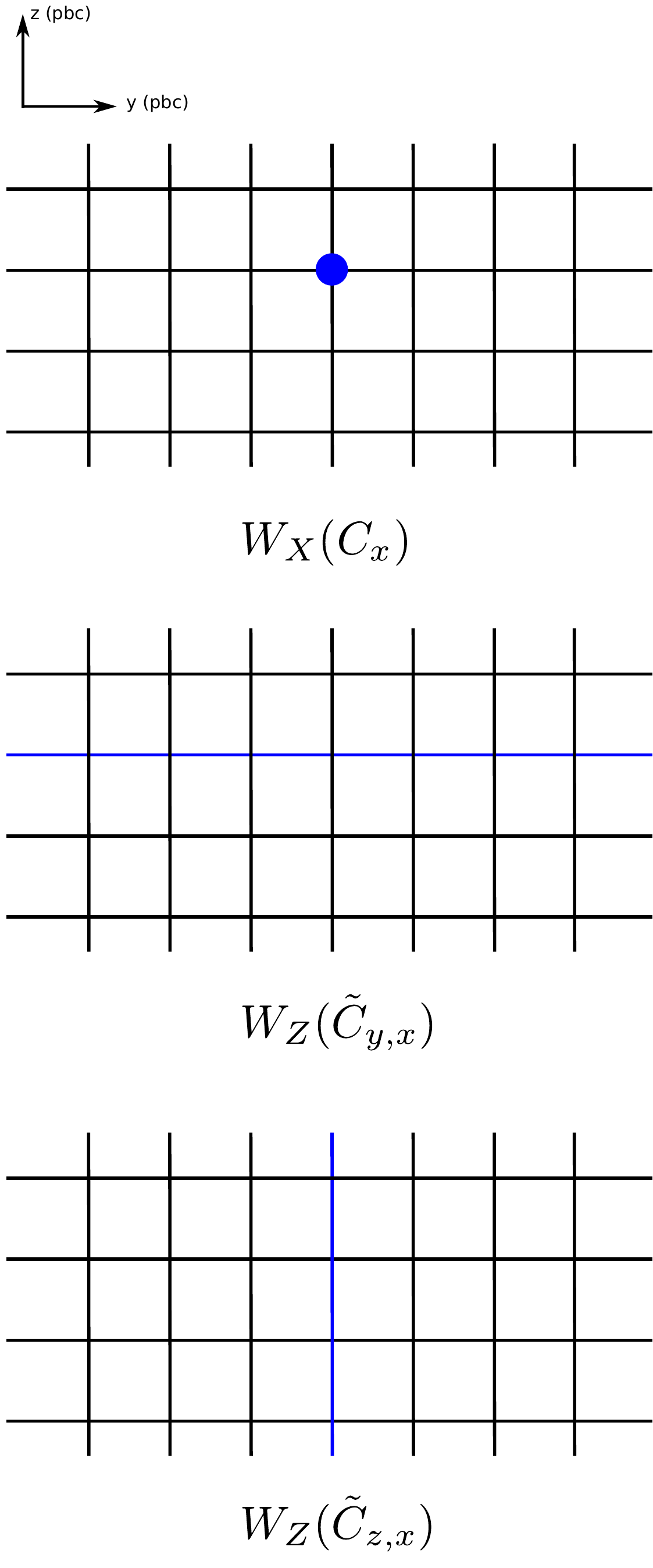}
\end{gathered},
\end{equation}
where the blue dot and the blue lines denote the projected operators on the $yz$-plane. There are $L_yL_z$ number of $W_X$ operators and $L_y+L_z$ number of $W_Z$ operators. 
The anti-commutation relations happens when they have overlaps:
\begin{equation}
\begin{gathered}
\includegraphics[width=0.5\columnwidth]{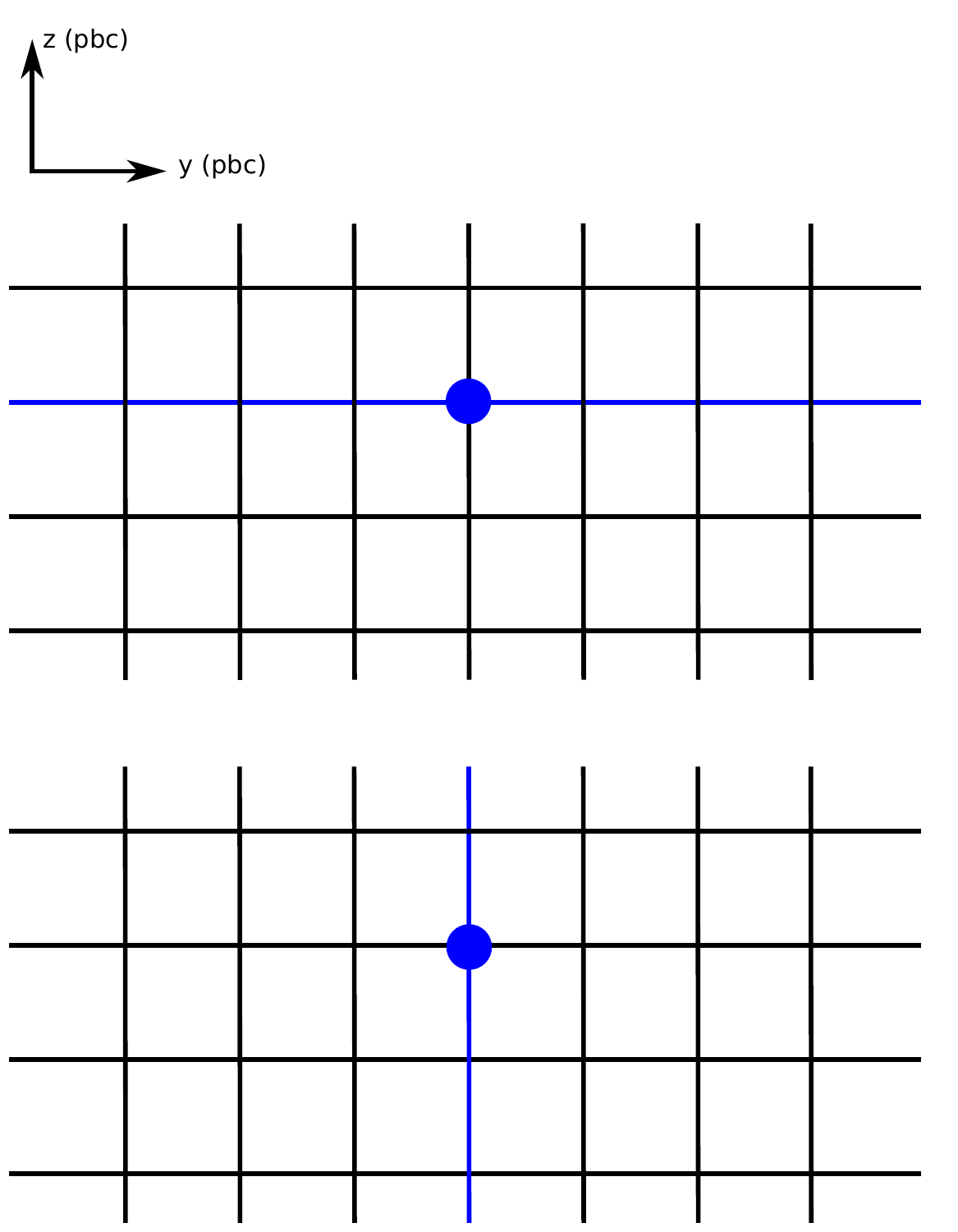}
\end{gathered}.
\end{equation}
Other combinations of operators commute. In a more pictorial language, the commutation relations are just that the blue point at the coordinate $(y,z)$ flips the vertical line and the horizontal line passing the blue point $(y,z)$. In this pictorial language, we can find that we can flip any pair of lines independently using the points. In the operator language, we can flip any pair of $W_Z$ operators using $W_X$ operators. For instance:
\begin{equation}
\begin{gathered}
\includegraphics[width=0.5\columnwidth]{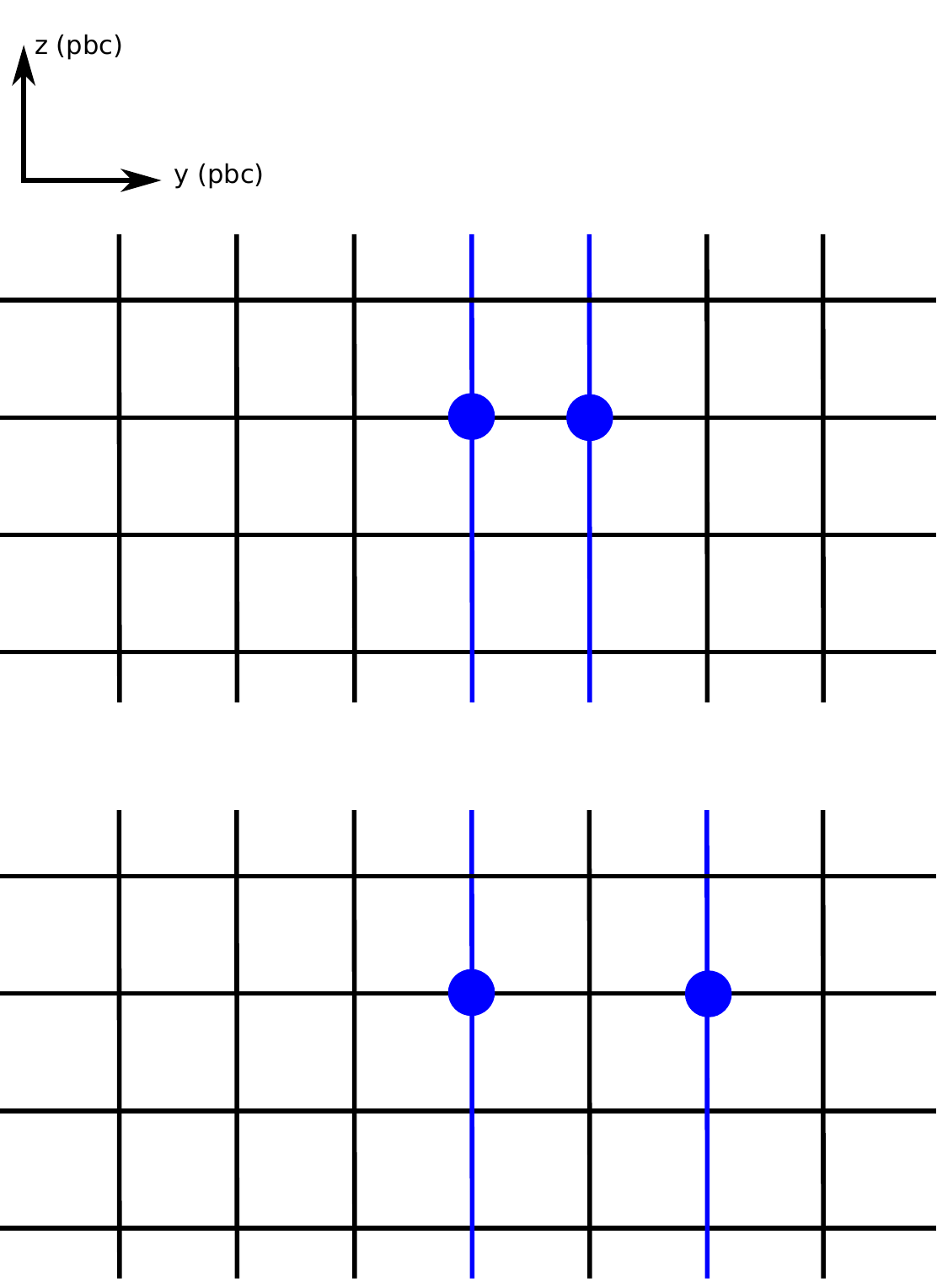}
\end{gathered}.
\end{equation}
Therefore, we set a ``reference" line and flip all other $L_y+L_z-1$ number of lines using the dots. In the operator language, we set a ``reference" $W_Z$ operator in this projected $yz$-plane, and flip all other $L_y+L_z-1$ number of $W_Z$ operators using $W_X$ operators.
Hence, we can generate $2^{L_y+L_z-1}$ dimensional Hilbert space using operators $W_X[C_x]$, $W_Z[\tilde{C}_{z,x}]$ and $W_Z[\tilde{C}_{y,x}]$. Similarly for other algebras below Eq.~\eqref{eq.figureoperatorsXcube}, we can generate $2^{L_x+L_z-1}$ and $2^{L_x+L_y-1}$ dimensional Hilbert space respectively. The ground state degeneracy is then their product: $2^{2L_x+2L_y+2L_z-3}$.

\begin{table}[t]
\begin{center}
 \begin{tabular}{||c | c ||} 
 \hline
 size $L_x\times L_y$~~~ & number of degenerate $1$'s \\ [0.5ex] 
 \hline\hline
 $1\times1$ & 2 \\ 
 \hline
 $1\times2$ & 2 \\
 \hline
 $1\times3$ & 2\\
 \hline
 $1\times4$ & 2\\
 \hline
 $1\times5$ & 2\\
 \hline
 $1\times6$ & 2\\
 \hline
 $2\times2$ & 2\\
 \hline
 $2\times3$ & 2\\
 \hline
 $2\times4$ & 2 \\ [1ex] 
 \hline
\end{tabular}
\end{center}
\caption{The number of degenerate $1$'s in the eigenvalues of the toric code transfer matrices of size $L_x \times L_y$. The other eigenvalues are all zeros.}
\label{table.ToricCodetransfermatrix}
\end{table}

\begin{table}[b]
\begin{center}
 \begin{tabular}{||c | c ||} 
 \hline
 size $L_x\times L_y$~~~ & number of degenerate $1$'s \\ [0.5ex] 
 \hline\hline
 $1\times1$ & 2 \\ 
 \hline
 $1\times2$ & 4 \\
 \hline
 $1\times3$ & 8\\
 \hline
 $1\times4$ & 16\\
 \hline
 $1\times5$ & 32\\
 \hline
 $1\times6$ & 64\\
 \hline
 $2\times2$ & 8\\
 \hline
 $2\times3$ & 16\\
 \hline
 $2\times4$ & 32 \\ [1ex] 
 \hline
\end{tabular}
\end{center}
\caption{The number of degenerate $1$'s in the eigenvalues of the X-cube transfer matrices of size $L_x \times L_y$. The other eigenvalues are all zeros.}
\label{table.Xcubetransfermatrix}
\end{table}

\section{TNS as a Projected Wave Function}\label{app.TNS_projected}

The ground states of the stabilizer codes can be written as projected wave functions. In this section, we will show that our TNSs are generically projected wave functions using the local Hamiltonian terms. In particular, we detail here the case of the 3D toric code model discussed in Sec.~\ref{sec.toriccode}. The proofs for other stabilizer codes are the same.

To begin with, we specify our projected wave function on the torus as:
\begin{equation}\label{Projectedwavefunction}
\ket{\psi} = \prod_{v} \left( \frac{1+A_v}{2}\right) \prod_{p} \left(\frac{1+B_p}{2}\right) \ket{0000\ldots}_x,
\end{equation}
where the local spin basis $\ket{0}_x$ satisfies:
\begin{eqnarray}
X \ket{0}_x = \ket{0}_x,
\end{eqnarray}
and $\ket{0000\ldots}_x$ is a product state. Hence, $B_p$ operators do not change $\ket{0000\ldots}_x$. The projected wave function then becomes:
\begin{equation}
\ket{\psi} = \prod_{v} \left(\frac{1+A_v}{2}\right) \ket{0000\ldots}_x.
\end{equation}
Now we show that:
\begin{equation}
\ket{\mathrm{TNS}} = \ket{\psi}
\end{equation}
for the 3D toric code model.

The TNS in Sec.~\ref{sec.toriccode} is constructed by two parts: (1) the projector $g$ tensors map the physical spins to the virtual spins; (2) the 6 virtual spins around a vertex are constrained by the local $T$ tensor at the vertex $v$. Equivalently, the TNS is a equal weight superposition of spin configurations in the Pauli $Z$ basis, such that $T$ tensors need to be nonzero.
Hence, our TNS for the toric code model can be written as:
\begin{equation}
\ket{\mathrm{TNS}} = \frac{1}{\mathcal{N}} \sum_{\{s\}} \prod_{v} \delta^v(\{s\}) \ket{\{s\}}_z,
\end{equation}
where $\ket{\{s\}}_z$ is the Pauli $Z$ basis, $\mathcal{N}$ is an overall normalization factor and $\delta^v(\{s\})$ enforces the 6 spins around the vertex $v$ that:
\begin{equation}
\sum_{s \in v} s = 0 \mod{2}, \quad s\in\{0,1\},
\end{equation}
otherwise the spin configuration has weight zero. We can write the $\delta^v$ using a projection operator:
\begin{equation}
\delta^v(\{s\}) \ket{\{s\}}_z = \left( \frac{1+A_v }{2} \right) \ket{\{s\}}_z.
\end{equation}
Hence, the TNS is:
\begin{equation}
\ket{\mathrm{TNS}} = \frac{1}{\mathcal{N}} \sum_{\{s\}} \prod_{v} \left( \frac{1+A_v}{2} \right) \ket{\{s\}}_z.
\end{equation}
We can switch the order of the projectors and the summation:
\begin{equation}\label{eq.toriccode_projection}
\begin{split}
\ket{\mathrm{TNS}} &= \prod_{v} \left( \frac{1+A_v}{2} \right) \sum_{\{s\}}  \frac{1}{\mathcal{N}}\ket{\{s\}}_z	\\
&= \prod_{v} \left(\frac{1+A_v}{2}\right)\ket{0000\ldots}_x	\\
&= \ket{\psi}.
\end{split}
\end{equation}
where we have used that:
\begin{equation}
\frac{1}{\sqrt{2}} \left( \ket{0}_z+\ket{1}_z \right) = \ket{0}_x,
\end{equation}
and the normalization factor $\mathcal{N}$ is canceled out.
Hence we have proved that for 3D toric code model, the TNS is a particular projected wave function. 

The proof can be repeated for the X-cube model and Haah code. In summary, the conclusions are:
\begin{equation}\label{eq.Xcube_projection}
\begin{split}
&\ket{\mathrm{TNS(\text{X-cube})}} \\
=& \prod_{v} \frac{1}{8} \left( 1+A_{v,xy}\right)\left( 1+A_{v,yz}\right)\left( 1+A_{v,xz}\right) \\
&\prod_{c} \frac{1}{2} \left( 1+B_{c}\right) \ket{000\ldots}_x
\end{split}
\end{equation}
and
\begin{equation}\label{eq.Haah_projection}
\begin{split}
\ket{\mathrm{TNS(\text{Haah})}} 
=& \prod_{c} \frac{1}{4} \left( 1+A_{c}\right) \left( 1+B_{c}\right) \ket{000\ldots}_x.
\end{split}
\end{equation}
Notice that our proof does not depend on the spatial manifold. The results in Eq.~\eqref{eq.toriccode_projection}, \eqref{eq.Xcube_projection} and \eqref{eq.Xcube_projection} also hold true on other closed manifolds.

\section{Numerics for Transfer Matrix Degeneracies for the Toric Code and X-cube Model}\label{app.transfermatrixnumerics}

The eigenvalues of the transfer matrix for the toric code model and the X-cube model have been numerically computed exactly and verified for small system sizes. We contract all the tensors by brute force evaluation of Eq.~\eqref{eq.TransferMatrix}, without using Eq.~\eqref{eq.projectorcontraction} or \eqref{eq.TransferMatrixContractT}. The data of the degenerate eigenvalues of the transfer matrices are listed in Table \ref{table.ToricCodetransfermatrix} and \ref{table.Xcubetransfermatrix}. They are consistent with the analytical results in Sec.~\ref{subsec.ToricCode_TransferMatrix} and \ref{subsec.Xcube_TransferMatrix}.

\section{Numerics for Entanglement Entropy and Entanglement Spectrum of the X-cube Model}
\label{AppNumericsXCube}

In this appendix, we present some numerical results of the entanglement entropies and entanglement spectra for the X-Cube model. The numerical procedure is described below.

\begin{figure}
\centering
\includegraphics[width=0.25\columnwidth]{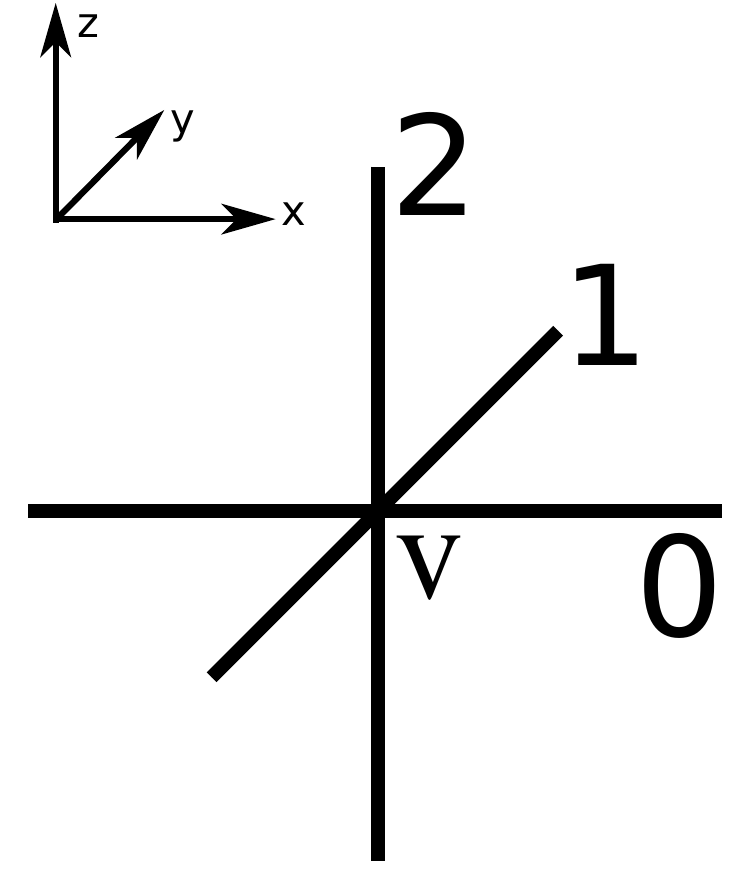}
\caption{The illustration for the ``Direction" catalog in Table \ref{TableXcubenumerics}. Since
the model has three spins per unit cell, we need an extra label to specify them. This extra label, the ``Direction" is the $0$, $1$ and $2$ in this figure. This is the same label that we use in the Table \ref{TableXcubenumerics} with respect to a vertex $v$.}
\label{fig.Direction}
\end{figure}

We start with the TNS of the X-cube defined on $\mathcal{T}^3$ of the size $L\times L\times L$. We choose a bipartition of the TNS $|\mathrm{TNS}\rangle=\sum_{\{t\}}|\{t\}\rangle_A\otimes |\{t\}\rangle_{\bar{A}}$
where $\{t\}$ is the set of open virtual indices. The entanglement cut is chosen such that in region $A$, $\{t\}$ connect directly with $g$ tensors, and in region $\bar{A}$, $\{t\}$ connect directly with $T$ tensors. For region A with small number of spins and small total system size $L$, we numerically trace over the degree of freedoms in region $\bar{A}$ and compute the reduced density matrix $\rho_A=\Tr_{\mathcal{H}_{\bar{A}}}\ket{\mathrm{TNS}}\bra{\mathrm{TNS}}$, and diagonalize it to find the entanglement spectrum, from which we can determine the entanglement entropy. Since in App.~\ref{app.TNS_projected} we have shown that the $\ket{\mathrm{TNS}}$ written in the $Z$-basis equals the projected wave function Eq.~\eqref{Projectedwavefunction}, we also directly generate the wave function in Eq.~\eqref{Projectedwavefunction} and perform the SVD to find the singular values (which equals the square root of the entanglement spectrum). For all the cases we computed, we find that the two methods agree with each other, and the non-zero eigenvalues of the reduced density matrix is fully degenerate. We denote the degenerate eigenvalue to be $\lambda$. When the reduced density matrix is diagonal (i.e., when the $\ket{\mathrm{TNS}}$ is SVD), we find that the rank of the reduced density matrix and the same entanglement entropy matches with our theoretical calculation. We list the numerical results in Table.~\ref{TableXcubenumerics}. The coordinates and the notion of ``Direction" are explained in Fig.~\ref{fig.Direction}. 

\begin{table}[t]
	\begin{center}
		\begin{tabular}{ c | c | c | c | c | c | c | c | c}
			\hline
			\multirow{2}*{System Size} &	\multicolumn{4}{c|}{Coordinate of Region $A$} & \multirow{2}*{$\lambda$}& \multirow{2}*{$S/\log 2$} & \multirow{2}*{SVD?} & \multirow{2}*{No.} \\ \cline{2-5}
			&	$x$ & $y$ & $z$ & Direction & & &\\ \cline{1-9}
			\multirow{21}*{$2\times 2\times 2$}&	0 & 0 & 0 & 0 & $\frac{1}{2}$& 1 & Yes & 1\\ \cline{2-9}
			&0 & 0 & 0 & 0 & \multirow{2}*{$\frac{1}{4}$}& \multirow{2}*{2} &  \multirow{2}*{Yes} & \multirow{2}*{2}\\\cline{2-5}
			&0 & 0 & 0 & 1 & & & &\\ \cline{2-9}
			&0 & 0 & 0 & 0 & \multirow{3}*{$\frac{1}{8}$} & \multirow{3}*{3} &  \multirow{3}*{Yes}& \multirow{3}*{3}\\\cline{2-5}
			&0 & 0 & 0 & 1 & & &\\\cline{2-5}
			&0 & 0 & 0 & 2 & & & \\ \cline{2-9}
			&0 & 0 & 0 & 0 & \multirow{4}*{$\frac{1}{8}$} & \multirow{4}*{3} &  \multirow{4}*{No} & \multirow{4}*{4} \\\cline{2-5}
			&0 & 0 & 0 & 1 & & &\\\cline{2-5}
			&0 & 0 & 0 & 2 & & &\\\cline{2-5}
			&1 & 0 & 0 & 0 & & &\\ \cline{2-9}
			&0 & 0 & 0 & 0 & \multirow{5}*{$\frac{1}{16}$} & \multirow{5}*{4} &  \multirow{5}*{No} & \multirow{5}*{5}\\\cline{2-5}
			&0 & 0 & 0 & 1 & & & \\\cline{2-5}
			&0 & 0 & 0 & 2 & & & \\\cline{2-5}
			&1 & 0 & 0 & 0 & & & \\\cline{2-5}
			&1 & 0 & 0 & 1 & &  &\\\cline{2-9}
			&0 & 0 & 0 & 0 & \multirow{6}*{$\frac{1}{32}$} & \multirow{6}*{5} &  \multirow{6}*{No} & \multirow{6}*{6}  \\\cline{2-5}
			&0 & 0 & 0 & 1 & & &\\\cline{2-5}
			&0 & 0 & 0 & 2 & & &\\\cline{2-5}
			&0 & 0 & 0 & 0 & & &\\\cline{2-5}
			&0 & 0 & 0 & 1 & & &\\\cline{2-5}
			&0 & 0 & 0 & 2 & & &\\\hline
			\multirow{6}*{$3\times 3\times 3$}&	0 & 0 & 0 & 0 & \multirow{6}*{$\frac{1}{16}$} & \multirow{6}*{4} & \multirow{6}*{Yes} & \multirow{6}*{7}\\ \cline{2-5}
			&1 & 0 & 0 & 0 & & &\\\cline{2-5}
			&1 & 0 & 0 & 1 & & &\\\cline{2-5}
			&1 & 2 & 0 & 1 & & &\\\cline{2-5}
			&1 & 0 & 0 & 2 & & &\\\cline{2-5}
			&1 & 0 & 2 & 2 & & &\\\hline
			\multirow{11}*{$4\times 4\times 4$}&	0 & 0 & 0 & 0 & \multirow{11}*{$\frac{1}{128}$} & \multirow{11}*{7} & \multirow{11}*{Yes} & \multirow{11}*{8}\\ \cline{2-5}
			& 1 & 0 & 0 & 0 & & &\\\cline{2-5}
			& 2 & 0 & 0 & 0 & & &\\\cline{2-5}
			& 1 & 0 & 0 & 1 & && \\\cline{2-5}
			& 1 & 3 & 0 & 1 & && \\\cline{2-5}
			& 2 & 0 & 0 & 1 & && \\\cline{2-5}
			& 2 & 3 & 0 & 1 & && \\\cline{2-5}
			& 1 & 0 & 0 & 2 & && \\\cline{2-5}
			& 1 & 0 & 3 & 2 & && \\\cline{2-5}
			& 2 & 0 & 0 & 2 & && \\\cline{2-5}
			& 2 & 0 & 3 & 2 & && \\\cline{1-9}
		\end{tabular}
		\caption{Entanglement entropies for various bipartitions of the $\ket{\mathrm{TNS}}$ of the X-cube model with system size $L\times L\times L$ ($L=2, 3, 4$). The second to fourth column list the coordinates of vertices in region $A$. The column of "Direction" labels one of the three bonds connecting with the vertex at position $(x,y,z)$ (in the positive direction), where $0,1$ and $2$ corresponds to the $x, y$ and $z$ direction respectively. }
		\label{TableXcubenumerics}
	\end{center}
\end{table}

We further explain how the numerical results match with our theoretical calculation when the bipartition is SVD. For the system size $2\times 2\times 2$,
when there is only one spin in region $A$, there are two open virtual indices connecting with the same $g$ tensor which identifies them. Hence there is only one independent open virtual index, and the entanglement entropy is $1\times \log 2$. 
When there are two spins in region $A$ whose coordinates are shown in Table.~\ref{TableXcubenumerics}, there are four virtual open indices, which are identified in pairs by two $g$ tensors. Hence there are two independent open virtual indices, and the entanglement entropy is $2\times \log 2$.
When there are three spins in region $A$ whose coordinates are shown in Table.~\ref{TableXcubenumerics}, there are six virtual open indices, which are identified in pairs by three $g$ tensors. Hence there are three independent open virtual indices, and the entanglement entropy is $3\times \log 2$.

For the system size $3\times 3\times 3$,  there are six spins in region $A$ whose coordinates are shown in Table.~\ref{TableXcubenumerics}. These six spins are connected with a $T$ tensor. There are six open virtual indices, and they subject to two independent constraints coming from $T$ tensor: the sum of four virtual indices in the $xy$ plane (which are $(x,y,z, \mathrm{Direction})=(0,0,0,0),(1,0,0,0),(1,0,0,1),(1,2,0,1)$) should be even, and the sum of four virtual indices in the $xz$ plane (which are $(x,y,z, \mathrm{Direction})=(0,0,0,0),(1,0,0,0),(1,0,0,2),(1,0,2,2)$) should be even. Hence there are $4=6-2$ independent open virtual indices, and the entanglement entropy is $4\times \log 2$.

For the system size $4\times 4\times 4$,  there are 11 spins in region $A$ whose coordinates are shown in Table.~\ref{TableXcubenumerics}. These 11 spins are connected with two contracted $T$ tensors. There are 10 open virtual indices, and they subject to four independent constraints coming from $T$ tensor: the sum of six virtual indices in the $xy$ plane (which are  $(x,y,z, \mathrm{Direction})=(0,0,0,0), (2, 0, 0, 0), (1,0,0,1), (1,3,0,1), (2,0,0,1), (2,3,0,1)$) should be even, the sum of six virtual indices in the $xz$ plane (which are $(x,y,z, \mathrm{Direction})=(0,0,0,0), (2,0,0,0), (1,0,0,2), (1,0,3,2), (2,0,0,2), (2,0,3,2)$) should be even, and the sum of four spins in the two parallel $yz$ planes respectively (which are $(x,y,z, \mathrm{Direction})=(1,0,0,1), (1,0,0,2), (1,3,0,1), (1,0,3,2)$ and $(x,y,z, \mathrm{Direction})=(2,0,0,1), (2,0,0,2), (2,3,0,1), (2,0,3,2)$) should be even respectively. Hence there are $7=11-4$ independent open virtual indices, and the entanglement entropy is $7\times \log 2$. 

\section{Numerics for Haah Code}
\label{app.HaahcodeBruteforce}

In this appendix, we present the results of numerical calculations for the entanglement entropies with various cuts. The numerical procedure is explained below.

We start with the TNS of the Haah code defined on $\mathcal{T}^3$ of the size $L_x\times L_y\times L_z$. We choose a bipartition of the TNS,  $|\mathrm{TNS}\rangle=\sum_{\{t\}}|\{t\}\rangle_A\otimes |\{t\}\rangle_{\bar{A}}$ where $\{t\}$ is the set of open virtual indices. For a given choice of region $A$, we then compute the reduced density matrix (RDM) $\rho_A=\Tr_{\bar{A}}|\mathrm{TNS}\rangle \langle \mathrm{TNS}|$, and diagonalize the RDM. Since in App.~\ref{app.TNS_projected} we shown that the $|\mathrm{TNS}\rangle$ in the $Z$ basis equals the projected wave function Eq.~\eqref{Projectedwavefunction} in the $X$ basis,  we also use Eq.~ \eqref{Projectedwavefunction} to compute the singular values. For all the cases we computed, the non-zero eigenvalues of the RDM is fully degenerate, which we denote as $\lambda$.

\begin{table}[H]
	\begin{center}
			\begin{tabular}{ c | c | c | c | c | c | c | c | c}
				\hline
				\multirow{2}*{System Size} &	\multicolumn{4}{c|}{Coordinate of Region $A$} & \multirow{2}*{$\lambda$}& \multirow{2}*{$S/\log 2$} & \multirow{2}*{SVD?} & \multirow{2}*{No.} \\ \cline{2-5}
				&	$x$ & $y$ & $z$ & Left/Right & & &\\ \cline{1-9}
					\multirow{8}*{$3\times 3\times 3$}&1 & 1 & 0 & 0 & \multirow{8}*{$\frac{1}{128}$} & \multirow{8}*{7} & \multirow{8}*{Yes} & \multirow{8}*{1}\\ \cline{2-5}
					&1 & 0 & 1 & 0 & & &\\\cline{2-5}
					&0 & 1 & 1 & 0 & & &\\\cline{2-5}
					&1 & 1 & 1 & 0 & & &\\\cline{2-5}
					&1 & 0 & 0 & 1 & & &\\\cline{2-5}
					&0 & 1 & 0 & 1 & & &\\\cline{2-5}
					&0 & 0 & 1 & 1 & & &\\\cline{2-5}
					&1 & 1 & 1 & 1 & & &\\\hline
					\multirow{15}*{$4\times 4\times 4$}&1 & 1 & 0 & 0 & \multirow{15}*{$\frac{1}{8192}$} & \multirow{15}*{13} & \multirow{15}*{Yes} & \multirow{15}*{2}\\ \cline{2-5}
					&1 & 0 & 1 & 0 & & &\\\cline{2-5}
					&0 & 1 & 1 & 0 & & &\\\cline{2-5}
					&1 & 1 & 1 & 0 & & &\\\cline{2-5}
					&1 & 2 & 0 & 0 & & &\\\cline{2-5}
					&1 & 2 & 1 & 0 & & &\\\cline{2-5}
					&0 & 2 & 1 & 0 & & &\\\cline{2-5}
					&1 & 0 & 0 & 1 & & &\\\cline{2-5}
					&0 & 1 & 0 & 1 & & &\\\cline{2-5}
					&0 & 0 & 1 & 1 & & &\\\cline{2-5}
					&1 & 1 & 1 & 1 & & &\\\cline{2-5}
					&0 & 2 & 0 & 1 & & &\\\cline{2-5}
					&1 & 2 & 1 & 1 & & &\\\cline{2-5}
					&1 & 1 & 0 & 1 & & &\\\cline{2-5}
					&0 & 1 & 1 & 1 & & &\\\hline
						\multirow{16}*{$3\times 3\times 3$}&0 & 0 & 0 & 0 & \multirow{16}*{$\frac{1}{16384}$} & \multirow{16}*{14} & \multirow{16}*{No} & \multirow{16}*{3}\\ \cline{2-5}
						&0 & 0 & 0 & 1 & & &\\\cline{2-5}
						&1 & 0 & 0 & 0 & & &\\\cline{2-5}
						&1 & 0 & 0 & 1 & & &\\\cline{2-5}
						&0 & 1 & 0 & 0 & & &\\\cline{2-5}
						&0 & 1 & 0 & 1 & & &\\\cline{2-5}
						&1 & 1 & 0 & 0 & & &\\\cline{2-5}
						&1 & 1 & 0 & 1 & & &\\\cline{2-5}
						&0 & 0 & 1 & 0 & & &\\\cline{2-5}
						&0 & 0 & 1 & 1 & & &\\\cline{2-5}
						&1 & 0 & 1 & 0 & & &\\\cline{2-5}
						&1 & 0 & 1 & 1 & & &\\\cline{2-5}
						&0 & 1 & 1 & 0 & & &\\\cline{2-5}
						&0 & 1 & 1 & 1 & & &\\\cline{2-5}
						&1 & 1 & 1 & 0 & & &\\\cline{2-5}
						&1 & 1 & 1 & 1 & & &\\\hline
			\end{tabular}
		\caption{Entanglement entropies for various bipartitions of the $\ket{\mathrm{TNS}}$ of the Haah code. The second to fourth column list the coordinates of vertices in region $A$. The column of "Left/Right" labels the spin on the left or right position on the vertex $(x,y,z)$, where $0$ and $1$ corresponds to the left and right position respectively. We used the coordinate frame as shown in Eq.~\ref{HaahHamiltonian} and Fig.~\ref{fig.HaahTNS}. }
		\label{TableHaahCodenumerics}
	\end{center}
\end{table}

The cases No.1 and No.2  in Table.~\ref{TableHaahCodenumerics} correspond to type 1 exact SVD regions with $l=2$ and $l=3$ respectively. We see that this is consistent with the general formula Eq.~\eqref{HaahexactSVDtype1}. When $l=2$, $6l-5=6\times 2-5=7$; when $l=3$, $6l-5=6\times 3-5=13$.  The case No.3 corresponds to the square region $A$ with size $2\times 2\times 2$. The TNS under such a cut is not an SVD (under the TNS basis), however, as we have shown in Sec. ~\ref{subsec.Haah_Entanglement}, we can make a change of basis such that in the new basis the wave function is an SVD. The counting of new basis gives the entanglement entropy of the cubic cut Eq.~\eqref{Haahlllentanglemententropy}. The brute force numerical calculation in case No.3 of Table. ~\ref{TableHaahCodenumerics} is consistent with the formula: $S(A)/\log2=6l^2-6l+2=6\times 2^2-6\times 2+2=14$. Notice that all these numerical results have been checked using the direct evaluation of the full GS wave function up to the system size $4 \times 3 \times 3$. 

\bibliography{TNS_Stabilizer}

\end{document}